# Massive migration from the steppe is a source for Indo-European languages in Europe

Wolfgang Haak[1,*], Iosif Lazaridis[2,3,*], Nick Patterson[3], Nadin Rohland[2,3], Swapan Mallick[2,3,4], Bastien Llamas[1], Guido Brandt[5], Susanne Nordenfelt[2,3], Eadaoin Harney[2,3,4], Kristin Stewardson[2,3,4], Qiaomei Fu[2,3,6,7], Alissa Mittnik[8], Eszter Bánffy[9,10], Christos Economou[11], Michael Francken[12], Susanne Friederich[13], Rafael Garrido Pena[14], Fredrik Hallgren[15], Valery Khartanovich[16], Aleksandr Khokhlov[17], Michael Kunst[18], Pavel Kuznetsov[17], Harald Meller[13], Oleg Mochalov[17], Vayacheslav Moiseyev[16], Nicole Nicklisch[5,13,19], Sandra L. Pichler[20], Roberto Risch[21], Manuel A. Rojo Guerra[22], Christina Roth[5], Anna Szécsényi-Nagy[5,9], Joachim Wahl[23], Matthias Meyer[6], Johannes Krause[8,12,24], Dorcas Brown[25], David Anthony[25], Alan Cooper[1], Kurt Werner Alt[5,13,19,20] and David Reich[2,3,4]

[*] Contributed equally to this work

[1] Australian Centre for Ancient DNA, School of Earth and Environmental Sciences & Environment Institute, University of Adelaide, Adelaide, South Australia, SA 5005, Australia

[2] Department of Genetics, Harvard Medical School, Boston, MA, 02115, USA

[3] Broad Institute of Harvard and MIT, Cambridge, MA, 02142, USA

[4] Howard Hughes Medical Institute, Harvard Medical School, Boston, MA, 02115, USA

[5] Institute of Anthropology, Johannes Gutenberg University of Mainz, D-55128 Mainz, Germany

[6] Max Planck Institute for Evolutionary Anthropology, Leipzig, 04103, Germany

[7] Key Laboratory of Vertebrate Evolution and Human Origins of Chinese Academy of Sciences, IVPP, CAS, Beijing, 100049, China

[8] Institute for Archaeological Sciences, University of Tübingen, Tübingen, 72074, Germany

[9] Institute of Archaeology, Research Centre for the Humanities, Hungarian Academy of Science, H-1014 Budapest, Hungary

[10] Römisch Germanische Kommission (RGK) Frankfurt, D-60325 Frankfurt, Germany

[11] Archaeological Research Laboratory, Stockholm University, 114 18, Sweden

[12] Department of Paleoanthropology, Senckenberg Center for Human Evolution and Paleoenvironment, University of Tübingen, Tübingen, D-72070, Germany

[13] State Office for Heritage Management and Archaeology Saxony-Anhalt and State Heritage Museum, D-06114 Halle, Germany

[14] Departamento de Prehistoria y Arqueología, Facultad de Filosofía y Letras, Universidad Autónoma de Madrid, E-28049 Madrid, Spain,

[15] The Cultural Heritage Foundation, Västerås, 722 12, Sweden

[16] Peter the Great Museum of Anthropology and Ethnography (Kunstkamera) RAS, St. Petersburg, Russia

[17] Volga State Academy of Social Sciences and Humanities, 443099 Russia, Samara, M. Gor'kogo, 65/67

[18] Deutsches Archaeologisches Institut, Abteilung Madrid, E-28002 Madrid, Spain

[19] Danube Private University, A-3500 Krems, Austria

[20] Institute for Prehistory and Archaeological Science, University of Basel, CH-4003 Basel, Switzerland
[21] Departamento de Prehistòria, Universitat Autonoma de Barcelona, E-08193 Barcelona, Spain
[22] Departamento de Prehistòria y Arqueolgia, Universidad de Valladolid, E-47002 Valladolid, Spain
[23] State Office for Cultural Heritage Management Baden-Württemberg, Osteology, Konstanz, D-78467, Germany
[24] Max Planck Institute for the Science of Human History, D-07745 Jena, Germany
[25] Anthropology Department, Hartwick College, Oneonta, NY

To whom correspondence should be addressed: David Reich
(reich@genetics.med.harvard.edu)

**We generated genome-wide data from 69 Europeans who lived between 8,000-3,000 years ago by enriching ancient DNA libraries for a target set of almost four hundred thousand polymorphisms. Enrichment of these positions decreases the sequencing required for genome-wide ancient DNA analysis by a median of around 250-fold, allowing us to study an order of magnitude more individuals than previous studies[1-8] and to obtain new insights about the past. We show that the populations of western and far eastern Europe followed opposite trajectories between 8,000-5,000 years ago. At the beginning of the Neolithic period in Europe, ~8,000-7,000 years ago, closely related groups of early farmers appeared in Germany, Hungary, and Spain, different from indigenous hunter-gatherers, whereas Russia was inhabited by a distinctive population of hunter-gatherers with high affinity to a ~24,000 year old Siberian[6]. By ~6,000-5,000 years ago, a resurgence of hunter-gatherer ancestry had occurred throughout much of Europe, but in Russia, the Yamnaya steppe herders of this time were descended not only from the preceding eastern European hunter-gatherers, but from a population of Near Eastern ancestry. Western and Eastern Europe came into contact ~4,500 years ago, as the Late Neolithic Corded Ware people from Germany traced ~3/4 of their ancestry to the Yamnaya, documenting a massive migration into the heartland of Europe from its eastern periphery. This steppe ancestry persisted in all sampled central Europeans until at least ~3,000 years ago, and is ubiquitous in present-day Europeans. These results provide support for the theory of a steppe origin[9] of at least some of the Indo-European languages of Europe.**

Genome-wide analysis of ancient DNA has emerged as a transformative technology for studying prehistory, providing information that is comparable in power to archaeology and linguistics. Realizing its promise, however, requires collecting genome-wide data from an adequate number of individuals to characterize population changes over time, which means not only sampling a succession of archaeological cultures[2], but also multiple individuals per culture. To make analysis of large numbers of ancient DNA samples practical, we used in-solution hybridization capture[10,11] to enrich next generation sequencing libraries for a target set of 394,577 single nucleotide polymorphisms (SNPs) ("390k capture"), 354,212 of which are autosomal SNPs that have also been genotyped using the Affymetrix Human Origins array in 2,345 humans from 203 populations[4,12]. This reduces the amount of sequencing required to obtain genome-wide data by a minimum of 45-fold and a median of 262-fold (Online Table 1). This strategy allows us to report genomic scale data on more than twice the number of ancient Eurasians as the entire preceding literature[1-8] (Extended Data Table 1).

We used this technology to study population transformations in Europe. We began by preparing 212 DNA libraries from 119 ancient samples in dedicated clean rooms, and testing these by light shotgun sequencing and mitochondrial genome capture (SI1, Online Table 1). We restricted to libraries with molecular signatures of authentic ancient DNA (elevated damage in the terminal nucleotide), negligible evidence of contamination based on mismatches to the mitochondrial consensus[13], and, where available, a mitochondrial DNA haplogroup that matched previous results using PCR[4,14,15] (SI2). For 123 libraries prepared in the presence of Uracil-DNA-glycosylase[16] to reduce errors due to ancient DNA damage[17], we performed 390k capture, carried out paired end sequencing, and mapped to the human genome. We restricted analysis to 95 libraries from 69 samples that had at least 0.06-fold average target coverage (average of 3.8-fold), and used majority rule to call an allele at each SNP covered at least once (Online Table 1). After combining our data (SI3) with 25 ancient samples from the literature — three Upper Paleolithic samples from Russia[1,7,6], seven people of European hunter gatherer ancestry[4,5,8,2], and fifteen European farmers[2,8,4,3], — we had data from 94 ancient Europeans. Geographically, these came from Germany (n=41), Spain (n=10), Russia (n=14), Sweden (n=12), Hungary (n=15), Italy (n=1) and Luxembourg (n=1) (Extended Data Table 2). Following the central European chronology, these included 19 hunter-gatherers (>5,500 BCE), 28 Early Neolithic farmers (EN: ~6,000-4,000 BCE), 11 Middle Neolithic farmers (MN: ~4,000-3,000 BCE) including the Tyrolean Iceman[3], 9 Late Copper/Early Bronze Age individuals (Yamnaya: ~3,300-2,700 BCE), 15 Late Neolithic individuals (LN: ~2,500-2,200 BCE), 9 Early Bronze Age individuals (~2,200-1,500 BCE), two Late Bronze Age individuals (~1,200-1,100 BCE) and one Iron Age individual (~900 BCE). Two individuals were excluded from analyses as they were related to others from the same population. The average number of SNPs covered at least once was 212,375 and the minimum was 22,869 (Fig. 1).

We determined that 34 of the 69 newly analyzed individuals were male and used 2,258 Y chromosome SNPs targets included in the capture to obtain high resolution Y chromosome haplogroup calls (SI4). Outside Russia, and before the Late Neolithic period, only a single R1b individual was found (early Neolithic Spain) in the combined literature (n=70). By contrast, haplogroups R1a and R1b were found in 60% of Late Neolithic/Bronze Age Europeans outside Russia (n=10), and in 100% of the samples from European Russia from all periods (7,500-2,700 BCE; n=9). R1a and R1b are the most common haplogroups in many European populations today[18,19], and our results suggest that they spread into Europe from the East after 3,000 BCE. Two hunter-gatherers from Russia included in our study belonged to R1a (Karelia) and R1b (Samara), the earliest documented ancient samples of either haplogroup discovered to date. These two hunter-gatherers did not belong to the derived lineages M417 within R1a and M269 within R1b that are predominant in Europeans today[18,19], but all 7 Yamnaya males did belong to the M269 subclade[18] of haplogroup R1b.

Principal components analysis (PCA) of all ancient individuals along with 777 present-day West Eurasians[4] (Fig. 2a, SI5) replicates the positioning of present-day Europeans between the Near East and European hunter-gatherers[4,20], and the clustering of early farmers from across Europe with present day Sardinians[3,4,27], suggesting that farming expansions across the Mediterranean to Spain and via the Danubian route to Hungary and Germany descended from a common stock. By adding samples from later periods and additional locations, we also observe several new patterns. All samples from Russia have affinity to the ~24,000 year old MA1[6], the type specimen for the Ancient North Eurasians (ANE) who contributed to both Europeans[4] and Native Americans[4,6,8]. The two hunter-gatherers from Russia (Karelia in the northwest of the country and Samara on the steppe near the Urals) form an "eastern European hunter-gatherer" (EHG) cluster at one end of a hunter-gatherer cline across Europe; people of hunter-gatherer ancestry from Luxembourg, Spain, and Hungary sit at the opposite "western European hunter-gatherer[4]" (WHG) end, while the hunter-gatherers from Sweden[4,8] (SHG) are intermediate. Against this background of differentiated European hunter-gatherers and homogeneous early farmers, multiple events transpired in all parts of Europe included in our study. Middle Neolithic Europeans from Germany, Spain, Hungary, and Sweden from the period ~4,000-3,000 BCE are intermediate between the earlier farmers and the WHG, suggesting an increase of WHG ancestry throughout much of Europe. By contrast, in Russia, the later Yamnaya steppe herders of ~3,000 BCE plot between the EHG and the Near East / Caucasus, suggesting a decrease of EHG ancestry during the same time period. The Late Neolithic and Bronze Age samples from Germany and Hungary[2] are distinct from the preceding Middle Neolithic and plot between them and the Yamnaya. This pattern is also seen in ADMIXTURE analysis (Fig. 2b, SI6), which implies that the Yamnaya have ancestry from populations related to the Caucasus and South Asia that is largely absent in 38 Early or Middle Neolithic farmers but present in all 25 Late Neolithic or Bronze Age individuals. This ancestry appears in Central Europe for the first time in our series with the Corded Ware around 2,500 BCE (SI6, Fig. 2b, Extended Data Fig. 1). The Corded Ware shared elements of material culture with steppe groups such as the Yamnaya although whether this reflects movements of people has been contentious[21]. Our genetic data provide direct evidence of migration and suggest that it was relatively sudden. The Corded Ware are genetically closest to the Yamnaya ~2,600 kilometers away, as inferred both from PCA and ADMIXTURE (Fig. 2) and $F_{ST}$ (0.011 ± 0.002) (Extended Data Table 3). If continuous gene flow from the east, rather than migration, had occurred, we would expect successive cultures in Europe to become increasingly differentiated from the Middle Neolithic, but instead, the Corded Ware are both the earliest and most strongly differentiated from the Middle Neolithic population.

"Outgroup" $f_3$-statistics[6] (SI7), which measure shared genetic drift between a pair of populations (Extended Data Fig. 2), support the clustering of hunter-gatherers, Early/Middle Neolithic, and Late Neolithic/Bronze Age populations into different groups as in the PCA (Fig. 2a). We also analyzed $f_4$-statistics, which allow us to test whether pairs of populations are consistent with descent from common ancestral populations, and to assess significance using a normally distributed Z-score. Early European farmers from the Early and Middle Neolithic were closely related but not identical. This is reflected in the fact that Loschbour shared more alleles with post-4,000 BCE European farmers from Germany, Spain, Hungary, Sweden, and Italy than with early farmers of Germany, Spain, and Hungary, documenting an increase of hunter-gatherer ancestry in multiple regions of Europe during the course of the Neolithic. The two EHG form a clade with respect to all other present-day and ancient populations ($|Z|<1.9$), and MA1 shares more alleles with them ($|Z|>4.7$) than with other ancient or modern populations, suggesting that they may be a source for the ANE ancestry in present Europeans[4,12,22] as they are geographically and temporally more proximate than Upper Paleolithic Siberians. The Yamnaya differ from the EHG by sharing fewer alleles with MA1 (|Z|=6.7) suggesting a dilution of ANE ancestry between 5,000-3,000 BCE on the European steppe. This was likely due to admixture of EHG with a population related to present-day Near Easterners, as the most negative $f_3$-statistic in the Yamnaya (giving unambiguous evidence of admixture) is observed when we model them as a mixture of EHG and present-day Near Eastern populations like Armenians (Z = -6.3; SI7). The Late Neolithic / Bronze Age groups of central Europe share more alleles with Yamnaya than the Middle Neolithic populations do (|Z|=12.4) and more alleles with the Middle Neolithic than the Yamnaya do (|Z|=12.5), and have a negative $f_3$-statistic with the Middle Neolithic and Yamnaya as references (Z=-20.7), proving that they were descended from a mixture of the local European populations and new migrants from the east. Moreover, the Yamnaya share more alleles with the Corded Ware (|Z|≥3.6) than with any other Late Neolithic/ Early Bronze age group with at least two individuals (SI7), proving that they had more eastern ancestry, consistent with the PCA and ADMIXTURE patterns (Fig. 2).

Modeling of the ancient samples shows that while Karelia is genetically intermediate between Loschbour and MA1, the topology that considers Karelia as a mixture of these two elements is not the only one that can fit the data (SI8). To avoid biasing our inferences by fitting an incorrect model, we developed new statistical methods that are substantial extensions of a previously reported approach[4], which allow us to obtain precise estimates of the proportion of mixture in later Europeans without requiring a formal model for the relationship among the ancestral populations. The method (SI9) is based on the idea that if a *Test* population has ancestry related to reference populations $Ref_1$, $Ref_2$, …, $Ref_N$ in proportions $\alpha_1$, $\alpha_2$, ...,$\alpha_N$, and the references are themselves differentially related to a triple of outgroup populations *A*, *B*, *C*, then $f_4(Test, A; B, C) = \sum_{i=1}^{N} \alpha_i \, f_4(Ref_i, A; B, C)$. By using a large number of outgroup populations we can fit the admixture coefficients $\alpha_i$ and estimate mixture proportions (SI9, Extended Data Fig. 3). Using 15 outgroups from Africa, South/East/North Asia, and the Americas, we obtain good fits as assessed by a formal test (SI10), and estimate that the Middle Neolithic populations of Germany and Spain have ~18-34% more WHG-related ancestry than Early Neolithic ones and that the Late Neolithic and Early Bronze Age populations of Germany have ~22-39% more EHG-related ancestry than the Middle

Neolithic ones (SI9). If we model them as mixtures of Yamnaya-related and Middle Neolithic populations, the inferred degree of population turnover is doubled to 48-80% (SI9, SI10).

To distinguish whether a Yamnaya or an EHG source fits the data better, we added ancient samples as outgroups (SI9). Adding any Early or Middle Neolithic farmer results in EHG-related genetic input into Late Neolithic populations being a poor fit to the data (SI9); thus, Late Neolithic populations have ancestry that cannot be explained by a mixture of EHG and Middle Neolithic. When using Yamnaya instead of EHG, however, we obtain a good fit (SI9, SI10). These results can be explained if the new genetic material that arrived in Germany was a composite of two elements: EHG and a type of Near Eastern ancestry different from that which was introduced by early farmers (also suggested by PCA and ADMIXTURE; Fig. 2, SI5, SI6). We estimate that these two elements each contributed about half the ancestry each of the Yamnaya (SI6, SI9), explaining why the population turnover inferred using Yamnaya as a source is about twice as high compared to the undiluted EHG. The estimate of Yamnaya-related ancestry in the Corded Ware is consistent when using either present populations or ancient Europeans as outgroups (SI9, SI10), and is 73.1 ± 2.2% when both sets are combined (SI10). The best proxies for ANE ancestry in Europe[4] were initially Native Americans[12,22], and then the Siberian MA1[6], but both are geographically and temporally too remote for what appears to be a recent migration into Europe[4]. We can now add three new pieces to the puzzle of how ANE ancestry was transmitted to Europe: first the EHG, then the Yamnaya formed by mixture between EHG and a Near Eastern related population, and then the Corded Ware who were formed by a mixture of the Yamnaya with Middle Neolithic Europeans. We caution that the sampled Yamnaya individuals from Samara might not be directly ancestral to Corded Ware individuals from Germany. It is possible that a more western Yamnaya population, or an earlier (pre-Yamnaya) steppe population may have migrated into central Europe, and future work may uncover more missing links in the chain of transmission of steppe ancestry.

By extending our model to a three way mixture of WHG, Early Neolithic and Yamnaya, we estimate that the ancestry of the Corded Ware was 79% Yamnaya-like, 4% WHG, and 17% Early Neolithic (Fig. 3). A small contribution of the first farmers is also consistent with uniparentally inherited DNA: *e.g.* mitochondrial DNA haplogroup N1a and Y chromosome haplogroup G2a, common in early central European farmers[14,23], almost disappear during the Late Neolithic and Bronze Age, when they are largely replaced by Y haplogroups R1a and R1b (SI4) and mtDNA haplogroups I, T1, U2, U4, U5a, W, and subtypes of H[14,23,24] (SI2).

The uniparental data not only confirm a link to the steppe populations but also suggest that both sexes participated in the migrations (SI2, SI4, Extended Data Table 3). The magnitude of the population turnover that occurred becomes even more evident if one considers the fact that the steppe migrants may well have mixed with eastern European agriculturalists on their way to central Europe. Thus, we cannot exclude a scenario in which the Corded Ware arriving in today's Germany had no ancestry at all from local populations.

Our results support a view of European pre-history punctuated by two major migrations: first, the arrival of first farmers during the Early Neolithic from the Near East, and second of Yamnaya pastoralists during the Late Neolithic from the steppe (Extended Data Fig. 5). Our data further show that both migrations were followed by resurgences of the previous inhabitants: first, during the Middle Neolithic, when hunter-gatherer ancestry rose again after its Early Neolithic decline, and then between the Late Neolithic and the present, when farmer and hunter-gatherer ancestry rose after its Late Neolithic decline. This second resurgence must have started during the Late Neolithic/Bronze Age period itself, as the Bell Beaker and Unetice groups had reduced Yamnaya ancestry compared to the earlier Corded Ware, and comparable levels to that in some present-day Europeans (Fig. 3). Today, Yamnaya related ancestry is lower in southern Europe and higher in northern Europe. Further data are needed to determine whether the steppe ancestry arrived in southern Europe at the time of the Late Neolithic / Bronze Age, or is due to migrations in historical times from northern Europe[25,26].

Our results provide new data relevant to debates on the origin and expansion of Indo-European languages in Europe (SI11). Although ancient DNA is silent on the question of the languages spoken by preliterate populations, it does carry evidence about processes of migration which are invoked by theories on Indo-European language dispersals. Such theories make predictions about movements of people to account for the spread of languages and material culture. The technology of ancient DNA makes it possible to reject or confirm the proposed migratory movements, as well as to identify new movements that were not previously known. The best argument for the "Anatolian hypothesis[27]" that Indo-European languages arrived in Europe from Anatolia ~8,500 years ago is that major language replacements are thought to require major migrations, and that after the Early Neolithic when farmers established themselves in Europe, the population base was likely to have been so large as to be impervious to subsequent turnover[27,28]. However, our study shows that a later major turnover did occur, and that steppe migrants replaced ~3/4 of the ancestry of central

Europeans. An alternative theory is the “Steppe hypothesis”, which proposes that early Indo-European speakers were pastoralists of the grasslands north of the Black and Caspian Seas, and that their languages spread into Europe after the invention of wheeled vehicles[9]. Our results make a compelling case for the steppe as a source of at least some of the Indo-European languages in Europe by documenting a massive migration ~4,500 years ago associated with the Yamnaya and Corded Ware cultures, which are identified by proponents of the Steppe hypothesis as vectors for the spread of Indo-European languages into Europe. These results challenge the Anatolian hypothesis by showing that not all Indo-European languages in Europe can plausibly derive from the first farmer migrations thousands of years earlier (SI11). We caution that the location of the Proto-Indo-European[9,27,29,30] homeland that also gave rise to the Indo-European languages of Asia, as well as the Indo-European languages of southeastern Europe, cannot be determined from the data reported here (SI11). Studying the mixture in the Yamnaya themselves, and understanding the genetic relationships among a broader set of ancient and present-day Indo-European speakers, may lead to new insight about the shared homeland.

**Acknowledgments**

We thank Peter Bellwood, Joachim Burger, Paul Heggarty, Mark Lipson, Colin Renfrew, Jared Diamond, Svante Pääbo, Ron Pinhasi and Pontus Skoglund for critical comments. We thank Svante Pääbo for support for establishing the ancient DNA facilities in Boston, and Pontus Skoglund for detecting the presence of two related individuals in our dataset. We thank Ludovic Orlando, Thorfinn S. Korneliussen, and Cristina Gamba for help in obtaining data. We thank Agilent Technologies and Götz Frommer for help in developing the capture reagents. We thank Clio Der Sarkissian, Guido Valverde, Luka Papac, and Birgit Nickel for wet lab support. We thank archaeologists Veit Dresely, Robert Ganslmeier, Oleg Balanvosky, José Ignacio Royo Guillén, Anett Osztás, Vera Majerik, Tibor Paluch, Krisztina Somogyi and Vanda Voicsek for sharing samples and discussion about archaeological context. This research was supported by an Australian Research Council grant to W.H. and B.L. (DP130102158), and German Research Foundation grants to K.W.A. (Al 287/7-1 and 7-3, Al 287/10-1 and Al 287/14-1) and to H.M. (Me 3245/1-1 and 1-3). D.R. was supported by U.S. National Science Foundation HOMINID grant BCS-1032255, U.S. National Institutes of Health grant GM100233, and the Howard Hughes Medical Institute.

**Author contributions**

WH, NP, NR, JK, KWA and DR supervised the study. WH, EB, CE, MF, SF, RGP, FH, VK, AK, MK, PK, HM, OM, VM, NN, SP, RR, MARG, CR, ASN, JW, JKr, DB, DA, AC, KWA and DR assembled archaeological material, WH, IL, NP, NR, SM, AM and DR analyzed genetic data. WH, NR, BL, GB, SN, EH, KS and AM performed wet laboratory ancient DNA work. NR, QF, MM and DR developed the 390k capture reagent. WH, IL and DR wrote the manuscript with help from all co-authors.

# Figure Legends

**Figure 1: Location and SNP coverage of samples included in this study.** (**a**) Geographic location and time-scale (central European chronology) of the 69 newly typed ancient individuals from this study (black outline) and 25 from the literature for which shotgun sequencing data was available (no outline). (**b**) Number of SNPs covered at least once in the analysis dataset of 94 individuals.

**Figure 2: Population transformations in Europe.** (**a**) PCA analysis, (**b**) ADMIXTURE analysis. The full ADMIXTURE analysis including present-day humans is shown in Extended Data Fig. 1.

**Figure 3: Admixture proportions.** We estimate mixture proportions using a method that gives unbiased estimates even without an accurate model for the relationships between the test populations and the outgroup populations (SI9). Population samples are grouped according to chronology (ancient) and Yamnaya ancestry (present-day humans).

## Online Methods

### Screening of libraries (shotgun sequencing and mitochondrial capture)

The 212 libraries screened in this study (SI1) from a total of 119 samples (SI3) were produced at Adelaide (n=151), Tübingen (n=16), and Boston (n=45) (Online Table 1).

The libraries from Adelaide and Boston had internal barcodes directly attached to both sides of the molecules from the DNA extract so that each sequence begins with the barcode[10]. The Adelaide libraries had 5 base pair (bp) barcodes on both sides, while the Boston libraries had 7 bp barcodes. Libraries from Tübingen had no internal barcodes, but were differentiated by the sequence of the indexing primer[31].

We adapted a reported protocol for enriching for mitochondrial DNA[10], with the difference that we adjusted the blocking oligonucleotides and PCR primers to fit our libraries with shorter adapters. Over the course of this project, we also lowered the hybridization temperature from 65°C to 60°C and performed stringent washes at 55°C instead of 60°C[32].

We used an aliquot of approximately 500ng of each library for target enrichment of the complete mitochondrial genome in two consecutive rounds[32], using a bait set for human mtDNA[32]. We performed enrichment in 96-well plates with one library per well, and used a liquid handler (Evolution P3, Perkin Elmer) for the capture and washing steps[33]. We used blocking oligonucleotides in hybridization appropriate to the adapters of the truncated libraries. After either of the two enrichment rounds, we amplified the enriched library molecules with the primer pair that keeps the adapters short (PreHyb) using Herculase Fusion II PCR Polymerase. We performed an indexing PCR of the final capture product using one or two indexing primers[31]. We cleaned up all PCR's using SPRI technology[34] and the liquid handler. Libraries from Tübingen were amplified with the primer pair IS5/IS6[31].

For libraries from Boston and Adelaide, we used a second aliquot of each library for shotgun sequencing after performing an indexing PCR[31]. We used unique index combinations for each library and experiment, allowing us to distinguish shotgun sequencing and mitochondrial DNA capture data, even if both experiments were in the same sequencing run. We sequenced shotgun libraries and mtDNA captured libraries from Tübingen in independent sequencing runs since the index was already attached at the library preparation stage.

We quantified the sequencing pool with the BioAnalyzer (Agilent) and/or the KAPA Library Quantification kit (KAPA biosystems) and sequenced on Illumina MiSeq, HiSeq2500 or NextSeq500 instruments for 2×75, 2×100 or 2×150 cycles along with the indexing read(s).

**Enrichment for 394,577 SNP targets ("390k capture")**

The protocol for enrichment for SNP targets was similar to the mitochondrial DNA capture, with the exception that we used another bait set (390k) and about twice as much library (up to 1000ng) compared to the mtDNA capture.

The specific capture reagent used in this study is described for the first time here. To target each SNP, we used a different oligonucleotide probe design compared to ref. 1. We used four 52 base pair probes for each SNP target. One probe ends just before the SNP, and one starts just after. Two probes are centered on the SNP, and are identical except for having the alternate alleles. This probe design avoids systematic bias toward one SNP allele or another. For the template sequence for designing the San and Yoruba panels baits, we used the sequence that was submitted for these same SNPs during the design of the Affymetrix Human Origins SNP array[12]. For SNPs that were both in the San and Yoruba panels, we used the Yoruba template sequence in preference. For all other SNPs, we used the human genome reference sequence as a template. Online Table 2 gives the list of SNPs that we targeted, along with details of the probes used. The breakdown of SNPs into different classes is:

124,106 "Yoruba SNPs": All SNPs in "Panel 5" of the Affymetrix Human Origins array (discovered as heterozygous in a Yoruba male: HGDP00927)[12] that passed the probe design criteria specified in ref. 11.

146,135 "San SNPs": All SNPs in "Panel 4" of the Affymetrix Human Origins array (discovered as heterozygous in a San male: HGDP01029)[12] that passed probe design criteria[11]. The full Affymetrix Human Origins array Panel 4 contains several tens of thousands of additional SNPs overlapping those from Panel 5, but we did not wish to redundantly capture Panel 5 SNPs.

98,166 "Compatibility SNPs": SNPs that overlap between the Affymetrix Human Origins the Affymetrix 6.0, and the Illumina 610 Quad arrays, which are not already included in the "Yoruba SNPs" or "San SNPs" lists[12] and that also passed the probe design design criteria[11].

26,170 "Miscellaneous SNPs": SNPs that did not overlap the Human Origins array. The subset analyzed in this study were 2,258 Y chromosome SNPs (http://isogg.org/tree/ISOGG_YDNA_SNP_Index.html) that we used for Y haplogroup determination.

**Processing of sequencing data**

We restricted analysis to read pairs that passed quality control according to the Illumina software ("PF reads").

We assigned read pairs to libraries by searching for matches to the expected index and barcode sequences (if present, as for the Adelaide and Boston libraries). We allowed no more than 1 mismatch per index or barcode, and zero mismatches if there was ambiguity in sequence assignment or if barcodes of 5 bp length were used (Adelaide libraries).

We used Seqprep (https://github.com/jstjohn/SeqPrep) to search for overlapping sequence between the forward and reverse read, and restricted to molecules where we could identify a minimum of 15 bp of overlap. We collapsed the two reads into a single sequence, using the consensus nucleotide if both reads agreed, and the read with higher base quality in the case of disagreement. For each merged nucleotide, we assigned the base quality to be the higher of the two reads. We further used Seqprep to search for the expected adapter sequences at either ends of the merged sequence, and to produce a trimmed sequence for alignment.

We mapped all sequences using BWA-0.6.1[35]. For mitochondrial analysis we mapped to the mitochondrial genome RSRS[36]. For whole genome analysis we mapped to the human reference genome *hg19*. We restricted all analyses to sequences that had a mapping quality of MAPQ≥37.

We sorted all mapped sequences by position, and used a custom script to search for mapped sequences that had the same orientation and start and stop positions. We stripped all but one of these sequences (keeping the best quality one) as duplicates.

**Mitochondrial sequence analysis and assessment of ancient DNA authenticity**

For each library for which we had average coverage of the mitochondrial genome of at least 10-fold after removal of duplicated molecules, we built a mitochondrial consensus sequence, assigning haplogroups for each library as described in SI2.

We used contamMix-1.0.9 to search for evidence of contamination in the mitochondrial DNA[13]. This software estimates the fraction of mitochondrial DNA sequences that match the consensus more closely than a comparison set of 311 worldwide mitochondrial genomes. This is done by taking the consensus sequence of reads aligning to the RSRS mitochondrial genome, and requiring a minimum coverage of 5 after filtering bases where the quality was <30. Raw reads are then realigned to this consensus. In addition, the consensus is multiply aligned with the other 311 mitochondrial genomes using kalign (2.0.4)[37] to build the necessary inputs for contamMix, trimming the first and last 5 bases of every read to mitigate against the confounding factor of ancient damage. This software had difficulty running on datasets with higher coverage, and for these datasets, we down-sampled to 50,000 reads, which we found produced adequate contamination estimation.

For all sequences mapping to the mitochondrial DNA that had a cytosine at the terminal nucleotide, we measured the proportion of sequences with a thymine at that position. For population genetic analysis, we only used partially UDG-treated libraries with a minimum of 3% C→T substitutions as recommended by ref. [33]. In cases where we used a fully UDG-treated library for 390k analysis, we examined mitochondrial capture data from a non-UDG-treated library made from the same extract, and verified that the non-UDG library had a minimum of 10% C→T at the first nucleotide as recommended by ref. [38]. Metrics for the mitochondrial DNA analysis on each library are given in Online Table 1.

**390k capture, sequence analysis and quality control**

For 390k analysis, we restricted to reads that not only mapped to the human reference genome *hg19* but that also overlapped the 354,212 autosomal SNPs genotyped on the Human Origins array[4]. We trimmed the last two nucleotides from each sequence because we found that these are highly enriched in ancient DNA damage even for UDG-treated libraries. We further restricted analyses to sites with base quality≥30.

We made no attempt to determine a diploid genotype at each SNP in each sample. Instead, we used a single allele – randomly drawn from the two alleles in the individual – to represent the individual at that site[20,39]. Specifically, we made an allele call at each target SNP using majority rule over all sequences overlapping the SNP. When each of the possible alleles was supported by an equal number of sequences, we picked an allele at random. We set the allele to "no call" for SNPs at which there was no read coverage.

We restricted population genetic analysis to libraries with a minimum of 0.06-fold average coverage on the 390k SNP targets, and for which there was an unambiguous sex determination based on the ratio of X to Y chromosome reads (SI4) (Online Table 1). For individuals for whom there were multiple libraries per sample, we performed a series of quality control analysis. First, we used the ADMIXTURE software[40,41] in supervised mode, using Kharia, Onge, Karitiana, Han, French, Mbuti, Ulchi and Eskimo as reference populations. We visually inspected the inferred ancestry components in each individual, and removed individuals with evidence of heterogeneity in inferred ancestry components across libraries. For all possible pairs of libraries for each sample, we also computed statistics of the form $D(Library_1, Library_2; Probe, Mbuti)$, where *Probe* is any of a panel of the same set of eight reference populations), to determine whether there was significant evidence of the *Probe* population being more closely related to one library from an ancient individual than another library from that same individual. None of the individuals that we used had strong evidence of ancestry heterogeneity across libraries. For samples passing quality control for which there were multiple libraries per sample, we merged the sequences into a single BAM.

We called alleles on each merged BAM using the same procedure as for the individual libraries. We used ADMIXTURE[41] as well as PCA as implemented in EIGENSOFT[42] (using the *lsqproject: YES* option to project the ancient samples) to visualize the genetic relationships of each set of samples with the same culture label with respect to 777 diverse present-day West Eurasians[4]. We visually identified outlier individuals, and renamed them for analysis either as outliers or by the name of the site at which they were sampled (Extended Data Table 1). We also identified two pairs of related individuals based on the proportion of sites covered in pairs of ancient samples from the same population that had identical allele calls using PLINK[43]. From each pair of related individuals, we kept the one with the most SNPs.

**Population genetic analyses**

We determined genetic sex using the ratio of X and Y chromosome alignments[44] (SI4), and mitochondrial haplogroup for all samples (SI2), and Y chromosome haplogroup for the male samples (SI4). We studied population structure (SI5, SI6). We used *f*-statistics to carry out formal tests of population relationships (SI6) and built explicit models of population history consistent with the data (SI7). We estimated mixture proportions in a way that was robust to uncertainty about the exact population history that applied (SI8). We estimated the minimum number of streams of migration into Europe needed to explain the data (SI9, SI10). The estimated mixture proportions shown in Fig. 3 were obtained using the *lsqlin* function of Matlab and the optimization method described in SI9 with 15 world outgroups.

**Figure 1**

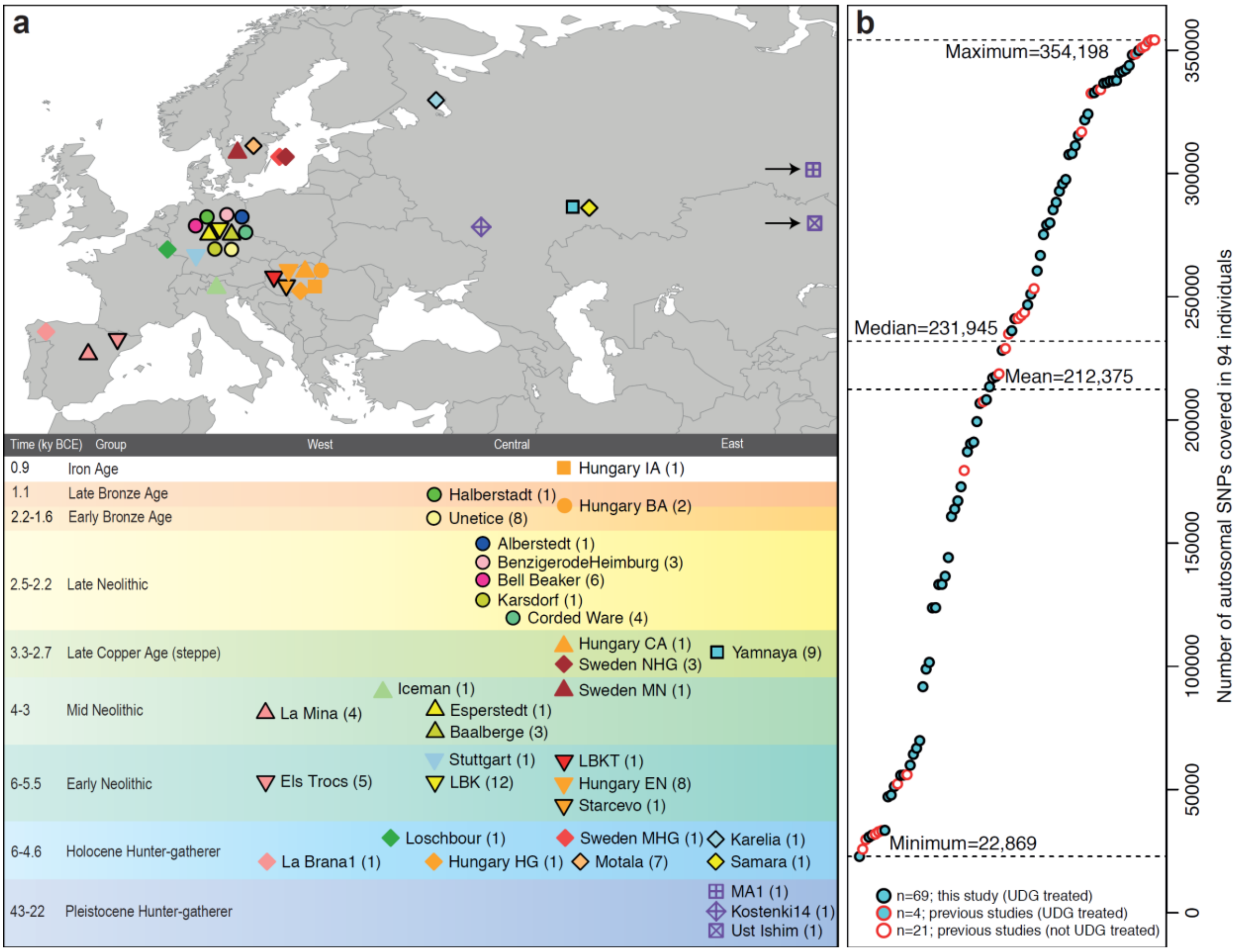
a
Time (ky BCE) Group
West
Central
East
0.9 Iron Age
Hungary IA (1)
1.1 Late Bronze Age
2.2-1.6 Early Bronze Age
Halberstadt (1)
Unetice (8)
Hungary BA (2)
2.5-2.2 Late Neolithic
Alberstedt (1)
BenzigerodeHeimburg (3)
Bell Beaker (6)
Karsdorf (1)
Corded Ware (4)
3.3-2.7 Late Copper Age (steppe)
Hungary CA (1)
Sweden NHG (3)
Yamnaya (9)
4-3 Mid Neolithic
La Mina (4)
Iceman (1)
Esperstedt (1)
Baalberge (3)
Sweden MN (1)
6-5.5 Early Neolithic
Els Trocs (5)
Stuttgart (1)
LBK (12)
LBKT (1)
Hungary EN (8)
Starcevo (1)
6-4.6 Holocene Hunter-gatherer
Loschbour (1)
Sweden MHG (1)
Karelia (1)
La Brana1 (1)
Hungary HG (1)
Motala (7)
Samara (1)
43-22 Pleistocene Hunter-gatherer
MA1 (1)
Kostenki14 (1)
Ust Ishim (1)
b
Maximum=354,198
Median=231,945
Mean=212,375
Minimum=22,869
n=69; this study (UDG treated)
n=4; previous studies (UDG treated)
n=21; previous studies (not UDG treated)
Number of autosomal SNPs covered in 94 individuals
0
50000
100000
150000
200000
250000
300000
350000

**Figure 2**

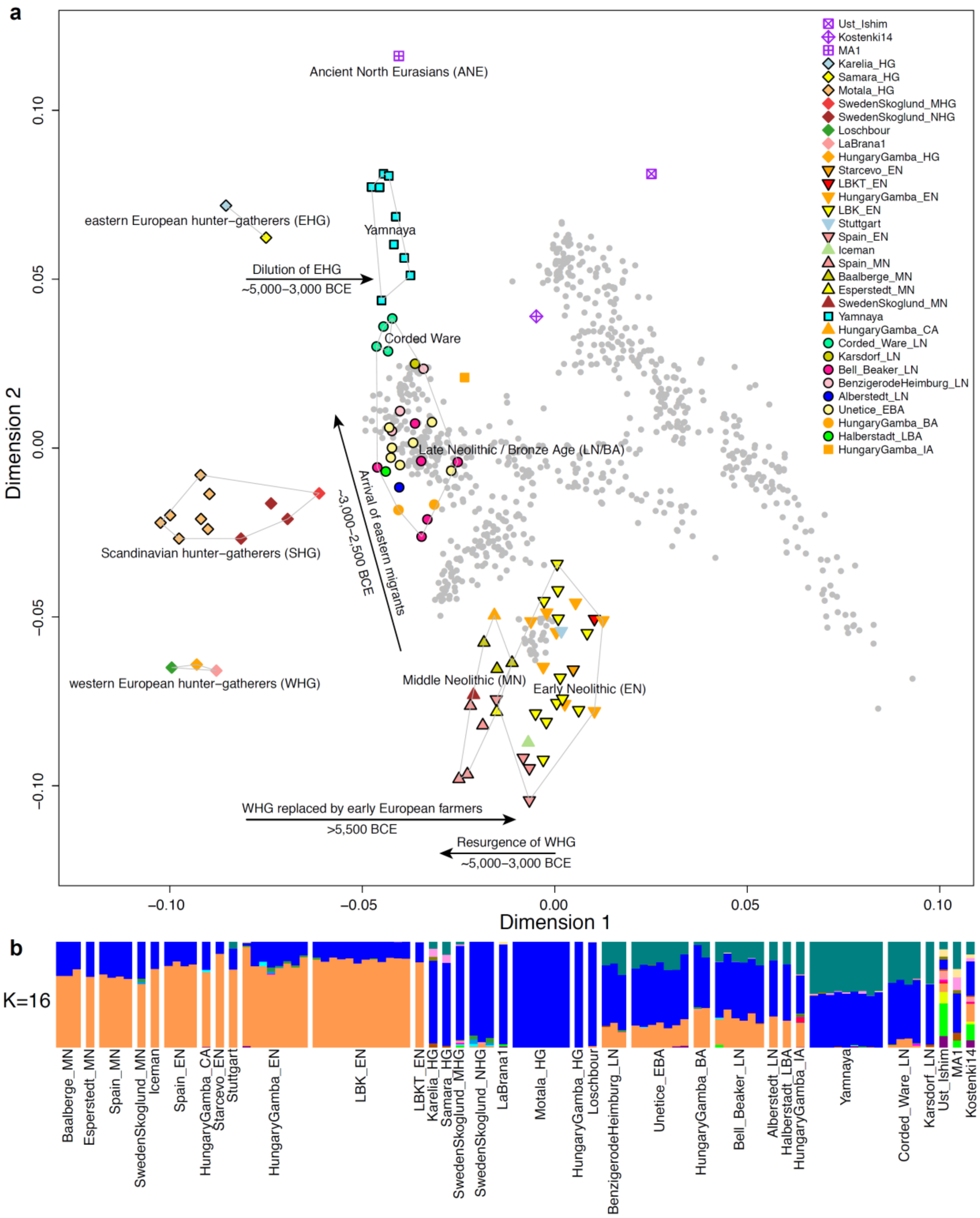
a
b
Ancient North Eurasians (ANE)
eastern European hunter–gatherers (EHG)
Yamnaya
Dilution of EHG
~5,000–3,000 BCE
Corded Ware
Late Neolithic / Bronze Age (LN/BA)
Arrival of eastern migrants
~3,000–2,500 BCE
Scandinavian hunter–gatherers (SHG)
western European hunter–gatherers (WHG)
Middle Neolithic (MN)
Early Neolithic (EN)
WHG replaced by early European farmers
>5,500 BCE
Resurgence of WHG
~5,000–3,000 BCE
Dimension 1
Dimension 2
Ust_Ishim
Kostenki14
MA1
Karelia_HG
Samara_HG
Motala_HG
SwedenSkoglund_MHG
SwedenSkoglund_NHG
Loschbour
LaBrana1
HungaryGamba_HG
Starcevo_EN
LBKT_EN
HungaryGamba_EN
LBK_EN
Stuttgart
Spain_EN
Iceman
Spain_MN
Baalberge_MN
Esperstedt_MN
SwedenSkoglund_MN
Yamnaya
HungaryGamba_CA
Corded_Ware_LN
Karsdorf_LN
Bell_Beaker_LN
BenzigerodeHeimburg_LN
Alberstedt_LN
Unetice_EBA
HungaryGamba_BA
Halberstadt_LBA
HungaryGamba_IA
K=16
Baalberge_MN
Esperstedt_MN
Spain_MN
SwedenSkoglund_MN
Iceman
Spain_EN
HungaryGamba_CA
Starcevo_EN
Stuttgart
HungaryGamba_EN
LBK_EN
LBKT_EN
Karelia_HG
Samara_HG
SwedenSkoglund_MHG
SwedenSkoglund_NHG
LaBrana1
Motala_HG
HungaryGamba_HG
Loschbour
BenzigerodeHeimburg_LN
Unetice_EBA
HungaryGamba_BA
Bell_Beaker_LN
Alberstedt_LN
Halberstadt_LBA
HungaryGamba_IA
Yamnaya
Corded_Ware_LN
Karsdorf_LN
Ust_Ishim
MA1
Kostenki14

**Figure 3**

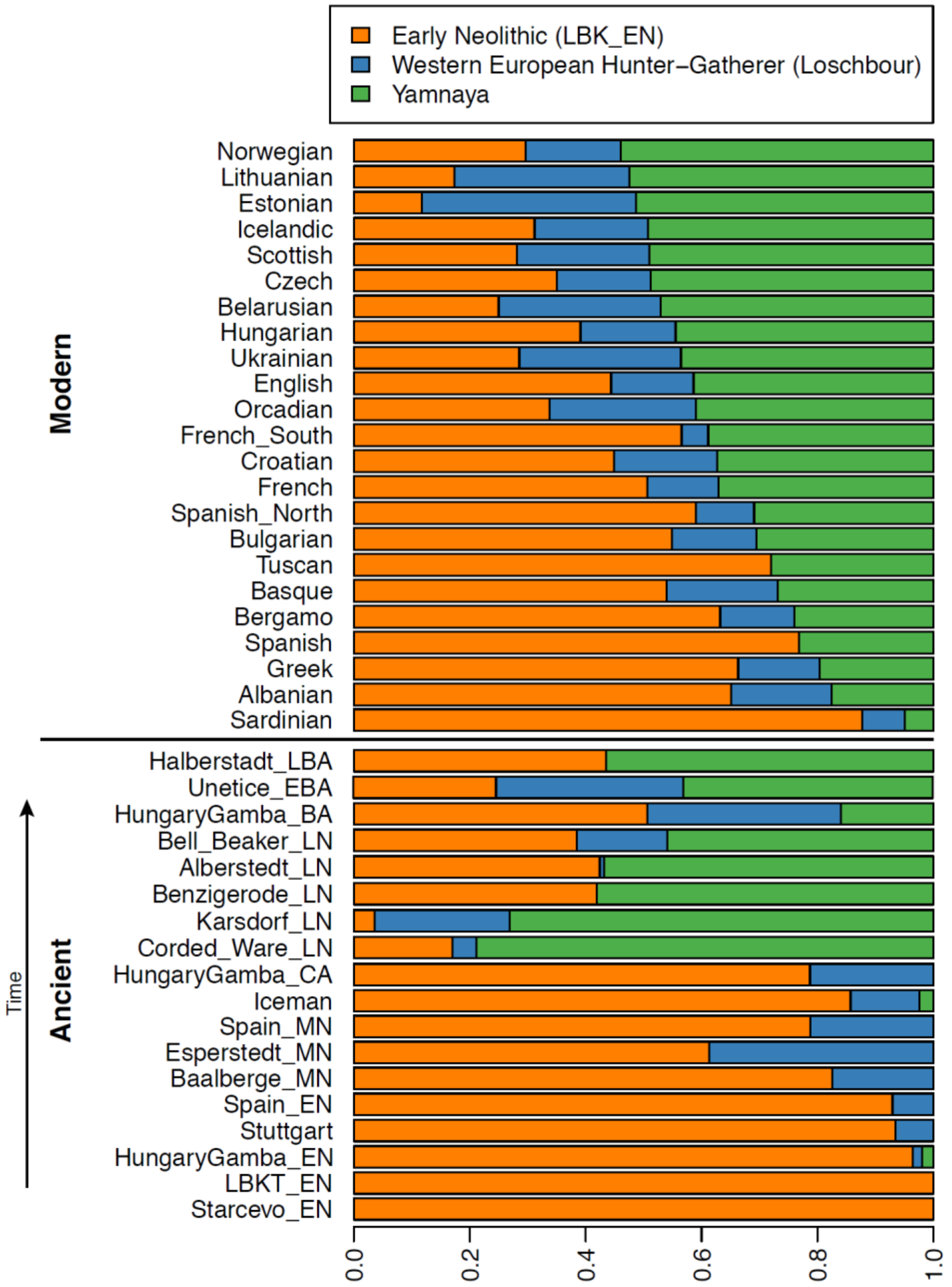

Early Neolithic (LBK_EN)
Western European Hunter–Gatherer (Loschbour)
Yamnaya
Modern
Norwegian
Lithuanian
Estonian
Icelandic
Scottish
Czech
Belarusian
Hungarian
Ukrainian
English
Orcadian
French_South
Croatian
French
Spanish_North
Bulgarian
Tuscan
Basque
Bergamo
Spanish
Greek
Albanian
Sardinian
Ancient
Time
Halberstadt_LBA
Unetice_EBA
HungaryGamba_BA
Bell_Beaker_LN
Alberstedt_LN
Benzigerode_LN
Karsdorf_LN
Corded_Ware_LN
HungaryGamba_CA
Iceman
Spain_MN
Esperstedt_MN
Baalberge_MN
Spain_EN
Stuttgart
HungaryGamba_EN
LBKT_EN
Starcevo_EN
0.0
0.2
0.4
0.6
0.8
1.0

**Extended Data Table 1: Number of ancient Eurasian modern human samples screened in genome-wide studies to date.**
Only studies that produced at least one sample at ≥0.05× coverage are listed.

| First author | Description | No. samples at ≥0.05× coverage (enough for Procrustes analysis) | No. samples at >0.25× coverage (enough to analyze in pairs) |
|---|---|---|---|
| Keller[3] | Tyrolean Iceman | 1 | 1 |
| Raghavan[6] | Upper Paleolithic Siberians | 2 | 1 |
| Olalde[5] | Mesolithic Iberian from LaBrana | 1 | 1 |
| Skoglund[8] | Farmers and hunter-gatherers from Sweden | 5 | 2 |
| Lazaridis[4] | Early European farmer from Germany & Mesolithic hunter-gatherers from Luxembourg and Sweden | 7 | 4 |
| Gamba[2] | Neolithic, Bronze Age, Iron Age Hungary | 13 | 9 |
| Fu[1] | Upper Paleolithic Siberian from Ust-Ishim | 1 | 1 |
| Seguin-Orlando[7] | Upper Paleolithic European from Kostenki | 1 | 1 |
| *Total before study* | | *31* | *20* |
| This study | Hunter-gatherers and pastoralists from Russia, Mesolithic hunter-gatherers from Sweden, Early Neolithic from Germany, Hungary, and Spain, Middle Neolithic from Germany & Spain, Late Neolithic / Bronze Age from Germany | 69 | 58 |

**Extended Data Table 2: The 69 samples newly reported on in this study.** Samples with direct radiocarbon dates are indicated by a calibrated date "cal BCE" along with associated laboratory numbers. Dates that are estimated based on faunal elements associated with the samples are not indicated with "cal" (although they are still calibrated, absolute dates).

| Reich ID | Pop Label for Analysis | Culture | Group | Location and sample details (e.g. sample, grave and museum ID) | Date (lab no.) | Country | Sex | mt-hg | Y-hg | Autosomal SNPs |
|---|---|---|---|---|---|---|---|---|---|---|
| **I0061** | Karelia_HG | Russian Mesolithic | EHG | Yuzhnyy Oleni Ostrov, Karelia, Russia; UzOO74, grave 142, MAE RAS 5773-74 | 5500-5000 BCE | Russia | M | C1g (formerly C1f) | R1a1 | 341554 |
| **I0124** | Samara_HG | Russian Neolithic HG | EHG | Sok River, Samara, Russia; SVP44 | 5650-5555 cal BCE (Beta – 392490) | Russia | M | U5a1d | R1b1a | 206748 |
| **I0011** | Motala_HG | Swedish Mesolithic | SHG | Motala, Sweden; Motala 1 | 5898-5531 cal BCE | Sweden | F | U5a1 | | 228271 |
| **I0012** | Motala_HG | Swedish Mesolithic | SHG | Motala, Sweden; Motala 2 | 5898-5531 cal BCE | Sweden | M | U2e1 | I2c2 | 292853 |
| **I0013** | Motala_HG | Swedish Mesolithic | SHG | Motala, Sweden; Motala 3 | 5898-5531 cal BCE | Sweden | M | U5a1 | I2a1b | 251108 |
| **I0014** | Motala_HG | Swedish Mesolithic | SHG | Motala, Sweden; Motala 4 | 5898-5531 cal BCE | Sweden | F | U5a2d | | 311299 |
| **I0015** | Motala_HG | Swedish Mesolithic | SHG | Motala, Sweden; Motala 6 | 5898-5531 cal BCE | Sweden | M | U5a2d | I2a1 | 285307 |
| **I0016** | Motala_HG | Swedish Mesolithic | SHG | Motala, Sweden; Motala 9 | 5898-5531 cal BCE | Sweden | M | U5a2 | I2a1 | 275233 |
| **I0017** | Motala_HG | Swedish Mesolithic | SHG | Motala, Sweden; Motala 12 | 5898-5531 cal BCE | Sweden | M | U2e1 | I2a1b | 337794 |
| **I0174** | Starcevo_EN | Starcevo | EN | Alsónyék-Bátaszék, Mérnöki telep, Hungary; BAM25a, feature 1532 | 5710-5550 cal BCE (MAMS 11939 ) | Hungary | M | N1a1a1b | H2 | 101653 |
| **I0176** | LBKT_EN | LBKT | EN | Szemely-Hegyes, Hungary: SZEH4b, feature 1001 | 5210-4940 cal BCE (Beta - 310038) | Hungary | | N1a1a1a3 | | 30718 |
| **I0046** | LBK_EN | LBK | EN | Halberstadt-Sonntagsfeld, Germany: HAL5, grave 2, feature 241.1 | 5206-5004 cal BCE (MAMS 21479) | Germany | F | T2c1d'e'f | | 266764 |
| **I0048** | LBK_EN | LBK | EN | Halberstadt-Sonntagsfeld, Germany: HAL25, grave 28, feature 861 | 5206-5052 cal BCE (MAMS 21482) | Germany | M | K1a | G2a2a | 123828 |
| **I0056** | LBK_EN | LBK | EN | Halberstadt-Sonntagsfeld, Germany; HAL14, grave 15, feature 430 | 5206-5052 cal BCE (MAMS 21480) | Germany | M | T2b(8) | G2a2a | 136578 |
| **I0057** | LBK_EN | LBK | EN | Halberstadt-Sonntagsfeld; HAL34, grave 38, feature 992 | 5207-5067 cal BCE (MAMS 21483) | Germany | F | N1a1a1 | | 55802 |
| **I0100** | LBK_EN | LBK | EN | Halberstadt-Sonntagsfeld; HAL4, grave 1, feature 139 | 5032-4946 cal BCE (KIA40341) | Germany | F | N1a1a1a | | 342342 |
| **I0659** | LBK_EN | LBK | EN | Halberstadt-Sonntagsfeld, Germany; HAL2, grave 35, feature 999 | 5079-4997 cal BCE (KIA40350) | Germany | M | N1a1a | G2a2a1 | 191007 |
| | | | | | 5066-4979 cal BCE (KIA30408) | | | | | |
| **0821** | LBK_EN | LBK | EN | Halberstadt-Sonntagsfeld, Germany; HAL24, grave 27, feature 867 | 5034-4942 cal BCE (KIA40348) | Germany | M | Pre-X2d1 | G2a2a1 | 55914 |
| **I0795** | LBK_EN | LBK | EN | Karsdorf, Germany; KAR6a, feature 170 | 5207-5070 cal BCE (MAMS 22823) | Germany | M | H1 | T1a | 47804 |
| **I0054** | LBK_EN | LBK | EN | Oberwiederstedt-Unterwiederstedt, UWS4, Germany, grave 6, feature 1 14 | 5209-5070 cal BCE (MAMS 21485) | Germany | F | J1c17 | | 337625 |
| **I0022** | LBK_EN | LBK | EN | Viesenhäuser Hof, Stuttgart-Mühlhausen, Germany; LBK1976 | 5500-4800 BCE | Germany | F | T2e | | 160852 |
| **I0025** | LBK_EN | LBK | EN | Viesenhäuser Hof, Stuttgart-Mühlhausen, Germany; LBK1992 | 5500-4800 BCE | Germany | F | T2b | | 307686 |
| **I0026** | LBK_EN | LBK | EN | Viesenhäuser Hof, Stuttgart-Mühlhausen, Germany; LBK2155 | 5500-4800 BCE | Germany | F | T2b | | 315484 |
| **I0409** | Spain_EN | Els_Trocs | EN | Els Trocs, Spain; Troc1 | 5311-5218 cal BCE (MAMS 16159) | Spain | F | J1c3 | | 172903 |
| **I0410** | Spain_EN | Els_Trocs | EN | Els Trocs, Spain; Troc3 | 5178-5066 cal BCE (MAMS 16161) | Spain | M | pre-T2c1d2 | R1b1 | 297595 |
| **I0411** | Spain_EN_relative_of_I0410 | Els_Trocs | EN | Els Trocs, Spain; Troc4 | 5177-5068 cal BCE (MAMS 16162) | Spain | M | K1a2a | F* | 31507 |
| **I0412** | Spain_EN | Els_Trocs | EN | Els Trocs, Spain; Troc 5 | 5310-5206 cal BCE (MAMS 16164) | Spain | M | N1a1a1 | I2a1b1 | 333940 |
| **I0413** | Spain_EN | Els_Trocs | EN | Els Trocs, Spain; Troc7 | 5303-5204 cal BCE (MAMS 16166) | Spain | F | V | | 295844 |
| **I0405** | Spain_MN | La_Mina | MN | La Mina, Spain; Mina3 | 3900-3600 BCE | Spain | M | K1a1b1 | I2a1a1/H2? | 133230 |
| **I0406** | Spain_MN | La_Mina | MN | La Mina, Spain; Mina4 | 3900-3600 BCE | Spain | M | H1 | I2a2a1 | 324169 |
| **I0407** | Spain_MN | La_Mina | MN | La Mina, Spain; Mina6b | 3900-3600 BCE | Spain | F | K1b1a1 | | 236225 |
| **I0408** | Spain_MN | La_Mina | MN | La Mina, Spain; Mina18a | 3900-3600 BCE | Spain | F | pre-U5b1i | | 321761 |
| **I0172** | Esperstedt_MN | Salzmünde/Bernburg | MN | Esperstedt, Germany; ESP24, feature 1841 | 3360-3086 cal BCE (Erl8699) | Germany | M | T2b | I2a1b1a | 279147 |
| **I0559** | Baalberge_MN | Baalberge | MN | Quedlinburg, Germany; QLB15D, feature 21033 | 3645-3537 cal BCE (MAMS 22818) | Germany | M | HV6'17 | R*? | 64304 |
| **I0560** | Baalberge_MN | Baalberge | MN | Quedlinburg, Germany; QLB18A, feature 21039 | 3640-3510 cal BCE (Er7856) | Germany | F | T2e1 | | 133305 |
| **I0807** | Baalberge_MN | Baalberge | MN | Esperstedt, Germany; ESP30, feature 6220 | 3887-3797 cal BCE (Er7784) | Germany | M | H1e1a | F* | 33481 |
| **I0231** | Yamnaya | Yamnaya | EBA | Ekaterinovka,_Southern Steppe, Samara , Russia, SVP3 | 2910-2875 cal BCE (Beta 392487) | Russia | M | U4a1 | R1b1a2a2 | 348142 |
| **I0357** | Yamnaya | Yamnaya | EBA | Lopatino I, Sok_River, Samara, Russia; SVP5 same sample as SVP37 | 3090-2910 BCE (Beta 392489) | Russia | F | W6 | | 163845 |
| **I0370** | Yamnaya | Yamnaya | EBA | Ishkinovka I, Eastern Orenburg, Pre-Ural steppe, Samara, Russia: SVP10 | 3300-2700 BCE | Russia | M | H13a1a1a | R1b1a2a2 | 199345 |
| **I0429** | Yamnaya | Yamnaya | EBA | Lopatino I, Sok River, Samara, Russia, SVP38 | 3339-2917 cal BCE (AA47804) | Russia | M | T2c1a2 | R1b1a2a2 | 217664 |
| **I0438** | Yamnaya | Yamnaya | EBA | Luzhki I, Samara River, Samara, Russia; SVP50 | 3021-2635 cal BCE (AA47807) | Russia | M | U5a1a1 | R1b1a2a2 | 213493 |
| **I0439** | Yamnaya | Yamnaya | EBA | Lopatino I, Sok River, Samara, Russia, SVP52 | 3305-2925 cal BCE (Beta 392491) | Russia | M | U5a1a1 | R1b1a | 98900 |
| **I0441** | Yamnaya | Yamnaya | EBA | Kurmanaevskii III, Buzuluk, Samara, Russia; SVP54 | 3010-2622 cal BCE (AA47805) | Russia | F | H2b | | 51326 |
| **I0443** | Yamnaya | Yamnaya | EBA | Lopatino II, Sok River, Samara, Russia; SVP57 | 3300-2700 BCE | Russia | M | W3a1a | R1b1a2a | 343890 |
| **I0444** | Yamnaya | Yamnaya | EBA | Kutuluk I, Kutuluk River, Samara, Russia; SVP58 | 3300-2700 BCE | Russia | M | H6a1b | R1b1a2a2 | 187126 |
| **I0550** | Karsdorf_LN | unknown | LN | Karsdorf, Germany; KAR22a, feature 191 | 3950-3400 BCE?, C14C pending | Germany | F | T1a1 | | 59907 |
| **I0103** | Corded_Ware_LN | Corded Ware | LN | Esperstedt, Germany; ESP16, feature 6236 | 2566-2477 cal BCE (MAMS 21488) | Germany | F | W6a | | 336918 |
| **I0049** | Corded_Ware_LN | Corded Ware | LN | Esperstedt, Germany; ESP22, feature 6140 | 2454-2291 cal BCE (MAMS 21489) | Germany | F | X2b4 | | 167170 |
| **I0106** | Corded_Ware_LN | Corded Ware | LN | Esperstedt, Germany; ESP26, feature 6233.1 | 2454-2291 cal BCE (MAMS 21490) | Germany | F | T2a1b1 | | 69886 |
| **I0104** | Corded_Ware_LN | Corded Ware | LN | Esperstedt, Germany; ESP11, feature 6216 | 2473-2348 cal BCE (MAMS 21487) | Germany | M | U4b1a1a1 | R1a1a1 | 336637 |
| **I0059** | BenzigerodeHeimburg_LN | Bell Beaker? | LN | Benzingerode-Heimburg, Germany; BZH6, grave 2, feature/find 1287/1036 | 2286-2153 cal BCE (MAMS 21486) | Germany | F | H1 /H1b'ad | | 241081 |
| **I0058** | BenzigerodeHeimburg_LN | Bell Beaker | LN | Benzingerode-Heimburg, Germany; BZH4, grave 7, feature 4607 | 2283-2146 cal BCE (MAMS 21491) | Germany | F | H1e | | 246728 |
| **I0171** | BenzigerodeHeimburg_LN | Bell Beaker? | LN | Benzingerode-Heimburg, Germany; BZH12, grave 3 ,feature 6256 | 2204-2136 cal BCE (KIA27952 ) | Germany | F | U5a1a2a | | 66800 |
| **I0112** | Bell_Beaker_LN | Bell Beaker | LN | Quedlinburg XII, Germany; QUEXII6, feature 6256 | 2340-2190 cal BCE (Er7038) | Germany | F | H13a1a2 | | 341003 |
| **I0113** | Bell_Beaker_LN | Bell Beaker | LN | Quedlinburg XII, Germany; QUEXII4, feature 6255.1 | 2290-2130 cal BCE (Er7283) | Germany | F | J1c5 | | 190352 |
| **I0108** | Bell_Beaker_LN | Bell Beaker | LN | Rothenschirmbach, Germany; ROT6, feature 10044 | 2497-2436 cal BCE (Er8710) | Germany | F | H5a3 | | 260528 |
| **I0111** | Bell_Beaker_LN | Bell Beaker | LN | Rothenschirmbach, Germany; ROT4, feature 10142 | 2414-2333 cal BCE (Er8712) | Germany | F | H3new | | 208256 |
| **I0060** | Bell_Beaker_LN | Bell Beaker | LN | Rothenschirmbach, Germany; ROT3, feature 10011 | 2294-2206 cal BCE (MAMS 22819) | Germany | F | K1a2c | | 47085 |
| **I0806** | Bell_Beaker_LN | Bell Beaker | LN | Quedlinburg VII 2, Germany; QLB28b, feature 19617 | 2296-2206 cal BCE (MAMS 22820) | Germany | M | H1 | R1b1a2a1a2 | 91757 |
| **I0118** | Alberstedt_LN | unknown | LN | Alberstedt, Germany; ALB3, feature 7144.2 | 2459-2345 cal BCE (MAMS 21492) | Germany | F | HV6'17 | | 349956 |
| **I0114** | Unetice_EBA_relative_of_I0117 | Unetice | EBA | Esperstedt, Germany; ESP2, feature 3340.1 | 2131-1979 cal BCE (MAMS 21493) | Germany | M | I3a | I2a2 | 217031 |
| **I0115** | Unetice_EBA | Unetice | EBA | Esperstedt, Germany; ESP3, feature 1559.1 | 1931-1780 cal BCE (MAMS 21494) | Germany | F | U5a1 | | 123744 |
| **I0116** | Unetice_EBA | Unetice | EBA | Esperstedt, Germany; ESP4, feature 3322/3323 | 2118-1961 cal BCE (MAMS 21495) | Germany | M | W3a1 | I2c2 | 308158 |
| **I0117** | Unetice_EBA | Unetice | EBA | Esperstedt, Germany; ESP29, feature 3332/3333 | 2199-2064 cal BCE (MAMS 21496) | Germany | F | I3a | | 279996 |
| **I0164** | Unetice_EBA | Unetice | EBA | Quedlinburg VIII, Germany; QUEVII6, feature 3580 | 2012-1919 cal BCE (MAMS 21497) | Germany | F | pre-U5b2a1b | | 332832 |
| **I0803** | Unetice_EBA | Unetice | EBA | Eulau, Germany; EUL41A, feature 882 | 2115-1966 cal BCE (MAMS 22822) | Germany | F | H4a1a1 | | 144186 |
| **I0804** | Unetice_EBA | Unetice | EBA | Eulau, Germany; EUL57B, feature1911.5 | 2131-1982 cal BCE (MAMS 22821) | Germany | M | H3 | I2 | 22869 |
| **I0047** | Unetice_EBA | Unetice | EBA | Halberstadt-Sonntagsfeld, Germany; HAL16, grave 19, feature 613.1 | 2022-1937 cal BCE (MAMS 21481) | Germany | F | V | | 288353 |
| **I0099** | Halberstadt_LBA | Late Bronze Age | LBA | Halberstadt-Sonntagsfeld, Germany; HAL36C, grave 40, feature 1114 | 1113-1021 cal BCE (MAMS 21484) | Germany | M | H23 | R1a1a1b1a2 | 337566 |

**Extended Data Table 3: Pairwise $F_{ST}$ for all ancient groups with ≥2 individuals, present-day Europeans with ≥10 individuals, and selected other groups.** $F_{ST}$ (below the diagonal), standard deviation (above the diagonal).

| | Armenian | Baalberge_MN | Basque | BedouinB | Belarusian | Bell_Beaker_LN | BenzigerodeHeimburg_LN | Bergamo | Bulgarian | Corded_Ware_LN | Croatian | Czech | EHG | English | Estonian | French | Greek | Han | Hungarian | HungaryGamba_BA | HungaryGamba_EN | Icelandic | Iraqi_Jew | Karitiana | LBK_EN | Lezgin | Lithuanian | Mala | Mordovian | Motala_HG | Norwegian | Onge | Orcadian | Papuan | Russian | Sardinian | Sicilian | Sindhi | Spain_EN | Spain_MN | Spanish | SwedenSkoglund_NHG | Turkish | Unetice_EBA | WHG | Yamnaya | Yoruba |
|---|---|---|---|---|---|---|---|---|---|---|---|---|---|---|---|---|---|---|---|---|---|---|---|---|---|---|---|---|---|---|---|---|---|---|---|---|---|---|---|---|---|---|---|---|---|---|---|
| Armenian | | 0.004 | 0.000 | 0.001 | 0.001 | 0.001 | 0.002 | 0.000 | 0.000 | 0.001 | 0.000 | 0.000 | 0.002 | 0.000 | 0.001 | 0.000 | 0.000 | 0.001 | 0.000 | 0.002 | 0.001 | 0.000 | 0.001 | 0.002 | 0.001 | 0.000 | 0.001 | 0.001 | 0.001 | 0.001 | 0.000 | 0.001 | 0.000 | 0.002 | 0.000 | 0.000 | 0.000 | 0.000 | 0.001 | 0.001 | 0.000 | 0.003 | 0.000 | 0.001 | 0.002 | 0.001 | 0.001 |
| Baalberge_MN | 0.023 | | 0.003 | 0.004 | 0.004 | 0.005 | 0.006 | 0.004 | 0.004 | 0.005 | 0.004 | 0.003 | 0.007 | 0.004 | 0.004 | 0.003 | 0.003 | 0.004 | 0.003 | 0.008 | 0.005 | 0.004 | 0.004 | 0.005 | 0.004 | 0.004 | 0.004 | 0.004 | 0.004 | 0.004 | 0.003 | 0.004 | 0.003 | 0.004 | 0.003 | 0.003 | 0.003 | 0.003 | 0.005 | 0.005 | 0.003 | 0.013 | 0.003 | 0.004 | 0.005 | 0.004 | 0.004 |
| Basque | 0.017 | 0.019 | | 0.001 | 0.000 | 0.001 | 0.002 | 0.000 | 0.000 | 0.001 | 0.000 | 0.000 | 0.002 | 0.000 | 0.000 | 0.000 | 0.000 | 0.001 | 0.000 | 0.002 | 0.001 | 0.000 | 0.001 | 0.002 | 0.001 | 0.000 | 0.000 | 0.001 | 0.000 | 0.001 | 0.000 | 0.001 | 0.000 | 0.002 | 0.000 | 0.000 | 0.000 | 0.000 | 0.001 | 0.001 | 0.000 | 0.003 | 0.000 | 0.001 | 0.002 | 0.001 | 0.001 |
| BedouinB | 0.025 | 0.047 | 0.041 | | 0.001 | 0.001 | 0.002 | 0.001 | 0.001 | 0.001 | 0.001 | 0.001 | 0.002 | 0.001 | 0.001 | 0.000 | 0.000 | 0.001 | 0.001 | 0.002 | 0.001 | 0.001 | 0.001 | 0.002 | 0.001 | 0.001 | 0.001 | 0.001 | 0.001 | 0.001 | 0.001 | 0.001 | 0.001 | 0.002 | 0.001 | 0.001 | 0.001 | 0.001 | 0.001 | 0.001 | 0.000 | 0.003 | 0.000 | 0.001 | 0.002 | 0.001 | 0.001 |
| Belarusian | 0.014 | 0.026 | 0.012 | 0.041 | | 0.001 | 0.002 | 0.000 | 0.000 | 0.001 | 0.000 | 0.000 | 0.002 | 0.000 | 0.000 | 0.000 | 0.000 | 0.001 | 0.000 | 0.002 | 0.001 | 0.000 | 0.001 | 0.002 | 0.001 | 0.000 | 0.000 | 0.001 | 0.000 | 0.001 | 0.000 | 0.001 | 0.000 | 0.002 | 0.000 | 0.000 | 0.000 | 0.000 | 0.001 | 0.001 | 0.000 | 0.003 | 0.000 | 0.001 | 0.002 | 0.001 | 0.001 |
| Bell_Beaker_LN | 0.017 | 0.020 | 0.012 | 0.042 | 0.008 | | 0.002 | 0.001 | 0.001 | 0.002 | 0.001 | 0.001 | 0.003 | 0.001 | 0.001 | 0.001 | 0.001 | 0.002 | 0.001 | 0.003 | 0.002 | 0.001 | 0.001 | 0.002 | 0.001 | 0.001 | 0.001 | 0.001 | 0.001 | 0.002 | 0.001 | 0.002 | 0.001 | 0.002 | 0.001 | 0.001 | 0.001 | 0.001 | 0.002 | 0.002 | 0.001 | 0.004 | 0.001 | 0.001 | 0.002 | 0.001 | 0.001 |
| BenzigerodeHeimburg_LN | 0.019 | 0.021 | 0.017 | 0.047 | 0.011 | 0.004 | | 0.002 | 0.002 | 0.003 | 0.002 | 0.002 | 0.004 | 0.002 | 0.002 | 0.002 | 0.002 | 0.002 | 0.002 | 0.004 | 0.003 | 0.002 | 0.002 | 0.003 | 0.002 | 0.002 | 0.002 | 0.002 | 0.002 | 0.002 | 0.002 | 0.003 | 0.002 | 0.003 | 0.002 | 0.002 | 0.002 | 0.002 | 0.003 | 0.003 | 0.002 | 0.006 | 0.002 | 0.002 | 0.003 | 0.002 | 0.002 |
| Bergamo | 0.007 | 0.015 | 0.008 | 0.032 | 0.007 | 0.008 | 0.012 | | 0.000 | 0.001 | 0.000 | 0.000 | 0.002 | 0.000 | 0.000 | 0.000 | 0.000 | 0.001 | 0.000 | 0.002 | 0.001 | 0.000 | 0.001 | 0.002 | 0.001 | 0.000 | 0.000 | 0.001 | 0.000 | 0.001 | 0.000 | 0.001 | 0.000 | 0.002 | 0.000 | 0.000 | 0.000 | 0.000 | 0.001 | 0.001 | 0.000 | 0.003 | 0.000 | 0.001 | 0.002 | 0.001 | 0.001 |
| Bulgarian | 0.006 | 0.019 | 0.009 | 0.031 | 0.003 | 0.008 | 0.010 | 0.002 | | 0.001 | 0.000 | 0.000 | 0.002 | 0.000 | 0.000 | 0.000 | 0.000 | 0.001 | 0.000 | 0.002 | 0.001 | 0.000 | 0.001 | 0.002 | 0.001 | 0.000 | 0.000 | 0.001 | 0.000 | 0.001 | 0.000 | 0.001 | 0.000 | 0.002 | 0.000 | 0.000 | 0.000 | 0.000 | 0.001 | 0.001 | 0.000 | 0.003 | 0.000 | 0.001 | 0.002 | 0.001 | 0.001 |
| Corded_Ware_LN | 0.023 | 0.032 | 0.025 | 0.052 | 0.015 | 0.008 | 0.007 | 0.019 | 0.017 | | 0.001 | 0.001 | 0.003 | 0.001 | 0.001 | 0.001 | 0.001 | 0.002 | 0.001 | 0.003 | 0.002 | 0.001 | 0.002 | 0.002 | 0.002 | 0.001 | 0.001 | 0.002 | 0.001 | 0.002 | 0.001 | 0.002 | 0.001 | 0.002 | 0.001 | 0.001 | 0.001 | 0.001 | 0.002 | 0.002 | 0.001 | 0.004 | 0.001 | 0.002 | 0.002 | 0.002 | 0.002 |
| Croatian | 0.009 | 0.019 | 0.010 | 0.035 | 0.002 | 0.007 | 0.011 | 0.004 | 0.001 | 0.015 | | 0.000 | 0.002 | 0.000 | 0.000 | 0.000 | 0.000 | 0.001 | 0.000 | 0.002 | 0.001 | 0.000 | 0.001 | 0.002 | 0.001 | 0.000 | 0.000 | 0.001 | 0.000 | 0.001 | 0.000 | 0.002 | 0.000 | 0.002 | 0.000 | 0.000 | 0.000 | 0.000 | 0.001 | 0.001 | 0.000 | 0.003 | 0.000 | 0.001 | 0.002 | 0.001 | 0.001 |
| Czech | 0.011 | 0.024 | 0.009 | 0.037 | 0.001 | 0.007 | 0.010 | 0.005 | 0.002 | 0.015 | 0.001 | | 0.002 | 0.000 | 0.000 | 0.000 | 0.000 | 0.001 | 0.000 | 0.002 | 0.001 | 0.000 | 0.001 | 0.002 | 0.001 | 0.000 | 0.000 | 0.001 | 0.000 | 0.001 | 0.000 | 0.002 | 0.000 | 0.002 | 0.000 | 0.000 | 0.000 | 0.000 | 0.001 | 0.001 | 0.000 | 0.003 | 0.000 | 0.001 | 0.002 | 0.001 | 0.001 |
| EHG | 0.067 | 0.082 | 0.060 | 0.093 | 0.042 | 0.041 | 0.038 | 0.057 | 0.051 | 0.034 | 0.048 | 0.043 | | 0.002 | 0.002 | 0.002 | 0.002 | 0.003 | 0.002 | 0.004 | 0.003 | 0.002 | 0.003 | 0.003 | 0.003 | 0.002 | 0.002 | 0.002 | 0.002 | 0.003 | 0.002 | 0.003 | 0.002 | 0.003 | 0.002 | 0.002 | 0.002 | 0.002 | 0.003 | 0.003 | 0.002 | 0.006 | 0.002 | 0.003 | 0.003 | 0.002 | 0.002 |
| English | 0.011 | 0.020 | 0.008 | 0.038 | 0.004 | 0.006 | 0.009 | 0.004 | 0.003 | 0.014 | 0.003 | 0.002 | 0.045 | | 0.000 | 0.000 | 0.000 | 0.001 | 0.000 | 0.002 | 0.001 | 0.000 | 0.001 | 0.002 | 0.001 | 0.000 | 0.000 | 0.001 | 0.000 | 0.001 | 0.000 | 0.001 | 0.000 | 0.002 | 0.000 | 0.000 | 0.000 | 0.000 | 0.001 | 0.001 | 0.000 | 0.003 | 0.000 | 0.001 | 0.002 | 0.001 | 0.001 |
| Estonian | 0.017 | 0.027 | 0.014 | 0.045 | 0.002 | 0.009 | 0.011 | 0.010 | 0.006 | 0.016 | 0.005 | 0.002 | 0.040 | 0.005 | | 0.000 | 0.000 | 0.001 | 0.000 | 0.002 | 0.001 | 0.000 | 0.001 | 0.002 | 0.001 | 0.001 | 0.000 | 0.001 | 0.000 | 0.001 | 0.000 | 0.001 | 0.000 | 0.002 | 0.000 | 0.000 | 0.000 | 0.000 | 0.001 | 0.001 | 0.000 | 0.003 | 0.000 | 0.001 | 0.002 | 0.001 | 0.001 |
| French | 0.009 | 0.018 | 0.006 | 0.035 | 0.004 | 0.006 | 0.008 | 0.002 | 0.002 | 0.015 | 0.002 | 0.001 | 0.048 | 0.001 | 0.005 | | 0.000 | 0.001 | 0.000 | 0.002 | 0.001 | 0.000 | 0.000 | 0.002 | 0.001 | 0.000 | 0.000 | 0.001 | 0.000 | 0.001 | 0.000 | 0.001 | 0.000 | 0.002 | 0.000 | 0.000 | 0.000 | 0.000 | 0.001 | 0.001 | 0.000 | 0.003 | 0.000 | 0.001 | 0.002 | 0.001 | 0.001 |
| Greek | 0.004 | 0.018 | 0.010 | 0.028 | 0.007 | 0.010 | 0.014 | 0.002 | 0.000 | 0.019 | 0.003 | 0.004 | 0.057 | 0.005 | 0.009 | 0.003 | | 0.001 | 0.000 | 0.002 | 0.001 | 0.000 | 0.000 | 0.002 | 0.001 | 0.000 | 0.000 | 0.001 | 0.000 | 0.001 | 0.000 | 0.001 | 0.000 | 0.002 | 0.000 | 0.000 | 0.000 | 0.000 | 0.001 | 0.001 | 0.000 | 0.003 | 0.000 | 0.001 | 0.002 | 0.001 | 0.001 |
| Han | 0.108 | 0.136 | 0.120 | 0.128 | 0.110 | 0.114 | 0.117 | 0.113 | 0.108 | 0.118 | 0.110 | 0.111 | 0.135 | 0.113 | 0.110 | 0.110 | 0.110 | | 0.001 | 0.003 | 0.002 | 0.001 | 0.001 | 0.002 | 0.001 | 0.001 | 0.001 | 0.001 | 0.001 | 0.002 | 0.001 | 0.001 | 0.001 | 0.002 | 0.001 | 0.001 | 0.001 | 0.001 | 0.002 | 0.002 | 0.001 | 0.003 | 0.001 | 0.001 | 0.002 | 0.001 | 0.001 |
| Hungarian | 0.009 | 0.019 | 0.009 | 0.036 | 0.001 | 0.007 | 0.009 | 0.003 | 0.001 | 0.014 | 0.002 | 0.000 | 0.046 | 0.001 | 0.003 | 0.001 | 0.003 | 0.108 | | 0.002 | 0.001 | 0.000 | 0.001 | 0.002 | 0.001 | 0.000 | 0.000 | 0.001 | 0.000 | 0.001 | 0.000 | 0.001 | 0.000 | 0.002 | 0.000 | 0.000 | 0.000 | 0.000 | 0.001 | 0.001 | 0.000 | 0.003 | 0.000 | 0.001 | 0.002 | 0.001 | 0.001 |
| HungaryGamba_BA | 0.016 | 0.021 | 0.010 | 0.039 | 0.005 | 0.008 | 0.010 | 0.008 | 0.005 | 0.016 | 0.005 | 0.005 | 0.046 | 0.007 | 0.006 | 0.005 | 0.007 | 0.115 | 0.005 | | 0.003 | 0.002 | 0.002 | 0.003 | 0.002 | 0.002 | 0.002 | 0.002 | 0.002 | 0.003 | 0.002 | 0.003 | 0.002 | 0.003 | 0.002 | 0.002 | 0.002 | 0.002 | 0.003 | 0.003 | 0.002 | 0.006 | 0.002 | 0.003 | 0.003 | 0.003 | 0.002 |
| HungaryGamba_EN | 0.016 | 0.009 | 0.018 | 0.037 | 0.021 | 0.020 | 0.025 | 0.009 | 0.012 | 0.037 | 0.015 | 0.018 | 0.081 | 0.018 | 0.026 | 0.014 | 0.010 | 0.129 | 0.016 | 0.014 | | 0.001 | 0.001 | 0.002 | 0.001 | 0.001 | 0.001 | 0.001 | 0.001 | 0.002 | 0.001 | 0.002 | 0.001 | 0.002 | 0.001 | 0.001 | 0.001 | 0.001 | 0.002 | 0.002 | 0.001 | 0.004 | 0.001 | 0.002 | 0.002 | 0.002 | 0.001 |
| Icelandic | 0.015 | 0.025 | 0.011 | 0.041 | 0.005 | 0.008 | 0.011 | 0.006 | 0.005 | 0.014 | 0.006 | 0.003 | 0.045 | 0.002 | 0.006 | 0.003 | 0.008 | 0.113 | 0.004 | 0.007 | 0.021 | | 0.001 | 0.002 | 0.001 | 0.000 | 0.000 | 0.001 | 0.000 | 0.001 | 0.000 | 0.001 | 0.000 | 0.002 | 0.000 | 0.000 | 0.000 | 0.000 | 0.001 | 0.001 | 0.000 | 0.003 | 0.000 | 0.001 | 0.002 | 0.001 | 0.001 |
| Iraqi_Jew | 0.005 | 0.031 | 0.021 | 0.026 | 0.020 | 0.022 | 0.024 | 0.011 | 0.010 | 0.029 | 0.014 | 0.017 | 0.073 | 0.016 | 0.023 | 0.014 | 0.009 | 0.112 | 0.015 | 0.018 | 0.020 | 0.021 | | 0.002 | 0.001 | 0.001 | 0.001 | 0.001 | 0.001 | 0.001 | 0.001 | 0.002 | 0.001 | 0.002 | 0.001 | 0.001 | 0.001 | 0.001 | 0.001 | 0.001 | 0.000 | 0.003 | 0.000 | 0.001 | 0.002 | 0.001 | 0.001 |
| Karitiana | 0.210 | 0.234 | 0.218 | 0.230 | 0.205 | 0.208 | 0.208 | 0.212 | 0.206 | 0.209 | 0.207 | 0.206 | 0.206 | 0.209 | 0.205 | 0.208 | 0.209 | 0.180 | 0.205 | 0.214 | 0.229 | 0.209 | 0.214 | | 0.002 | 0.002 | 0.002 | 0.002 | 0.002 | 0.002 | 0.002 | 0.002 | 0.002 | 0.002 | 0.002 | 0.002 | 0.002 | 0.002 | 0.002 | 0.002 | 0.002 | 0.004 | 0.002 | 0.002 | 0.003 | 0.002 | 0.001 |
| LBK_EN | 0.023 | 0.009 | 0.024 | 0.043 | 0.029 | 0.021 | 0.025 | 0.017 | 0.018 | 0.035 | 0.022 | 0.024 | 0.084 | 0.024 | 0.033 | 0.020 | 0.016 | 0.137 | 0.022 | 0.018 | 0.006 | 0.029 | 0.026 | 0.237 | | 0.001 | 0.001 | 0.001 | 0.001 | 0.001 | 0.001 | 0.002 | 0.001 | 0.002 | 0.001 | 0.001 | 0.001 | 0.001 | 0.001 | 0.001 | 0.001 | 0.003 | 0.001 | 0.001 | 0.002 | 0.001 | 0.001 |
| Lezgin | 0.005 | 0.033 | 0.018 | 0.034 | 0.012 | 0.015 | 0.015 | 0.010 | 0.007 | 0.017 | 0.009 | 0.010 | 0.052 | 0.011 | 0.014 | 0.009 | 0.007 | 0.103 | 0.009 | 0.017 | 0.025 | 0.013 | 0.011 | 0.201 | 0.032 | | 0.001 | 0.001 | 0.000 | 0.001 | 0.000 | 0.001 | 0.000 | 0.002 | 0.000 | 0.000 | 0.000 | 0.000 | 0.001 | 0.001 | 0.000 | 0.003 | 0.000 | 0.001 | 0.002 | 0.001 | 0.001 |
| Lithuanian | 0.019 | 0.030 | 0.014 | 0.046 | 0.001 | 0.010 | 0.010 | 0.011 | 0.006 | 0.016 | 0.005 | 0.003 | 0.043 | 0.006 | 0.002 | 0.006 | 0.010 | 0.114 | 0.004 | 0.007 | 0.026 | 0.007 | 0.024 | 0.209 | 0.034 | 0.015 | | 0.001 | 0.000 | 0.001 | 0.000 | 0.002 | 0.000 | 0.002 | 0.000 | 0.000 | 0.000 | 0.001 | 0.001 | 0.001 | 0.000 | 0.003 | 0.000 | 0.001 | 0.002 | 0.001 | 0.001 |
| Mala | 0.045 | 0.074 | 0.061 | 0.068 | 0.051 | 0.055 | 0.057 | 0.053 | 0.048 | 0.058 | 0.051 | 0.051 | 0.081 | 0.053 | 0.053 | 0.051 | 0.049 | 0.070 | 0.050 | 0.056 | 0.070 | 0.055 | 0.049 | 0.187 | 0.077 | 0.043 | 0.056 | | 0.001 | 0.001 | 0.001 | 0.001 | 0.001 | 0.002 | 0.001 | 0.001 | 0.001 | 0.000 | 0.002 | 0.001 | 0.001 | 0.003 | 0.001 | 0.001 | 0.002 | 0.001 | 0.001 |
| Mordovian | 0.014 | 0.028 | 0.014 | 0.041 | 0.002 | 0.010 | 0.011 | 0.009 | 0.005 | 0.013 | 0.005 | 0.003 | 0.040 | 0.005 | 0.003 | 0.005 | 0.008 | 0.100 | 0.003 | 0.007 | 0.025 | 0.006 | 0.020 | 0.196 | 0.032 | 0.011 | 0.004 | 0.047 | | 0.001 | 0.000 | 0.001 | 0.000 | 0.002 | 0.000 | 0.000 | 0.000 | 0.000 | 0.001 | 0.001 | 0.000 | 0.003 | 0.000 | 0.001 | 0.002 | 0.001 | 0.001 |
| Motala_HG | 0.088 | 0.083 | 0.072 | 0.113 | 0.060 | 0.057 | 0.053 | 0.073 | 0.069 | 0.060 | 0.067 | 0.062 | 0.048 | 0.063 | 0.058 | 0.064 | 0.075 | 0.166 | 0.062 | 0.058 | 0.091 | 0.062 | 0.095 | 0.254 | 0.093 | 0.080 | 0.058 | 0.112 | 0.060 | | 0.001 | 0.002 | 0.001 | 0.002 | 0.001 | 0.001 | 0.001 | 0.001 | 0.002 | 0.002 | 0.001 | 0.004 | 0.001 | 0.001 | 0.002 | 0.001 | 0.001 |
| Norwegian | 0.014 | 0.024 | 0.010 | 0.041 | 0.004 | 0.007 | 0.009 | 0.006 | 0.005 | 0.014 | 0.003 | 0.002 | 0.044 | 0.002 | 0.005 | 0.002 | 0.007 | 0.112 | 0.003 | 0.009 | 0.021 | 0.002 | 0.020 | 0.207 | 0.028 | 0.012 | 0.006 | 0.053 | 0.006 | 0.062 | | 0.002 | 0.000 | 0.002 | 0.000 | 0.000 | 0.000 | 0.000 | 0.001 | 0.001 | 0.000 | 0.003 | 0.000 | 0.001 | 0.002 | 0.001 | 0.001 |
| Onge | 0.168 | 0.191 | 0.181 | 0.186 | 0.171 | 0.174 | 0.177 | 0.174 | 0.170 | 0.180 | 0.171 | 0.172 | 0.197 | 0.174 | 0.173 | 0.172 | 0.170 | 0.155 | 0.171 | 0.176 | 0.187 | 0.175 | 0.172 | 0.283 | 0.194 | 0.167 | 0.177 | 0.129 | 0.166 | 0.227 | 0.174 | | 0.001 | 0.002 | 0.001 | 0.001 | 0.001 | 0.001 | 0.002 | 0.002 | 0.001 | 0.004 | 0.001 | 0.002 | 0.002 | 0.002 | 0.001 |
| Orcadian | 0.014 | 0.021 | 0.011 | 0.041 | 0.006 | 0.009 | 0.010 | 0.007 | 0.006 | 0.016 | 0.006 | 0.004 | 0.047 | 0.003 | 0.007 | 0.004 | 0.008 | 0.114 | 0.004 | 0.008 | 0.021 | 0.005 | 0.020 | 0.209 | 0.029 | 0.013 | 0.008 | 0.054 | 0.007 | 0.063 | 0.004 | 0.175 | | 0.002 | 0.000 | 0.000 | 0.000 | 0.000 | 0.001 | 0.001 | 0.000 | 0.003 | 0.000 | 0.001 | 0.002 | 0.001 | 0.001 |
| Papuan | 0.186 | 0.206 | 0.199 | 0.202 | 0.190 | 0.194 | 0.194 | 0.193 | 0.187 | 0.199 | 0.190 | 0.191 | 0.218 | 0.192 | 0.192 | 0.191 | 0.189 | 0.180 | 0.189 | 0.192 | 0.204 | 0.195 | 0.190 | 0.304 | 0.212 | 0.185 | 0.196 | 0.151 | 0.185 | 0.244 | 0.193 | 0.232 | 0.193 | | 0.002 | 0.002 | 0.002 | 0.002 | 0.002 | 0.002 | 0.002 | 0.004 | 0.002 | 0.002 | 0.002 | 0.002 | 0.002 |
| Russian | 0.015 | 0.027 | 0.014 | 0.042 | 0.002 | 0.010 | 0.010 | 0.009 | 0.005 | 0.015 | 0.005 | 0.003 | 0.039 | 0.005 | 0.002 | 0.005 | 0.008 | 0.099 | 0.003 | 0.007 | 0.024 | 0.006 | 0.021 | 0.196 | 0.032 | 0.012 | 0.004 | 0.047 | 0.002 | 0.058 | 0.005 | 0.166 | 0.007 | 0.185 | | 0.000 | 0.000 | 0.000 | 0.001 | 0.001 | 0.000 | 0.003 | 0.000 | 0.001 | 0.002 | 0.001 | 0.001 |
| Sardinian | 0.014 | 0.014 | 0.013 | 0.036 | 0.017 | 0.016 | 0.022 | 0.007 | 0.009 | 0.033 | 0.011 | 0.013 | 0.074 | 0.012 | 0.020 | 0.009 | 0.009 | 0.123 | 0.012 | 0.014 | 0.010 | 0.016 | 0.017 | 0.225 | 0.015 | 0.021 | 0.021 | 0.064 | 0.020 | 0.085 | 0.016 | 0.183 | 0.016 | 0.201 | 0.020 | | 0.000 | 0.000 | 0.001 | 0.001 | 0.000 | 0.003 | 0.000 | 0.001 | 0.002 | 0.001 | 0.001 |
| Sicilian | 0.003 | 0.017 | 0.011 | 0.027 | 0.009 | 0.011 | 0.013 | 0.003 | 0.002 | 0.022 | 0.004 | 0.006 | 0.059 | 0.006 | 0.012 | 0.004 | 0.002 | 0.109 | 0.005 | 0.010 | 0.010 | 0.010 | 0.008 | 0.209 | 0.016 | 0.009 | 0.013 | 0.048 | 0.010 | 0.078 | 0.007 | 0.169 | 0.009 | 0.187 | 0.010 | 0.008 | | 0.000 | 0.001 | 0.001 | 0.000 | 0.003 | 0.000 | 0.001 | 0.002 | 0.001 | 0.001 |
| Sindhi | 0.016 | 0.047 | 0.032 | 0.043 | 0.023 | 0.026 | 0.028 | 0.023 | 0.020 | 0.029 | 0.022 | 0.022 | 0.057 | 0.023 | 0.025 | 0.022 | 0.020 | 0.081 | 0.021 | 0.027 | 0.042 | 0.025 | 0.020 | 0.185 | 0.049 | 0.014 | 0.028 | 0.014 | 0.021 | 0.088 | 0.024 | 0.142 | 0.026 | 0.162 | 0.021 | 0.035 | 0.021 | | 0.001 | 0.001 | 0.000 | 0.003 | 0.000 | 0.001 | 0.002 | 0.001 | 0.001 |
| Spain_EN | 0.034 | 0.021 | 0.030 | 0.053 | 0.037 | 0.028 | 0.035 | 0.024 | 0.028 | 0.044 | 0.030 | 0.032 | 0.092 | 0.032 | 0.040 | 0.028 | 0.027 | 0.144 | 0.031 | 0.028 | 0.018 | 0.036 | 0.037 | 0.245 | 0.015 | 0.042 | 0.041 | 0.085 | 0.039 | 0.099 | 0.036 | 0.202 | 0.036 | 0.219 | 0.041 | 0.021 | 0.026 | 0.058 | | 0.002 | 0.001 | 0.004 | 0.001 | 0.002 | 0.002 | 0.002 | 0.002 |
| Spain_MN | 0.033 | 0.014 | 0.025 | 0.054 | 0.032 | 0.022 | 0.026 | 0.023 | 0.026 | 0.035 | 0.027 | 0.029 | 0.077 | 0.027 | 0.036 | 0.024 | 0.025 | 0.138 | 0.028 | 0.028 | 0.023 | 0.031 | 0.037 | 0.241 | 0.016 | 0.040 | 0.035 | 0.079 | 0.035 | 0.081 | 0.030 | 0.196 | 0.031 | 0.214 | 0.035 | 0.021 | 0.025 | 0.052 | 0.009 | | 0.001 | 0.004 | 0.001 | 0.002 | 0.002 | 0.002 | 0.002 |
| Spanish | 0.008 | 0.014 | 0.005 | 0.032 | 0.006 | 0.007 | 0.011 | 0.002 | 0.002 | 0.018 | 0.004 | 0.003 | 0.054 | 0.003 | 0.008 | 0.001 | 0.003 | 0.111 | 0.003 | 0.006 | 0.011 | 0.005 | 0.013 | 0.210 | 0.017 | 0.011 | 0.009 | 0.052 | 0.008 | 0.069 | 0.005 | 0.172 | 0.005 | 0.190 | 0.008 | 0.007 | 0.003 | 0.024 | 0.024 | 0.020 | | 0.003 | 0.000 | 0.001 | 0.002 | 0.001 | 0.001 |
| SwedenSkoglund_NHG | 0.071 | 0.066 | 0.058 | 0.094 | 0.047 | 0.050 | 0.048 | 0.056 | 0.053 | 0.054 | 0.052 | 0.047 | 0.058 | 0.050 | 0.050 | 0.049 | 0.058 | 0.153 | 0.050 | 0.041 | 0.068 | 0.051 | 0.076 | 0.244 | 0.078 | 0.066 | 0.049 | 0.101 | 0.047 | 0.034 | 0.049 | 0.211 | 0.053 | 0.230 | 0.047 | 0.067 | 0.061 | 0.074 | 0.084 | 0.074 | 0.052 | | 0.003 | 0.004 | 0.004 | 0.003 | 0.003 |
| Turkish | 0.001 | 0.023 | 0.014 | 0.026 | 0.010 | 0.013 | 0.016 | 0.005 | 0.004 | 0.021 | 0.006 | 0.008 | 0.058 | 0.009 | 0.013 | 0.006 | 0.003 | 0.095 | 0.007 | 0.013 | 0.016 | 0.012 | 0.006 | 0.198 | 0.022 | 0.005 | 0.015 | 0.040 | 0.010 | 0.081 | 0.011 | 0.160 | 0.012 | 0.178 | 0.011 | 0.013 | 0.003 | 0.013 | 0.032 | 0.030 | 0.006 | 0.064 | | 0.001 | 0.002 | 0.001 | 0.001 |
| Unetice_EBA | 0.016 | 0.016 | 0.014 | 0.042 | 0.007 | 0.002 | 0.003 | 0.008 | 0.007 | 0.005 | 0.007 | 0.005 | 0.034 | 0.005 | 0.006 | 0.005 | 0.009 | 0.112 | 0.006 | 0.007 | 0.022 | 0.007 | 0.021 | 0.206 | 0.022 | 0.013 | 0.007 | 0.053 | 0.007 | 0.051 | 0.005 | 0.173 | 0.007 | 0.190 | 0.008 | 0.019 | 0.011 | 0.024 | 0.030 | 0.023 | 0.007 | 0.043 | 0.013 | | 0.002 | 0.001 | 0.001 |
| WHG | 0.086 | 0.073 | 0.062 | 0.106 | 0.054 | 0.055 | 0.056 | 0.066 | 0.065 | 0.070 | 0.061 | 0.056 | 0.078 | 0.058 | 0.053 | 0.058 | 0.070 | 0.162 | 0.058 | 0.049 | 0.080 | 0.057 | 0.091 | 0.260 | 0.091 | 0.082 | 0.053 | 0.110 | 0.056 | 0.053 | 0.057 | 0.220 | 0.059 | 0.239 | 0.054 | 0.074 | 0.072 | 0.087 | 0.092 | 0.070 | 0.060 | 0.050 | 0.078 | 0.057 | | 0.002 | 0.002 |
| Yamnaya | 0.030 | 0.054 | 0.034 | 0.060 | 0.020 | 0.016 | 0.013 | 0.028 | 0.023 | 0.011 | 0.022 | 0.020 | 0.028 | 0.021 | 0.020 | 0.022 | 0.026 | 0.118 | 0.020 | 0.023 | 0.050 | 0.020 | 0.037 | 0.204 | 0.052 | 0.019 | 0.022 | 0.056 | 0.018 | 0.063 | 0.021 | 0.178 | 0.022 | 0.197 | 0.019 | 0.044 | 0.030 | 0.028 | 0.060 | 0.053 | 0.027 | 0.062 | 0.026 | 0.012 | 0.076 | | 0.001 |
| Yoruba | 0.142 | 0.163 | 0.155 | 0.141 | 0.150 | 0.151 | 0.153 | 0.148 | 0.145 | 0.158 | 0.148 | 0.149 | 0.179 | 0.150 | 0.152 | 0.147 | 0.145 | 0.177 | 0.148 | 0.151 | 0.156 | 0.152 | 0.141 | 0.270 | 0.163 | 0.143 | 0.155 | 0.139 | 0.148 | 0.200 | 0.151 | 0.222 | 0.152 | 0.231 | 0.148 | 0.153 | 0.139 | 0.133 | 0.169 | 0.167 | 0.144 | 0.185 | 0.137 | 0.151 | 0.195 | 0.158 | |

# Extended Data Figure Legends

**Extended Data Figure 1: ADMIXTURE analysis.**

**Extended Data Figure 2: Outgroup $f_3$-statistic $f_3$(Dinka; X, Y), measuring the degree of shared drift among pairs of ancient individuals.**

**Extended Data Figure 3: Modeling Corded Ware as a mixture of *N*=1, 2, or 3 ancestral populations.**

(**a**) We present in the left column a histogram of raw $f_4$-statistic residuals and on the right Z-scores for the best-fitting (lowest squared 2-norm of the residuals, or *resnorm*) model at each *N*. (**b**), we show on the left how *resnorm* and on the right how the maximum |Z| score change for different *N*. (**c**) *resnorm* of different *N*=2 models. The set of outgroups used in this analysis, in the terminology of SI9, is "World Foci 15 + Ancients".

**Extended Data Fig. 4: Modeling Europeans as mixtures of increasing complexity: *N*=1 (EN), *N*=2 (EN, WHG), *N*=3 (EN, WHG, Yamnaya), *N*=4 (EN, WHG, Yamnaya, Nganasan), *N*=5 (EN, WHG, Yamnaya, Nganasan, BedouinB).** The residual norm of the fitted model (SI9) and its changes are indicated.

**Extended Data Fig. 5: Geographic distribution of archaeological cultures and graphic illustration of proposed population movements / turnovers discussed in the main text (symbols of samples are identical to Figure 1):**

(**a**) proposed routes of migration by early farmers into Europe ~9,000-7000 years ago, (**b**) resurgence of hunter-gatherer ancestry during the Middle Neolithic 7,000-5,000 years ago, (**c**) arrival of steppe ancestry in central Europe during the Late Neolithic ~4,500 years ago. White arrows indicate the two possible scenarios of the arrival of Indo-European language groups.

**Extended Data Figure 1 (high-resolution version uploaded separately).**

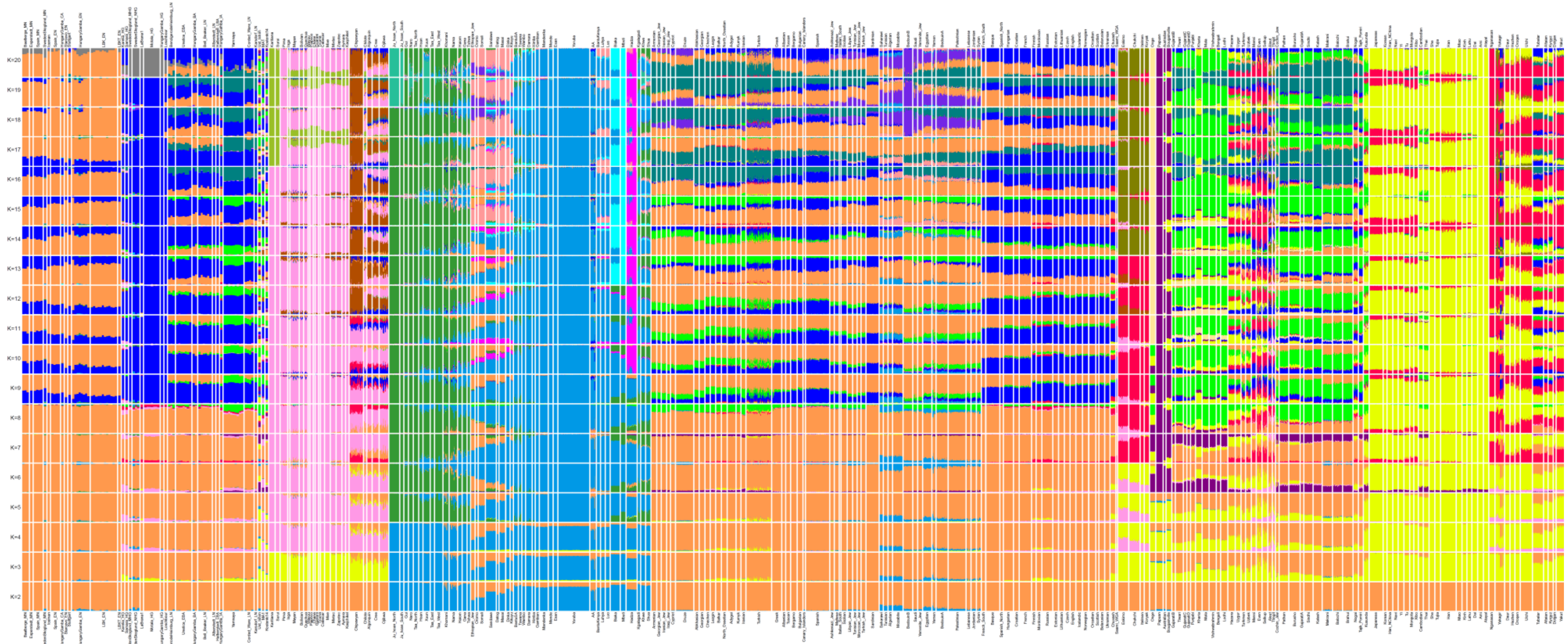

**Extended Data Figure 2**

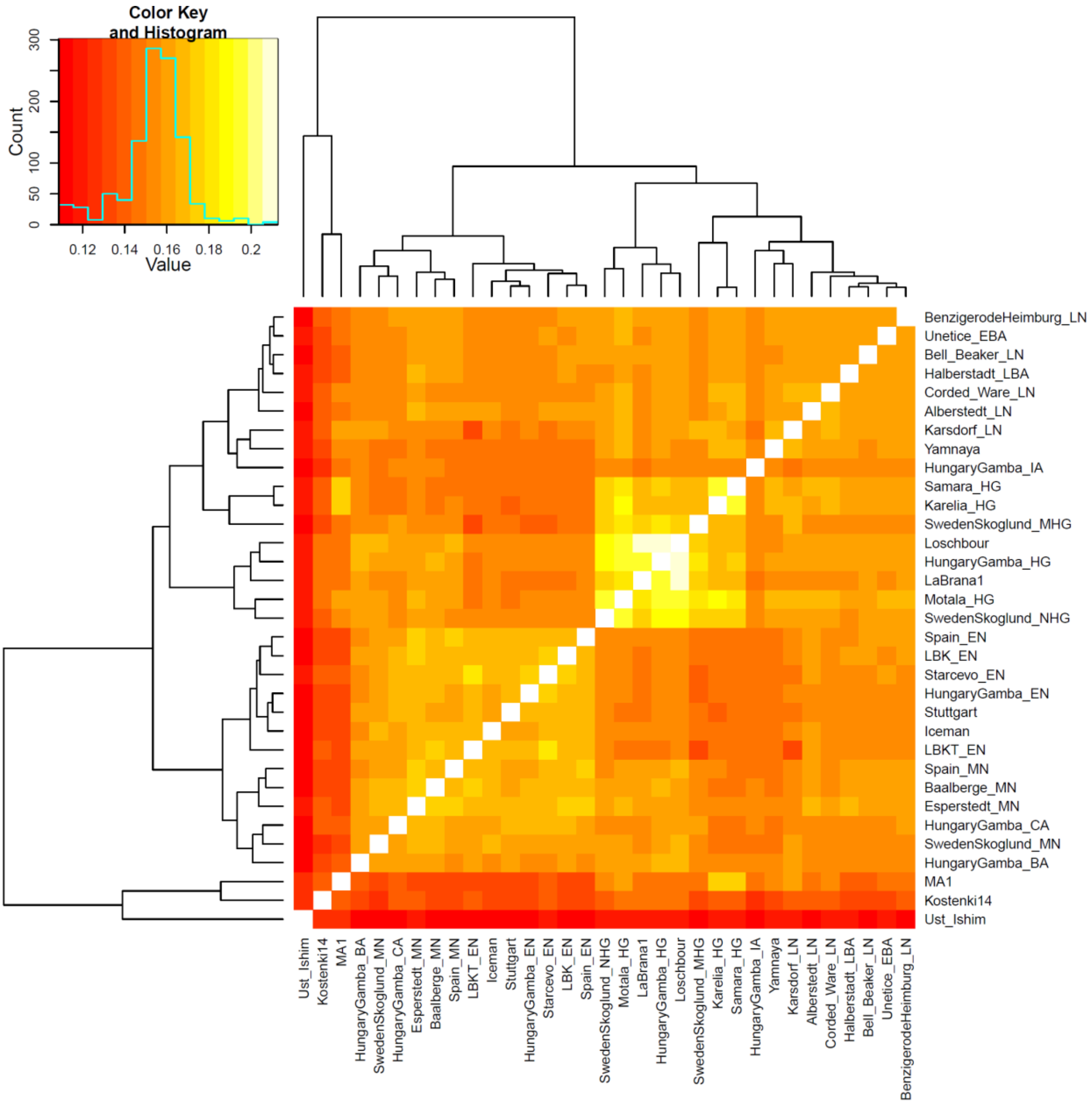

**Extended Data Figure 3**

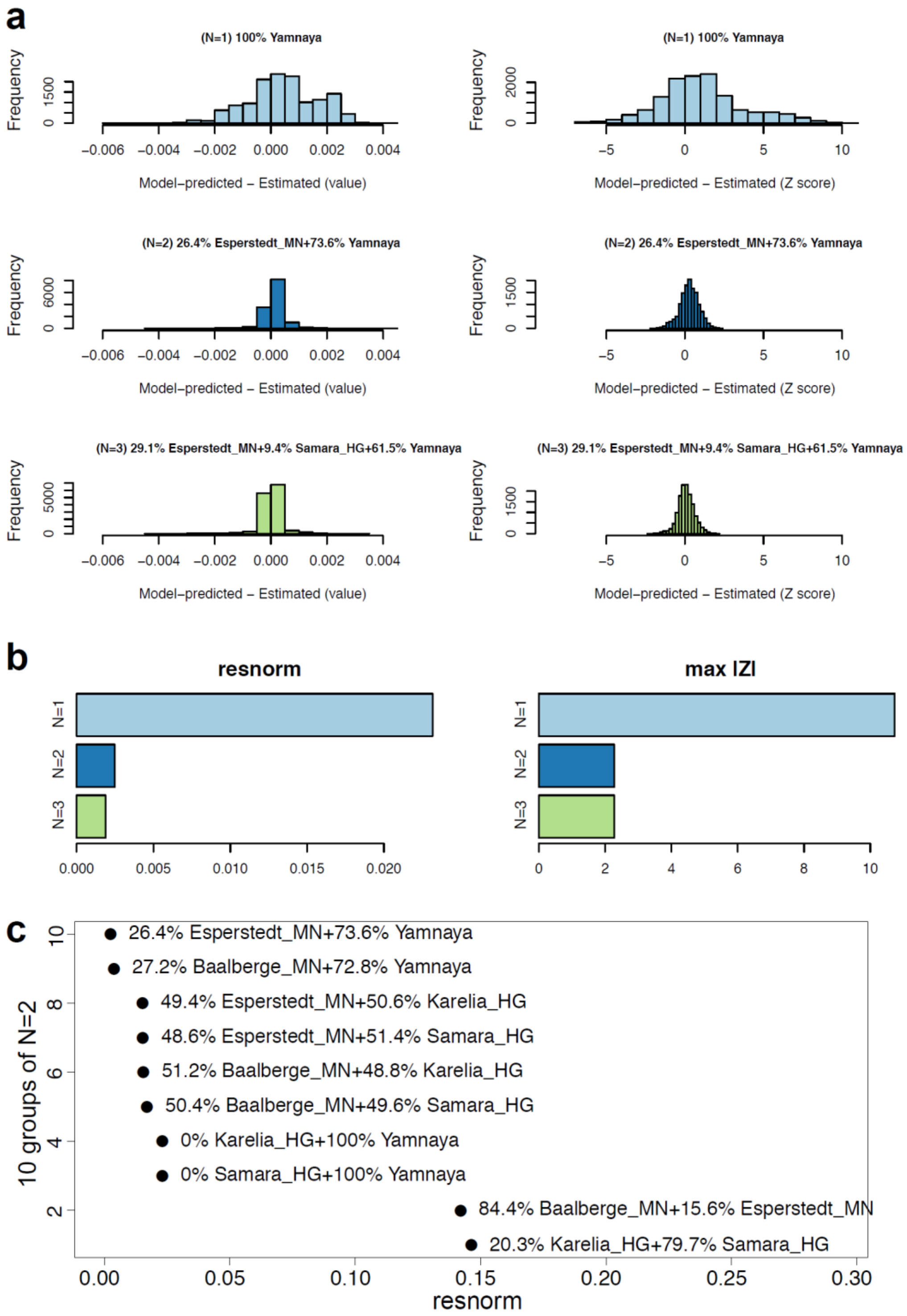

# Extended Data Figure 4

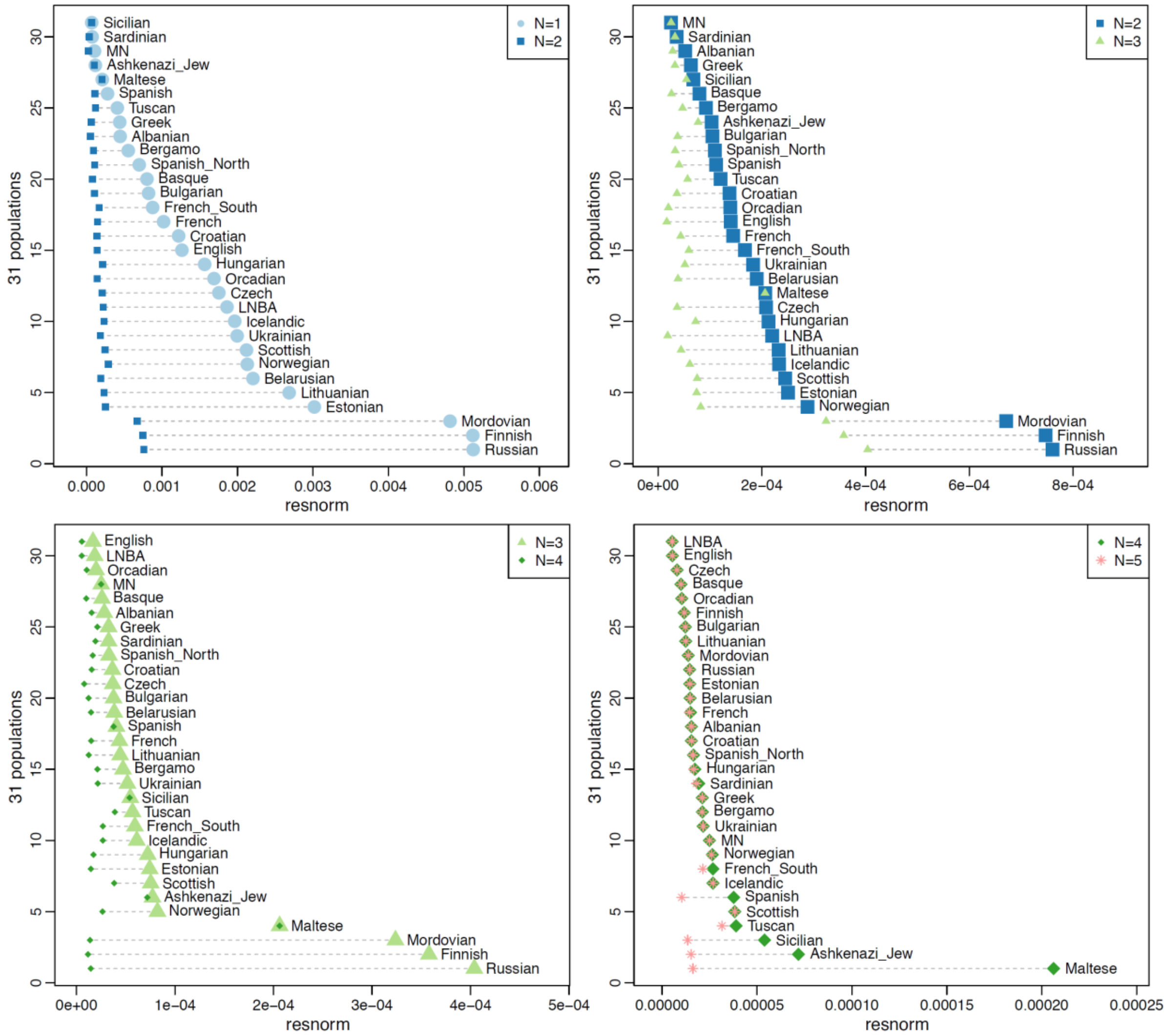

**Extended Data Figure 5**

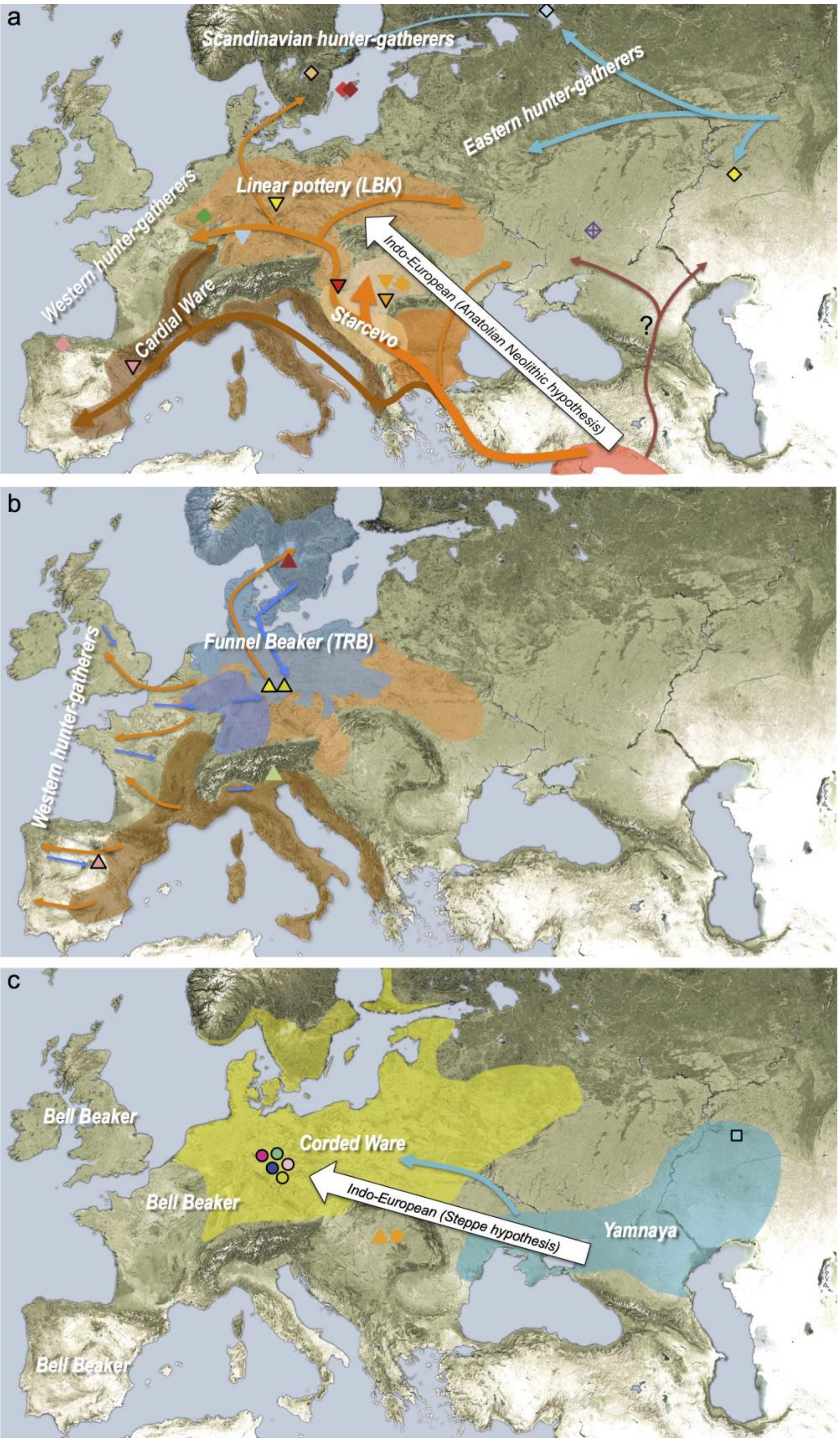

# Supplementary Information

# Massive migration from the steppe is a source for Indo-European languages in Europe

# Supplementary Information 1

## Protocols for DNA extraction and library preparation

Nadin Rohland*, Wolfgang Haak*, Alissa Mittnik, Bastien Llamas, Eadaoin Harney, Susanne Nordenfelt, Kristin Stewardson, David Reich

* To whom correspondence should be addressed (wolfgang.haak@adelaide.edu.au or nrohland@genetics.med.harvard.edu)

### Strategy

The 69 samples newly reported in this study reflect data from 95 ancient DNA libraries. They are a subset of 119 samples and 212 libraries that were screened in the course of this study (Online Table 1).

Briefly, our screening strategy included the following steps:

- Silica-based DNA extraction[1,2]
- Double-stranded library preparation with truncated barcoded Illumina adapters, with either no DNA damage repair[3], full USER-treatment[4] or partial UDG repair[5]
- Shallow shotgun sequencing
- Mitochondrial DNA (mtDNA) capture and sequencing

We then selected samples for 390k SNP capture based on the following criteria:

- DNA damage patterns characteristic of ancient DNA
- Low estimates of contamination based on the mtDNA data[6]
- Evidence of substantial complexity in the library

Libraries with no UDG treatment were only used for screening (shallow shotgun sequencing and mtDNA capture) to assess the authenticity of the DNA extracted from these samples. We then prepared partial or full UDG treated libraries for the 390k SNP capture. All such libraries also were subjected to their own round of screening (mtDNA capture and shotgun) prior to 390k capture.

In what follows we present details on the methods used for DNA extraction and library preparation at the University of Adelaide's Australian Centre for Ancient DNA (ACAD) (n=151 libraries), the University of Tübingen Germany (n=16 libraries), and the Harvard Medical School Boston USA (n=45 libraries).

### Adelaide lab (n=151 libraries)

DNA extraction was carried out in the clean-room facilities at ACAD, using an in-solution silica-based protocol[2]. Briefly, 200 mg of tooth or bone powder were demineralized in 4 mL 0.5M EDTA and 60 μL of Proteinase K at 37°C overnight, followed by an additional incubation with 60 μL of Proteinase K at 55°C for 1 h. DNA was subsequently bound to silica particles using a custom binding buffer [90% Qiagen QG buffer (i.e. 5 M guanidinium thiocyanate), 220 mM sodium acetate, 25 mM sodium chloride, 1% Triton X 100] at a 1:4 ratio of lysate: binding buffer (vol:vol). DNA binding for 1h was followed by two washes in 80% ethanol and a final elution in 200 μL TE buffer (10mM Tris, 0.1mM EDTA, pH8.0) containing 0.05% Tween 20.

DNA libraries were prepared in the clean-room facility following the preparation of the extract. For a subset of samples, double-stranded libraries were prepared with truncated barcoded adapters using the library preparation protocol by Meyer and Kircher (2010)[3]. This protocol started with 20 μL of extract and did not involve UDG treatment to remove characteristic ancient DNA damage. For libraries that had characteristics of authentic ancient DNA as determined by shallow shotgun sequencing and mtDNA capture, we prepared one or more double-stranded libraries with truncated barcoded adapters and a UDG and endonuclease VIII damage repair treatment (USER enzyme, NEB)[4].

DNA libraries were amplified by PCR in quintuplicates for an initial 12-13 cycles (AmpliTaq Gold, Life Technologies), and PCR replicates were pooled and purified with the Agencourt AMPure XP system. DNA libraries were then re-amplified for 12-13 cycles in quintuplicates or sextuplicates, followed by pooling and purification, visual inspection on a 3.5% agarose gel, and final quantification using a NanoDrop 2000c spectrophotometer.

This protocol evolved over the course of the study. For the last set of ancient DNA libraries built for this study, partial DNA damage removal[5] was performed using truncated barcoded adapters. Briefly, this method removes DNA damage except for the first and last positions of the reads, making it possible to test for the presence of characteristic DNA damage patterns while making the DNA libraries more robust for population genetic analysis. These partially UDG-treated libraries were amplified using isothermal amplifications using the commercial TwistAmp® Basic kit (TwistDx Ltd). The amplification followed the manufacturer's recommendations and used 13.4 μL of libraries after the Bst amplification step, as well as an increased incubation time of the isothermal reaction to 44 min at 37°C, followed by gel electrophoresis and Nanodrop quantification. Approximately 1-2 μg DNA was shipped to Harvard Medical School in Boston for further analysis.

**Tübingen lab (n=16 libraries)**

The preparation of Motala samples was described in Lazaridis et al. (2014)[7].

For the 9 Stuttgart *Linienbandkeramik* (LBK) samples newly reported in this study, 100 μL DNA extracts were prepared from 39-69 mg of dentin powder in the clean room facilities of the University of Tübingen using the silica-based protocol by Dabney et al. (2013)[1]. A 40 μL aliquot of each of these extracts was turned into a double-stranded library using the protocol of Meyer and Kircher (2010)[3], and indices were then added to both ends of the library molecules[8]. An aliquot of each LBK extract was also sent to Boston for preparation of partial UDG-treated libraries.

**Boston lab (n=45 libraries)**

DNA was extracted from approximately 75mg of bone powder following the extraction method by Dabney et al. 2013[1] with the exception that the MinElute column Zymo extender assembly was replaced by the High Pure Extender Assembly from the Roche High Pure Viral Nucleic Acid Large Volume Kit (M. Meyer, personal communication). The elution volume varied between 32 and 90 uL.

Library preparation was performed starting from between 20 and 30 uL of extract. We used a double stranded library preparation that is an adaption of Meyer and Kircher (2010)[3] and Kircher, Sawyer and Meyer (2012)[8], resulting in barcoded libraries with truncated adapters following (Rohland et al. 2014)[5]. This library preparation, involving different types of UDG-treatment, was use for all extracts prepared in Boston (from the Samara Valley Project samples) as well as on nine DNA extracts prepared in Tübingen (from Stuttgart LBK samples). Online Table 1 shows which library preparation method was used (i.e. no [minus] or partial [half] or full [plus] UDG treatment); UDG minus was only used to assess the authenticity of extracted DNA from the samples; nuclear data only originates from plus or half libraries

A subset of libraries from Adelaide were amplified further in Boston to achieve sufficient material for enrichment (mtDNA and 390k). We used Herculase II Fusion DNA Polymerase (Agilent) and the primers PreHyb-F and PreHyb-R[5] for these re-amplifications, and cleaned the products with the MinElute PCR Purification System.

Summary information on all libraries made for this study is given in Online Table 1.

# Supplementary Information 2

## Mitochondrial genome analysis

Wolfgang Haak*, Swapan Mallick, Guido Brandt and Bastien Llamas

* To whom correspondence should be addressed (wolfgang.haak@adelaide.edu.au),

We used an automated procedure to obtain an initial mitochondrial DNA haplogroup determination. Following this, we double-checked reads for each sample visually. The final set of haplogroup determinations is given in Online Table 1.

We imported bam files (deduplicated and quality filtered) into Geneious (Geneious version 7.1.3 created by Biomatters. Available from http://www.geneious.com/) and re-assembled the reads against the Revised Sapiens Reference Sequence (RSRS)[1]. SNPs were called in Geneious for all polymorphisms with a minimum coverage of 2 and a minimum variant frequency of 0.7. The assembly and the resulting list of SNPs were double-checked against SNPs reported in phylotree.org (mtDNA tree Build 16 (19 Feb 2014))[2]. Following recommendations in van Oven et al. 2010, we excluded common indels at nucleotide positions 309.1C(C), 315.1C, AC indels at 515-522, 16182C, 16183C, 16193.1C(C), but report C16519T. Haplotype calls are given in Table S2.1, reporting all SNPs compared to RSRS and following the hierarchical structure until the most derived SNP shared with an existing sequence. Additional SNPs were considered to be private mutations and are reported in a separate column.

We report 105 complete mitochondrial genomes from the 119 individuals that were analysed for this study.

Overall, we found no inconsistencies between consensus calls from different library architectures (UDG treated and untreated), between libraries from different extracts from the same individual, or from mitochondrial sequences that had been generated previously using a different technique[3,4].

Sample HAL14 appears to harbor two potentially heteroplasmic sites at np 3338, where we find that 60% of reads show a T and 40% a C in a UDG-treated library and 70/30% T/C ratio in an untreated library, and at 16093 (51/49% C/T ratio UDG-treated and 17/83% C/T ratio untreated). Since we observe both alleles of both sites in the UDG-treated library, DNA damage is unlikely to be an explanation. The fact that all other SNP calls are unambiguous, i.e. the variant is observed in >99.9% of all reads, also suggests that contamination is an unlikely explanation. Of note, T3338C has been reported as a sub-haplogroup defining SNP in T2b8, immediately derived from T2b, while 16093 has been described as recurrent mutation[5]. Another heteroplasmic site was found in sample La Mina 2 at np 16274, where 54% of the reads show and A and 46% a G in a UDG-treated libraries, whereas all neighbouring D-loop SNP are consistent throughout all reads, again rendering contamination unlikely.

The majority of samples presented here were previously typed by sequencing of the hypervariable segment I (HVS I) and/or by typing of 22 haplogroup-defining coding region

SNPs[6] as part of large-scale diachronic studies[7-9]. The fact that previous PCR-based mitochondrial analysis of these samples was successful likely explains the high success rate of our mtDNA capture (88%). Haplogroup calls from PCR-based typing techniques were consistent in all cases with our results from complete sequencing of the mitochondrial genomes, expect for individual BENZ18 due to a sample mixup with another sample from the same site. Haplogroup calls were also consistent with previously reported results using a Mitochip v2 typing assay[3]. The full mitochondrial sequencing provided higher resolution haplogroup determination, but did not alter our previous findings about the history of the Mittelelbe-Saale sample series. Briefly[10]:

1. Hunter-gatherers carry predominantly U-haplotypes.
2. Early farmers predominantly carry haplotypes T2, K, HV, H, V, X and N1a.
3. Middle Neolithic farmers exhibit a similar mitochondrial profile as early farmers and show an increase in the frequency of U-haplotypes, providing mitochondrial DNA based confirmation of the resurgence of hunter-gatherer lineages that is also documented from our autosomal data.
4. Late Neolithic individuals mark a second major turnover, highlighted by the appearance of I, T1, and increase of U2, U4, U5a, H and W haplotypes.

We used the mitochondrial haplotype frequencies in Brandt et al. 2014[10], and in a recent study of the Yamnaya and Catacomb cultures[11] together with the new data (10 Yamnaya) reported here to explore the maternal relationships between the Late Cooper Age/early Bronze Age groups from the steppes and Late Neolithic groups from Central Europe (Table S2.2). We added newly reported mitochondrial genomes from this study to existing population data sets, notably eight novel sequences to the early Neolithic LBK set, and also removed two samples that radiocarbon dating in this study showed were from more recent time periods (HAL36 and HAL16). We added the Samara hunter-gatherer to the Eastern hunter-gather group (HGE) and two new samples to the Middle Neolithic Baalberge group (BAC). The ten new samples from Els Trocs (5) and La Mina (5) were pooled with the Epi-Cardial (CAR) and Middle Neolithic (MNS) data from Spain, respectively. We did not include the three N1a genomes from Hungary as these were chosen non-randomly for a phylogeographic study.

We performed principal component analysis (PCA) based on frequencies from the 22 most common West Eurasia mitochondrial haplogroups (Table S2.2; Figure S2.1). The results show that Bronze Age Kurgan samples from the steppe (BAK) and Siberia (BAS) as well as the Catacomb culture (CAT) are linked to the hunter-gatherer samples from eastern Europe (HGE) and northern Europe (PWC), and separated from central European early and Middle Neolithic samples by the first PC (20%).

The Late Copper Age/Early Bronze Age Yamnaya fall closest to the central European Late Neolithic Corded Ware (CWC) and Bell Beaker groups (BBC), and in particular to samples from the early Bronze Age Unetice culture (UC). This supports a genetic contribution of eastern groups to central Europe during the Late Neolithic ~4,500 BCE, consistent with the autosomal SNP data. These results indicate that the Late Neolithic migration into central Europe was not entirely male in origin.

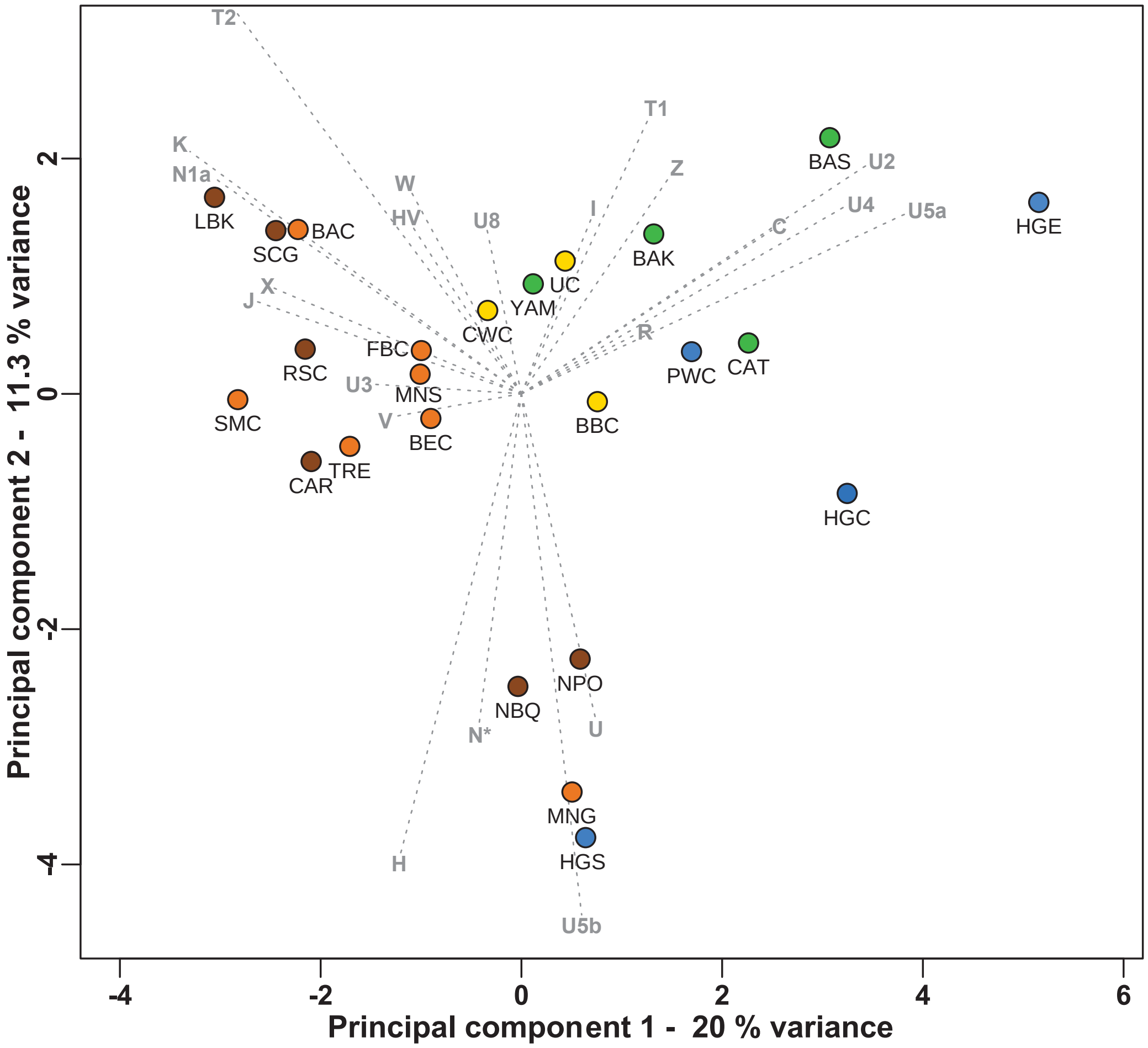

**Figure S2.1. PCA based on mitochondrial DNA haplogroup frequencies from 24 prehistoric populations.** Temporal groups are color-coded: hunter-gatherer (blue), Early Neolithic (brown), Middle Neolithic (orange), Late Neolithic/Early Bronze Age (yellow), and Late Neolithic/Early Bronze Age East (green). For full length names and details of samples and populations please refer to Table S2.2.

**Table S2.1. Mitochondrial sequence haplotypes of the 105 ancient samples in this study.**

| Lab ID | Individual ID | mtDNA-hg call | Private mutations | SNPs against RSRS |
|---|---|---|---|---|
| I0061 | UzOO74 | C1g (formerly C1f) | A8577G, A11605t, A12217G, T16189C! | C146T, C152T, C195T, A247G, A249d, 290-291d, T489C, A769G, A825t, A1018G, A2758G, C2885T, T3552a, T3594C, G4104A, T4312C, A4715G, G7146A, C7196a, T7256C, A7521G, T8468C, A8577G, G8584A, T8655C, A9545G, C10400T, T10664C, A10688G, C10810T, C10915T, A11605t, A12217G, G13105A, A13263G, G13276A, T13506C, T13650C, T14318C, T14783C, G15043A, G15301A, A15487t, A16129G, T16187C, G16230A, T16278C, T16298C, C16311T, T16325C, C16327T, C16519T |
| I0124 | SVP44 | U5a1d | A16241C, C16519T | C146T, C152T, C195T, A247G, A769G, A825t, A1018G, A2758G, C2885T, T3027C, T3197C, T3594C, G4104A, T4312C, G7146A, T7256C, A7521G, T8468C, T8655C, G8701A, G9477A, C9540T, G10398A, T10664C, A10688G, C10810T, C10873T, C10915T, A11467G, A11914G, A12308G, G12372A, T12705C, G13105A, G13276A, T13506C, T13617C, T13650C, A14793G, A15218G, A16129G, T16187C, C16189T, C16192T, T16223C, G16230A, A16241C, C16256T, C16270T, T16278C, C16311T, A16399G, C16519T |
| I0011 | Motala1 | U5a1 | G5460A, T16093C, C16519T | C146T, C152T, C195T, A247G, A769G, A825t, A1018G, A2758G, C2885T, T3197C, T3594C, G4104A, T4312C, G5460A, G7146A, T7256C, A7521G, T8468C, T8655C, G8701A, G9477A, C9540T, G10398A, T10664C, A10688G, C10810T, C10873T, C10915T, A11467G, A11914G, A12308G, G12372A, T12705C, G13105A, G13276A, T13506C, T13617C, T13650C, A14793G, A15218G, T16093C, A16129G, T16187C, C16189T, C16192T, T16223C, G16230A, C16256T, C16270T, T16278C, C16311T, A16399G, C16519T |
| I0017 | Motala12 | U2e1 | C16527T | C146T, C195T, T217C, A247G, C340T, A508G, A769G, A825t, A1018G, A1811G, A2758G, C2885T, T3594C, A3720G, G4104A, T4312C, A5390G, T5426C, C6045T, T6152C, G7146A, T7256C, A7521G, T8468C, T8655C, G8701A, C9540T, G10398A, T10664C, A10688G, C10810T, C10873T, A10876G, C10915T, A11467G, A11914G, A12308G, G12372A, T12705C, T13020C, G13105A, G13276A, T13506C, T13650C, T13734C, A15907G, A16051G, G16129c, T16187C, T16223C, G16230A, T16278C, C16311T, T16362C, C16527T |
| I0012 | Motala2 | U2e1 | C16527T | C146T, C195T, T217C, A247G, C340T, A508G, A769G, A825t, A1018G, A1811G, A2758G, C2885T, T3594C, A3720G, G4104A, T4312C, A5390G, T5426C, C6045T, T6152C, G7146A, T7256C, A7521G, T8468C, T8655C, G8701A, C9540T, G10398A, T10664C, A10688G, C10810T, C10873T, A10876G, C10915T, A11467G, A11914G, A12308G, G12372A, T12705C, T13020C, G13105A, G13276A, T13506C, T13650C, T13734C, A15907G, A16051G, G16129c, T16187C, T16223C, G16230A, T16278C, C16311T, T16362C, C16527T |
| I0013 | Motala3 | U5a1 | G5460A, G8860A, A9389G, C16519T | C146T, C152T, C195T, A247G, A769G, A825t, A1018G, A2758G, C2885T, T3197C, T3594C, G4104A, T4312C, G5460A, G7146A, T7256C, A7521G, T8468C, T8655C, G8701A, G8860A, A9389G, G9477A, C9540T, G10398A, T10664C, A10688G, C10810T, C10873T, C10915T, A11467G, A11914G, A12308G, G12372A, T12705C, G13105A, G13276A, T13506C, T13617C, T13650C, A14793G, A15218G, A16129G, T16187C, C16189T, C16192T, T16223C, G16230A, C16256T, C16270T, T16278C, C16311T, A16399G, C16519T |
| I0014 | Motala4 | U5a2d | A13158G, C16519T | C146T, C152T, C195T, A247G, A769G, A825t, A1018G, A2758G, C2885T, T3197C, T3594C, G4104A, T4312C, G7146A, T7256C, A7521G, A7843G, T8468C, T8655C, G8701A, G9477A, C9540T, G10398A, T10664C, A10688G, C10810T, C10873T, C10915T, A11467G, A11914G, A12308G, G12372A, T12705C, G13105A, A13158G, G13276A, T13506C, T13617C, T13650C, A14793G, A16129G, T16187C, C16189T, C16192T, T16223C, G16230A, C16256T, C16270T, T16278C, C16311T, C16519T, G16526A |
| I0015 | Motala6 | U5a2d | T152C!, G6480A, C16519T | C146T, C195T, A247G, A769G, A825t, A1018G, A2758G, C2885T, T3197C, T3594C, G4104A, T4312C, G6480A, G7146A, T7256C, A7521G, A7843G, T8468C, T8655C, G8701A, G9477A, C9540T, G10398A, T10664C, A10688G, C10810T, C10873T, C10915T, A11467G, A11914G, A12308G, G12372A, T12705C, G13105A, G13276A, T13506C, T13617C, T13650C, A14793G, A16129G, T16187C, C16189T, C16192T, T16223C, G16230A, C16256T, C16270T, T16278C, C16311T, C16519T, G16526A |
| I0016 | Motala9 | U5a2 | T195C!, G228A, C514T, G1888A, A2246G, A3756T, G6917A, G8854A, A9531G, C16519T | C146T, C152T, G228A, A247G, C514T, A769G, A825t, A1018G, G1888A, A2246G, A2758G, C2885T, T3197C, T3594C, A3756T, G4104A, T4312C, G6917A, G7146A, T7256C, A7521G, T8468C, T8655C, G8701A, G8854A, G9477A, A9531G, C9540T, G10398A, T10664C, A10688G, C10810T, C10873T, C10915T, A11467G, A11914G, A12308G, G12372A, T12705C, G13105A, G13276A, T13506C, T13617C, T13650C, A14793G, A16129G, T16187C, C16189T, C16192T, T16223C, G16230A, C16256T, C16270T, T16278C, C16311T, C16519T, G16526A |

| Lab ID | Individual ID | mtDNA-hg call | Private mutations | SNPs against RSRS |
|---|---|---|---|---|
| I0101 | DEB36 | U5a1a'g | C150T, G1007A, T16093C, C16519T | C146T, C150T, C152T, C195T, A247G, A769G, A825t, G1007A, A1018G, A2758G, C2885T, T3197C, T3594C, G4104A, T4312C, G7146A, T7256C, A7521G, T8468C, T8655C, G8701A, G9477A, C9540T, G10398A, T10664C, A10688G, C10810T, C10873T, C10915T, A11467G, A11914G, A12308G, G12372A, T12705C, G13105A, G13276A, T13506C, T13617C, T13650C, A14793G, A15218G, T16093C, A16129G, T16187C, C16189T, T16223C, G16230A, C16256T, C16270T, T16278C, C16311T, A16399G, C16519T |
| I0056 | HAL14 | T2b(8?) | heteroplasmy at np 3338 (60% T, 40% C) and 16093 (51% C, 49% T) | C146T, C152T, C195T, A247G, G709A, A769G, A825t, G930A, A1018G, G1888A, A2758G, C2885T, T3338C (40%), T3594C, G4104A, T4216C, T4312C, A4917G, G5147A, G7146A, T7256C, A7521G, T8468C, T8655C, G8697A, G8701A, C9540T, G10398A, T10463C, T10664C, A10688G, C10810T, C10873T, C10915T, A11251G, A11812G, A11914G, T12705C, G13105A, G13276A, G13368A, T13506C, T13650C, A14233G, G14905A, C15452a, A15607G, G15928A, T16126C, A16129G, T16187C, C16189T, T16223C, G16230A, T16278C, C16294T, C16296T, T16304C, C16311T |
| I0102 | HAL15 | N1a1a1a3 | T4772C, G16129A! | C146T, C195T, T199C, T204C, A247G, 573.XC, T669C, A769G, A825t, A1018G, G1719A, G2702A, A2758G, C2885T, T3336C, T3594C, G4104A, T4312C, T4772C, A5315G, G7146A, T7256C, A7521G, T8468C, T8655C, G8701A, A8901G, C9540T, T10238C, T10664C, A10688G, C10810T, C10873T, C10915T, A11914G, G12501A, G13105A, G13276A, T13506C, T13650C, A13780G, G15043A, C16147A, T16154C, T16172C, T16187C, C16189T, G16230A, C16248T, T16278C, C16311T, C16320T, C16355T |
| I0821 | HAL24 | pre-X2d1 | Topologically missing: T204C, A5186G | C146T, C152T, G207A, A247G, A769G, A825t, A1018G, G1719A, A2758G, C2885T, T3594C, G4104A, T4312C, T6221C, C6371T, A6791G, G7146A, T7256C, A7521G, T8468C, T8503C, T8655C, G8701A, C9540T, G10398A, T10664C, A10688G, C10810T, C10873T, C10915T, T11878C, A11914G, G13105A, G13276A, T13506C, T13650C, G13708A, A13966G, T14470C, A16129G, T16187C, G16230A, C16311T |
| I0048 | HAL25 | K1a | 195C!, T8433g, T16093C | C146T, C152T, A247G, C497T, A769G, A825t, A1018G, T1189C, A1811G, A2758G, C2885T, A3480G, T3594C, G4104A, T4312C, G7146A, T7256C, A7521G, T8433g, T8468C, T8655C, G8701A, G9055A, C9540T, T9698C, A10550G, T10664C, A10688G, C10810T, C10873T, C10915T, T11299C, A11467G, A11914G, A12308G, G12372A, T12705C, G13105A, G13276A, T13506C, T13650C, C14167T, T14798C, T16093C, A16129G, T16187C, C16189T, T16223C, T16224C, G16230A, T16278C |
| I0659 | HAL2A | N1a1a1a2 | Topologically missing: T15299C, C16172T! | C146T, C195T, T199C, T204C, G207A, A247G, 573.XC, T669C, A769G, A825t, A1018G, G1719A, G2702A, A2758G, C2885T, T3336C, T3594C, G4104A, T4312C, A5315G, G7146A, T7256C, A7521G, T8468C, G8485A, T8655C, G8701A, A8901G, C9540T, T10238C, T10664C, A10688G, C10810T, C10873T, C10915T, A11914G, G12501A, G13105A, G13276A, T13506C, T13650C, A13780G, G15043A, T16086C, A16129G, C16147A, T16172C, T16187C, C16189T, G16230A, C16248T, T16278C, C16311T, C16320T, C16355T |
| I0057 | HAL34 | N1a1a1 | C152T!!, A11884G, G14259A | C146T, C152T!!, C195T, T199C, T204C, A247G, 573.XC, T669C, A769G, A825t, A1018G, G1719A, G2702A, A2758G, C2885T, T3336C, T3594C, G4104A, T4312C, A5315G, G7146A, T7256C, A7521G, T8468C, T8655C, G8701A, A8901G, C9540T, T10238C, T10664C, A10688G, C10810T, C10873T, C10915T, A11884G, A11914G, G12501A, G13105A, G13276A, T13506C, T13650C, A13780G, G14259A, G15043A, A16129G, C16147A, T16172C, T16187C, C16189T, G16230A, C16248T, T16278C, C16311T, C16355T |
| I0820 | HAL37 | W1c'i | none | T119C, C146T, C152T, A189G, T204C, G207A, A247G, G709A, A769G, A825t, A1018G, T1243C, A2758G, C2885T, A3505G, T3594C, G4104A, T4312C, G5046A, G5460A, G7146A, T7256C, A7521G, C7864T, G8251A, T8468C, T8655C, G8701A, G8994A, C9540T, G10398A, T10664C, A10688G, C10810T, C10873T, C10915T, C11674T, A11914G, A11947G, T12414C, G13105A, G13276A, T13506C, T13650C, G15884c, A16129G, T16187C, C16189T, G16230A, T16278C, C16292T, C16311T |
| I0100 | HAL4 | N1a1a1a | T15900C | C146T, C152T!!, C195T, T199C, T204C, A247G, 573.XC, T669C, A769G, A825t, A1018G, G1719A, G2702A, A2758G, C2885T, T3336C, T3594C, G4104A, T4312C, A5315G, G7146A, T7256C, A7521G, T8468C, T8655C, G8701A, A8901G, C9540T, T10238C, T10664C, A10688G, C10810T, C10873T, C10915T, A11914G, G12501A, G13105A, G13276A, T13506C, T13650C, A13780G, G15043A, T15900C, A16129G, C16147A, T16172C, T16187C, C16189T, G16230A, C16248T, T16278C, C16311T, C16320T, C16355T |
| I0046 | HAL5 | T2c1d'e'f | C9651a | C152T, C195T, A247G, G709A, A769G, A825t, A1018G, G1888A, A2758G, C2885T, T3594C, G4104A, T4216C, T4312C, A4917G, G6261A, G7146A, T7256C, A7521G, T8468C, T8655C, G8697A, G8701A, C9540T, C9651a, G10398A, T10463C, T10664C, A10688G, C10810T, C10822T, C10873T, C10915T, A11251G, A11812G, A11914G, T12705C, G13105A, G13276A, G13368A, T13506C, |

| Lab ID | Individual ID | mtDNA-hg call | Private mutations | SNPs against RSRS |
|---|---|---|---|---|
| | | | | T13650C, A14233G, G14905A, C15452a, A15607G, G15928A, T16126C, A16129G, T16187C, C16189T, T16223C, G16230A, T16278C, C16292T, C16294T, C16296T, C16311T |
| I0796 | KAR11B | H | T152C! | G73A, C146T, C195T, A247G, A769G, A825t, A1018G, G2706A, A2758G, C2885T, T3594C, G4104A, T4312C, T7028C, G7146A, T7256C, A7521G, T8468C, T8655C, G8701A, C9540T, G10398A, T10664C, A10688G, C10810T, C10873T, C10915T, A11719G, A11914G, T12705C, G13105A, G13276A, T13506C, T13650C, T14766C, A16129G, T16187C, C16189T, T16223C, G16230A, T16278C, C16311T |
| I0797 | KAR16A | H46b | C152T!! | G73A, C146T, C152T!!, C195T, A247G, A769G, A825t, A1018G, G2706A, A2758G, C2772T, C2885T, T3594C, G4104A, T4312C, T7028C, G7146A, T7256C, A7521G, T8468C, T8655C, G8701A, C9540T, G10398A, T10664C, A10688G, C10810T, C10873T, C10915T, A11719G, A11893G, A11914G, T12705C, G13105A, G13276A, T13506C, T13650C, T14766C, A16129G, T16187C, C16189T, T16223C, G16230A, T16278C, C16311T |
| I0795 | KAR6A | H1 | G1719A, C14380T | G73A, C146T, C152T, C195T, A247G, A769G, A825t, A1018G, G1719A, G2706A, A2758G, C2885T, G3010A, T3594C, G4104A, T4312C, T7028C, G7146A, T7256C, A7521G, T8468C, T8655C, G8701A, C9540T, G10398A, T10664C, A10688G, C10810T, C10873T, C10915T, A11719G, A11914G, T12705C, G13105A, G13276A, T13506C, T13650C, C14380T, T14766C, A16129G, T16187C, C16189T, T16223C, G16230A, T16278C, C16311T |
| I0019 | LBK1254 | HV6'17 | C16519T | G73A, C146T, C152T, C195T, A247G, A769G, A825t, A1018G, A2758G, C2885T, T3594C, G4104A, T4312C, G7146A, T7256C, A7521G, T8468C, T8655C, G8701A, C9540T, G10398A, T10664C, A10688G, C10810T, C10873T, C10915T, A11719G, A11914G, T12705C, G13105A, G13276A, T13506C, T13650C, T14766C, A16129G, T16187C, C16189T, T16223C, G16230A, T16278C, C16519T |
| I0020 | LBK1577 | T2e | T195C! | C146T, C150T, C152T, A247G, G709A, A769G, A825t, A1018G, G1888A, A2758G, C2885T, T3594C, G4104A, T4216C, T4312C, A4917G, G7146A, T7256C, A7521G, T8468C, T8655C, G8697A, G8701A, C9540T, G10398A, T10463C, T10664C, A10688G, C10810T, C10873T, C10915T, A11251G, A11812G, A11914G, T12705C, G13105A, G13276A, G13368A, T13506C, T13650C, A14233G, G14905A, C15452a, A15607G, G15928A, T16126C, A16129G, G16153A, T16187C, C16189T, T16223C, G16230A, T16278C, C16294T, C16296T, C16311T |
| I0021 | LBK1581 | T2b | none | C146T, C152T, C195T, A247G, G709A, A769G, A825t, G930A, A1018G, G1888A, A2758G, C2885T, T3594C, G4104A, T4216C, T4312C, A4917G, G5147A, G7146A, T7256C, A7521G, T8468C, T8655C, G8697A, G8701A, C9540T, G10398A, T10463C, T10664C, A10688G, C10810T, C10873T, C10915T, A11251G, A11812G, A11914G, T12705C, G13105A, G13276A, G13368A, T13506C, T13650C, A14233G, G14905A, C15452a, A15607G, G15928A, T16126C, A16129G, T16187C, C16189T, T16223C, G16230A, T16278C, C16294T, T16304C, C16311T |
| I0022 | LBK1976 | T2e | none | C146T, C150T, C152T, C195T, A247G, G709A, A769G, A825t, A1018G, G1888A, A2758G, C2885T, T3594C, G4104A, T4216C, T4312C, A4917G, G7146A, T7256C, A7521G, T8468C, T8655C, G8697A, G8701A, C9540T, G10398A, T10463C, T10664C, A10688G, C10810T, C10873T, C10915T, A11251G, A11812G, A11914G, T12705C, G13105A, G13276A, G13368A, T13506C, T13650C, A14233G, G14905A, C15452a, A15607G, G15928A, T16126C, A16129G, G16153A, T16187C, C16189T, T16223C, G16230A, T16278C, C16294T, C16296T, C16311T |
| I0023 | LBK1979 | H | C16186T | G73A, C146T, C152T, C195T, A247G, A769G, A825t, A1018G, G2706A, A2758G, C2885T, T3594C, G4104A, T4312C, T7028C, G7146A, T7256C, A7521G, T8468C, T8655C, G8701A, C9540T, G10398A, T10664C, A10688G, C10810T, C10873T, C10915T, A11719G, A11914G, T12705C, G13105A, G13276A, T13506C, T13650C, T14766C, A16129G, C16186T, T16187C, C16189T, T16223C, G16230A, T16278C, C16311T |
| I0024 | LBK1988 | W1c'i | none | T119C, C146T, C152T, A189G, T204C, G207A, A247G, G709A, A769G, A825t, A1018G, T1243C, A2758G, C2885T, A3505G, T3594C, G4104A, T4312C, G5046A, G5460A, G7146A, T7256C, A7521G, C7864T, G8251A, T8468C, T8655C, G8701A, G8994A, C9540T, G10398A, T10664C, A10688G, C10810T, C10873T, C10915T, C11674T, A11914G, A11947G, T12414C, G13105A, G13276A, T13506C, T13650C, G15884c, A16129G, T16187C, C16189T, G16230A, T16278C, C16292T, C16311T |
| I0025 | LBK1992 | T2b | none | C146T, C152T, C195T, A247G, G709A, A769G, A825t, G930A, A1018G, G1888A, A2758G, C2885T, T3594C, G4104A, T4216C, T4312C, A4917G, G5147A, G7146A, T7256C, A7521G, T8468C, T8655C, G8697A, G8701A, C9540T, G10398A, T10463C, T10664C, |

| Lab ID | Individual ID | mtDNA-hg call | Private mutations | SNPs against RSRS |
|---|---|---|---|---|
| | | | | A10688G, C10810T, C10873T, C10915T, A11251G, A11812G, A11914G, T12705C, G13105A, G13276A, G13368A, T13506C, T13650C, A14233G, G14905A, C15452a, A15607G, G15928A, T16126C, A16129G, T16187C, C16189T, T16223C, G16230A, T16278C, C16294T, C16296T, T16304C, C16311T |
| I0026 | LBK2155 | T2b | none | C146T, C152T, C195T, A247G, G709A, A769G, A825t, G930A, A1018G, G1888A, A2758G, C2885T, T3594C, G4104A, T4216C, T4312C, A4917G, G5147A, G7146A, T7256C, A7521G, T8468C, T8655C, G8697A, G8701A, C9540T, G10398A, T10463C, T10664C, A10688G, C10810T, C10873T, C10915T, A11251G, A11812G, A11914G, T12705C, G13105A, G13276A, G13368A, T13506C, T13650C, A14233G, G14905A, C15452a, A15607G, G15928A, T16126C, A16129G, T16187C, C16189T, T16223C, G16230A, T16278C, C16294T, C16296T, T16304C, C16311T |
| I0027 | LBK2172 | H40 | none | G73A, C146T, C152T, C195T, A247G, A769G, A825t, A1018G, G2706A, A2758G, C2885T, T3594C, G4104A, T4312C, T7028C, G7146A, T7256C, A7521G, T7621C, T8468C, T8655C, G8701A, C9540T, G10398A, T10664C, A10688G, C10810T, C10873T, C10915T, A11719G, A11914G, T12705C, G13105A, G13276A, T13506C, T13650C, T14766C, A16129G, T16187C, C16189T, T16223C, G16230A, T16278C, C16311T |
| I0054 | UWS4 | J1c17 | C9302T, A15467G | C146T, C152T, G185A, C195T, G228A, A247G, C295T, C462T, T489C, A769G, A825t, A1018G, A2758G, C2885T, G3010A, T3594C, T3847C, G4104A, T4216C, T4312C, G7146A, T7256C, A7521G, T8468C, T8655C, G8701A, C9302T, C9540T, T10664C, A10688G, C10810T, C10873T, C10915T, A11251G, A11914G, A12612G, T12705C, G13105A, G13276A, T13506C, T13650C, G13708A, T14798C, C15452a, A15467G, C16069T, T16126C, A16129G, T16187C, C16189T, T16223C, G16230A, T16278C, C16311T |
| I0176 | SZEH4b | N1a1a1a3 | A16300G | C146T, C195T, T199C, T204C, A247G, 573.XC, T669C, A769G, A825t, A1018G, G1719A, G2702A, A2758G, C2885T, T3336C, T3594C, G4104A, T4312C, A5315G, G7146A, T7256C, A7521G, T8468C, T8655C, G8701A, A8901G, C9540T, T10238C, T10664C, A10688G, C10810T, C10873T, C10915T, A11914G, G12501A, G13105A, G13276A, T13506C, T13650C, A13780G, G15043A, A16129G, C16147A, T16154C, T16172C, T16187C, C16189T, G16230A, C16248T, T16278C, A16300G, C16311T, C16320T, C16355T |
| I0409 | Troc1 | J1c3 | C16519T | C146T, C152T, G185A, C195T, G228A, A247G, C295T, C462T, T489C, A769G, A825t, A1018G, A2758G, C2885T, G3010A, T3594C, G4104A, T4216C, T4312C, G7146A, T7256C, A7521G, T8468C, T8655C, G8701A, C9540T, T10664C, A10688G, C10810T, C10873T, C10915T, A11251G, A11914G, A12612G, T12705C, G13105A, G13276A, T13506C, T13650C, G13708A, C13934T, T14798C, C15452a, C16069T, T16126C, A16129G, T16187C, C16189T, T16223C, G16230A, T16278C, C16311T, C16519T |
| I0410 | Troc3 | pre-T2c1d2 | none | C195T, A247G, T279C, G709A, A769G, A825t, A1018G, G1888A, A2758G, C2885T, T3594C, G4104A, T4216C, T4312C, A4917G, C5187T, G6261A, G7146A, T7256C, A7521G, C7873T, T8468C, T8655C, G8697A, G8701A, C9540T, G10398A, T10463C, T10664C, A10688G, C10810T, C10822T, C10873T, C10915T, A11251G, A11812G, A11914G, T12705C, G13105A, G13276A, G13368A, T13506C, T13650C, A14233G, G14905A, C15452a, A15607G, G15928A, T16126C, A16129G, T16187C, C16189T, T16223C, G16230A, T16278C, C16292T, C16294T, C16311T |
| I0412 | Troc5 | N1a1a1 | G5460A | C146T, C152T!!, C195T, T199C, T204C, A247G, 573.XC, T669C, A769G, A825t, A1018G, G1719A, G2702A, A2758G, C2885T, T3336C, T3594C, G4104A, T4312C, A5315G, G5460A, G7146A, T7256C, A7521G, T8468C, T8655C, G8701A, A8901G, C9540T, T10238C, T10664C, A10688G, C10810T, C10873T, C10915T, A11884G, A11914G, G12501A, G13105A, G13276A, T13506C, T13650C, A13780G, G15043A, A16129G, C16147A, T16172C, T16187C, C16189T, G16230A, C16248T, T16278C, C16311T, C16355T |
| I0413 | Troc7 | V | C16519T | T72C, G73A, C146T, C152T, C195T, A247G, A769G, A825t, A1018G, A2758G, C2885T, T3594C, G4104A, T4312C, G4580A, G7146A, T7256C, A7521G, T8468C, T8655C, G8701A, C9540T, G10398A, T10664C, A10688G, C10810T, C10873T, C10915T, A11719G, A11914G, T12705C, G13105A, G13276A, T13506C, T13650C, T14766C, C15904T, A16129G, T16187C, C16189T, T16223C, G16230A, T16278C, T16298C, C16311T, C16519T |
| I0411 | Troc4 | K1a2a | none | C146T, C152T, C195T, A247G, C497T, A769G, A825t, A1018G, T1189C, A1811G, A2758G, C2885T, A3480G, T3594C, G4104A, T4312C, G5773A, G7146A, T7256C, A7521G, T8468C, T8655C, G8701A, G9055A, C9540T, T9698C, A10550G, T10664C, A10688G, C10810T, C10873T, C10915T, T11025C, T11299C, A11467G, A11914G, A12308G, G12372A, T12705C, G13105A, G13276A, T13506C, T13650C, C14167T, T14798C, A16129G, T16187C, C16189T, T16223T, T16224C, G16230A, T16278C |
| I0174 | BAM25a | N1a1a1 | C16193T | C146T, C152T!!, C195T, T199C, T204C, A247G, 573.XC, T669C, A769G, A825t, A1018G, G1719A, G2702A, A2758G, C2885T, |

| Lab ID | Individual ID | mtDNA-hg call | Private mutations | SNPs against RSRS |
|---|---|---|---|---|
| | | | | T3336C, T3594C, G4104A, T4312C, A5315G, G7146A, T7256C, A7521G, T8468C, T8655C, G8701A, A8901G, C9540T, T10238C, T10664C, A10688G, C10810T, C10873T, C10915T, A11884G, A11914G, G12501A, G13105A, G13276A, T13506C, T13650C, A13780G, G15043A, A16129G, C16147A, T16172C, C16189T, C16193T, G16230A, C16248T, T16278C, C16311T, C16355T |
| I0166 | HAL13 | V1a | C7331T, T16311C!, C16519T | T72C, G73A, C146T, C152T, C195T, A247G, A769G, A825t, A1018G, A2758G, C2885T, T3594C, G4104A, T4312C, G4580A, T4639C, G7146A, T7256C, C7331T, A7521G, T8468C, T8655C, G8701A, A8869G, C9540T, G10398A, T10664C, A10688G, C10810T, C10873T, C10915T, A11719G, A11914G, T12705C, G13105A, G13276A, T13506C, T13650C, T14766C, C15904T, A16129G, T16187C, C16189T, T16223C, G16230A, T16278C, T16298C, C16519T |
| I0162 | OSH1 | H16a'c'd | none | G73A, C146T, C195T, A247G, A769G, A825t, A1018G, G2706A, A2758G, C2885T, T3594C, G4104A, T4312C, T7028C, G7146A, T7256C, A7521G, T8468C, T8655C, G8701A, C9540T, C10394T, G10398A, T10664C, A10688G, C10810T, C10873T, C10915T, A11719G, A11914G, T12705C, G13105A, G13276A, T13506C, T13650C, T14766C, A16129G, T16187C, C16189T, T16223C, G16230A, T16278C, C16311T |
| I0163 | OSH7 | H5b | none | G73A, C146T, C152T, C195T, A247G, C456T, A769G, A825t, A1018G, G2706A, A2758G, C2885T, T3594C, G4104A, T4312C, G5471A, T7028C, G7146A, T7256C, A7521G, T8468C, T8655C, G8701A, C9540T, G10398A, T10664C, A10688G, C10810T, C10873T, C10915T, A11719G, A11914G, T12705C, G13105A, G13276A, T13506C, T13650C, T14766C, A16129G, T16187C, C16189T, T16223C, G16230A, T16278C, T16304C, C16311T, C16519T |
| I0165 | OSH9 | U5b1b | T152C!, C6644T, A14053G, T16192C! | C146T, C150T, C195T, A247G, A769G, A825t, A1018G, A2758G, C2885T, T3197C, T3594C, G4104A, T4312C, A5656G, C6644T, G7146A, T7256C, A7521G, A7768G, T8468C, T8655C, G8701A, G9477A, C9540T, G10398A, T10664C, A10688G, C10810T, C10873T, C10915T, A11467G, A11914G, A12308G, G12372A, G12618A, T12705C, G13105A, G13276A, T13506C, T13617C, T13650C, A14053G, T14182C, A16129G, T16187C, T16223C, G16230A, C16270T, T16278C, C16311T |
| I0170 | OAW1 | HV6'17 | C16519T | G73A, C146T, C152T, C195T, A247G, A769G, A1018G, A825t, A2758G, C2885T, T3594C, G4104A, T4312C, G7146A, T7256C, A7521G, T8468C, T8655C, G8701A, C9540T, G10398A, T10664C, A10688G, C10810T, C10873T, C10915T, A11719G, A11914G, T12705C, G13105A, G13276A, T13506C, T13650C, T14766C, A16129G, T16187C, C16189T, T16223C, G16230A, T16278C, C16519T |
| I0807 | ESP30 | H1e1a | C5960T, G8865A | G73A, C146T, C152T, C195T, A247G, A769G, A825t, A1018G, G2706A, A2758G, C2885T, G3010A, T3594C, G4104A, T4312C, G5460A, C5960T, T7028C, G7146A, T7256C, A7521G, T8468C, A8512G, T8655C, G8701A, G8865A, C9540T, G10398A, T10664C, A10688G, C10810T, C10873T, C10915T, A11719G, A11914G, T12705C, G13105A, G13276A, T13506C, T13650C, T14766C, C14902T, A16129G, T16187C, C16189T, T16223C, G16230A, T16278C, C16311T |
| I0822 | HQU3 | K1e | G1018A!, C10757T | C146T, C195T, A247G, A769G, A825t, T1189C, A1811G, T1819C, A2758G, C2885T, A3480G, T3594C, G4104A, T4312C, G7146A, T7256C, A7521G, G8251A, T8468C, T8655C, G8701A, G9055A, C9540T, T9698C, C10478T, A10550G, T10664C, A10688G, C10757T, C10810T, C10873T, C10915T, T11299C, A11467G, A11914G, A12308G, G12372A, T12705C, G13105A, G13276A, T13506C, T13650C, C14142g, C14167T, T14798C, A16129G, T16187C, C16189T, T16223T, T16224C, G16230A, T16278C, T16362C |
| I0808 | HQU4 | H7d | np 304-311, 3569-3589, 16129-16156, 16301-16309, 16380-16389 no coverage | G73A, C146T, C152T, C195T, A247G, A769G, A825t, A1018G, G2706A, A2758G, C2885T, T3594C, G4104A, T4312C, A4793G, T7028C, G7146A, T7256C, A7521G, T8468C, T8655C, G8701A, C9540T, G10398A, T10664C, A10688G, C10810T, C10873T, C10915T, A11719G, A11914G, T12705C, G13105A, G13276A, T13506C, T13650C, T14766C, C15409T, A16129G, T16187C, C16189T, T16223C, G16230A, T16278C, C16311T |
| I0212 | HQU5 | T2c1d1 | T152C!, C4017T, C6340T, G8860A, G15110A | C195T, A247G, T279C, G709A, A769G, A825t, A1018G, G1888A, A2758G, C2885T, T3594C, C4017T, G4104A, T4216C, T4312C, A4917G, C5187T, G6261A, C6340T, G7146A, T7256C, A7521G, C7873T, T8468C, T8655C, G8697A, G8701A, G8860A, C9540T, G10398A, T10463C, T10664C, A10688G, C10810T, C10822T, C10873T, C10915T, A11251G, A11812G, T12705C, G13105A, G13276A, G13368A, T13506C, T13650C, A14233G, G14905A, G15110A, C15452a, A15607G, G15928A, T16126C, A16129G, T16187C, C16189T, T16223C, G16230A, T16278C, C16292T, C16294T, C16311T |
| I0559 | QLB15D | HV6'17 | C16519T | G73A, C146T, C152T, C195T, A247G, A769G, A825t, A1018G, A2758G, C2885T, T3594C, G4104A, T4312C, G7146A, T7256C, A7521G, T8468C, T8655C, G8701A, C9540T, G10398A, T10664C, A10688G, C10810T, C10873T, C10915T, A11719G, A11914G, |

| Lab ID | Individual ID | mtDNA-hg call | Private mutations | SNPs against RSRS |
|---|---|---|---|---|
| | | | | T12705C, G13105A, G13276A, T13506C, T13650C, T14766C, A16129G, T16187C, C16189T, T16223C, G16230A, T16278C, C16519T |
| I0560 | QLB18A | T2e1 | none | C41T, C146T, C150T, C152T, C195T, A247G, G709A, A769G, A825t, A1018G, G1888A, A2758G, C2885T, T3594C, G4104A, T4216C, T4312C, A4917G, G7146A, T7256C, A7521G, T8468C, T8655C, G8697A, G8701A, C9540T, G10398A, T10463C, T10664C, A10688G, C10810T, C10873T, C10915T, A11251G, A11812G, A11914G, T12705C, G13105A, G13276A, G13368A, T13506C, T13650C, A14233G, G14905A, C15452a, A15607G, G15928A, T16126C, A16129G, G16153A, T16187C, C16189T, T16223C, G16230A, T16278C, C16294T, C16311T |
| I0556 | QLB2A | U8a1 | T7660C, T16342C, T16342C | C146T, C152T, C195T, A247G, T282C, A769G, A825t, A1018G, A1811G, A2758G, C2885T, T3594C, C3738T, G4104A, T4312C, T6392C, C6455T, A7055G, G7146A, T7256C, A7521G, T7660C, T8468C, T8655C, G8701A, C9365T, C9540T, T9698C, G10398A, T10664C, A10688G, C10810T, C10873T, C10915T, A11467G, A11914G, A12308G, G12372A, T12705C, G13105A, G13145A, G13276A, T13506C, T13650C, A16129G, T16187C, C16189T, T16223T, G16230A, T16278C, C16311T, T16342C, C16519T |
| I0557 | QLB6B | U5b2a2 | T16192C!, C16519T | C146T, C150T, C152T, C195T, A247G, A769G, A825t, A1018G, C1721T, A2758G, C2885T, T3197C, C3212T, T3594C, G4104A, T4312C, A4732G, G7146A, T7256C, A7521G, A7768G, T8468C, T8655C, G8701A, G9477A, C9540T, G10398A, T10664C, A10688G, C10810T, C10873T, C10915T, A11467G, A11914G, A12308G, G12372A, T12705C, G13105A, G13276A, T13506C, T13617C, A13637G, T13650C, T14182C, A16129G, T16187C, T16223C, G16230A, C16270T, T16278C, C16311T, G16398A, C16519T, |
| I0548 | BENZ14 | U5a2b4 | C16519T | C146T, C152T, C195T, A247G, A769G, A825t, A1018G, A2758G, C2885T, T3197C, T3594C, G4104A, T4312C, G7146A, T7256C, A7521G, T8468C, T8655C, G8701A, G9477A, C9540T, G9548A, G10398A, T10664C, A10688G, C10810T, C10873T, C10915T, A11467G, A11914G, A12308G, G12372A, T12705C, G13105A, G13276A, T13506C, T13617C, T13650C, A14793G, G15301A!!, A16129G, T16187C, C16189T, C16192T, T16223C, G16230A, C16256T, C16270T, T16278C, C16311T, C16519T, G16526A |
| I0549 | BENZ18/BENZ15? | W1c'i | none | T119C, C146T, C152T, A189G, T204C, G207A, A247G, G709A, A769G, A825t, A1018G, T1243C, A2758G, C2885T, A3505G, T3594C, G4104A, T4312C, G5046A, G5460A, G7146A, T7256C, A7521G, C7864T, G8251A, T8468C, T8655C, G8701A, G8994A, C9540T, G10398A, T10664C, A10688G, C10810T, C10873T, C10915T, C11674T, A11914G, A11947G, T12414C, G13105A, G13276A, T13506C, T13650C, G15884c, A16129G, T16187C, C16189T, G16230A, T16278C, C16292T, C16311T |
| I0172 | ESP24 | T2b | T12811C | C146T, C152T, C195T, A247G, G709A, A769G, A825t, G930A, A1018G, G1888A, A2758G, C2885T, T3594C, G4104A, T4216C, T4312C, A4917G, G5147A, G7146A, T7256C, A7521G, T8468C, T8655C, G8697A, G8701A, C9540T, G10398A, T10463C, T10664C, A10688G, C10810T, C10873T, C10915T, A11251G, A11812G, A11914G, T12705C, T12811C, G13105A, G13276A, G13368A, T13506C, T13650C, A14233G, G14905A, C15452a, A15607G, G15928A, T16126C, A16129G, T16187C, C16189T, T16223C, G16230A, T16278C, C16294T, C16296T, T16304C, C16311T |
| I0175 | BAL16 | N1a1a1 | C152T!!, T8227C, G10143A, A11884G, T16355C! | C146T, C152T!!, C195T, T199C, T204C, A247G, 573.XC, T669C, A769G, A825t, A1018G, G1719A, G2702A, A2758G, C2885T, T3336C, T3594C, G4104A, T4312C, A5315G, G7146A, T7256C, A7521G, T8227C, T8468C, T8655C, G8701A, A8901G, C9540T, G10143A, T10238C, T10664C, A10688G, C10810T, C10873T, C10915T, A11884G, A11914G, G12501A, G13105A, G13276A, T13506C, T13650C, A13780G, G15043A, A16129G, C16147A, T16172C, T16187C, C16189T, G16230A, C16248T, T16278C, C16311T |
| I0551 | SALZ3B | U3a1 | none | C146T, C150T, C152T, C195T, A247G, A769G, A825t, A1018G, A1811G, A2294G, A2758G, C2885T, G3010A, T3594C, G4104A, T4312C, T4703C, C6518T, G7146A, T7256C, A7521G, T8468C, T8655C, G8701A, G9266A, C9540T, G10398A, A10506G, T10664C, A10688G, C10810T, C10873T, C10915T, A11467G, A11914G, A12308G, G12372A, T12705C, G13105A, G13276A, T13506C, T13650C, C13934T, A14139G, T15454C, A16129G, T16187C, C16189T, T16223C, G16230A, T16278C, C16311T, A16343G, G16390A |
| I0800 | SALZ57A | H3 | T152C!, A5515G | G73A, C146T, C195T, A247G, A769G, A825t, A1018G, G2706A, A2758G, C2885T, T3594C, G4104A, T4312C, A5515G, T6776C, T7028C, G7146A, T7256C, A7521G, T8468C, T8655C, G8701A, C9540T, G10398A, T10664C, A10688G, C10810T, C10873T, C10915T, A11719G, A11914G, T12705C, G13105A, G13276A, T13506C, T13650C, T14766C, A16129G, T16187C, C16189T, T16223C, G16230A, T16278C, C16311T |
| I0802 | SALZ77A | H3 | none | G73A, C146T, C152T, C195T, A247G, A769G, A825t, A1018G, G2706A, A2758G, C2885T, T3594C, G4104A, T4312C, T6776C, T7028C, G7146A, T7256C, A7521G, T8468C, T8655C, G8701A, C9540T, G10398A, T10664C, A10688G, C10810T, C10873T, C10915T, A11719G, A11914G, T12705C, G13105A, G13276A, T13506C, T13650C, T14766C, A16129G, T16187C, C16189T, |

| Lab ID | Individual ID | mtDNA-hg call | Private mutations | SNPs against RSRS |
|---|---|---|---|---|
| | | | | T16223C, G16230A, T16278C, C16311T |
| I0552 | SALZ7A | H5 | C5993T, A14566G, C16519T | G73A, C146T, C152T, C195T, A247G, C456T, A769G, A825t, A1018G, G2706A, A2758G, C2885T, T3594C, G4104A, T4312C, C5993T, T7028C, G7146A, T7256C, A7521G, T8468C, T8655C, G8701A, C9540T, G10398A, T10664C, A10688G, C10810T, C10873T, C10915T, A11719G, A11914G, T12705C, G13105A, G13276A, T13506C, T13650C, A14566G, T14766C, A16129G, T16187C, C16189T, T16223C, G16230A, T16278C, T16304C, C16311T, C16519T |
| I0554 | SALZ88A | J1c | A235G, T4083C, A7049G, A15799G, C16519T | C146T, C152T, C195T, G228A, A235G, A247G, C295T, C462T, T489C, A769G, A825t, A1018G, A2758G, C2885T, G3010A, T3594C, T4083C, G4104A, T4216C, T4312C, A7049G, G7146A, T7256C, A7521G, T8468C, T8655C, G8701A, C9540T, T10664C, A10688G, C10810T, C10873T, C10915T, A11251G, A11914G, A12612G, T12705C, G13105A, G13276A, T13506C, T13650C, G13708A, T14798C, C15452a, A15799G, C16069T, T16126C, A16129G, T16187C, C16189T, T16223C, G16230A, T16278C, C16311T, C16519T, |
| I0798 | SALZ18A | H10e'f'g | C13503T | G73A, C146T, C152T, C195T, A247G, A769G, A825t, A1018G, G2706A, A2758G, C2885T, T3594C, G4104A, T4312C, T7028C, G7146A, T7256C, A7521G, T8468C, T8655C, G8701A, C9540T, G10398A, T10664C, A10688G, C10810T, C10873T, C10915T, A11719G, A11914G, T12705C, G13105A, G13276A, C13503T, T13506C, T13650C, T14470a, T14766C, T16093C, A16129G, T16187C, C16189T, T16223C, G16230A, T16278C, C16311T |
| I0799 | SALZ21B | H1e | T1766C | G73A, C146T, C152T, C195T, A247G, A769G, A825t, A1018G, T1766C, G2706A, A2758G, C2885T, G3010A, T3594C, G4104A, T4312C, G5460A, T7028C, G7146A, T7256C, A7521G, T8468C, T8655C, G8701A, C9540T, G10398A, T10664C, A10688G, C10810T, C10873T, C10915T, A11719G, A11914G, T12705C, G13105A, G13276A, T13506C, T13650C, T14766C, A16129G, T16187C, C16189T, T16223C, G16230A, T16278C, C16311T |
| I0408 | Mina18a | pre-U5b1i | T13656C, T16209C, A16399G (topologically missing C3498T, T6674G, G15777A, T16311C, T16356C | C146T, C150T, C152T, C195T, A247G, A769G, A825t, A1018G, A2758G, C2885T, A3105G, T3197C, T3594C, G4104A, T4312C, A5656G, G7146A, T7256C, A7521G, A7768G, T8468C, T8655C, G8701A, G9477A, C9540T, G10398A, T10664C, A10688G, C10810T, C10873T, C10915T, A11467G, A11914G, A12308G, G12372A, T12705C, G13105A, G13276A, T13506C, T13617C, T13650C, T13656C, T14182C, A16129G, C16167T, T16187C, C16189T, C16192T, T16209C, T16223C, G16230A, C16270T, T16278C, C16311T, A16399G |
| I0404 | Mina2 | J2a1a1 | T195C!, A10499G, G10586A, G11377A, C16519T; heteroplasmy at np 16274 (54% A, 46% G) | C146T, C150T, A215G, A247G, C295T, T319C, T489C, G513A, A769G, A825t, A1018G, A2758G, C2885T, T3594C, G4104A, T4216C, T4312C, G7146A, T7256C, C7476T, A7521G, G7789A, T8468C, T8655C, G8701A, C9540T, A10499G, G10586A, T10664C, A10688G, C10810T, C10873T, C10915T, A11251G, G11377A, A11914G, A12612G, T12705C, G13105A, G13276A, T13506C, T13650C, G13708A, A13722G, A14133G, G15257A, C15452a, C16069T, T16126C, A16129G, G16145A, T16187C, C16189T, T16223C, G16230A, T16231C, C16261T, T16278C, C16311T, C16519T |
| I0405 | Mina3 | K1a1b1 | C16301T; | C114T, C146T, C152T, C195T, A247G, C497T, A769G, A825t, A1018G, T1189C, A1811G, A2758G, C2885T, A3480G, T3594C, G4104A, T4312C, G7146A, T7256C, A7521G, T8468C, T8655C, G8701A, G9055A, C9540T, T9698C, A10550G, T10664C, A10688G, C10810T, C10873T, C10915T, T11299C, A11467G, A11470G, A12308G, G12372A, T12705C, G13105A, G13276A, T13506C, T13650C, C14167T, T14798C, A15924G, T16093C, A16129G, T16187C, C16189T, T16223T, T16224C, G16230A, T16278C, C16301T |
| I0406 | Mina4 | H1 | C722T | G73A, C146T, C152T, C195T, A247G, C722T, A769G, A825t, A1018G, G2706A, A2758G, C2885T, G3010A, T3594C, G4104A, T4312C, T7028C, G7146A, T7256C, A7521G, T8468C, T8655C, G8701A, C9540T, G10398A, T10664C, A10688G, C10810T, C10873T, C10915T, A11719G, A11914G, T12705C, G13105A, G13276A, T13506C, T13650C, T14766C, A16129G, T16187C, C16189T, T16223C, G16230A, T16278C, C16311T |
| I0407 | Mina6b | K1b1a1 | none | C146T, C195T, A247G, A769G, A825t, A1018G, T1189C, A1811G, A2758G, C2885T, A3480G, T3594C, G4104A, T4312C, G5913A, G7146A, T7256C, A7521G, T8468C, T8655C, G8701A, G9055A, C9540T, T9698C, G9962A, A10289G, A10550G, T10664C, A10688G, C10810T, C10873T, C10915T, T11299C, A11467G, A11914G, A11923G, A12308G, G12372A, T12705C, G13105A, G13276A, T13506C, T13650C, C13967T, C14167T, T14798C, G15257A, C15946T, T16093C, A16129G, T16187C, C16189T, T16223T, T16224C, |

| Lab ID | Individual ID | mtDNA-hg call | Private mutations | SNPs against RSRS |
|---|---|---|---|---|
| | | | | G16230A, T16278C, G16319A, A16463G |
| I0370 | SVP10 | H13a1a1 | C14809T, C16261T, C16519T | G73A, C146T, C152T, C195T, A247G, A769G, A825t, A1018G, C2259T, G2706A, A2758G, C2885T, T3594C, G4104A, T4312C, A4745G, T7028C, G7146A, T7256C, G7337A, A7521G, T8468C, T8655C, G8701A, C9540T, G10398A, T10664C, A10688G, C10810T, C10873T, C10915T, A11719G, A11914G, T12705C, G13105A, G13276A, T13326C, T13506C, T13650C, C13680T, T14766C, C14809T, C14872T, A16129G, T16187C, C16189T, T16223C, G16230A, C16261T, T16278C, C16311T, C16519T |
| I0355 | SVP2 | K1b2a | none | C152T, A247G, A769G, A825t, A1018G, T1189C, A1811G, A2758G, C2885T, A3480G, T3594C, G4104A, T4312C, G5913A, G7146A, T7256C, A7521G, T8468C, T8655C, G8701A, G9055A, C9540T, T9698C, A10550G, T10664C, A10688G, C10810T, C10873T, C10915T, T11299C, A11467G, A11914G, A12308G, G12372A, T12705C, T12738g, G12771A, G13105A, G13276A, T13506C, T13650C, C14167T, T14798C, A16129G, T16187C, C16189T, T16223T, T16224C, G16230A, T16278C |
| I0231 | SVP3 | U4a1 | T7153C | C146T, A247G, G499A, A769G, A825t, A1018G, A1811G, A2758G, C2885T, T3594C, G4104A, T4312C, T4646C, T5999C, A6047G, G7146A, T7153C, T7256C, A7521G, T8468C, T8655C, G8701A, C8818T, C9540T, G10398A, T10664C, A10688G, C10810T, C10873T, C10915T, C11332T, A11467G, A11914G, A12308G, G12372A, T12705C, A12937G, G13105A, G13276A, T13506C, T13650C, C14620T, T15693C, A16129G, C16134T, T16187C, C16189T, T16223C, G16230A, T16278C, C16311T, T16356C |
| I0429 | SVP38 | T2c1a2 | none | C146T, C152T, C195T, A247G, 573.XC, G709A, A769G, A825t, A1018G, G1888A, A2758G, C2885T, T3594C, G4104A, T4216C, T4312C, A4917G, C5817T, G6261A, G7146A, T7256C, A7521G, C8455T, T8468C, T8655C, G8697A, G8701A, C9540T, G10398A, T10463C, T10664C, A10688G, C10810T, C10822T, C10873T, C10915T, A11251G, A11812G, A11914G, T12705C, G13105A, G13276A, G13368A, T13506C, T13650C, A13973t, A14233G, G14905A, C15452a, A15607G, G15928A, T16126C, A16129G, T16187C, C16189T, T16223C, G16230A, T16278C, C16292T, C16294T, C16311T |
| I0357 | SVP5 | W6c | G7852A | C146T, C152T, A189G, C194T, T204C, G207A, A247G, G709A, A769G, A825t, A1018G, T1243C, A2758G, C2885T, A3505G, T3594C, A4093G, G4104A, T4312C, G5046A, G5460A, G7146A, T7256C, A7521G, G7852A, C8002T, G8251A, T8468C, T8614C, T8655C, C8658T, G8701A, G8994A, C9540T, G10398A, T10664C, A10688G, C10810T, C10873T, C10915T, C11674T, A11914G, A11947G, T12414C, G13105A, G13276A, T13506C, T13650C, G15884c, A16129G, T16187C, C16189T, C16192T, G16230A, T16278C, C16292T, C16311T, T16325C |
| I0438 | SVP50 | U5a1a1 | C16519T | C146T, C152T, C195T, A247G, A769G, A825t, A1018G, T1700C, A2758G, C2885T, T3197C, T3594C, G4104A, T4312C, T5495C, G7146A, T7256C, A7521G, T8468C, T8655C, G8701A, G9477A, C9540T, G10398A, T10664C, A10688G, C10810T, C10873T, C10915T, A11467G, A11914G, A12308G, G12372A, T12705C, G13105A, G13276A, T13506C, T13617C, T13650C, A14793G, A15218G, A15924G, A16129G, T16187C, C16189T, T16223C, G16230A, C16256T, C16270T, T16278C, C16311T, A16399G, C16519T |
| I0439 | SVP52 | U5a1a1 | C16519T | C146T, C152T, C195T, A247G, A769G, A825t, A1018G, T1700C, A2758G, C2885T, T3197C, T3594C, G4104A, T4312C, T5495C, G7146A, T7256C, A7521G, T8468C, T8655C, G8701A, G9477A, C9540T, G10398A, T10664C, A10688G, C10810T, C10873T, C10915T, A11467G, A11914G, A12308G, G12372A, T12705C, G13105A, G13276A, T13506C, T13617C, T13650C, A14793G, A15218G, A15924G, A16129G, T16187C, C16189T, T16223C, G16230A, C16256T, C16270T, T16278C, C16311T, A16399G, C16519T |
| I0441 | SVP54 | H2b | C16519T | G73A, C146T, C195T, A247G, A769G, A825t, A1018G, G1438A, G2706A, A2758G, C2885T, T3594C, G4104A, T4312C, T7028C, G7146A, T7256C, A7521G, T8468C, T8598C, T8655C, G8701A, C9540T, G10398A, T10664C, A10688G, C10810T, C10873T, C10915T, A11719G, A11914G, T12705C, G13105A, G13276A, T13506C, T13650C, T14766C, A16129G, T16187C, C16189T, T16223C, G16230A, T16278C, C16519T |
| I0443 | SVP57 | W3a1a | A15951G | C146T, C152T, A189G, C194T, T204C, G207A, A247G, G709A, A769G, A825t, A1018G, T1243C, T1406C, A2758G, C2885T, A3505G, T3594C, G4104A, T4312C, G5046A, G5460A, G7146A, C7151T, T7256C, A7521G, G8251A, T8468C, T8655C, G8701A, G8994A, C9540T, G10398A, T10664C, A10688G, C10810T, C10873T, C10915T, C11674T, A11914G, A11947G, T12414C, G13105A, A13263G, G13276A, T13506C, T13650C, T15784C, G15884c, A15951G, A16129G, T16187C, C16189T, G16230A, T16278C, C16292T, C16311T |
| I0444 | SVP58 | H6a1b | C16519T | G73A, C146T, C152T, C195T, T239C, A247G, A769G, A825t, A1018G, G2706A, A2758G, C2885T, T3594C, G3915A, G4104A, T4312C, A4727G, T7028C, G7146A, T7256C, A7521G, T8468C, T8655C, G8701A, G9380A, C9540T, G10398A, G10589A, T10664C, |

| Lab ID | Individual ID | mtDNA-hg call | Private mutations | SNPs against RSRS |
|---|---|---|---|---|
| | | | | A10688G, C10810T, C10873T, C10915T, A11719G, A11914G, T12705C, G13105A, G13276A, T13506C, T13650C, T14766C, A16129G, T16187C, C16189T, T16223C, G16230A, T16278C, C16311T, T16362C, A16482G, C16519T |
| I0051 | ALB2 | H3b | np 2118-2125, 4771, 10026-10069 no coverage | G73A, C146T, C152T, C195T, A247G, A769G, A825t, A1018G, A2581G, G2706A, A2758G, C2885T, T3594C, G4104A, T4312C, T6776C, T7028C, G7146A, T7256C, A7521G, T8468C, T8655C, G8701A, C9540T, G10398A, T10664C, A10688G, C10810T, C10873T, C10915T, A11719G, A11914G, T12705C, G13105A, G13276A, T13506C, T13650C, T14766C, A16129G, T16187C, C16189T, T16223C, G16230A, T16278C, C16311T |
| I0118 | ALB3 | HV6'17 | C12112T, C16519T, G16526A | G73A, C146T, C152T, C195T, A247G, A769G, A825t, A1018G, A2758G, C2885T, T3594C, G4104A, T4312C, G7146A, T7256C, A7521G, T8468C, T8655C, G8701A, C9540T, G10398A, T10664C, A10688G, C10810T, C10873T, C10915T, A11719G, A11914G, C12112T, T12705C, G13105A, G13276A, T13506C, T13650C, T14766C, A16129G, T16187C, C16189T, T16223C, G16230A, T16278C, C16519T, G16526A |
| I0805 | QLB26 | H1 | none | G73A, C146T, C152T, C195T, A247G, A769G, A825t, A1018G, G2706A, A2758G, C2885T, G3010A, T3594C, G4104A, T4312C, T7028C, G7146A, T7256C, A7521G, T8468C, T8655C, G8701A, C9540T, G10398A, T10664C, A10688G, C10810T, C10873T, C10915T, A11719G, A11914G, T12705C, G13105A, G13276A, T13506C, T13650C, T14766C, A16129G, T16187C, C16189T, T16223C, G16230A, T16278C, C16311T |
| I0806 | QLB28 | H1 | none | G73A, C146T, C152T, C195T, A247G, A769G, A825t, A1018G, G2706A, A2758G, C2885T, G3010A, T3594C, G4104A, T4312C, T7028C, G7146A, T7256C, A7521G, T8468C, T8655C, G8701A, C9540T, G10398A, T10664C, A10688G, C10810T, C10873T, C10915T, A11719G, A11914G, T12705C, G13105A, G13276A, T13506C, T13650C, T14766C, A16129G, T16187C, C16189T, T16223C, G16230A, T16278C, C16311T |
| I0113 | QUEXII4 | J1c5 | T14470C, C16519T | C146T, C152T, G185A, C195T, G228A, A247G, C295T, C462T, T489C, A769G, A825t, A1018G, A2758G, C2885T, G3010A, T3594C, G4104A, T4216C, T4312C, A5198G, G7146A, T7256C, A7521G, T8468C, T8655C, G8701A, C9540T, T10664C, A10688G, C10810T, C10873T, C10915T, A11251G, A11914G, A12612G, T12705C, G13105A, G13276A, T13506C, T13650C, G13708A, T14470C, T14798C, C15452a, C16069T, T16126C, A16129G, T16187C, C16189T, T16223C, G16230A, T16278C, , C16311T, C16519T, |
| I0112 | QUEXII6 | H13a1a2 | G9025A, C16519T | G73A, C146T, C152T, C195T, A247G, A769G, A825t, A1018G, C2259T, G2706A, A2758G, C2885T, T3594C, G4104A, T4312C, A4745G, T7028C, G7146A, T7256C, A7521G, T8468C, T8655C, G8701A, G9025A, C9540T, G10398A, T10664C, A10688G, C10810T, C10873T, C10915T, A11719G, A11914G, T12705C, G13105A, G13276A, T13506C, A13542G, T13650C, C13680T, T14766C, C14872T, A16129G, T16187C, C16189T, T16223C, G16230A, T16278C, C16311T, C16519T |
| I0060 | ROT3 | K1a2c | T146C!, T152C!, A214G, G14305A, T14311C, T16086C | C195T, A214G, A247G, C497T, A769G, A825t, A1018G, A1811G, T1189C, A2758G, C2885T, A3480G, T3594C, G4104A, T4216C, T4312C, G7146A, T7256C, A7521G, G7775A, T8468C, T8655C, G8701A, G9055A, C9540T, T9698C, A10550G, T10664C, A10688G, C10810T, C10873T, C10915T, T11025C, T11299C, A11467G, A11914G, A12308G, G12372A, T12705C, G13105A, G13276A, T13506C, T13650C, C14167T, G14305A, T14311C, T14798C, T16086C, A16129G, T16187C, C16189T, T16223C, T16224C, G16230A, T16278C |
| I0111 | ROT4 | H3new | C4577T, C16256T | G73A, C146T, C152T, C195T, A247G, A769G, A825t, A1018G, G2706A, A2758G, C2885T, T3594C, G4104A, T4312C, C4577T, T6776C, T7028C, G7146A, T7256C, A7521G, T8468C, T8655C, G8701A, C9540T, G10398A, T10664C, A10688G, C10810T, C10873T, C10915T, A11719G, A11914G, T12705C, G13105A, G13276A, T13506C, T13650C, T14766C, A16129G, T16187C, C16189T, T16223C, G16230A, C16256T, T16278C, C16311T |
| I0108 | ROT6 | H5a3 | C16519T | G73A, C146T, C152T, C195T, A247G, C456T, G513A, A769G, A825t, A1018G, G2706A, A2758G, C2885T, T3594C, G4104A, T4312C, T4336C, T7028C, G7146A, T7256C, A7521G, T8468C, T8655C, G8701A, C9540T, G10398A, T10664C, A10688G, C10810T, C10873T, C10915T, A11719G, A11914G, T12705C, G13105A, G13276A, T13506C, T13650C, T14766C, G15884A, A16129G, T16187C, C16189T, T16223C, G16230A, T16278C, T16304C, C16311T, C16519T |
| I0171 | BZH12 | U5a1a2a | C16519T | C146T, C152T, C195T, A247G, 573.XC, A769G, A825t, A1018G, T1700C, A2758G, C2885T, T3197C, T3594C, G4104A, T4312C, A5319G, A6629G, T6719C, G7146A, T7256C, A7521G, T8468C, T8655C, G8701A, G9477A, C9540T, G10398A, T10664C, A10688G, C10810T, C10873T, C10915T, A11467G, A11914G, A12308G, C12346T, G12372A, T12705C, G13105A, G13276A, T13506C, |

| Lab ID | Individual ID | mtDNA-hg call | Private mutations | SNPs against RSRS |
|---|---|---|---|---|
| | | | | T13617C, T13650C, A14793G, A15218G, A16129G, T16187C, C16189T, T16223C, G16230A, C16256T, C16270T, T16278C, C16311T, A16399G, C16519T |
| I0058 | BZH4 | H1e | A15220G, A15401G, A16293G | G73A, C146T, C152T, C195T, A247G, A769G, A825t, A1018G, G2706A, A2758G, C2885T, G3010A, T3594C, G4104A, T4312C, G5460A, T7028C, G7146A, T7256C, A7521G, T8468C, T8655C, G8701A, C9540T, G10398A, T10664C, A10688G, C10810T, C10873T, C10915T, A11719G, A11914G, T12705C, G13105A, G13276A, T13506C, T13650C, T14766C, A15220G, A15401G, A16129G, T16187C, C16189T, T16223C, G16230A, T16278C, A16293G, C16311T |
| I0059 | BZH6 | H1/H1b'ad | A8149G, A9377G, T9467C, A13671G, T14319C, T16189C! | G73A, C146T, C152T, C195T, A247G, A769G, A825t, A1018G, G2706A, A2758G, C2885T, G3010A, T3594C, G4104A, T4312C, T7028C, G7146A, T7256C, A7521G, A8149G, T8468C, T8655C, G8701A, A9377G, T9467C, C9540T, G10398A, T10664C, A10688G, C10810T, C10873T, C10915T, A11719G, A11914G, T12705C, G13105A, G13276A, T13506C, T13650C, A13671G, T14319C, T14766C, A16129G, T16187C, T16223C, G16230A, T16278C, C16311T |
| I0104 | ESP11 | U4b1a1a1 | none | C146T, C152T, A247G, A499G, A769G, A825t, A1018G, A1811G, T2083C, A2758G, C2885T, T3594C, A3672G, G4104A, T4312C, T4646C, T5999C, A6047G, G7146A, T7256C, A7521G, T7705C, T8468C, A8642G, T8655C, G8701A, C9540T, G10398A, T10664C, A10688G, C10810T, C10873T, C10915T, C11332T, T11339C, A11467G, A11914G, T12297C, A12308G, G12372A, T12705C, G13105A, G13276A, T13506C, T13650C, C14620T, T15693C, C15789T, A16129G, T16187C, C16189T, T16223C, G16230A, T16278C, C16311T, T16356C, T16362C |
| I0103 | ESP16 | W6a | none | C146T, C152T, A189G, C194T, T204C, G207A, A247G, G709A, A769G, A825t, A1018G, T1243C, A2758G, C2885T, A3505G, T3594C, A4093G, G4104A, T4312C, G5046A, G5460A, G7146A, T7256C, A7521G, G8251A, T8468C, T8610C, T8614C, T8655C, G8701A, G8994A, C9540T, G10398A, T10664C, A10688G, C10810T, C10873T, C10915T, C11674T, A11914G, A11947G, T12414C, G13105A, G13276A, T13506C, T13650C, G15884c, A16129G, T16187C, C16189T, C16192T, G16230A, T16278C, C16292T, C16311T, T16325C |
| I0049 | ESP22 | X2b4 | none | C146T, C152T, A153G, G225A, T226C, A247G, A769G, A825t, A1018G, G1719A, A2758G, C2885T, T3594C, G3705A, G4104A, T4312C, T6221C, C6371T, G7146A, T7256C, A7521G, C8393T, T8468C, T8655C, G8701A, C9540T, G10398A, T10664C, A10688G, C10810T, C10873T, C10915T, A11914G, G13105A, G13276A, T13506C, T13650C, G13708A, A13966G, T14470C, G15927A, A16129G, T16187C, G16230A, C16311T |
| I0106 | ESP26 | T2a1b1 | none | C146T, C152T, C195T, A247G, G709A, A769G, A825t, A1018G, G1888A, T2141C, A2758G, C2885T, T3594C, G4104A, T4216C, T4312C, A4917G, G7146A, T7256C, A7521G, T8468C, T8655C, G8697A, G8701A, C9540T, G10398A, T10463C, T10664C, A10688G, C10810T, C10873T, C10915T, A11251G, A11812G, A11914G, T12705C, G13105A, G13276A, G13368A, T13506C, T13650C, T13965C, A13966G, A14233G, A14687G, G14905A, C15452a, A15607G, G15928A, T16126C, A16129G, T16187C, C16189T, T16223C, G16230A, T16278C, C16294T, C16311T, T16324C |
| I0050 | ESP5 | U5a2d | T5999C, A9389G, T13474C, G14544A, A14587G, C14854T, C16519T | C146T, C152T, C195T, A247G, A769G, A825t, A1018G, A2758G, C2885T, T3197C, T3594C, G4104A, T4312C, T5999C, G7146A, T7256C, A7521G, A7843G, T8468C, T8655C, G8701A, A9389G, G9477A, C9540T, G10398A, T10664C, A10688G, C10810T, C10873T, C10915T, A11467G, A11914G, A12308G, G12372A, T12705C, G13105A, G13276A, T13474C, T13506C, T13617C, T13650C, G14544A, A14587G, A14793G, C14854T, A16129G, T16187C, C16189T, C16192T, T16223C, G16230A, C16256T, C16270T, T16278C, C16311T, C16519T |
| I0550 | KAR22A | T1a1 | none | C146T, A247G, G709A, A769G, A825t, A1018G, G1888A, A2758G, C2885T, T3594C, G4104A, T4216C, T4312C, A4917G, G7146A, T7256C, A7521G, T8468C, T8655C, G8697A, G8701A, C9540T, T9899C, G10398A, T10463C, T10664C, A10688G, C10810T, C10873T, C10915T, A11251G, A11914G, C12633a, T12705C, G13105A, G13276A, G13368A, T13506C, T13650C, G14905A, C15452a, A15607G, G15928A, T16126C, A16129G, A16163G, C16186T, T16187C, T16223C, G16230A, T16278C, C16294T, C16311T |
| I0117 | ESP29 | I3a | none | C146T, C195T, T199C, T204C, G207A, T239C, A247G, T250C, 573.XC, A769G, A825t, A1018G, G1719A, A2758G, C2885T, T3594C, G4104A, T4312C, A4529t, G7146A, T7256C, A7521G, G8251A, T8468C, T8655C, G8701A, C9540T, T10034C, T10238C, T10664C, A10688G, C10810T, C10873T, C10915T, A11914G, G12501A, G13105A, G13276A, T13506C, T13650C, A13780G, G15043A, A15924G, T16086C, T16187C, C16189T, G16230A, T16278C, C16311T, G16391A |

| Lab ID | Individual ID | mtDNA-hg call | Private mutations | SNPs against RSRS |
|---|---|---|---|---|
| I0115 | ESP3 | U5a1 | C4796T, C12103A, C14003T, A14893G, T14971C, C16519T | C146T, C152T, C195T, A247G, A769G, A825t, A1018G, A2758G, C2885T, T3197C, T3594C, G4104A, T4312C, C4796T, G7146A, T7256C, A7521G, T8468C, T8655C, G8701A, G9477A, C9540T, G10398A, T10664C, A10688G, C10810T, C10873T, C10915T, A11467G, A11914G, C12103A, A12308G, G12372A, T12705C, G13105A, G13276A, T13506C, T13617C, T13650C, C14003T, A14793G, A14893G, T14971C, A15218G, A16129G, T16187C, C16189T, C16192T, T16223C, G16230A, C16256T, C16270T, T16278C, C16311T, A16399G, C16519T |
| I0116 | ESP4 | W3a1 | C5211T, G6267A, T14025C, C16147G | C146T, C152T, A189G, C194T, T204C, G207A, A247G, G709A, A769G, A825t, A1018G, T1243C, T1406C, A2758G, C2885T, A3505G, T3594C, G4104A, T4312C, G5046A, C5211T, G5460A, G6267A, G7146A, T7256C, A7521G, G8251A, T8468C, T8655C, G8701A, G8994A, C9540T, G10398A, T10664C, A10688G, C10810T, C10873T, C10915T, C11674T, A11914G, A11947G, T12414C, G13105A, A13263G, G13276A, T13506C, T13650C, T14025C, T15784C, G15884c, A16129G, C16147G, T16187C, C16189T, G16230A, T16278C, C16292T, C16311T |
| I0803 | EUL41A | H4a1a1 | C13545T, C16519T | C146T, C152T, C195T, A247G, A769G, A825t, A1018G, G2706A, A2758G, C2885T, T3594C, C3992T, A4024G, G4104A, T4312C, T5004C, T7028C, G7146A, T7256C, A7521G, G8269A, T8468C, T8655C, G8701A, G9123A, C9540T, A10044G, G10398A, T10664C, A10688G, C10810T, C10873T, C10915T, A11719G, A11914G, T12705C, G13105A, G13276A, T13506C, C13545T, T13650C, C14365T, A14582G, T14766C, A16129G, T16187C, C16189T, T16223C, G16230A, T16278C, C16311T, C16519T |
| I0804 | EUL57B | H3 | T152C! | G73A, T152C!, C146T, C195T, A247G, A769G, A825t, A1018G, G2706A, A2758G, C2885T, T3594C, G4104A, T4312C, T6776C, T7028C, G7146A, T7256C, A7521G, T8468C, T8655C, G8701A, C9540T, G10398A, T10664C, A10688G, C10810T, C10873T, C10915T, A11719G, A11914G, T12705C, G13105A, G13276A, T13506C, T13650C, T14766C, A16129G, T16187C, C16189T, T16223C, G16230A, T16278C, C16311T |
| I0047 | HAL16 | V | G207A, A14841G | T72C, G73A, C146T, C152T, C195T, G207A, A247G, A769G, A825t, A1018G, A2758G, C2885T, T3594C, G4104A, T4312C, 4580A, G7146A, T7256C, A7521G, T8468C, T8655C, G8701A, C9540T, G10398A, T10664C, A10688G, C10810T, C10873T, C10915T, A11719G, A11914G, T12705C, G13105A, G13276A, T13506C, T13650C, T14766C, A14841G, C15904T, A16129G, T16187C, C16189T, T16223C, G16230A, T16278C, T16298C, C16311T, C16519T, |
| I0164 | QUEVIII6 | U5b2a1 (pre-U5b2a1b) | T5918C, T16192C!, C16519T | C146T, C150T, C152T, C195T, A247G, A769G, A825t, A1018G, C1721T, A2758G, C2885T, T3197C, T3594C, G4104A, T4312C, A4732G, T5918C, G7146A, T7256C, A7521G, A7768G, T8468C, T8655C, G8701A, G9477A, C9540T, G10398A, T10664C, A10688G, C10810T, C10873T, C10915T, A11467G, A11914G, A12308G, G12372A, T12705C, G13105A, G13276A, T13506C, T13617C, A13637G, T13650C, T14182C, A16129G, T16187C, T16223C, G16230A, T16278C, C16311T, C16519T |
| I0114 | ESP2 | I3a | none | C146T, C195T, T199C, T204C, G207A, T239C, A247G, T250C, 573.XC, A769G, A825t, A1018G, G1719A, A2758G, C2885T, T3594C, G4104A, T4312C, A4529t, G7146A, T7256C, A7521G, G8251A, T8468C, T8655C, G8701A, C9540T, T10034C, T10238C, T10664C, A10688G, C10810T, C10873T, C10915T, A11914G, G12501A, G13105A, G13276A, T13506C, T13650C, A13780G, G15043A, A15924G, T16086C, T16187C, C16189T, G16230A, T16278C, C16311T, G16391A |
| I0099 | HAL36C | H23 | none | G73A, C146T, C152T, C195T, A247G, A769G, A825t, A1018G, G2706A, A2758G, C2885T, T3594C, G4104A, T4312C, T7028C, G7146A, T7256C, A7521G, T8468C, T8655C, G8701A, C9540T, C10211A, G10398A, T10664C, A10688G, C10810T, C10873T, C10915T, A11719G, A11914G, T12705C, G13105A, G13276A, T13506C, T13650C, T14766C, A16129G, T16187C, C16189T, T16223C, G16230A, T16278C, C16311T |

## Table S2.2. Haplogroup frequencies of 24 ancient populations used for principal component analysis.

| Population | abbr. | n | C | Z | N* | N1a | I | W | X | R | HV | V | H | T1 | T2 | J | U | U2 | U3 | U4 | U5a | U5b | U8 | K | Publication (number of samples) |
|---|---|---|---|---|---|---|---|---|---|---|---|---|---|---|---|---|---|---|---|---|---|---|---|---|---|
| Hunter-gatherer central | HGC | 28 | 0.0 | 0.0 | 0.0 | 0.0 | 0.0 | 0.0 | 0.0 | 0.0 | 0.0 | 0.0 | 0.0 | 0.0 | 0.0 | 0.0 | 14.3 | 10.7 | 0.0 | 7.1 | 32.1 | 32.1 | 3.6 | 0.0 | Bramanti et al. 2009 (11)[12], Fu et al. 2013 (5)[13], Bollongino et al. 2013 (5)[14], Lazaridis et al. 2014 (7)[15] |
| Linear Pottery culture | LBK | 115 | 0.0 | 0.0 | 0.0 | 11.3 | 0.0 | 3.5 | 0.9 | 0.0 | 5.2 | 3.5 | 16.5 | 0.0 | 25.2 | 11.3 | 0.0 | 0.0 | 0.9 | 0.0 | 1.7 | 0.9 | 0.0 | 19.1 | Haak et al. 2005 (14)[16], Bramanti 2008[17], Haak et al. 2010 (22)[6], Brandt et al. 2013 (64)[8], Lazaridis et al. 2014 (1)[15], this study (8) |
| Rössen culture | RSC | 17 | 0.0 | 0.0 | 0.0 | 5.9 | 0.0 | 0.0 | 5.9 | 0.0 | 23.5 | 5.9 | 29.4 | 0.0 | 11.8 | 0.0 | 0.0 | 0.0 | 0.0 | 0.0 | 0.0 | 5.9 | 0.0 | 11.8 | Brandt et al. 2013 (11)[8], Lee et al. 2013[18] |
| Schöningen group | SCG | 33 | 0.0 | 0.0 | 0.0 | 3.0 | 0.0 | 9.1 | 3.0 | 0.0 | 3.0 | 0.0 | 15.2 | 0.0 | 12.1 | 15.2 | 0.0 | 0.0 | 0.0 | 0.0 | 0.0 | 6.1 | 3.0 | 30.3 | Brandt et al. 2013 (33)[8] |
| Baalberge culture | BAC | 21 | 0.0 | 0.0 | 0.0 | 4.8 | 0.0 | 0.0 | 9.5 | 0.0 | 4.8 | 0.0 | 23.8 | 4.8 | 23.8 | 4.8 | 0.0 | 0.0 | 0.0 | 0.0 | 0.0 | 4.8 | 4.8 | 14.3 | Brandt et al. 2013 (19)[8], this study (2) |
| Salzmünde culture | SMC | 29 | 0.0 | 0.0 | 0.0 | 6.9 | 0.0 | 0.0 | 3.4 | 0.0 | 3.4 | 3.4 | 31.0 | 0.0 | 6.9 | 20.7 | 0.0 | 0.0 | 10.3 | 0.0 | 0.0 | 3.4 | 0.0 | 10.3 | Brandt et al. 2013 (29)[8] |
| Bernburg culture | BEC | 17 | 0.0 | 0.0 | 0.0 | 0.0 | 0.0 | 5.9 | 5.9 | 0.0 | 0.0 | 5.9 | 23.5 | 0.0 | 11.8 | 0.0 | 0.0 | 0.0 | 0.0 | 0.0 | 11.8 | 17.6 | 0.0 | 17.6 | Brandt et al. 2013 (17)[8] |
| Corded Ware culture | CWC | 44 | 0.0 | 0.0 | 0.0 | 0.0 | 2.3 | 2.3 | 6.8 | 0.0 | 2.3 | 0.0 | 22.7 | 6.8 | 11.4 | 9.1 | 0.0 | 2.3 | 0.0 | 6.8 | 9.1 | 4.5 | 0.0 | 13.6 | Haak et al. 2008 (9)[19], Brandt et al. 2013 (35)[8] |
| Bell Beaker culture | BBC | 35 | 0.0 | 0.0 | 0.0 | 0.0 | 2.9 | 8.6 | 0.0 | 0.0 | 0.0 | 0.0 | 40.0 | 5.7 | 5.7 | 2.9 | 0.0 | 2.9 | 0.0 | 5.7 | 14.3 | 5.7 | 0.0 | 5.7 | Brandt et al. 2013 (29)[8] |
| Unetice culture | UC | 94 | 0.0 | 0.0 | 0.0 | 0.0 | 12.8 | 4.3 | 5.3 | 1.1 | 2.1 | 3.2 | 21.3 | 2.1 | 6.4 | 6.4 | 2.1 | 6.4 | 0.0 | 1.1 | 12.8 | 2.1 | 3.2 | 7.4 | Brandt et al. 2013 (94)[8] |
| Middle Neolithic Germany | MNG | 16 | 0.0 | 0.0 | 0.0 | 0.0 | 0.0 | 0.0 | 0.0 | 0.0 | 0.0 | 0.0 | 37.5 | 0.0 | 0.0 | 6.3 | 6.3 | 0.0 | 0.0 | 0.0 | 0.0 | 50.0 | 0.0 | 0.0 | Bollongino et al. 2013 (16)[14] |
| Funnel Beaker culture | FBC | 10 | 0.0 | 0.0 | 0.0 | 0.0 | 0.0 | 0.0 | 0.0 | 0.0 | 0.0 | 0.0 | 10.0 | 0.0 | 30.0 | 20.0 | 0.0 | 0.0 | 0.0 | 0.0 | 10.0 | 20.0 | 0.0 | 10.0 | Malmström et al. 2009[20], Skoglund et al. 2012 (3)[21], Bramanti et al. 2009 (7)[12] |
| Pitted Ware culture | PWC | 19 | 0.0 | 0.0 | 0.0 | 0.0 | 0.0 | 0.0 | 0.0 | 0.0 | 10.5 | 5.3 | 0.0 | 0.0 | 5.3 | 0.0 | 0.0 | 0.0 | 0.0 | 42.1 | 15.8 | 15.8 | 0.0 | 5.3 | Malmström et al. 2009[20], Skoglund et al. 2012 (19)[21] |
| Hunter-Gatherer south | HGS | 13 | 0.0 | 0.0 | 15.4 | 0.0 | 0.0 | 0.0 | 0.0 | 0.0 | 0.0 | 0.0 | 38.5 | 0.0 | 0.0 | 0.0 | 0.0 | 0.0 | 0.0 | 7.7 | 0.0 | 38.5 | 0.0 | 0.0 | Chandler 2003[22], Chandler et al. 2005 (7)[23], Hervella et al. 2012 (4)[24], Sánchez-Quinto et al. 2012 (2)[25] |
| (Epi)Cardial | CAR | 23 | 0.0 | 0.0 | 13 | 4.3 | 0.0 | 0.0 | 4.3 | 0.0 | 0.0 | 0.0 | 26.1 | 0.0 | 13 | 4.3 | 0.0 | 0.0 | 0.0 | 0.0 | 0.0 | 4.3 | 0.0 | 30.4 | Gamba et al. 2011 (11)[26], Lacan 2011, Lacan et al. 2011b (7)[27], this study (5) |
| Neolithic Portugal | NPO | 17 | 0.0 | 0.0 | 0.0 | 0.0 | 0.0 | 0.0 | 0.0 | 0.0 | 0.0 | 5.9 | 70.6 | 0.0 | 0.0 | 0.0 | 0.0 | 0.0 | 0.0 | 0.0 | 17.6 | 5.9 | 0.0 | 0.0 | Chandler 2003, Chandler et al. 2005 (17)[22,23] |
| Neolithic Basque Country & Navarre | NBQ | 43 | 0.0 | 0.0 | 0.0 | 0.0 | 2.3 | 0.0 | 2.3 | 0.0 | 2.3 | 0.0 | 44.2 | 0.0 | 2.3 | 4.7 | 25.6 | 0.0 | 0.0 | 0.0 | 0.0 | 7.0 | 0.0 | 9.3 | Hervella 2010, Hervella et al. 2012 (43)[24] |
| Middle Neolithic Spain | MNS | 16 | 0.0 | 0.0 | 0.0 | 0.0 | 6.3 | 6.3 | 0.0 | 0.0 | 0.0 | 0.0 | 31.3 | 0.0 | 12.5 | 18.8 | 0.0 | 0.0 | 0.0 | 6.3 | 0.0 | 6.3 | 0.0 | 12.5 | Sampietro et al. 2007 (11)[28], this study (5) |
| Treilles | TRE | 29 | 0.0 | 0.0 | 0.0 | 0.0 | 0.0 | 0.0 | 13.8 | 0.0 | 6.9 | 3.4 | 20.7 | 0.0 | 6.9 | 20.7 | 3.4 | 0.0 | 0.0 | 0.0 | 3.4 | 13.8 | 0.0 | 6.9 | Lacan 2011, Lacan et al. 2011a (29)[29] |
| Hunter-gatherer east | HGE | 15 | 20 | 0.0 | 0.0 | 0.0 | 0.0 | 0.0 | 0.0 | 0.0 | 0.0 | 0.0 | 6.7 | 0.0 | 0.0 | 0.0 | 0.0 | 20 | 0.0 | 26.7 | 26.7 | 0.0 | 0.0 | 0.0 | Bramanti et al. 2009 (2)[12], Krause et al. 2010 (1)[30], Der Sarkissian et al. in 2013 (11)[7], this study (1) |
| Bronze Age Siberia | BAS | 11 | 0.0 | 9.1 | 0.0 | 0.0 | 0.0 | 0.0 | 0.0 | 0.0 | 0.0 | 0.0 | 9.1 | 9.1 | 9.1 | 0.0 | 0.0 | 9.1 | 0.0 | 27.3 | 18.2 | 0.0 | 0.0 | 9.1 | Keyser et al. 2009 (11)[31] |
| Bronze Age Kazakhstan | BAK | 8 | 0.0 | 0.0 | 0.0 | 0.0 | 12.5 | 0.0 | 0.0 | 0.0 | 12.5 | 0.0 | 12.5 | 25.0 | 12.5 | 0.0 | 0.0 | 0.0 | 0.0 | 0.0 | 12.5 | 12.5 | 0.0 | 0.0 | Lalueza-Fox et al. 2004 (8)[32] |
| Catacomb culture | CAT | 24 | 0.0 | 0.0 | 0.0 | 0.0 | 4.2 | 0.0 | 0.0 | 8.3 | 4.2 | 0.0 | 25 | 8.3 | 0.0 | 8.3 | 0.0 | 0.0 | 0.0 | 29.2 | 12.5 | 0.0 | 0.0 | 0.0 | Wilde et al. 2014 (24)[11] |
| Bronze Age Yamnaya | YAM | 36 | 0.0 | 0.0 | 0.0 | 2.8 | 2.8 | 8.3 | 2.8 | 0.0 | 0.0 | 0.0 | 25 | 5.6 | 11.1 | 5.6 | 2.8 | 2.8 | 0.0 | 5.6 | 16.7 | 0.0 | 0.0 | 8.3 | Wilde et al. 2014 (26)[11], this study (10) |

# Supplementary Information 3

## Description of ancient samples and archaeological context

Wolfgang Haak*, Guido Brandt, Christina Roth, Anna Szécsényi-Nagy, Susanne Friederich, Harald Meller, Michael Francken, Joachim Wahl, Roberto Risch, Michael Kunst, Manuel A. Rojo Guerra, Rafael Garrido-Pena, Eszter Bánffy, Kurt W. Alt, Dorcas Brown, David Anthony and David Reich

* To whom correspondence should be addressed (wolfgang.haak@adelaide.edu.au)

**Overview**

The archaeology of prehistoric Europe has revealed numerous material cultures that differed from each other over space and time. Some cultures left behind artifacts similar to those that preceded them. Other cultures left behind artifacts that mark a clear break from preceding local traditions.

The largest single group of samples analyzed in this study was from the transect through time study from the Early Neolithic to the Early Bronze Age reported in Brandt et al. 2013[1]. We selected samples from that study based on two criteria: We predominantly chose samples based on overall DNA quality, which we assessed based on performance in our previous study (largely PCR success rates, replication of results, and completeness of SNP typing). From the subset of samples that performed well, we chose ones that could be assigned most confidently to an archaeological culture via accompanying grave goods/context, orientation of the burials, available radiocarbon dates, and seemingly typical mtDNA haplotype. In addition, we added to this study some individuals for which the cultural attribution was less clear, in that there were no characteristic grave goods or context. These individuals were included to test the ability of the 390k SNP capture strategy to assign a potentially unknown or unprovenanced sample to a candidate population in combination with absolute dating. For the latter set of samples, we obtained 19 new radiocarbon dates at the Curt-Engelholm-Centre for Archaeometry, Mannheim, Germany.

In what follows we provide archaeological context information for the 69 samples for which genome-wide data is newly presented in this study (a total of 119 individuals were screened as detailed in Online Table 1, but here we discuss only the subset that produced genome-wide data). We present the samples in chronological order, largely following the central European terminology, i.e. Mesolithic (>6000-5000 BCE), Early Neolithic (6000-4000 BCE), Middle Neolithic (4000-2800 BCE), Late Neolithic (2800-2200 BCE), and Early Bronze Age (2200-1600 BCE; Figure 1a). We note that there are different traditional terminology for samples from the Russian steppe and regions of southern Europe.

We use “BCE” (Before Common Era) to refer to a calibrated date that is estimated based on archaeological context (e.g. radiocarbon dates from associated faunal remains), and “calBCE” to refer to a calibrated radiocarbon date range (1σ) directly from human skeletal remains, followed by the respective dating laboratory number.

**Hunter-gatherer samples**

We generated data on two hunter-gatherer samples from western Russia (the easternmost part of Europe) to bridge the geographical gap between the West European site Loschbour (WHG)[2] and the Siberian site Mal'ta in Siberia, Russia[3], from which hunter-gatherer genetic data are available from the literature.

The individual we refer to as 'Karelia' in this study is

- UZ0074/I0061 (MAE RAS collection number 5773-74, grave number 142)

from the ~5500 BCE Mesolithic site Yuzhnyy Oleni Ostrov (island in Lake Onega) in Karelia, Western Russia, for which a complete mitochondrial genome was published recently[4]. Mitochondrial HVS I data from eight other individuals from the same site have also been described[5].

The individual we refer to as 'Samara hunter-gatherer'

- I0124/SVP44 (5640-5555 calBCE, Beta-392490)

is an adult male from grave 1 in a Neolithic-Eneolithic settlement producing artifacts from the Elshanka, Samara, and Repin cultures. The specific site is Lebyazhinka IV, on the Sok River, Samara oblast, Russia. ('Neolithic' here refers to the presence of ceramics, not to domesticated animals or plants.) The radiocarbon date of this individual, based on a femur, is centuries before the appearance of domesticated animals in the middle Volga region. Lebyazhinka IV and the neighboring Lebyazhinka V site were occupied seasonally by multiple cultures between 7000-3500 BCE; a few graves were found in the settled areas[6].

**Early Neolithic**

The Early Neolithic in Europe in this study is represented by new samples reported from sites in Hungary, Germany and Spain.

The central European Neolithic (6200–3950 BCE) is first manifested in the Starčevo culture, the Transdanubian *Linienbandkeramik* (LBKT), and the central European distribution of the *Linienbandkeramik* (LBK); these were the first people in the region to exploit agriculture and animal husbandry[7]. The *Linienbandkeramik* appears earliest in the archaeological record of western Hungary (Transdanubia), where it incorporated novel technologies and cultural elements from the preceding, southeastern Starčevo culture, which in turn also shows similarities in material culture to early farming groups further southeast, including Anatolia, the Levant and the Near East. During the Neolithic transition, the LBK expanded relatively rapidly along the major waterways and fertile Loess plains towards central Europe, extending as far as the Paris Basin in the West and Ukraine in the East[8,9].

*Starčevo in Hungary: Alsónyék-Bátaszék, Mérnöki telep*

The Alsónyék-Bátaszék, Mérnöki telep site was excavated by the Institute of Archaeology, Hungarian Academy of Sciences, between 2006 and 2009 (excavators: A. Osztás, I. Zalai-Gaál). This site is a part of the site complex "Alsónyék", which forms the largest agglomeration of Neolithic settlements and cemeteries in Hungary. Alsónyék is situated along the Tolna county part of the M6 motorway track in south Hungary. At Mérnöki telep, traces of the Early Neolithic Starčevo culture and other Neolithic features, including LBK and the Late Neolithic Lengyel culture, came to light across a total of 80 hectares. Out of the 1568 archaeological features, more than 400 could be associated with the Starčevo culture including 26 burials. The majority of features were pits, in various shapes and sizes; however, ditches and various oven types were also found. The skeletal remains were all found inside ovens or in pits, and were all in flexed positions. Only one burial contained grave goods[10]. All 26 Starčevo skeletons were investigated as part of the German Research Foundation (DFG) funded bioarchaeological (ancient DNA and stable isotopes) project entitled "*The population history of the Carpathian Basin in the Neolithic period and its influence on the colonization of central Europe*". Skeletons in uncertain stratigraphic positions, and buried without any grave goods, were radiocarbon dated at the CEZA laboratory in Mannheim, Germany. Of these, individual

- BAM25a/I0174 (feature 1532, 5710-5550 calBCE, MAMS 11939)

was included in this study.

*Transdanubian LBK in Hungary: Szemely-Hegyes*

The south-Hungarian site Szemely-Hegyes is located in Baranya county along the M6 motorway. The site was excavated between 2006 and 2007 by a commercial archaeology contractor (Ásatárs Ltd; excavators T. Paluch and K. Somogyi). Approximately 1400 archaeological features were documented across 4 hectares, including finds from the LBK, Sopot culture, Copper Age Balaton-Lasinja, and Furchenstich groups[11] (T. Paluch, K. Somogyi, J. Jakucs personal communication), including ten Neolithic graves. An LBK settlement with 20 houses, several pits, ditches and ovens was also uncovered. Six out of the ten Neolithic graves could be assigned to either the Transdanubian LBK or to the Sopot period, and the skeletons were radiocarbon dated at Beta Analytic in Miami, USA. Sample

- SZEH4b/I0176 (feature 1001, 5210-4940 calBCE, Beta - 310038)

associated with the Transdanubian LBK was analyzed here.

*LBK in Germany: Halberstadt-Sonntagsfeld*

The most intensively studied region in this study is Mittelelbe-Saale (MES) in Saxony-Anhalt, Germany. MES is situated within a tight network of waterways along ancient established trade routes. The region is climatically and geologically well characterized and benefits from its location in the rain-shadow of the nearby Harz Mountains, its deposits of ores and salts, and its fertile black and para-brown soils[12-14]. The earliest MES culture sampled in this study is the LBK.

The LBK Halberstadt-Sonntagsfeld site was discovered during construction of a new development and was excavated between 1999 and 2002, uncovering 1324 archaeological features across an area of 9947m$^2$. The majority of the finds could be attributed to the LBK, with contours and remnants of seven long houses, several pits and 42 graves[15]. The remaining finds could be attributed to the Middle Neolithic Bernburg culture as well as the Unetice and Urnfield cultures of the Bronze Age. The site is a classical example of LBK settlement burials, where the majority of graves were grouped around five long houses, and one group of six graves located inside the central yard area of the settlement. Skeletal material from the site is well preserved and led to mitochondrial DNA results for 39 individuals[1]. We included seven individuals clearly associated with an LBK house:

- HAL5/I0046 (grave 2, feature 241.1, 5206-5004 calBCE, MAMS 21479)
- HAL25/I0048 (grave 28, feature 861, 5206-5052 calBCE, MAMS 21482)
- HAL14/I0056 (grave 15, feature 430, 5206-5052 calBCE, MAMS 21480)
- HAL34/I0057 (grave 38, feature 992, 5207-5067 calBCE, MAMS 21483)
- HAL4/I0100 (grave 1, feature 139, 5032-4946 calBCE, KIA40341)

- HAL2/I0659 (grave 35, feature 999, 5079-4997 calBCE, KIA40350 and 5066-4979 calBCE, KIA30408; Fig. S3.1)
- HAL24/I0821 (grave 27, feature 867, 5034-4942 calBCE, KIA40348)

Radiocarbon dates obtained from these samples confirmed the attribution to the older and intermediate phase of the LBK period.

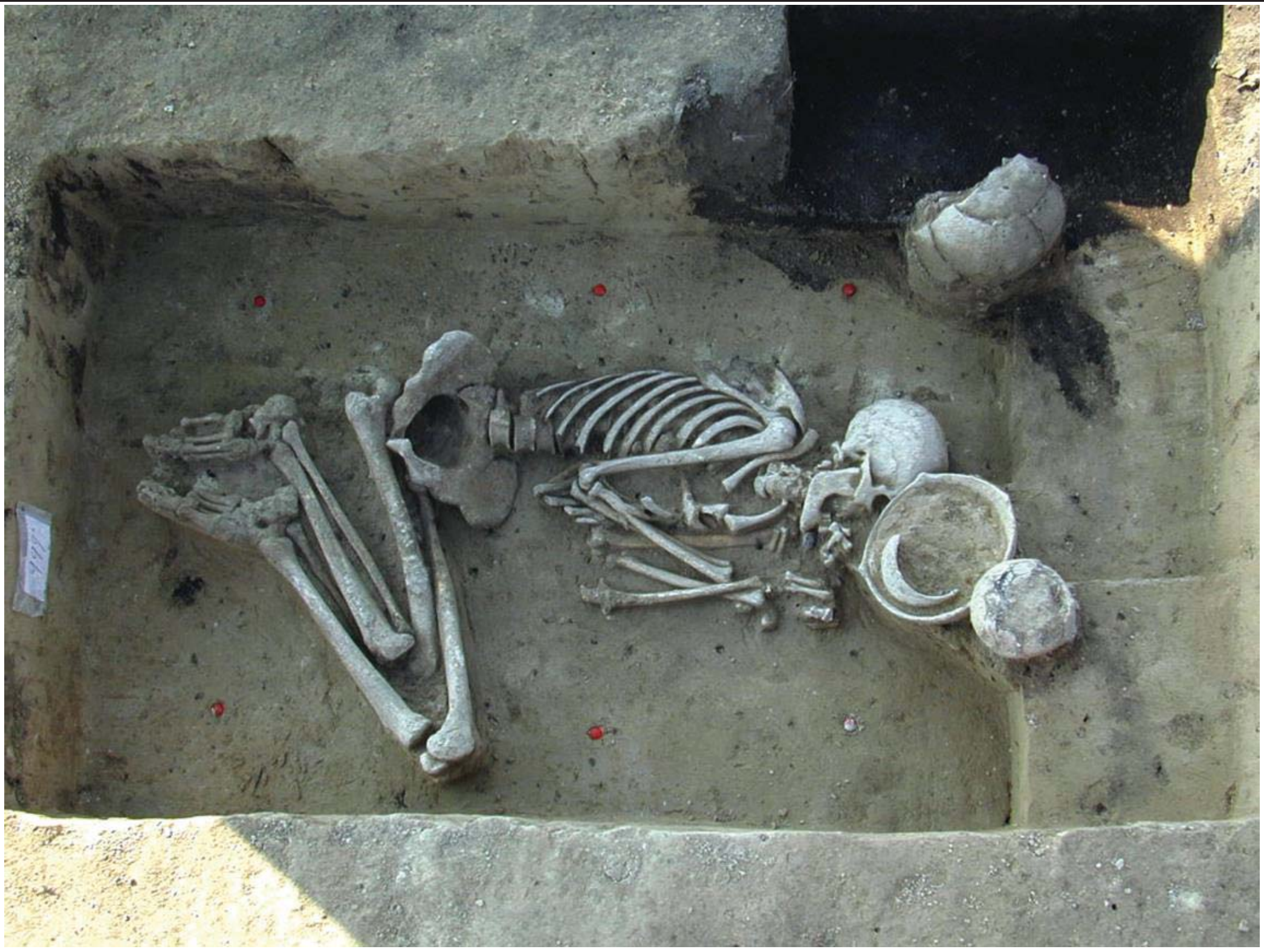

**Figure S3.1. Skeleton HAL2 (I0659 in this study) from grave 35 at the Halberstadt-Sonntagsfeld site** (Photo: J. Lipták, LDA Sachsen-Anhalt, Germany).

*LBK in Germany: Oberwiederstedt-Unterwiederstedt*

Several sites around Wiederstedt, Mansfeld-Südharz, and Saxony-Anhalt were excavated in the late 20th century. Promontory areas above the Wipper River in the Harz region in Hettstedt were repeatedly used during LBK times, as reflected in traces of settlements. In addition, excavations in 2000 revealed a mass grave of 12 individuals of which individual

- UWS4/I0054 (grave 6, feature 1 14, 5209-5070 calBCE, MAMS 21485)

was included in this study[16].

*LBK in Germany: Karsdorf*

The site of Karsdorf is located in the valley of Unstrut, Burgenlandkreis, Saxony-Anhalt, Germany. The slope on which Karsdorf is located is characterized by alluvial loess. The place itself was settled intensively since the earliest phase of the LBK in the region. The settlement area is at least 50 acres in size and nearly 30 houses have been excavated. So-called 'settlement burials' were regularly found in pits in the center of the settlement area, of which individual

- KAR6/I0795 (feature 170, 5207-5070 calBCE, MAMS 22823)

was sampled for this study.

*LBK in Germany: Stuttgart-Mühlhausen*

The site Stuttgart-Mühlhausen "Viesenhäuser Hof" was excavated in several campaigns between 1931 and 1993. It reflects a long period of habitation starting from the Early Neolithic to the Iron Age. The Early Neolithic is represented by a large number of well-preserved burials from the LBK, separated into two areas including burials from the early phases of the LBK (area-2; 5500-5300 BCE) and the middle and later phases (area-1). Three individuals from area-2 are newly reported in this study (a fourth LBK sample from the Viesenhäuser Hof site was previously sequenced to high coverage and is co-analyzed with the newly sequenced LBK in this study[2]). Based on morphology, individual

- LBK1976/I0022 (grave II-106, area-2, 5500-5300 BCE)

is a female who died at an estimated age of 50-65 years. The skeleton derives from a grave that was excavated among 90 others from area-2 of the cemetery. The woman was buried in the characteristic way of the LBK, lying in flexed position on the right side, but lacks any grave goods. The burial was oriented from SW-NE with the skull facing NW. Most of the body parts were represented and in anatomical order.

The second well-preserved skeleton

- LBK1992/I0025 (grave II-113, area-2, 5500-5300 BCE)

is a male who died at an estimated age of 50-70 years. The skeleton was found prone in the grave with the position of the left leg indicated a flexed position. The burial was oriented from SE-NW with the skull facing south. The grave only contained a single pottery sherd found in the filling of the grave.

The third skeleton is a female

- LBK2155/I0026 (grave II-125, area-2, 5500-5300 BCE)

who died at 35-50 years. This individual was found prone with flexed legs imitating a seated position. The grave was oriented from W-E with the skull facing north. The only associated finding was a pottery fragment from the filling.

*Epicardial Neolithic in Spain: Els Trocs*

The Pyrenean cave site of Els Trocs (close to San Feliu de Veri, Bisaurri (Huesca), Spain) is located in the Upper Ribagorza of Aragón at 1561 meters above sea level, inside a conical hill, which dominates the surrounding 'Selvaplana' plain, traditionally used for pasture and cultivation. The entrance to the cave (UTM coordinates: x-298.198, y-4.702.955, z-1561) is oriented to the south and hidden by abundant vegetation and huge stone boulders, which partially close the entrance (currently 2.3 m high and 1.8 m wide). After a steep access ramp, a chamber of 15 m of length and 6 m of maximum width opens up, in which the temperature remains stable at 6 and 8°C throughout the year. The excavations carried out inside the cave have uncovered a stratigraphic sequence that in combination with a series of radiocarbon dates characterize the following four phases of human occupation[17]:

TROCS I: the first occupation of the cave, dated to 5300-5000 BCE.
TROCS II: the Middle Neolithic horizon, dated around 4500 BCE.
TROCS III: intensive use spanning >1000 years between 4000-2900 BCE.
TROCS IV: upper layer of the cave from historical times (100 BCE) to today.

Human bones were discovered from all four phases, but especially from Trocs I during which they appear to be associated with a marked and complex ritual behavior, which involved manipulation of skeletal elements and the deposition of abundant faunal remains (mainly young lambs) that had been consumed.

The first occupation phase of the cave (Trocs I) is related to the so-called Epicardial tradition. However, this attribution is currently under re-consideration[18] since both the incised-impressed-grooved decorative techniques and the Cardial pottery types are assumed to be two variants of the same archaeological culture responsible of the initial Neolithization of Iberia. The samples successfully analyzed for this study are

- Troc1/I0409 (5311-5218 cal BCE, MAMS 16159)
- Troc3/I0410 (5178-5066 cal BCE, MAMS 16161)
- Troc4/I0411 (5177-5068 cal BCE, MAMS 16162)
- Troc5/I0412 (5310-5206 cal BCE, MAMS 16164)
- Troc7/I0413 (5303-5204 cal BCE, MAMS 16166)

The genetic data indicates that Troc4 was a close relative of Troc3. We excluded the data from Troc4 from many genome-wide analyses since Troc3 was higher coverage.

**Middle Neolithic**

*Baalberge, Salzmuende, and Bernburg in Germany: Esperstedt*

Esperstedt is located in the Merseburg-Querfurt district, Saxony-Anhalt, and is situated in the centre of the Querfurter Platte, a plain that is primarily loess.

The single grave of individual

- ESP30/I0807 (feature 6220; 3887-3797 calBCE, Er7784)

discovered in 2004 during investigations in advance to major roadworks could be assigned to the Baalberge group.

Site 4b is located east of the Weida near Esperstedt and forms part of large-scale excavations that were initiated in 2005 in the context of major infrastructural roadworks in Saxony-Anhalt, Germany. Site 4b revealed mostly finds from the Middle Neolithic Salzmünde and Bernburg cultures, viewed as regional forms of the Funnel Beaker tradition (or *Trichterbecherkultur* TRB), which defines the Middle Neolithic in central Europe. The genetic results from individual

- ESP24/I0172 (3360-3086 calBCE, Erl8699)

were retrieved from skeletal remains found in a settlement pit that was attributed to the Bernburg culture, but had no associated grave goods. Radiocarbon dating from the >60 year-old male individual matched both the temporally overlapping phases of the Salzmünde (3400-3025 BCE) and Bernburg (3100-2650 BCE) cultures[19].

*Baalberge in Germany: Quedlinburg*

The site Quedlinburg, Harzkreis, is situated in the fertile foothills of the northern Harz, a region characterized by rich loess soils. The specific site Lehofsberg (Reference site 9) was excavated during major roadworks for highway B6n, at which archaeological excavations revealed a small graveyard with a dozen graves (without trapezoidal enclosure) from the Baalberge culture and a total 20 burials spread over a length of 200 meters. The individuals

- QLB15/I0559 (feature 21033, 3645-3537 cal BCE, MAMS 22818)
- QLB18/I0560 (feature 21039, 3640-3510 calBCE, Er7856)

that were sampled for this study are likely to be burials of commoners as more elaborate grave architectural elements such as cists are missing.

*Middle Neolithic in Spain: La Mina*

The site La Mina (Alcubilla de las Peñas, Soria, Spain) includes a classic passage grave with a corridor longer than five meters oriented south-southeast. The site suffered from a systematic process of closure and dismantling of its stone structure with all orthostats of the chamber and the passage being removed. However, the burial chamber, from which the remains of approximately 20 buried individuals have been recovered, remained intact. At the time of the burials, the appearance of the tomb was modified by increasing the diameter and height of the original mound, placing a menhir higher than three meters possibly on top of the monument, and building a wall of orthostats parallel to the old burial surrounding the entire perimeter wall. As a result, the original collective grave had been transformed into a ceremonial center, and given its unique topographic location the mound became a territorial landmark. A radiocarbon date from a human bone from the ossuary (3780-3700 calBCE, Beta 316132) places the tomb at the beginning of Megalithism in the inner Iberian Peninsula, at the early 4$^{th}$ millennium.

The samples of La Mina originate from two seasons. The samples of Mina1-10 are from season 2010 and Mina 11-18 from 2012. In both seasons, samples were taken directly at the excavation with gloves and facemask and immediately stored under cool conditions. In 2010, samples were taken by geneticist Sarah Lang and in 2011 by archaeologist Manuel Rojo-Guerra. The samples that produced 390k data and that are included in our analysis are:

- Mina3/I0405 (3900-3600 BCE)
- Mina5/I0406 (3900-3600 BCE)
- Mina6/I0407 (3900-3600 BCE)
- Mina18/I0408 (3900-3600 BCE)

**Late Copper/Early Bronze Age steppe chronology**

The Yamnaya horizon was the first cultural complex that spread across all of the Pontic-Caspian-Ural steppes, beginning about 3300 BCE. It is divided into at least six regional groups and at least two chronological stages (up to four phases have been suggested), and the earliest ceramics included at least two distinct types (originating on the lower Don and lower Volga), both of which were decorated with cord impressions and usually were round-bottomed. The unifying traits were: a mobile pastoral economy that operated from newly invented wheeled vehicles using horseback-mounted herders; the introduction of new metal types (shafthole axes, tanged daggers, and spiral hair-rings or Lockenringe), new metal techniques, and copper mining on a significant scale, perhaps inspired by the Maikop culture; the burial of a select few adult males (who made up 75-80% of the graves under kurgans in the Samara region) under kurgans that usually covered 1 to 3 graves (although in the Kuban and NW Pontic steppes a wider range of people was buried under kurgans); the burial of wheels or whole wagons above or beside elite graves; deposition of animal sacrifices (usually sheep-goat) in about 15% of graves; and erection of grave-associated stone stelae, in the Ukrainian steppes. A few stratified settlements are found in the Ukrainian steppes west of the Don but none are known east of the Don in the Volga-Ural region. The Mikhailovka settlement on the Dnieper River is the standard stratigraphic guide to Yamnaya chronology[20,21]. Pre-Yamnaya level 1 began about 3800 BCE, early Yamnaya level 2 began about 3500 BCE, and late Yamnaya level 3 began about 2800 BCE. At Mikhailovka the Yamnaya phase (early level 2) began about 3500-3300 BCE[22], but the majority of early Yamnaya kurgans began later, closer to 3300 BCE, from the Volga to the Dnieper, so 3300 BCE is a safer date for the general beginning of the Yamnaya era[23]. EBA Yamnaya evolved into the MBA Poltavka culture in the Volga-Ural steppes and into the MBA Catacomb culture in the Dnieper-Don-Caucasus steppes beginning about 2800 BCE. In the Dniester-Bug steppes, graves of the late Yamnaya style continued as late as 2200 BCE.

*Yamnaya in Russia: Ekaterinovka*

The site Ekaterinovka is located in the Southern Steppe on the left bank of the Volga 128 km southwest of the city of Samara. The >45 years old mature male individual included in this study is

- SVP3/I0231 (Ekaterinovka, grave 1, 2910-2875 calBCE, Beta 392487).

*Yamnaya in Russia: Lopatino I*

A large cemetery of 39 kurgans was located on a low terrace beside the Sok River, Samara oblast, Russia (N53°38'24"/E50°39'18"). Eight kurgans were excavated in different years by various teams. Five were constructed in the Yamnaya period and three were added by the MBA Poltavka culture. We included three individuals from this site:

- SVP5/I0357 (kurgan 35, central grave 1, 3090-2910 calBCE, Beta 39248)
  was from a grave that contained the remains of two Yamnaya individuals, including an adult woman and a child, an unusual pair, because 75-80% of individuals in Yamnaya kurgans in the Samara region were adult males.
- SVP38/I0429 (kurgan 31, central grave 1, 3339-2917 calBCE, AA47804)
  was from a grave of an adult male 35-45 years old, 175 cm tall, supine, with a triangular flint projectile point beside him.
- SVP52/I0439 (kurgan 1, central grave 1, 3305-2925 calBCE, Beta 392491)
  was from an adult male 25-35 years old, 178.5 cm tall.

*Yamnaya in Russia: Ishkinovka I*

Ishkinovka (or Ishkinino) is a kurgan cemetery located 30 km north of Novotroitsk on a right-bank tributary of the Ural River, among the easternmost Yamnaya sites, 570 km southeast of Lopatino I. The region contains copper ores exploited by miners who were active throughout the Bronze Age beginning in the EBA, which might explain the Yamnaya occupation of this eastern region, 250 km east of the Yamnaya mines at the Kargaly copper ore field[24]. The male individual included in this study is

- SVP10/I0370 (Kurgan 3, grave 7, 3300-2700 BCE)

*Yamnaya in Russia: Luzhki I*

This unique cemetery, 1-2 km distant from Lopatino I, might represent the undocumented majority of the Yamnaya population that was not buried under kurgans. Individual,

- SVP50/I0438 (3021-2635 calBCE, AA47807)

an adult male 25-35 years old was found in a cemetery with no obvious kurgan, surrounded by the graves of 5 children. He was buried face down with his hands behind his back. He was crippled by a blow with a heavy blunt instrument to his right hip that crushed his femur just below the trochanter, with no healing evident, and his skull was gouged pre-mortem in six places with a serrated tool. His wounds and burial position suggest that he was tortured and executed.

*Yamnaya in Russia: Kurmanaevka III*

This cemetery of three Yamnaya kurgans was erected on the first terrace overlooking the floodplain of the Buzuluk River, a left-bank tributary of the upper Samara River, in western Orenburg oblast, Russia, 170 km southeast of Lopatino I.

- SVP54/I0441 (Kurgan 3, 3010-2622 calBCE, AA47805)
  The sample is taken from the femur of an adult woman 35-45 years old whose bones were heavily stained with red ochre. Her grave is a peripheral grave (burial 2), and is located above the central grave of a male (burial 1).

*Yamnaya in Russia: Lopatino II*

A cemetery of three kurgans was located on the first terrace above the floodplain 2 km northeast of the village of Lopatino on the Sok River in Samara oblast, Russia. One kurgan (3) was assigned to the EBA Yamnaya culture, the second (kurgan 1) to the MBA Poltavka culture, and the third (kurgan 2) to the final MBA II Potapovka culture, related culturally to Sintashta east of the Urals.

- SVP57/I0443 (Grave 1, kurgan 3, 3300-2700 BCE)
  contained a male aged 15-17, apparently killed by an unhealed blunt-force trauma to his right parietal by a hammer-like weapon. Unusually for Yamnaya kurgans, none of the three graves under kurgan 3 were located under the center of the mound, but grave 1 was farthest from the center, near the northern margin.

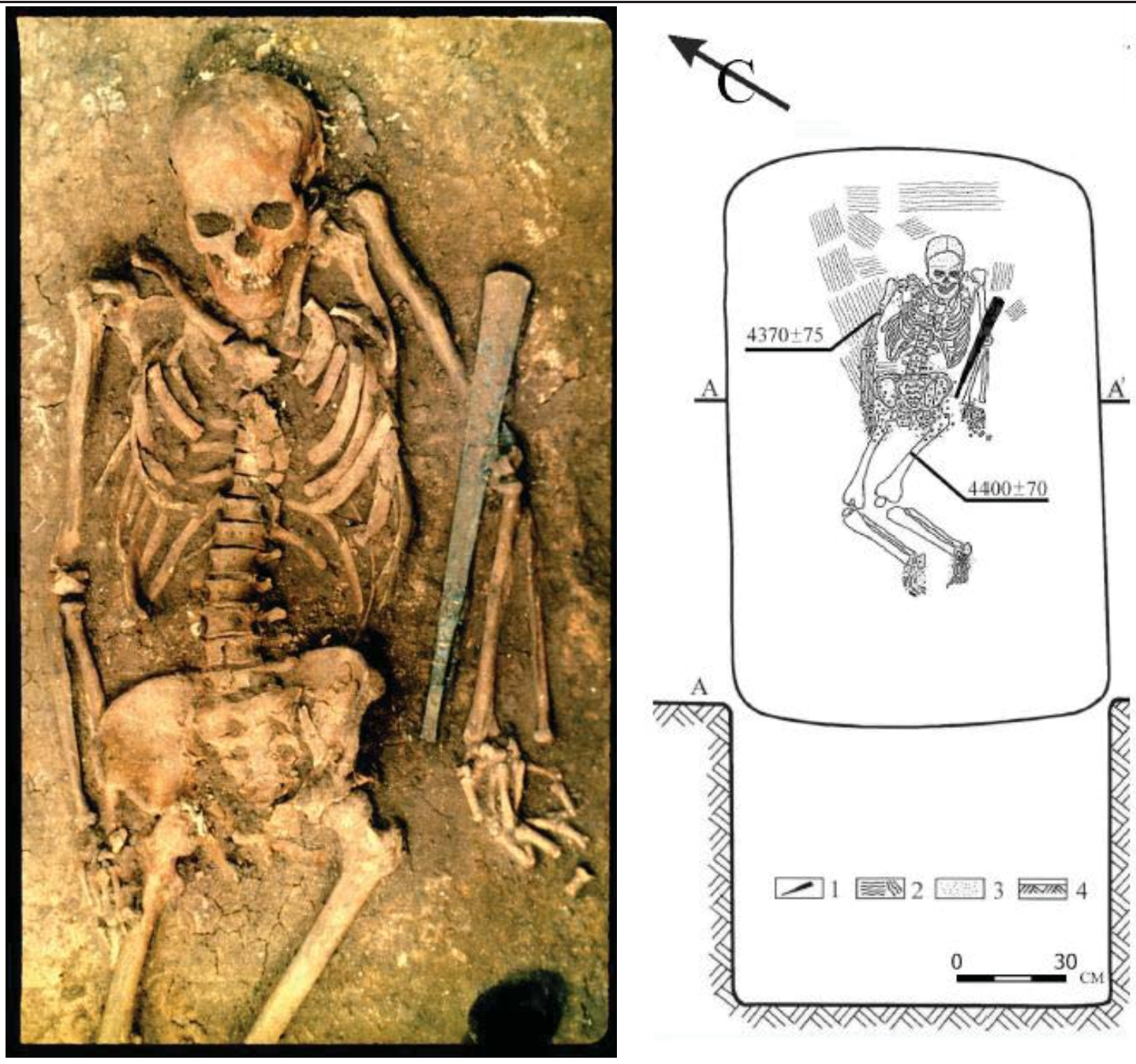

**Figure S3.2. Skeleton SVP58 from grave 1 at Kutuluk kurgan cemetery I[25,26].**

*Yamnaya in Russia: Kutuluk*

Kutuluk kurgan cemetery I, located 60 km east of the city of Samara, contained:

- SVP58/I0444 (central grave 1, kurgan 4, 3335-2881 calBCE, AA12570)

The remains are of male aged 25-35 years (Fig. S3.2), estimated height 176 cm, with no obvious injury or disease, and buried with the largest metal object found in a Yamnaya grave anywhere[26]. The object was a blunt mace 48 cm long, 767 g in weight, cast/annealed and made of pure copper, like most Yamnaya metal objects.

**Late Neolithic**

Since most of our samples were collected in present-day Germany, we follow the German/north-central European chronology for the Late Neolithic. Here, the Late Neolithic horizon in central Europe is defined as 2725–2200 BCE, even though earliest signs of the Corded Ware culture can be found around 2800 BCE, whereas remains of Bell Beaker cultures prevail until 2050 BCE. The Bell Beaker horizon is evident in the archaeological record from 2500 BCE onward in the southern Mittelelbe-Saale (MES), when it starts to move into settlement areas previously occupied by Corded Ware people, the latter of which have cultural affinities to archaeological groups further east such as the Globular Amphora and the Yamnaya culture in the North Pontic and Russian steppes. The settlement density increases during later Bell Beaker phases (2300-2050 BCE), when the Corded Ware is superseded by Bell Beaker elements, often at the same sites. This Late Neolithic Bell Beaker phenomenon is of interest, since the earliest archaeological evidence has been described in the Tagus region of Western Iberia around 2800-2700 BCE and it has been traced archaeologically over large parts of Western Europe but is also found as far as Hungary, Ireland, and southern Scandinavia, and in smaller enclaves in North Africa[27]. Here, earlier Neolithic cultures were overlain by discernable Bell Beaker elements, which became visible in rich grave goods (including the eponymous bell-shaped ceramic beakers). During the transition to the Bronze Age, Early Bronze Age cultural elements of the Únětice culture appear contemporaneously to late elements of the Bell Beaker culture, again sometimes also at the same site. A number of sites, such as Eulau and Quedlinburg show the presence of both Corded Ware and Bell Beaker cultures, and later of Únětice culture, and highlight the cultural diversity and dynamic at the time.

*Corded Ware in Germany: Esperstedt*

The site Esperstedt forms part of large-scale excavations initiated in 2005 in the context of major infrastructural roadworks in Saxony-Anhalt, Germany, to build motorway A38. Individuals from Esperstedt reference site 4 could be unambiguously assigned to the Corded Ware, both by accompanying pottery and characteristic orientation of the burials[28]. Males were usually buried in a right-hand side flexed position with head in the west and facing south, while females were buried on their left-hand side with their head in the east. We studied four individuals from this site:

- ESP11/I0104 (feature 6216, 2473-2348 calBCE, MAMS 21487)
- ESP16/I0103 (feature 6236, 2566-2477 calBCE, MAMS 21488)
- ESP22/I0049 (feature 6233.1, 2454-2291 calBCE, MAMS 21489)
- ESP 26/I0106 (feature 6216, 2454-2291 calBCE, MAMS 21490)

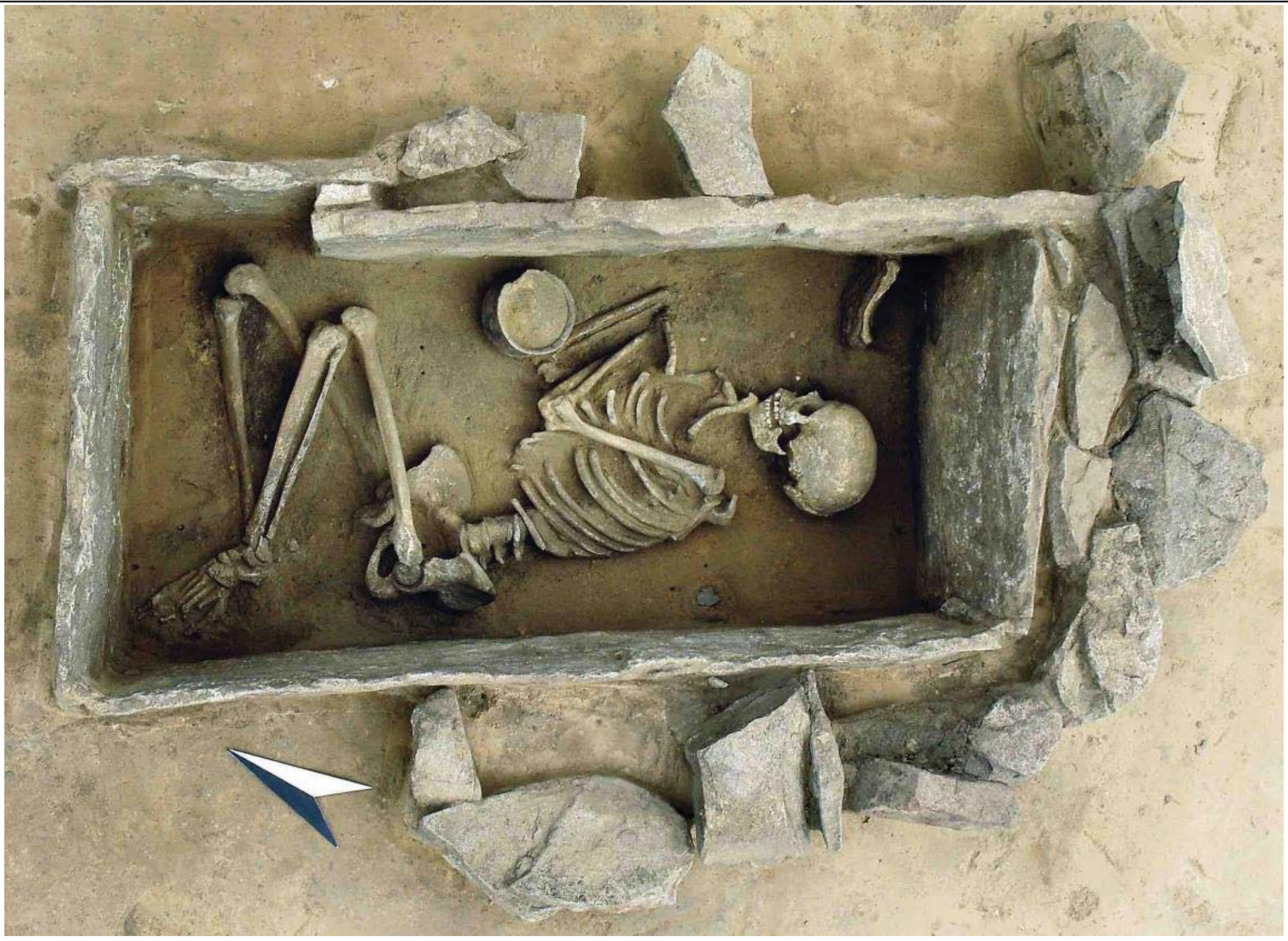

**Figure S3.3. Feature 10049 from Rothenschirmbach, Germany, highlighting the complex grave architecture at the site (Photo: H. Arnold, LDA Sachsen-Anhalt, Germany).**

*Bell Beaker in Germany: Rothenschirmbach*

The site Rothenschirmbach in the district Merseburg-Querfurt is also located on the loess plain of the 'Querfurter Platte'. The site features a densely occupied cemetery from the Bell Beaker period including 14 burials. Females of the Bell Beaker culture were usually buried on their right side with the head in the south and facing east, while males were buried with their heads in the north, facing west. Key features of this cemetery are a high proportion of children's burials and the large number of rich and embellished graves with partially complex architecture (cists, menhirs, e.g. Fig. S3.3), and the rare finding of gold in a man's grave (the oldest gold find in central Germany)[29]. We sampled three individuals from this site: Individual

- ROT3/I0060 (feature 10011, 2294-2206 calBCE, MAMS 22819)
  is a 4-6 year old child (infans I) buried in a stone cist.
- ROT4/I0111 (feature 10142, 2414-2333 calBCE, Er8712)
  featured remnants of a strong wooden cist covered by a 400kg menhir stone of 1.5 x 0.55m in size.
- ROT6/I0108 (feature 10044, 2497-2436 calBCE, Er8710)
  is an adult buried in a open stone cist.

*Bell Beaker and Late Neolithic in Germany: Benzingerode-Heimburg*

- BZH6/I0059 (grave 2, feature 1287, 2286-2153 calBCE, MAMS 21486)
  was attributed to the Bell Beaker period by excavator Tanja Autze as the pit overlaps with a neighbouring grave harboring BBC features. The orientation matches BBC customs, and the radiocarbon date supports assignment to this period.

- BZH4/I0058 (grave 7, feature 4607, 2283-2146 calBCE, MAMS 21491)
  had no grave goods, and the skeleton was oriented SE-NW, rightside-flexed position, facing north.
- BZH12/I0171 (grave 3, feature 2350, 2204-2136 calBCE, KIA27952)
  was buried in SE-NW orientation, in a right-sided flexed position, facing west. While the burial customs do not match standard Bell Beaker or the Corded Ware rites, the radiocarbon date supports a Late Neolithic assignment.

Interestingly, radiocarbon dates of all three individuals fall in the overlapping period of the late Corded Ware, Bell Beaker and early Únětice period in this region[30].

*Bell Beaker and Late Neolithic in Germany: Quedlinburg VII and XIIa*

Reference site XII-West is located northeast of Quedlinburg, Saxony-Anhalt, and forms part of a small graveyard with mixed grave goods from Late Neolithic cultures[31]. The grave

- QUEXII6/I0112 (feature 6256, 2340-2190 calBCE, Er7038)
  was archaeologically assigned to Bell Beaker based on the orientation, however grave goods include both a Bell Beaker 'Füsschenschale' and a 'Corded Ware' cup with ornaments characteristic for the single grave culture. The individual was an outlier from other Bell Beaker samples in PCA, and so we marked it as such for genome-wide analysis.
- QUEXII4/I0113 (feature 6255.1, 2290-2130 cal BCE (Er7283)
  was also a female whose single grave good, a small cup, was uninformative. The orientation of her body is typical for the Bell beaker period.

Another group of six graves was discovered at Quedlinburg reference site VII and attributed to the Bell beaker culture based on form and orientation of the burials[32]. We included a >50 year old male individual buried in an extreme flexed position.

- QLB28/I0806 (feature 19617, 2296-2206 calBCE, MAMS 22820)

*Other Late Neolithic samples in Germany: Karsdorf*

We also included a sample from an unusual burial at the Karsdorf site in Germany

- KAR22/I0550 (feature 00191, date unknown)

The original archaeological attribution to the Baalberge culture is ambiguous and while the grave appears prehistoric, the determination was entirely based its position inside the eastern gate of a burial complex from the Baalberge culture. The mitochondrial genome of KAR22 could be determined as T1a1, matching previous HVS-I based results[1]. By analyzing genomic data KAR22 clusters with Late Neolithic Corded Ware individuals (Figure 2a), while the admixture proportions in Figure 3 reveal a complete absence of an Early Neolithic component (even less than other Corded Ware individuals). Given the large proportion of Yamnaya-related ancestry, a broader group assignment to the Late Neolithic appears likely. Radiocarbon dating results are pending.

*Other Late Neolithic samples in Germany: Alberstedt*

The site Alberstedt in Merseburg-Querfurt, Saxony-Anhalt, on the loess-bearing Querfurter Platte, is located on a hilltop. This promontory site was used by the Corded Ware and Bell Beaker people as a burial ground. The single grave of individual

- ALB3/I0118 (feature 7144.2, 2459-2345 calBCE, MAMS 21492)

was uncovered within a strip of 20m width in preparation for major roadworks. The grave was initially ascribed to the Bell Beaker culture but is rather an unusual burial complex strewn with cattle bones as well as a few sherds of Corded Ware-like pottery in the back filling[28]. The radiocarbon date falls in line with both the Bell Beaker and Corded Ware occupation phases of this region. Given the ambiguous archaeological classification we have decided to use the location and date to classify this sample (Late Neolithic Alberstedt). The intermediate position of this sample on the PCA plot (Figure 2a) between unambiguously assigned Corded Ware (Esperstedt) and Bell Beaker (Rothenschirmbach) individuals, and >50% Yamnaya ancestry are consistent with an individual who has mixed Corded Ware and Bell Beaker ancestry.

**Early Bronze Age**

*Únětice in Germany: Esperstedt*

The reference site 4e at Esperstedt in Merseburg-Querfurt, Saxony-Anhalt/Germany includes a small graveyard of 10 burials that could be assigned to the Early Bronze Age Únětice culture. All individuals are buried on the their right-hand side in flexed position, which is typical for this time period, with the head in the south and facing west. The graves at Esperstedt are slightly tilted towards SE, facing NE. Individuals

- ESP2/I0114 (feature 3340.1, 2131-1979 calBCE, MAMS 21493) and
- ESP29/I0117 (feature 3332/3333, 2199-2064 calBCE, MAMS 21496)

form a small group and appear to be genetically closely related.

- ESP4/I0116 (feature 3322/3323, 2118-1961 calBCE, MAMS 21495)

was also buried as part of a small group of three graves nearby, while individual

- ESP3/I0115 (feature 1559.1, 1931-1780 calBCE, MAMS 21494),

a female adult from a more recent phase at reference site 4b, formed part of a double inhumation with a ~10 year-old child at approximately 100 meters distance from the older graveyard[33].

*Únětice in Germany: Quedlinburg VIII*

The site in Quedlinburg, located in the northern Harz region, in the foreland of the Harz mountains, harbors several graves from the Early Bronze Age. Three adjacent grave groups were found next to the stream Sülze and could be assigned to the Únětice culture. From these, we sampled individual

- QUEVIII6/I0164 (feature 3580, 2012-1919 calBCE, MAMS 21497)

whose otherwise ordinary grave was sealed with a stone cover.

*Únětice in Germany: Eulau*

The site Eulau, Burgenlandkreis, is located in the valley of the Saale river. The promontory that forms "Eulau" is a loess above-gravel formation and was intensively populated during the Early Bronze Age. Individual

- EUL41A/I0803 (feature 882, 2115-1966 calBCE, MAMS 22822)

was buried on the eastern edge of the promontory in a small burial ground of less than 10 burials attributed to the the Early Bronze Age. Individual

- EUL57B/I0804 (feature 1911.5, 2131-1982 calBCE, MAMS 22821)

was buried 200 meters distance away on the western edge of the promontory in another burial ground with approximately 20 features. Feature 1911 is an unusual grave context, including skeletal remains from three juvenile males in the bottom, and a cist in which a child and a neonate had been laid down. The samples taken here are from one of the three male individuals. The graves have been dated to 2200-1550 BCE based on archaeological context.

*Bronze Age in Germany: Halberstadt-Sonntagsfeld, please see also N.B. below*

- I0047/HAL16 (grave 19, feature 613.1, 2022-1937 calBCE, MAMS 21481)

features a skeleton buried in right-handed flexed position, with head in the west, facing south. The grave contained one bone artifact, but no pottery. This individual was originally attributed to the LBK based on the presence of an LBK settlement with associated burials nearby[15], but direct radiocarbon dating newly generated for this study revealed a younger date overlapping with the Bronze Age Únětice culture.

- I0099/HAL36C (grave 40, feature 1114, 1113-1021 calBCE, MAMS 21484)

was buried in right-handed flexed position, head SSW, facing SE. Two decorated LBK pots and two undecorated globular pots were found above the grave but it was not clear whether they were part of the burial or the back filling. Thus, the skeleton was also originally thought to be part of the LBK burial series found at the same site, but subsequent radiocarbon dating performed for this study indicated a much younger date, placing this individual within the Late Bronze Age Urnfield culture of the Mittelelbe-Saale region.

*Nota bene*: To our knowledge, these results are the first published case in which ancient DNA data have been used to identify outlier individuals: individuals who are genetically distinct from others at the same site that have been classified as being from the same archaeological culture (in this case, the Early Neolithic LBK). The fact that this genetic outlier status is consistent with the recent radiocarbon dates and the position of the individuals at the periphery of the site – not directly associated with one of the grave groups that accompany each of the LBK houses (Figure S3.4) – indicates that the genetic analysis is likely to be accurately identifying individuals that are outliers. These results suggest that in the future, genetic analysis may be useful for classifying burials when grave goods and other context are missing or unclear.

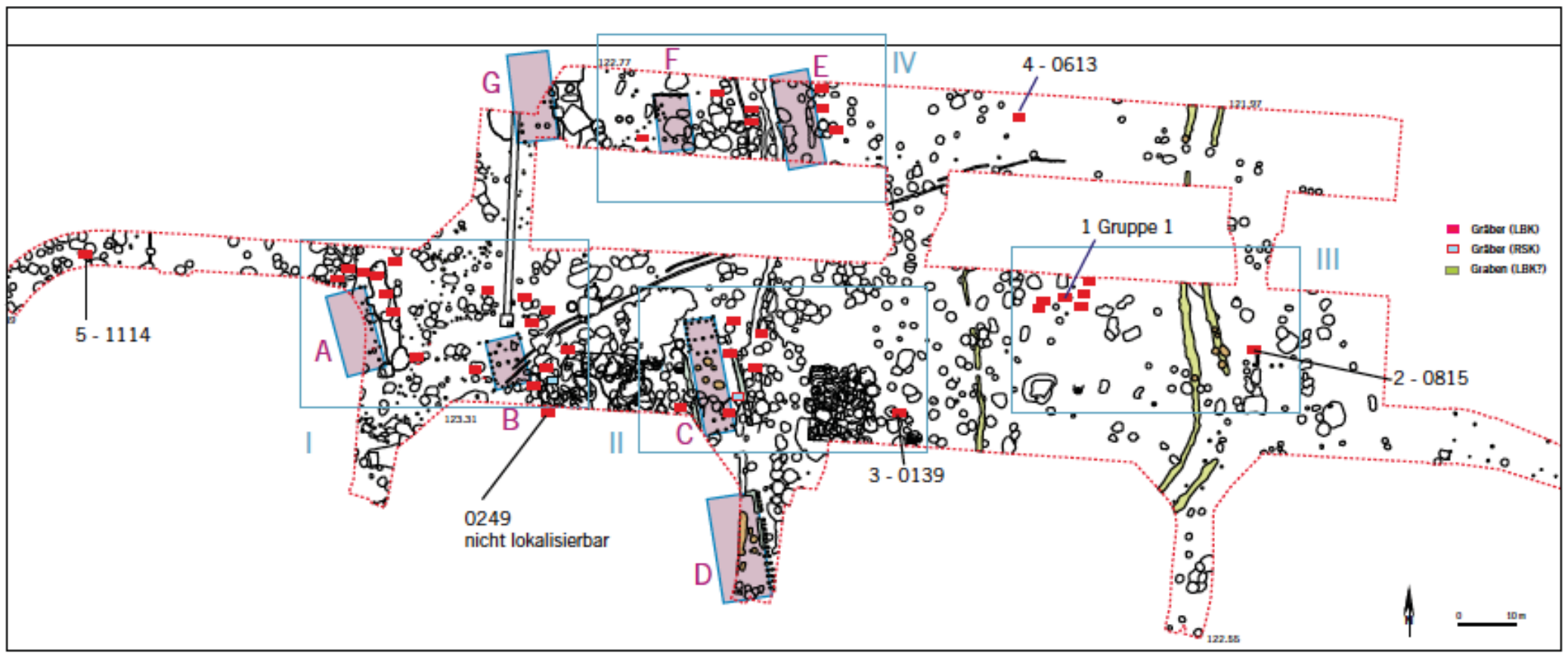

**Figure S3.4. Map of the Halberstadt settlement with groups of graves (red rectangles) associated to LBK houses (purple).** The graves dating to the Bronze Age (features 1114 and 0613) are clearly marked as outliers with 20-30m distances to the LBK graves (red rectangles) (source: LDA Sachsen-Anhalt, Germany).

# Supplementary Information 4

## Sex determination and Y chromosome analysis

Iosif Lazaridis*, Swapan Mallick, Nick Patterson, and David Reich

* To whom correspondence should be addressed (lazaridis@genetics.med.harvard.edu)

### Sex determination

We determined the sex of the 69 individuals newly analyzed in this study by examining the number of reads overlapping targets on 390k capture reagent on the X chromosome (n=1,829) and Y chromosome (n=2,258) SNPs targeted by the 390k capture reagent.

We applied the read mapping and filtering procedure described in the Online Methods section. Table S4.1 gives the number of non-duplicated reads mapping to the X and Y chromosome targets for each individual.

The ratio of Y/(Y+X) reads[1] (Fig. S4.1) shows a bimodal distribution, with 35 individuals having a ratio that is 0.0011 ± 0.0012 (1 standard deviation) who we interpret as female, and 34 individuals having a ratio of 0.508 ± 0.025 who we interpret as male. The remainder of this note focuses on Y chromosome haplogroups for the 34 individuals determined to be male.

**Fig. S4.1: Ratio of Y/(Y+X) alignments.** Samples are sorted on the value of this ratio, demonstrating a discontinuity between females (where the ratio is ~0) and males (where it is ~0.5).

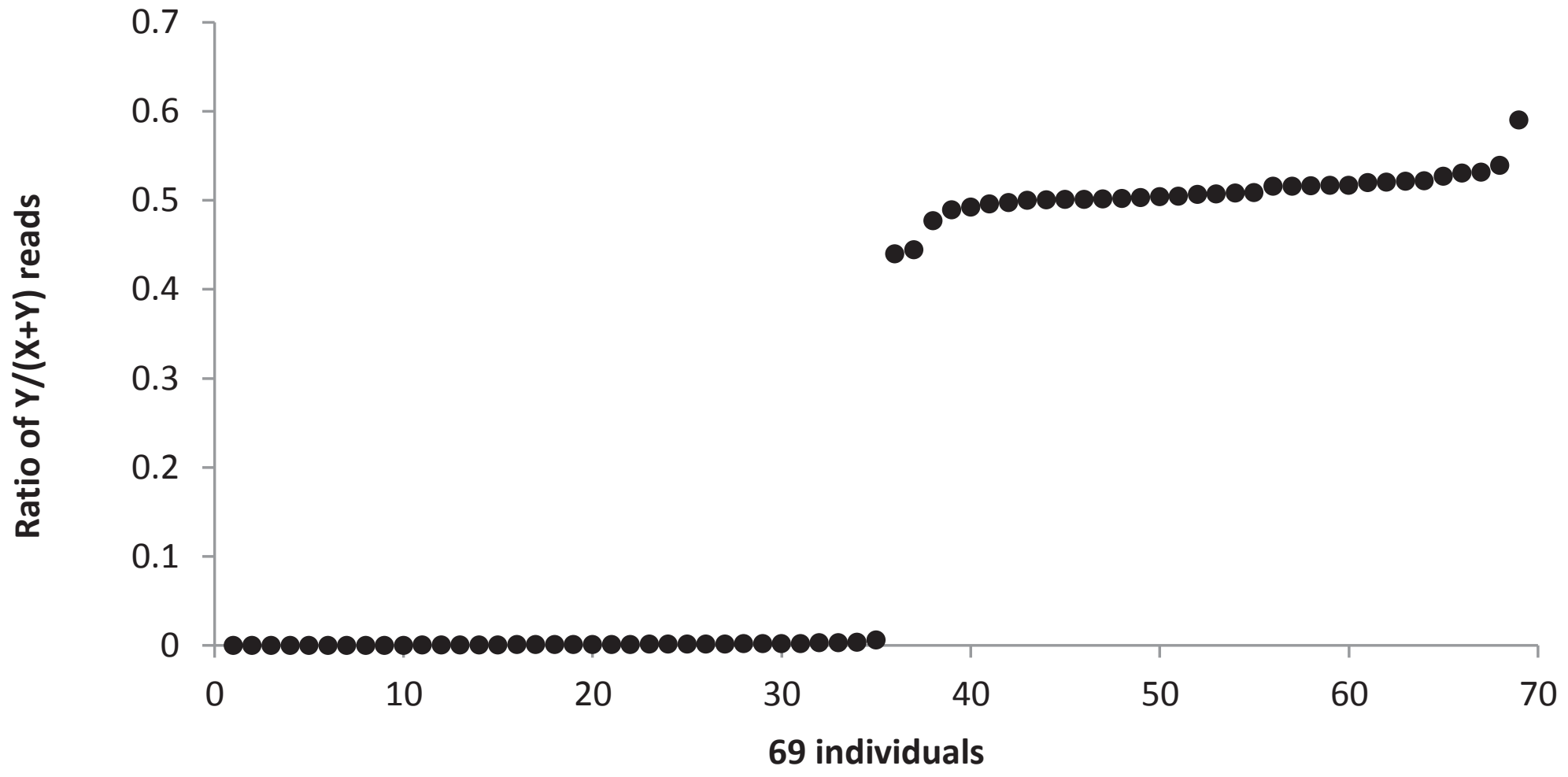

### Y chromosome haplogroup determination.

The 390k capture reagent targeted all SNPs present in the Y-DNA SNP index of the International Society of Genetic Genealogy (ISOGG) version 8.22 as of April 22, 2013 (http://isogg.org/tree/ISOGG_YDNA_SNP_Index.html). At each SNP, we represent the individual using the majority allele (breaking ties randomly) for all non-duplicated reads overlapping the SNP, requiring MAPQ≥30, base quality≥30, and trimming 2 bases at the ends of reads.

**Table S4.1: Number of reads aligning to the X and Y chromosome targets**

| ID | Xreads | Yreads | Sex |
|---|---|---|---|
| I0011 | 1091 | 3 | XX |
| I0012 | 1151 | 1225 | XY |
| I0013 | 898 | 977 | XY |
| I0014 | 1581 | 3 | XX |
| I0015 | 1173 | 1252 | XY |
| I0016 | 1057 | 1145 | XY |
| I0017 | 1621 | 1725 | XY |
| I0022 | 831 | 0 | XX |
| I0025 | 1606 | 3 | XX |
| I0026 | 1662 | 1 | XX |
| I0046 | 2403 | 2 | XX |
| I0047 | 1826 | 1 | XX |
| I0048 | 398 | 401 | XY |
| I0049 | 1083 | 0 | XX |
| I0054 | 3584 | 5 | XX |
| I0056 | 410 | 374 | XY |
| I0057 | 307 | 1 | XX |
| I0058 | 1384 | 1 | XX |
| I0059 | 1506 | 1 | XX |
| I0060 | 257 | 0 | XX |
| I0061 | 2477 | 2490 | XY |
| I0099 | 1620 | 1673 | XY |
| I0100 | 1748 | 1 | XX |
| I0103 | 5231 | 4 | XX |
| I0104 | 3069 | 3081 | XY |
| I0106 | 354 | 0 | XX |
| I0108 | 1738 | 0 | XX |
| I0111 | 1071 | 2 | XX |
| I0112 | 1737 | 2 | XX |
| I0113 | 1050 | 2 | XX |
| I0114 | 752 | 773 | XY |
| I0115 | 665 | 2 | XX |
| I0116 | 1301 | 1321 | XY |
| I0117 | 1933 | 1 | XX |
| I0118 | 3449 | 3 | XX |
| I0124 | 688 | 680 | XY |
| I0164 | 1722 | 1 | XX |
| I0171 | 347 | 0 | XX |
| I0172 | 5264 | 5621 | XY |
| I0174 | 282 | 282 | XY |
| I0176 | 143 | 0 | XX |
| I0231 | 4596 | 4752 | XY |
| I0357 | 1011 | 6 | XX |
| I0370 | 627 | 670 | XY |
| I0405 | 353 | 359 | XY |
| I0406 | 1470 | 1472 | XY |
| I0407 | 1054 | 1 | XX |
| I0408 | 4245 | 5 | XX |
| I0409 | 831 | 1 | XX |
| I0410 | 1172 | 1176 | XY |
| I0411 | 62 | 69 | XY |
| I0412 | 4164 | 4272 | XY |
| I0413 | 1483 | 1 | XX |
| I0429 | 776 | 763 | XY |
| I0438 | 701 | 758 | XY |
| I0439 | 293 | 332 | XY |
| I0441 | 276 | 0 | XX |
| I0443 | 2869 | 2900 | XY |
| I0444 | 590 | 572 | XY |
| I0550 | 277 | 0 | XX |
| I0559 | 214 | 168 | XY |
| I0560 | 692 | 1 | XX |
| I0659 | 642 | 614 | XY |
| I0795 | 111 | 121 | XY |
| I0803 | 714 | 0 | XX |
| I0804 | 50 | 72 | XY |
| I0806 | 243 | 284 | XY |
| I0807 | 105 | 84 | XY |
| I0821 | 155 | 175 | XY |

We performed Y-haplogroup determination by examining the state of SNPs present in ISOGG version 9.129 (accessed Dec 08, 2014); we used this later version—even though it includes many more SNPs than were present in version 8.22 used during the design of the 390k capture reagent—in order to obtain up-to-date Y-haplogroup nomenclature. We determined Y chromosome haplogroups by identifying the most derived Y chromosome SNP in each individual.

.**Table S4.2: Y-Haplogroup assignments for 34 ancient European males.**

| ID | Mean coverage on SNP targets (genome-wide) | Haplogroup | Population |
|---|---|---|---|
| I0559 | 0.20 | R*? | Baalberge_MN |
| I0807 | 0.10 | F* | Baalberge_MN |
| I0806 | 0.31 | R1b1a2a1a2 | Bell_Beaker_LN |
| I0104 | 6.21 | R1a1a1 | Corded_Ware_LN |
| I0172 | 58.03 | I2a1b1a | Esperstedt_MN |
| I0099 | 8.90 | R1a1a1b1a2 | Halberstadt_LBA |
| I0061 | 8.01 | R1a1 | Karelia_HG |
| I0048 | 0.45 | G2a2a | LBK_EN |
| I0056 | 0.53 | G2a2a | LBK_EN |
| I0659 | 0.82 | G2a2a1 | LBK_EN |
| I0795 | 0.15 | T1a | LBK_EN |
| I0821 | 0.18 | G2a2a1 | LBK_EN |
| I0012 | 2.76 | I2c2 | Motala_HG |
| I0013 | 1.89 | I2a1b | Motala_HG |
| I0015 | 3.41 | I2a1 | Motala_HG |
| I0016 | 2.35 | I2a1 | Motala_HG |
| I0017 | 8.79 | I2a1b | Motala_HG |
| I0124 | 0.96 | R1b1 | Samara_HG |
| I0410 | 3.29 | R1b1 | Spain_EN |
| I0412 | 28.17 | I2a1b1 | Spain_EN |
| I0411 | 0.11 | F* | Spain_EN_relative_of_I0410 |
| I0405 | 0.56 | I2a1a1/H2? | Spain_MN |
| I0406 | 9.28 | I2a2a1 | Spain_MN |
| I0174 | 0.36 | H2 | Starcevo_EN |
| I0116 | 3.86 | I2c2 | Unetice_EBA |
| I0804 | 0.07 | I2 | Unetice_EBA |
| I0114 | 1.07 | I2a2 | Unetice_EBA_relative_of_I0117 |
| I0231 | 10.02 | R1b1a2a2 | Yamnaya |
| I0370 | 0.97 | R1b1a2a2 | Yamnaya |
| I0429 | 1.06 | R1b1a2a2 | Yamnaya |
| I0438 | 1.00 | R1b1a2a2 | Yamnaya |
| I0439 | 0.33 | R1b1a | Yamnaya |
| I0443 | 9.10 | R1b1a2a | Yamnaya |
| I0444 | 0.79 | R1b1a2a2 | Yamnaya |

We discuss the specific results for individual samples below.

**I0559 (Baalberge_MN)**

An assignment to haplogroup R is possible based on P224:17285993C→T. This may represent ancient DNA damage, so the assignment should be viewed with caution. The sample can be assigned to the upstream haplogroup P1 (the newly defined parent node of haplogroups Q and R) based on P230:17470112G→A. Downstream haplogroups that could be excluded were R1a1a (M515:14054623T→A), and R1b1a2a1a (L151:16492547C→T), so it is R*(xR1a1a, R1b1a2a1a).

**I0807 (Baalberge_MN)**
The assignment to haplogroup F* is based on a single mutation P316:16839641A→T. We could exclude haplogroups G (Page94:2846401C→T, PF2956:14993358A→G, PF3134:15275200C→G), H1a2a (Z14683:17431284A→C), H1b1 (Z14006:7786084G→T), H1b2a (Z14385:22458616G→C), H3a1 (Z13118:17493503T→C, Z13181:23121729G→A), H3a2 (Z12794:22191274A→C), I1a3 (S243:14401486C→T), I2a1b (M423:19096091G→A), J1a2b3a (L818:19136821A→G), J2a1h2 (L25:19136822T→C), J2b2a1a1 (Z631:2819161G→A), L (M11:21730647A→G), NO (P194:16202980C→G) and P1 (P237:8334875A→G, P240:14598808T→C, P282:18028661A→G).

**I0806 (Bell_Beaker_LN)**
The individual was assigned to haplogroup R1b1a2a1a2 based on mutation P312:22157311C→A. Two Bell Beaker individuals from Kromsdorf, Germany were previously determined[2] to belong to haplogroup R1b.

The individual also has upstream mutations for R1 (P236:17782178C→G), R1b1 (L278:18914441C→T), R1b1a2 (F1794:14522828G→A), and R1b1a2a1 (L51:8502236G→A). Its haplotype is ancestral for R1b1a2a1a2a1a1a (S1217:7193830C→G, Z262:16320197C→T), R1b1a2a1a2c1a (DF49:22735599G→A), R1b1a2a1a2c1a1 (DF23:17774409G→A), R1b1a2a1a2c1f1 (L554:15022777A→G), R1b1a2a1a2c1f2 (S868:19033817T→C), R1b1a2a1a2c1i (CTS6581:16992602T→C) and R1b1a2a1a2c1l1a1 (CTS2457.2:14313081C→T).

**I0104 (Corded_Ware_LN)**
This individual was assigned to haplogroup R1a1a1 based on mutation M417:8533735G→A, and also had six upstream mutations placing it in haplogroup R1a1a (M515:14054623T→A, M198:15030752C→T, L168:16202177A→G, M512:16315153C→T, M514:19375294C→T, and L449:22966756C→T) and four mutations placing it in haplogroup R1a1 (L145:14138745C→A, L62:17891241A→G and L63:18162834T→C, L146:23473201T→A).

We could further exclude lineages R1a1a1a (CTS7083:17275047T→G) and R1a1a1b (S441:7683058G→A, S224:8245045C→T), so I0104 could be more precisely described as R1a1a1*(x R1a1a1a, R1a1a1b).

Three related individuals belonging to haplogroup R1a were previously described from Corded Ware individuals from Eulau, Germany[3]. The R-M417 haplogroup has an estimated TMRCA of ~5,800 years ago[4], which indicates that I0104 lived about 1.5ky after the foundation of this lineage from which the vast majority of modern R1a-related chromosomes from across Eurasia are descended. More than 96% of modern European R-M417 Y-chromosomes belong[4] to lineage R1a1a1b1a-Z282 which can be excluded for individual I0104, both indirectly, based on its ancestral state for the upstream mutation defining haplogroup R1a1a1b, and directly, as the individual was ancestral for the Z282 polymorphism itself (15588401T→C). Thus, I0104 was related to the modern European set of R1a Y-chromosomes but did not belong to the more dominant group within this set.

**I0172 (Esperstedt_MN)**

This individual was assigned to haplogroup I2a1b1a based on mutation L1498:18668472C→T, and could also be assigned to upstream haplogroups I2a1b1 (L161.1:22513718C→T) and I2a1b (CTS1293:7317227G→A, CTS1802:14074218A→T, L178:15574052G→A, CTS8239:17893806A→G, M423:19096091G→A, CTS11030:22905944G→C).

Thus, I0172 belonged to a more derived clade than the ~8,000-year old Loschbour male[5] from Luxembourg, and may represent a hunter-gatherer Y-chromosomal lineage that was incorporated in the population of Middle Neolithic farmers from Germany. Haplogroup I2a lineages were also detected in Swedish hunter-gatherers[5,6] from 7-5 thousand years ago, an early Hungarian individual (~5,700 years cal BC) with a "hunter-gatherer" autosomal makeup that belonged to an early farmer community[7], as well as later ~5,000 year old individuals from Treilles, France[8], while haplogroup I lineages were observed in two early Neolithic farmers from Hungary belonging to the early Neolithic Trans-Danubian Linear Pottery (LBKT) and Starcevo cultures[9]. It thus appears that there was gene flow from male hunter-gatherers into the Early and Middle Neolithic farmers across Europe.

**I0099 (Halberstadt_LBA)**

This single individual from the Late Bronze Age could be assigned to haplogroup R1a1a1b1a2 based on S204:16474793G→A, and to upstream R1a1a1b1a (S198:15588401T→C), and R1a1a1b1 (PF6217:21976303T→A), and R1a1a1b (S224:8245045C→T, S441:7683058G→A). It is thus more derived than the earlier Corded Ware I0104 individual and belongs firmly within the present-day European variation of R1a Y-chromosomes.

**I0061 (Karelia_HG)**

In contrast to I0104 and I0099, the hunter-gatherer from Karelia could only be assigned to haplogroup R1a1 (M459:6906074A→G, Page65.2:2657176C→T) and the upstream haplogroup R1a (L145:14138745C→A, L62:17891241A→G, L63:18162834T→C, L146:23473201T→A). It was ancestral for the downstream clade R1a1a (M515:14054623T→A, M198:15030752C→T, M512:16315153C→T, M514:19375294C→T, L449:22966756C→T). Thus, it can be designated as belonging to haplogroup R1a1*(xR1a1a) and it occupied a basal position to the vast majority of modern Eurasian R1a-related Y-chromosomes[4], although more basal (R1a-M420*) Y-chromosomes have been detected in Iran and eastern Turkey[4]. Overall, our detection of haplogroup R1a1 in a northwest Russian hunter-gatherer establishes the early presence of this lineage in eastern Europe, and is consistent with a later migration from eastern Europe into central Europe which contributed such haplogroups to the Corded Ware population.

**I0048 (LBK_EN)**

This individual could be assigned to haplogroup G2a2a (PF3185:22894488C→T) and to upstream haplogroup G2a (L31:14028148C→A). It was ancestral for haplogroup G2a2a1b (L91:21645555G→C), so it could be designated G2a2a*(xG2a2a1b).

Haplogroup G2a has been found in early Neolithic farmers from Germany[10], Hungary[9], the Tyrolean Iceman[11], ~5,000 year old farmers from France[8], and ~7,000 year old ones from Spain[12]. It it thus a link between Early Neolithic farmers of central Europe and the Mediterranean, as its presence[13] in modern Sardinians, a population with known links to the early European farmers[5,11,14,15] also suggests.

**I0056 (LBK_EN)**
This individual also belonged to haplogroup G2a2a (PF3147:7738069G→A, PF3175:18962113C→T, PF3181:21808944C→A), but not to G2a2a1a (M286:22741799G→A) or G2a2a1b1 (FGC5668:22467833A→G), so it could be designated G2a2a*(xG2a2a1a, G2a2a1b1).

**I0659 (LBK_EN)**
This individual also belonged to haplogroup G2a2a1 (PF3170:18090604G→A), and to upstream haplogroup G2a2a (PF3151:9785736A→G, PF3161:15702713A→C, PF3175:18962113C→T, PF3184:22576860C→T, PF3185:22894488C→T), but not to downstream haplogroup G2a2a1 (PF3177:21327198C→T). Thus, this individual carried the derived state for one of the SNPs defining haplogroup G2a2a1 (PF3170), and the ancestral state for another (PF3177), suggesting that the first of these mutations occurred before the second.

**I0795 (LBK_EN)**
This individual belonged to haplogroup T1a (PF5604:7890461C→T, M70:21893881A→C). This is the first instance of this haplogroup in an ancient individual that we are aware of and strengthens the case for the early Neolithic origin of this lineage in modern Europeans[16], rather than a more recent introduction from the Near East where it is more abundant today.

**I0821 (LBK_EN)**
This individual belonged to haplogroup G2a2a1 (PF3155:14006343T→C) and also to upstream haplogroup G2a2a (PF3166:16735582T→G), and G2a2 (CTS4367:15615340C→G). We could exclude downstream haplogroup G2a2a1b (L91:21645555G→C), so it could be designated G2a2a1*(xG2a2a1b).

**I0012 (Motala_HG)**
This is the Motala2 individual whose shotgun data was previously analyzed[5]. It belonged to haplogroup I2c2 (PF3827:22444389T→A), with the upstream haplogroup I2c (L597:18887888T→A) also supported. Our higher coverage capture data defines its haplogroup more precisely than the I haplogroup previously reported[5].

**I0013 (Motala_HG)**
This is the Motala3 individual whose shotgun data was previous analyzed[5]. It belonged to haplogroup I2a1b (M423:19096091G→A), consistent with the previous analysis. Haplogroup I2a1b1 (L161.1:22513718C→T) could be excluded, and it could thus be designated I2a1b*(xI2a1b1). The analysis of the shotgun data[5] could reject M359.2 (previously listed as I2a1b1, but currently designated as under "Investigation" by ISOGG), and I2a1b2-L621 (previously listed as I2a1b3).

**I0015 (Motala_HG)**
This is the Motala6 individual whose shotgun data was previous analyzed[5]. It belonged to haplogroup I2a1 (P37.2:14491684T→C) and the upstream haplogroup I2a (L460:7879415A→C). Its phylogenetic position could not be determined in the previous analysis of the shotgun data. We also find that it was ancestral for I2a1a (L159.1:15810964T→G, M26:21865821G→A and L158:23496560G→A), I2a1b (M423:19096091G→A), I2a1c (L233:14487362G→A), and I2a1e (L1294:2887401T→C), so it could be designated I2a1*(xI2a1a, I2a1b, I2a1c, I2a1e).

**I0016 (Motala_HG)**
This is the Motala9 individual whose shotgun data was previous analyzed[5]. It belonged to haplogroup I2a1a1a (L672:22228628T→A) and the upstream haplogroups I2a1 (P37.2:14491684T→C) and I2a (L460:7879415A→C), so it could be better resolved than in the analysis of the shotgun data which could only designated it as I*(xI1). It was also ancestral for I2a1a1a1a1b (Z118:20834727A→G), and could be designated I2a1a1a*(xI2a1a1a1a1b).

**I0017 (Motala_HG)**

This is the Motala12 individual whose shotgun data was previous analyzed[5]. It could be assigned to haplogroup I2a1b2a1 (L147.2:6753258T→C) and also the upstream haplogroup I2a1b (L178:15574052G→A, M423:19096091G→A). However, in the shotgun data haplogroup I2a1b2-L621 (previously known as I2a1b3) could be excluded based on mutation L621:1876008G→A, which is inconsistent with this individual carrying the derived state for haplogroup I2a1b2a1. We do not have a call for L621 in the capture data, so we are certain only of this individual's assignment to haplogroup I2a1b.

To summarize the data from the five Motala males, we could assign 4 of 5 males to haplogroup I2a1 and its subclades and one (Motala2/I0012) to haplogroup I2c2.

**I0124 (Samara_HG)**

The hunter-gatherer from Samara belonged to haplogroup R1b1 (L278:18914441C→T), with upstream haplogroup R1b (M343:2887824C→A) also supported. However, he was ancestral for both the downstream haplogroup R1b1a1 (M478:23444054T→C) and R1b1a2 (M269:22739367T→C) and could be designated as R1b1*(xR1b1a1, R1b1a2). Thus, this individual was basal to most west Eurasian R1b individuals which belong to the R-M269 lineage as well as to the related R-M73/M478 lineage that has a predominantly non-European distribution[17]. The occurrence of chromosomes basal to the most prevalent lineages within haplogroups R1a and R1b in eastern European hunter-gatherers, together with the finding of basal haplogroup R* in the ~24,000-year old Mal'ta (MA1) boy[18] suggests the possibility that some of the differentiation of lineages within haplogroup R occurred in north Eurasia, although we note that we do not have ancient DNA data from more southern regions of Eurasia. Irrespective of the more ancient origins of this group of lineages, the occurrence of basal forms of R1a and R1b in eastern European hunter-gatherers provide a geographically plausible source for these lineages in later Europeans where both lineages are prevalent[4,17,19].

**I0410 (Spain_EN)**

We determined that this individual belonged to haplogroup R1b1 (M415:9170545C→A), with upstream haplogroup R1b (M343:2887824C→A) also supported. However, the individual was ancestral for R1b1a1 (M478:23444054T→C), R1b1a2 (PF6399:2668456C→T, L265:8149348A→G, L150.1:10008791C→T and M269:22739367T→C), R1b1c2 (V35:6812012T→A), and R1b1c3 (V69:18099054C→T), and could thus be designated R1b1*(xR1b1a1, R1b1a2, R1b1c2, R1b1c3).

The occurrence of a basal form of haplogroup R1b1 in both western Europe and R1b1a in eastern Europe (I0124 hunter-gatherer from Samara) complicates the interpretation of the origin of this lineage. We are not aware of any other western European R1b lineages reported in the literature before the Bell Beaker period (ref. [2] and this study). It is possible that either (i) the Early Neolithic Spanish individual was a descendant of a Neolithic migrant from the Near East that introduced this lineage to western Europe, or (ii) there was a very sparse distribution of haplogroup R1b in European hunter-gatherers and early farmers, so the lack of its detection in the published literature may reflect its occurrence at very low frequency.

The occurrence of a basal form of R1b1 in western Europe logically raises the possibility that present-day western Europeans (who belong predominantly to haplogroup R1b1a2-M269) may trace their origin to early Neolithic farmers of western Europe. However, we think this is not likely given the existence of R1b1a2-M269 not only in western Europe but also in the Near East; such a distribution implies migrations of M269 males from western Europe to the Near East which do not seem archaeologically plausible. We prefer the explanation that R-M269 originated in the eastern end of its distribution, given its first appearance in the Yamnaya males (below) and in the Near East[17].

**I0412 (Spain_EN)**
We determined that this individual belonged to haplogroup I2a1b1 (L161.1:22513718C→T), with upstream haplogroup I2a1b also supported (CTS1293:7317227G→A, L178:15574052G→A, M423:19096091G→A). Haplogroup I-L161.1 has not been studied in representative samples of modern Europeans to our knowledge. A project devoted to this haplogroup in the genetic genealogy community suggests a relatively high (but not exclusive) occurrence in the present-day British Isles (https://www.familytreedna.com/public/I2a-L161/; administered by Robert Gabel (Ulrich); accessed Dec. 09, 2014). This may be a hunter-gatherer lineage that was absorbed by early farmers of western Europe, as its present-day distribution and discovery in an early Neolithic Iberian suggest.

**I0411 (Spain_EN_relative_of_I0410)**
This individual is not included in the Spain_EN sample as we that determined it was a relative to individual I0410 and we retained I0410 because of the latter's better quality. We could only assign it to haplogroup F based on mutation P135:21618856C→T. Of the known subclades of F, it was found to be ancestral for haplogroup G (F1551:9448354A→G), I1 (M450:7548915G→A), I2a (S247:15224591G→A), J (CTS26:2675457A→T, YSC0000228:22172960G→T), L1b2 (M274:22737801C→T), T (PF5607:8459278G→A, CTS5268:16174116C→T, CTS7749:17644174C→T), O2b (M176:2655180G→A), Q1a2a (L475:18146921G→A), Q1b1 (FGC1861:21365952G→A), R1a1a (L449:22966756C→T) and R1b1c2 (V35:6812012T→A).

**I0405 (Spain_MN)**
This individual was assigned to haplogroup I2a1a1 (L672:22228628T→A). We note, however, that haplogroup H2 is also supported (L279:6932824G→T, L285:21869856C→T). Given the occurrence of haplogroup I2a chromosomes in many individuals from prehistoric Europe (this study and ref.[5-9]), assignment to I2a1a1 seems plausible. Haplogroup I-L672 is nested within haplogroup I-M26, which is rare in Europe today except in Sardinians (40.9%) and other populations from Southwestern Europe[20]. Haplogroup H2 is detected in an early Neolithic Starcevo individual (I0174, below), so we cannot determine this individual's haplogroup with certainty.

**I0406 (Spain_MN)**
This individual was assigned to haplogroup I2a2a1 (CTS9183:18732197A→G) with upstream haplogroup I2a2a also supported (L368:6931594C→T, L34:7716262A→C, P221:8353707C→A, P223:16699334C→G, P222:18888200C→G, M223:21717307G→A and P220:24475669G→T). We could exclude downstream haplogroups I2a2a1a1a (L1195:18865320G→A), I2a2a1b1 (L702:7629205C→T), I2a2a1b2a (L801:21763755A→C), and I2a2a1b2b (L147.3:6753258T→C), so this individual could be designated I2a2a1*(x I2a2a1a1a, I2a2a1b1, I2a2a1b2a, I2a2a1b2b).

Haplogroup I2a2a1 is nested within haplogroup I2a2a-M223, previously designated I1c, which occurs at low frequency throughout Europe[20], and represents another European hunter-gatherer lineage in the Middle Neolithic farmers of Spain.

**I0174 (Starcevo_EN)**
This individual was assigned to haplogroup H2 (L281:8353840T→G). Upstream haplogroup F was also supported (P142:7218079G→A, P145:8424089G→A, P138:14199284T→C,

P316:16839641A→T, P14:17398598C→T, P159:18097251C→A). An individual bearing mutation P96 which also defines haplogroup H2 was found in the Netherlands[21]; while haplogroup H is rare in present-day Europeans, its discovery in I0174 suggests that it was present in Neolithic Europe.

**I0116 (Unetice_EBA)**
This individual was assigned to haplogroup I2c2 (PF3827:22444389T→A) and upstream haplogroups I2c (L597:18887888T→A), I2 (M438:16638804A→G) were also supported.

**I0804 (Unetice_EBA)**
This individual was assigned to haplogroup I2 (L68:18700150C→T) with upstream haplogroup F also supported (P158:17493513C→T).

**I0114 (Unetice_EBA_relative_of_I0117)**
This individual was initially assigned to haplogroup I2a2a (L368:6931594C→T) with upstream haplogroups I2a2 (L181:19077754G→T, P218:17493630T→G, P217:7628484C→T) and I2 (M438:16638804A→G) also supported. However, the sample was ancestral for other mutations defining haplogroup I2a2a (L34:7716262A→C, P223:16699334C→G, M223:21717307G→A), so it is possible that either it represents a branch of the Y-chromosome phylogeny that possessed the L368 but not the L34, P223, and M223 mutations, or that the derived L368C→T represents ancient DNA damage. We thus assign it only to haplogroup I2a2.

**I0231 (Yamnaya)**
This individuals was assigned to haplogroup R1b1a2a2 (CTS1078:7186135G→C, Z2105:15747432C→A) with upstream haplogroups R1b1a2a (L23:6753511G→A), and R1b1a2 (PF6399:2668456C→T, L265:8149348A→G, PF6434:8411202A→G, L150.1:10008791C→T, PF6482:18381735A→G, M269:22739367T→C) also supported. It was ancestral for R1b1a2a2a (L584:28731917C→T), so it could be designated R1b1a2a2*(xR1b1a2a2a).

Notably, the individual did not belong to haplogroup R1b1a2a1-M412 (8502236G→A), and it has been observed that R-L23*(xM412) chromosomes "often exceed 10% frequency in the Caucasus, Turkey and some SE Europe and Circum-Uralic populations" but "they typically display frequencies ≤5% in Western Europe (except for an instance of 27% in Switzerland's Upper Rhone Valley) in contrast to the prominent spread of derived M412 varieties in West Europe (Figure 1f)." (ref. [17]). A study of modern Armenians[22] reports a frequency of ~28% of L23 in modern Armenians while noting that "The derived M412 allele, which is found in nearly all haplogroup R1b1b1*-L23 chromosomes in Europe, is absent in the sampled Armenians, which also exhibit a scarcity of haplotype sharing with Europeans, suggesting a limited role for Armenians in the introduction of R1b into Europe." Moreover, results from the Armenian DNA Project (https://www.familytreedna.com/public/ArmeniaDNAProject/default.aspx?section=ysnp; administered by Hovann Simonian, Mark Arslan, and Peter Hrechdakian; accessed Dec 09, 2014) indicate the presence of the Z2103 derived state in many modern Armenians.

It is not possible to determine whether the appearance of R-Z2103 in the Yamnaya individual is due to (i) gene flow from the south to the steppe and related to the autosomal signal of "dilution" of Eastern European hunter-gatherers, or (ii) gene flow from the steppe to the south. Modern Armenians have a signal of admixture from the Yamnaya, as when we test $f_3$-statistics of the form $f_3$(Armenian; Yamnaya, $X$) we find the lowest Z-score for $f_3$(Armenian; Yamnaya, BedouinB) = -0.00296 (Z=-7.1). However, the lowest Z-score of statistics of the form $f_3$(Armenian; $X$, $Y$) involves the ($X$, $Y$) = (LBK_EN, Sindhi) pair (value -0.00575, Z=-15.3), so the signal of admixture from the Yamnaya is not the strongest one for Armenians. Moreover, as shown in SI 7, the Yamnaya have a negative $f_3$-statistic with ($X$, $Y$) = (Karelia_HG, Armenian). A negative statistic for both Armenians and Yamnaya with each other as a reference population may suggest that a third (unsampled) population admixed into both the Yamnaya and to Armenians. The question of directionality can only be furthered elucidated by the study of additional ancient samples from the Caucasus, Near East and the steppe.

**I0370 (Yamnaya)**
This individual was assigned to haplogroup R1b1a2a2 (CTS1078/Z2103:7186135G→C), with upstream haplogroups R1b1a2 (M269:22739367T→C, L150.1:10008791C→T), R1b1a (L320:4357591C→T) also supported.

**I0429 (Yamnaya)**
This individual was assigned to haplogroup R1b1a2a2 (Z2105:15747432C→A) and to the upstream haplogroups R1b1a2a (L23:6753511G→A) and R1b1a2 (L150.1:10008791C→T, M269:22739367T→C). It was ancestral for R1b1a2a2a (L584:28731917C→T) and so could be designated R1b1a2a2*(x R1b1a2a2a).

**I0438 (Yamnaya)**
This individual could also be assigned to haplogroup R1b1a2a2 (Z2105:15747432C→A). It could also be assigned to the upstream haplogroups R1b1a2a (L23:6753511G→A), R1b1a (L320:4357591C→T). It was ancestral for R1b1a2a2a (L584:28731917C→T), and R1b1a2a2c (CTS7822:17684699A→T), so it could be designated R1b1a2a2*(xR1b1a2a2a, R1b1a2a2c).

**I0439 (Yamnaya)**
This individual could be assigned to haplogroup R1b1a (P297:18656508G→C), with upstream haplogroup R1 (M173:15026424A→C, M306:22750583C→A) also supported. It was ancestral for haplogroup R1b1a2a1 (L51:8502236G→A) and so could be designated R1b1a*(xR1b1a2a1).

**I0443 (Yamnaya)**
This individual could only be assigned to haplogroup R1b1a2a (L49.1:2842212T→A, L23:6753511G→A). It could also be assigned to the upstream haplogroups R1b1a2 (PF6399:2668456C→T, L150.1:10008791C→T, L1353:19179540G→A, PF6509:22190371A→G, M269:22739367T→C, CTS12478:28590278G→A). The individual was ancestral for haplogroup R1b1a2a1 (L51/M412:8502236G→A) and, unlike I0231, I0370 and I0438 also for R1b1a2a2 (Z2105:15747432C→A). Thus, it could be designated as R1b1a2a*(xR1b1a2a1, R1b1a2a2).

**I0444 (Yamnaya)**
The individual could be assigned to haplogroup R1b1a2a2 (CTS1078/Z2103:7186135G→C) and also to the upstream haplogroups R1b1a2 (L150.1:10008791C→T) and R1b1 (M415:9170545C→A).

Summarizing the results from the Yamnaya males, all seven belonged to haplogroup R1b1a. Six of these could be further assigned to haplogroup R1b1a2a, and five of these to haplogroup R1b1a2a2. The uniformity of R1b Y-chromosomes in this sample suggests a patrilineal organization of the Yamnaya, or at least of the people who were given expensive Kurgan burials. We cannot exclude the presence of other haplogroups in the general population, or in other individuals located elsewhere in the expansive Yamnaya horizon[23]. We also emphasize the absence of M412 (the dominant lineage within haplogroup R-M269 in Europe) in this sample, as well as the absence of the R1a haplogroup which was detected in the Corded Ware and Late Bronze Age Halberstadt individual from central Europe. A survey of other European steppe groups may reveal the more immediate patrilineal kin of the major founding lineages of modern European R1a and R1b chromosomes.

**Discussion**

In Table S4.3 we summarize results from previouslys of 61 ancient European and 25 ancient Siberian/Central Asian Y-chromosomes >1,000BCE. In combination with our own results (Table S4.2), this summary makes it clear that only a single R1b Y-chromosome has been found in 70 ancient Europeans outside Russia (1.4%) before the Late Neolithic period. In contrast, all 9 ancient individuals (100%) from Russia belonged to haplogroups R1a and R1b, and 18/23 (78%) Bronze Age individuals from Central Asia/Siberia belonged to haplogroup R1a. In Europe except Russia, both R1a and R1b have been found in the Late Neolithic and Bronze Age periods for a combined frequency of 6/10 (60%). Present-day Europeans have high frequencies[4,17] of haplogroups R1a and R1b.

Thus, it appears that before ~4,500 years ago, the frequency of R1a and R1b in Europe outside Russia was very low, and it rose in the Late Neolithic/Bronze Age period. The young, star-like phylogenies of these two haplogroups[24] also suggest relatively recent expansions. The ubiquity of these haplogroups in Russia, Siberia, and Central Asia suggest that their rise in Europe was likely to have been due to a migration from the east, although more work is needed to trace these migrations and also to correlate them with regions of the world that have not yet been studied with ancient DNA (such as southern Europe, the Caucasus, the Near East, Iran, and Central and South Asia). Nonetheless, the Y-chromosome results suggest the same east-to-west migration as our analysis of autosomal DNA.

**Table S4.3: Previous Y-chromosome studies on ancient Europeans, Siberians and Central Asians >1,000BCE.** We do not include the Motala individuals from ref.[5] in this tabulation as they are included in Table S4.2. We use haplogroup nomenclature from ISOGG version 9.129 and note the terminal derived mutation assessed for each sample.

| First author | Year | N | Country | Period | Haplogroup |
|---|---|---|---|---|---|
| EUROPE | | | | | |
| Seguin-Orlando[25] | 2014 | 1 | Russia | Upper Paleolithic H/G | C1- F3393 |
| Haak[3],[+] | 2008 | 1 | Germany | Late Neolithic | R1a1-SRY10831.2 |
| Haak[10] | 2010 | 1 | Germany | Early Neolithic | G2a2b-S126 |
| | | 2 | Germany | Early Neolithic | F-M89 |
| Lacan[12],[*] | 2011 | 5 | Spain | Early Neolithic | G2a-P15 |
| | | 1 | Spain | Early Neolithic | E1b1b1a1b1a-V13 |
| Lacan[8],[**] | | 20 | France | Middle Neolithic | G2a-P15 |
| | | 2 | France | Middle Neolithic | I2a1-P37.2 |
| Lacan[26],[***] | 2011 | 2 | France | Middle Neolithic | I2a1 |
| Lee[2] | 2012 | 1 | Germany | Late Neolithic | R1b1a2-M269 |
| | | 1 | Germany | Late Neolithic | R1b-M343 |
| Lazaridis[5] | 2014 | 1 | Luxembourg | Mesolithic H/G | I2a1b-L178 |
| Olalde[27] | 2014 | 1 | Spain | Early Neolithic | C1a2-V20 |
| Skoglund[6] | 2014 | 1 | Sweden | Neolithic H/G | I2a1-P37.2 |
| Gamba[7] | 2014 | 2 | Hungary | Early Neolithic | I2a-L460 |
| | | 2 | Hungary | Early Neolithic | C1a2-V20/V184 |
| | | 1 | Hungary | Late Bronze Age | J2a1-S57 |
| Szécsényi-Nagy[9] | 2014 | 5 | Hungary | Early Neolithic | G2a-P15 |
| | | 4 | Hungary | Early Neolithic | G2a2b-S126 |
| | | 3 | Hungary | Early Neolithic | F-M89 |
| | | 1 | Hungary | Early Neolithic | I2a1-P37.2 |
| | | 1 | Hungary | Early Neolithic | I-M170 |
| | | 1 | Hungary | Early Neolithic | I1-M253 |
| Keller[11] | 2012 | 1 | Italy | Middle Neolithic | G2a2a1b-L91 |
| SIBERIA & CENTRAL ASIA | | | | | |
| Fu | 2014 | 1 | Russia | Upper Paleolithic H/G | K |
| Raghavan[18] | 2014 | 1 | Russia | Upper Paleolithic H/G | R |
| Keyser[28] | 2009 | 7 | Russia | Bronze Age | R1a1a-M17 |
| | | 1 | Russia | Bronze Age | C |
| Hollard[29] | 2014 | 4 | Mongolia | Bronze Age | R1a1a1b2-Z93 |
| | | 3 | Mongolia | Bronze Age | Q1a2a1-L54 |
| | | 1 | Mongolia | Bronze Age | C-M130 |
| Li[30] | 2010 | 7 | China | Bronze Age | R1a1a-M198 |

[+] Three individuals belonged to this haplogroup but the individuals were related to each other.

[*] One of these 5 individuals was inferred to belong to haplogroup G2a based on its Y-STR profile.

[**] Three of these 20 individuals were inferred to belong to haplogroup G2a based on Y-STR profile.

[***] These individuals from the dolmen of la Pierre Fritte were inferred to belong to haplogroup I2a1 based on Y-STR profile.

# Supplementary Information 5
## Principal Component Analysis

Iosif Lazaridis* and David Reich

* To whom correspondence should be addressed (lazaridis@genetics.med.harvard.edu)

In Fig. 2a we show the PCA analysis that was conducted using *smartpca* from the EIGENSOFT[1] software package on 777 present-day West Eurasians[2] and 92 ancient individuals at 66,225 transversion SNPs.

We excluded transition SNPs in this analysis as non-UDG-treated ancient DNA libraries from the literature had high rates of damage causing apparent C→T and G→A substitutions, which potentially leads to misleading results, for example by causing ancient DNA samples to appear as extreme outliers, dominating one the first two principal components.

To avoid visual clutter in Fig. 2a, we represented modern individuals as grey dots, and used colored and labeled symbols to represent the ancient individuals.

In Figure S5.1 we show the complement of Fig. 2a by color coding modern individuals and representing ancient ones as grey dots.

In Fig. S5.2 we show a PCA over all 354,212 SNPs, with ancient individuals projected (using the *lsqproject: YES* option) onto the variation of the modern ones. Topologically, this is very similar to Fig. 2a. An interesting aspect of Fig. S5.2 is that while most clusters are preserved, there is a spurious clustering of MA1 with the Yamnaya. However, MA1 and the Yamnaya are clearly not a clade, as, for example, the statistic $f_4$(MA1, Yamnaya; Karitiana, Chimp) is significantly negative both on all SNPs (Z=-6.3), or limiting to transversions (Z=-3.9).

We chose to show Fig. 2a in the main paper as limiting to transversions did not seem to impede our ability to distinguish among populations, did not produce any spurious visual clustering that contradicted the analysis of *f*-statistics (SI7) and allowed us to treat ancient and modern individuals in the same way.

**Figure S5.1: Complement of PCA of Fig. 2a, showing ancient individuals as grey dots and color coding modern ones.**

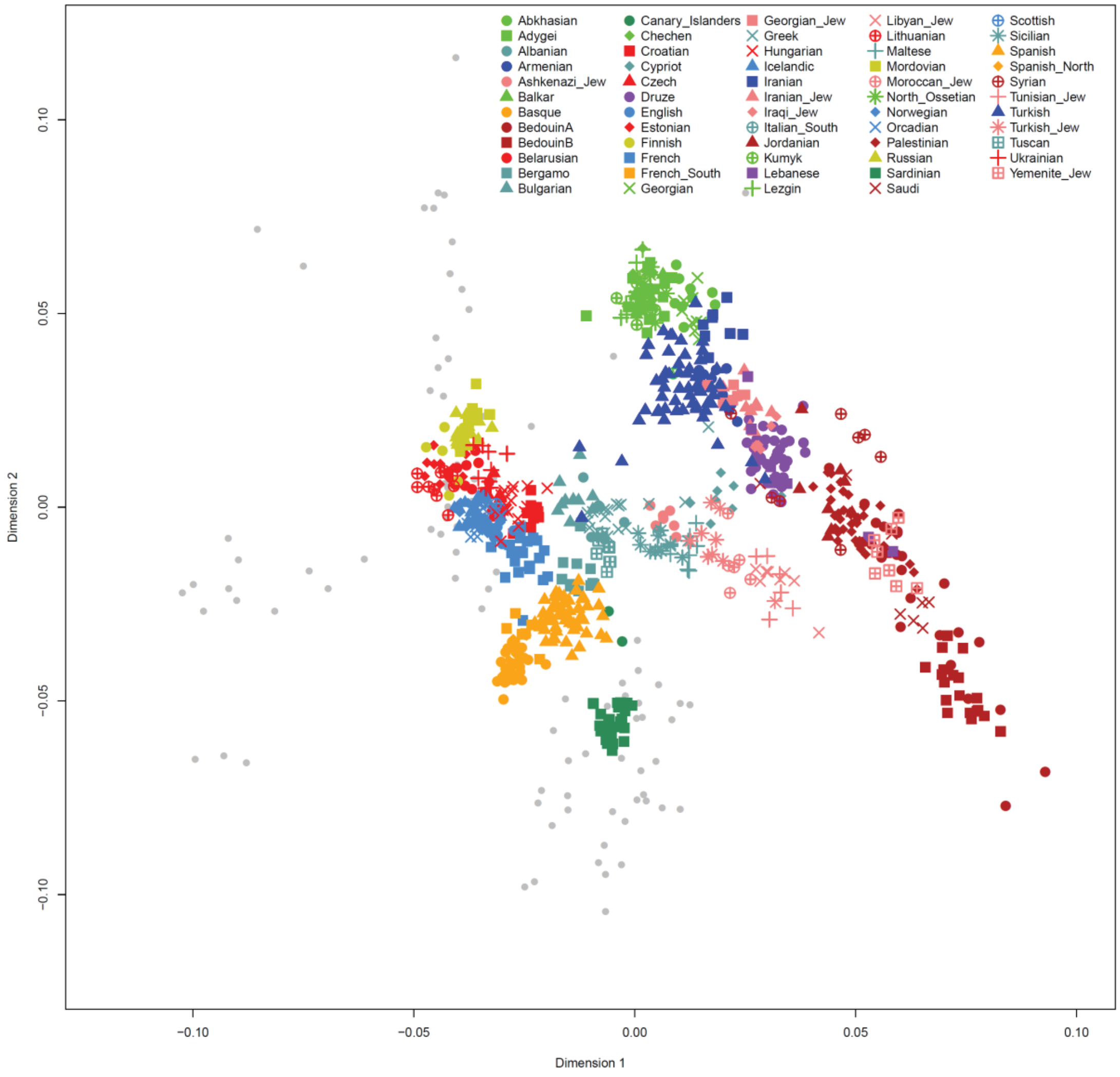

**Figure S5.2: PCA analysis with ancient individuals projected onto the variation of the present-day ones.**

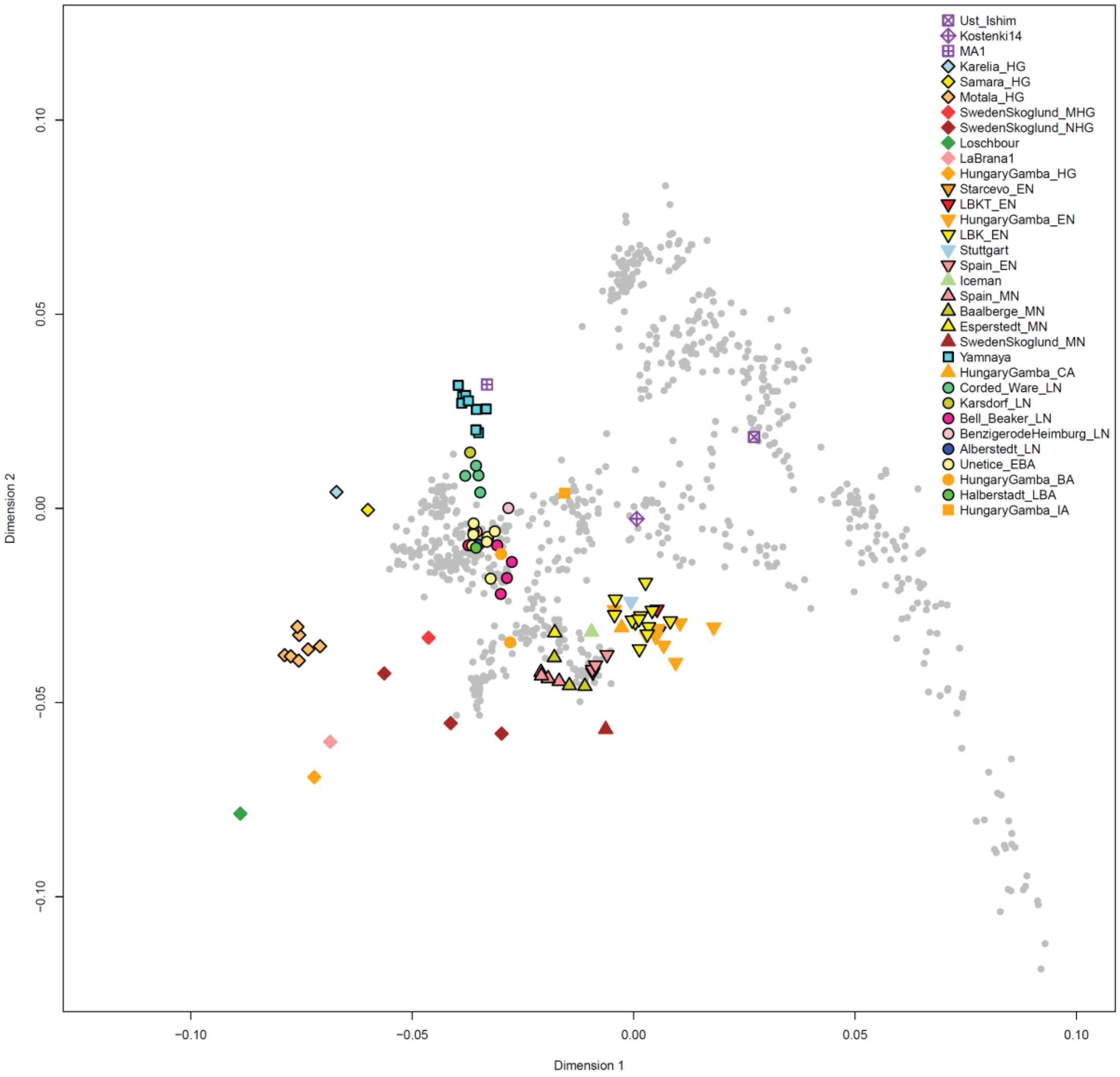

# Supplementary Information 6

## ADMIXTURE analysis

Iosif Lazaridis* and David Reich

* To whom correspondence should be addressed (lazaridis@genetics.med.harvard.edu)

We carried out admixture analysis using ADMIXTURE[1] (v. 1.23), including 2,437 individuals (2,345 modern humans[2] and 92 ancient ones. This includes Ust'-Ishim[3], Kostenki14[4], MA1[5], LaBrana1[6], the Iceman[7], Loschbour and Stuttgart[2], Neolithic, Copper, Bronze, and Iron Age individuals from Hungary[8], a farmer and hunter-gatherers from Sweden[9], and 67 new individuals from the present study. We excluded individuals I0114 and I0411 that appear to be related to individuals I0117 and I0410 respectively (Online methods), as in our experience, unsupervised ADMIXTURE clustering analysis often infers spurious clusters when pairs of relatives are included in the analysis dataset.

The initial set of 354,212 SNPs was pruned for linkage disequilibrium in PLINK[10] (v. 1.07) using parameters --indep-pairwise 200 25 0.5, resulting in a pruned set of 230,538 SNPs used for analysis. (LD-pruning is recommended by the authors of ADMIXTURE, and we chose a relatively high $r^2$=0.5 in order to retain a large number of SNPs, as more aggressive pruning would reduce the SNPs available for analysis for the individuals with most missing data.) ADMIXTURE was run with cross-validation (--cv) and with 100 different seeds (chosen at random from the set 1…1000). The number of ancestral populations (*K*) was varied between 2 and 20.

**Figure S6.1: Log likelihood of best of 100 replicates as a function of *K***

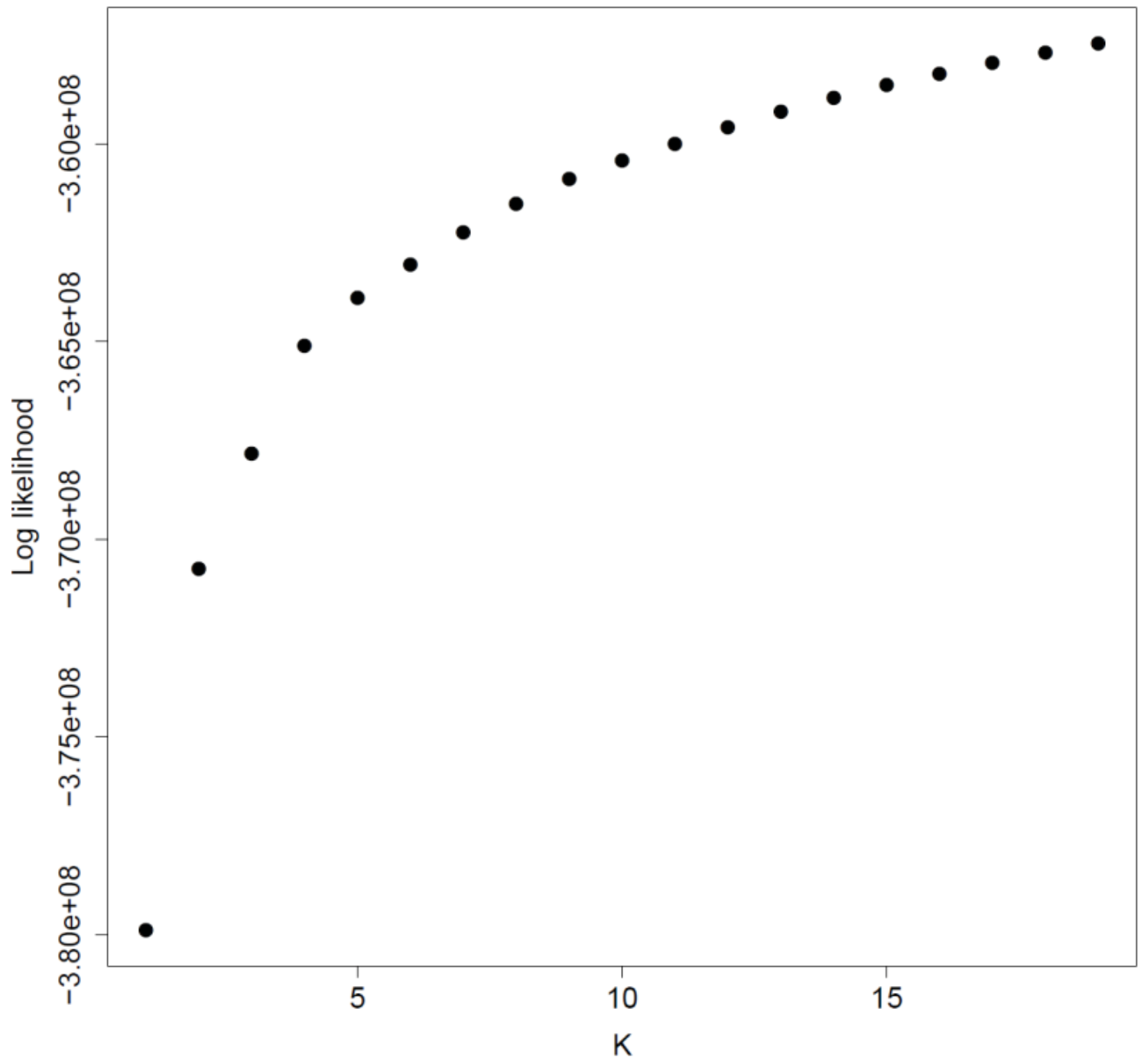

In Extended Data Fig. 1 we show the admixture proportions for 236 populations, including 33 ancient ones. The 33 ancient populations are placed at the beginning of the figure and individuals are given a higher line width for added clarity. The populations are clustered (using the function *hclust* in R) using their mixture proportions across all *K* values, with the goal of visually clustering populations that have similar mixture proportions. We visualize in Extended Data Fig. 1 the runs (for each value of *K*) that had the highest loglikelihood. In Fig. S6.1 we show the change in log likelihood with *K*, showing a steady increase with increasing *K*.

In Fig. S6.2 we show a box-and-whiskers plot of the cross-validation error for *K*=2 to *K*=20. This seems to plateau as *K* increases. The lowest median CV error is attained for K=16.

**Figure S6.2: Cross-validation error for 100 replicates as a function of *K***

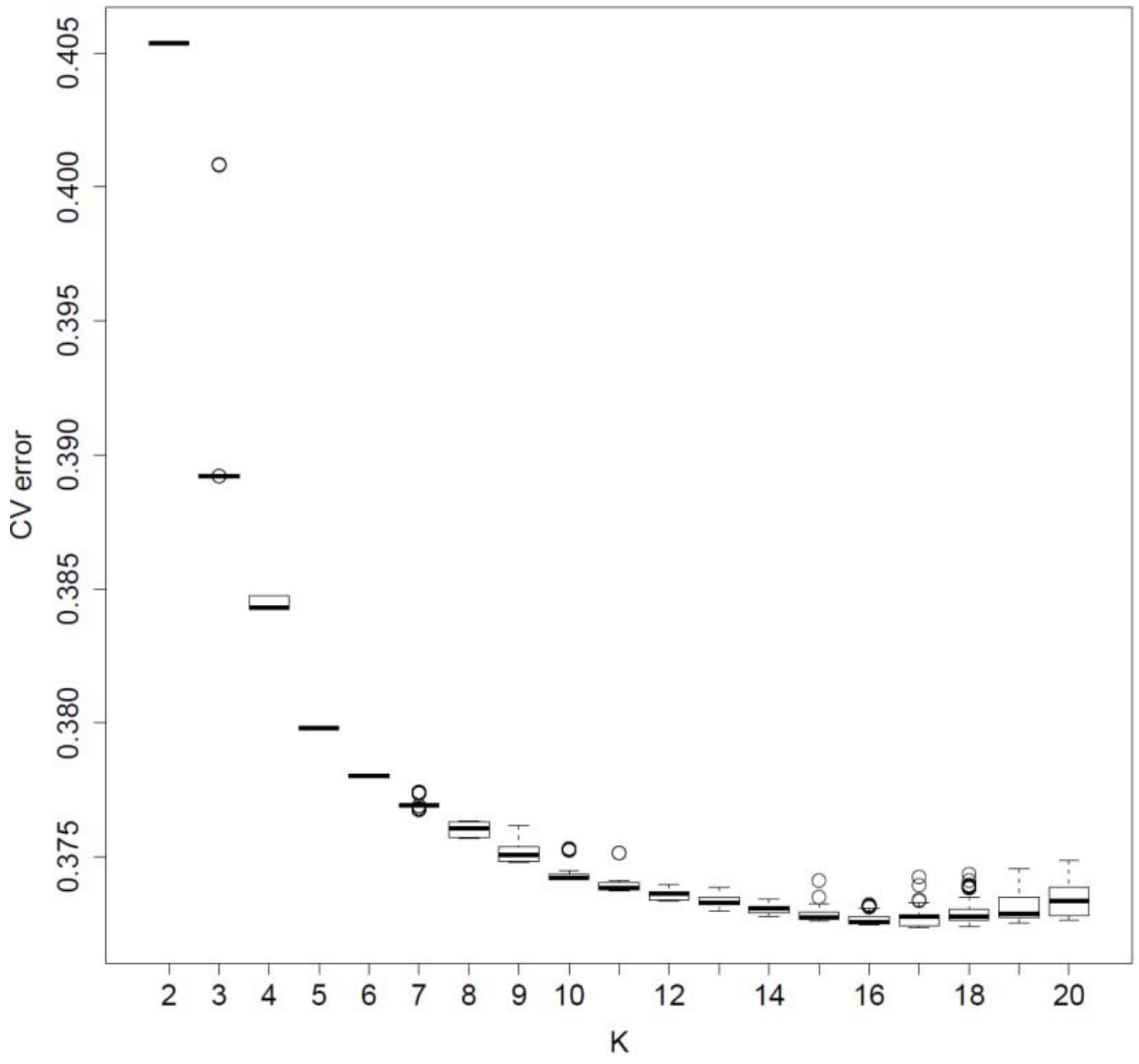

Fig. S6.3 shows a detail of Extended Data Fig. 1 showing only the ancient populations. We observe that none of the populations with at least two individuals have any major outliers, thus justifying our decision to pool individuals from each archaeologically defined population for the main analyses of our paper. The relative homogeneity of the populations also suggests that the number of SNPs available for analysis, even after LD-pruning, is adequate to correctly cluster individuals.

The Early/Middle Neolithic European populations belong almost entirely to the "orange" ancestral population from K=2 to K=8, while hunter-gatherers show a relationship to eastern non-Africans from K=3 to K=8, consistent with sharing more genetic drift with these populations due to their lack of "Basal Eurasian" ancestry[2]. From K=4 to K=6, the hunter-gatherers and late Neolithic/Bronze Age (LN/BA) groups possess some of the "pink" component that is dominant in Native Americans; this may reflect either the presence of west Eurasian-related "Ancient North Eurasian" ancestry in Native Americans[5] or of the same type of ancestry in European hunter-gatherers. An interesting pattern occurs at K=8, with all the late LN/BA groups from central Europe and the Yamnaya having some of the "light green" component that is lacking in earlier European farmers and hunter-gatherers; this component is found at high frequencies in South Asian populations and its co-occurrence in late

Neolithic/Bronze Age Europeans (but not earlier ones) and South Asians might reflect a degree of common ancestry associated with late Neolithic migratory movements (e.g., the ~5,800-year old TMRCA of Y-chromosome haplogroup R1a-M417 suggests some gene flow affecting both Europe and South Asia in this time frame[11], although this date is subject to uncertainty due to poor estimates of the human mutation rate.)

At K=9 a European hunter-gatherer ancestral population ("dark blue") appears; this was not present in an earlier analysis of the Human Origins modern populations and a much smaller number of ancient individuals[2]. The inclusion of a large number of ancient hunter-gatherers has probably caused such an ancestral population to appear in this analysis. European farmers now appear to be mixture of a Near Eastern (orange) and European hunter-gatherer (dark blue) ancestral populations, with an increase in the hunter-gatherer ancestry during the Middle Neolithic (reflecting the "resurgence" of such ancestry shown in PCA, Fig. 2a) and also during the Late Neolithic. Note, also, the persistent presence of the "light green" component that ties LN/EBA groups to South Asia between K=9 and K=15. A similar (darker green) component also distinguishes LN/EBA groups from earlier ones at K=16; this component appears to be highly represented in groups from South Asia, the Near East, and the Caucasus. The existence of this component may correspond to the evidence for "dilution" of EHG ancestry in the Yamnaya (SI7), showing them to have evenly split ancestry between the "dark blue" hunter-gatherer and "dark green" component; the analysis of SI9 also suggests an even split between an EHG and a Near Eastern component in the ancestry of the Yamnaya. The "dark green" component seems to have been carried from a Yamnaya-related population to the Corded Ware and other Late Neolithic and Bronze Age populations of central Europe. A useful topic for future work is to study the relationship of LN/BA populations to contemporary South Asians, Caucasian and Near Eastern populations and to see if this affinity (in contrast to earlier Europeans) may be related to the dispersal of Indo-European languages.

**Figure S6.3: Ancient populations ADMIXTURE results.** For the full set of populations, see Extended Data Fig. 1.

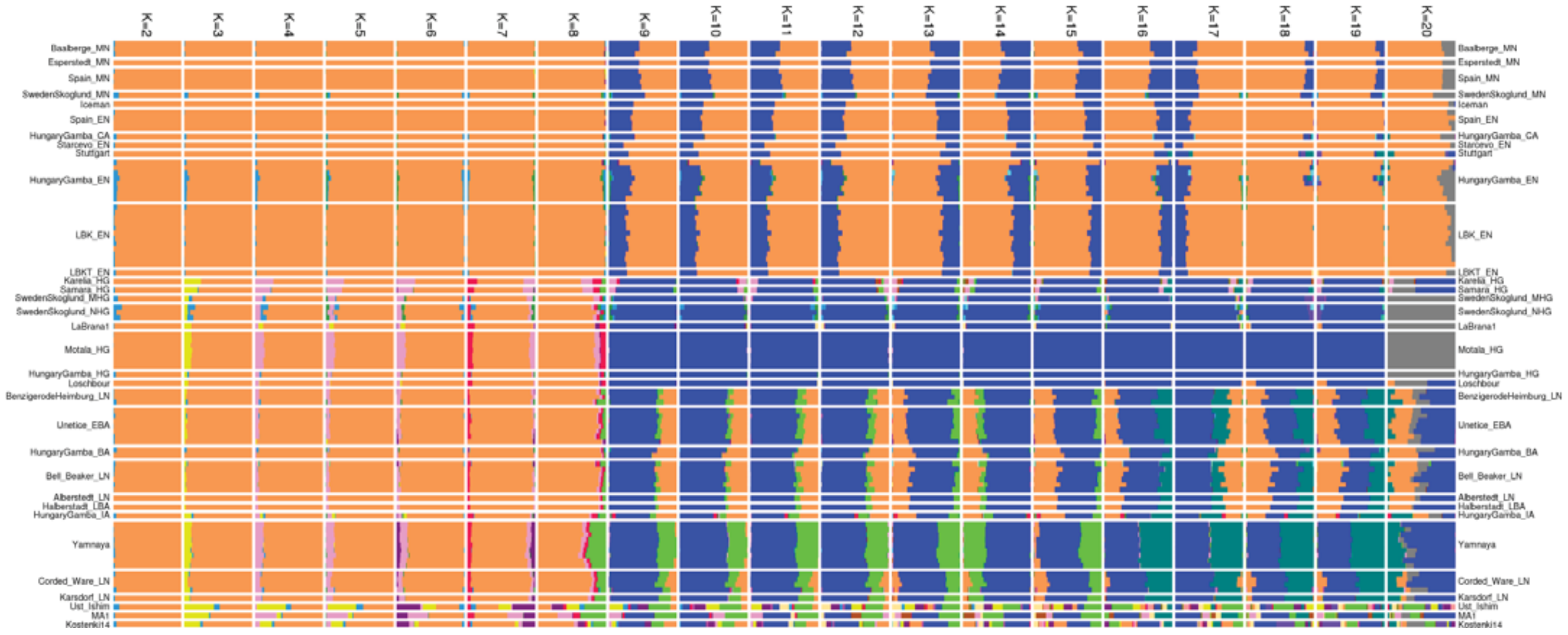

# Supplementary Information 7

## Population relationships and admixture using *f*-statistics

Iosif Lazaridis*, Nick Patterson, and David Reich

* To whom correspondence should be addressed (lazaridis@genetics.med.harvard.edu)

We carried out analysis of population relationships with *f*-statistics[1,2] to confirm formally the qualitative patterns observed via Principal Components Analysis[3,4] (PCA) (SI5) and ADMIXTURE[5] analysis (SI6). We use italics to refer to a variable (that is iterated over different populations, e.g. *MN* is Spain_MN, Esperstedt_MN, or Baalberge_MN) and non-italics to refer to a concrete population (e.g., MN is a meta-population consisting of all Middle Neolithic individuals). Significance of statistics is assessed using a block jackknife[6] with a 5cM block size, using the ADMIXTOOLS suite of programs[1].

**Figure S7.1: "Outgroup" statistic $f_3$(Dinka; $X$, $Y$).** Panel (a) shows that population $Y_2$ shares more genetic drift (indicated in blue) with $X$ than $Y_1$ does. In panel (b), an admixed population $X$ shares genetic drift with $Y_1$ and $Y_2$ that is weighted by the admixture proportions $\alpha_1$ and $\alpha_2$.

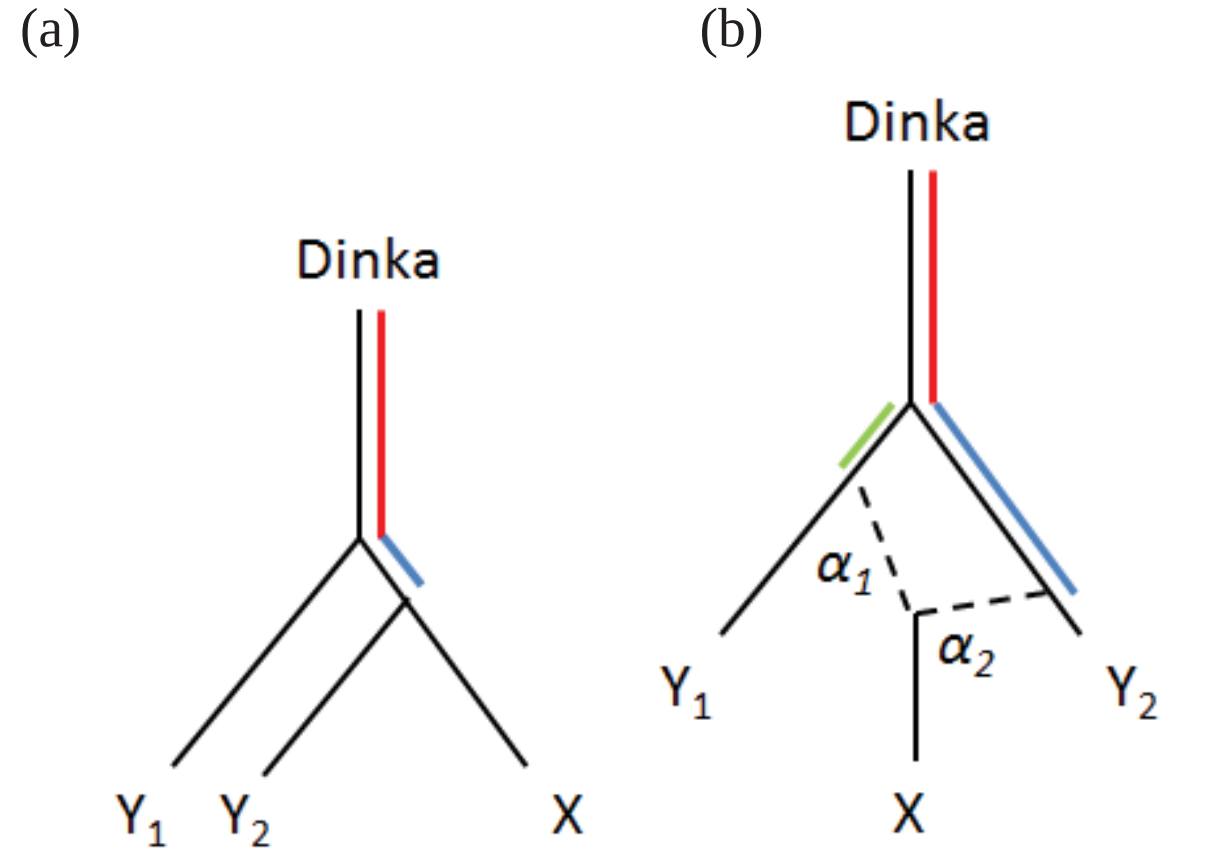

### Outgroup $f_3$-statistics

We first examined "outgroup" $f_3$-statistics[7] of the form $f_3$(Dinka; $X$, $Y$) (Table S7.1). These measure the amount of genetic drift shared by two populations $X$ and $Y$ after their divergence from the Dinka, an African outgroup population that is phylogenetically close to the Out-of-Africa bottleneck[8] and has no detectible archaic Eurasian admixture[9] (results with other African outgroups such as Yoruba or Ju_hoan_North (San) are qualitatively similar). Fig. S7.1 shows this statistic. We note that in Fig. S7.1b an admixed population $X$ may share more genetic drift with $Y_2$ than with $Y_1$ even if it has less ancestry related to $Y_2$ than to $Y_1$; this is due to the fact that $Y_2$ may have experienced more genetic drift than $Y_1$ since their separation from Dinka. Thus, $f_3$(Dinka; $X$, $Y_2$)> $f_3$(Dinka; $X$, $Y_1$) does not imply that $\alpha_2$>$a_1$. It does, however, imply that $\alpha_2$>0 (that $X$ shares at least some genetic drift with $Y_2$ that is not shared with $Y_1$).

These observations are consistent with the groupings of Fig. 2a (PCA plot) but show only the population that shares most genetic drift with each population $X$. In Extended Data Fig. 2 we show the pairwise heat matrix of this statistic for all populations under study. All Holocene ancient European group together at the exclusion of Upper Paleolithic Eurasians MA1[7], Kostenki14[10], and Ust_Ishim[11], suggesting that they share common genetic drift as a result of at least some shared history. Two major groupings of Holocene ancient Europeans involve Early/Middle Neolithic farmers on the bottom-left quadrant, and hunter-gatherers, Yamnaya, and Late Neolithic/Bronze Age central Europeans on the top-right quadrant. The strong relationship between European hunter-gatherers is clearly seen in the middle of the plot, with finer WHG (Loschbour, LaBrana1, and HungaryGamba_HG) and EHG (Karelia_HG and Samara_HG) groupings also evident. MA1 clearly shares less genetic drift with Europeans than Europeans do with each other. However, MA1 shares more genetic drift with EHG than Europeans shared with EHG. Early and Middle Neolithic farmers group with each other, with EN and MN sub-groupings also evident; the Middle Neolithic group also includes Copper and Bronze Age individuals from Hungary[12]. Importantly, the Late Neolithic/Bronze Age samples group with the hunter-gatherers and Yamnaya, and not the Early/Middle Neolithic ones, as might be expected if they were a direct continuation of the EN/MN population of Europe.

**Table S7.1: Populations maximizing $f_3$(Dinka; *X*, *Y*) for each ancient population *X*.**

| *X* | *Y* | $f_3$(Dinka; *X*, *Y*) | std. error |
|---|---|---|---|
| Alberstedt_LN | Esperstedt_MN | 0.167986 | 0.002435 |
| Baalberge_MN | LBKT_EN | 0.173694 | 0.006135 |
| Bell_Beaker_LN | HungaryGamba_HG | 0.163192 | 0.001971 |
| BenzigerodeHeimburg_LN | Motala_HG | 0.165990 | 0.001716 |
| Corded_Ware_LN | Karsdorf_LN | 0.166848 | 0.003064 |
| Esperstedt_MN | Spain_MN | 0.175997 | 0.001956 |
| Halberstadt_LBA | Esperstedt_MN | 0.166007 | 0.002412 |
| HungaryGamba_BA | HungaryGamba_HG | 0.165637 | 0.002290 |
| HungaryGamba_CA | Starcevo_EN | 0.169771 | 0.003655 |
| HungaryGamba_EN | Starcevo_EN | 0.171040 | 0.002473 |
| HungaryGamba_HG | Loschbour | 0.212255 | 0.002640 |
| HungaryGamba_IA | SwedenSkoglund_NHG | 0.156723 | 0.002325 |
| Iceman | Starcevo_EN | 0.170761 | 0.003149 |
| Karelia_HG | Samara_HG | 0.193494 | 0.002698 |
| Karsdorf_LN | SwedenSkoglund_MHG | 0.170276 | 0.011495 |
| Kostenki14 | Loschbour | 0.149636 | 0.002404 |
| LaBrana1 | Loschbour | 0.212582 | 0.002587 |
| LBK_EN | Esperstedt_MN | 0.175495 | 0.001809 |
| LBKT_EN | Starcevo_EN | 0.182378 | 0.008451 |
| Loschbour | LaBrana1 | 0.212582 | 0.002587 |
| MA1 | Samara_HG | 0.173080 | 0.002898 |
| Motala_HG | Loschbour | 0.194857 | 0.002037 |
| Samara_HG | Karelia_HG | 0.193494 | 0.002698 |
| Spain_EN | Spain_MN | 0.174572 | 0.001652 |
| Spain_MN | Esperstedt_MN | 0.175997 | 0.001956 |
| Starcevo_EN | LBKT_EN | 0.182378 | 0.008451 |
| Stuttgart | LBK_EN | 0.168289 | 0.001642 |
| SwedenSkoglund_MHG | HungaryGamba_HG | 0.181851 | 0.005697 |
| SwedenSkoglund_MN | Baalberge_MN | 0.165744 | 0.002777 |
| SwedenSkoglund_NHG | Motala_HG | 0.192765 | 0.001963 |
| Unetice_EBA | Motala_HG | 0.165904 | 0.001468 |
| Ust_Ishim | Chane | 0.125962 | 0.002174 |
| Yamnaya | Samara_HG | 0.169258 | 0.001955 |

We also plot the value of the $f_3$(Dinka; *X*, *Y*) for different populations *X* of particular interest (Fig. S7.3-7). These outgroup $f_3$-statistics strongly suggest that the ancestry of the Late Neolithic/Bronze Age groups was not entirely drawn from the local Early/Middle Neolithic population. In particular, the Corded Ware (Fig. S7.7a) shares more genetic drift with both the hunter-gatherers from Russia, the hunter-gatherers from Sweden, and the Yamnaya group from Russia than with any earlier (MN or EN) farmer groups from Europe. This strongly suggests that this group, which occupies the northernmost position in the PCA plot (Fig. 2a), does indeed possess some ancestry that is not rooted in its local central European context.

**Figure S7.3 [Western European hunter-gatherers]: Populations that share most genetic drift with a population *X* according to the statistic $f_3$(Dinka; *X*, *Y*). The estimated value and ±3 standard errors are indicated.** Three hunter-gatherers from Spain (LaBrana1), Luxembourg (Loschbour), and Hungary (HungaryGamba_HG, or KO1 of ref.[12]) share more genetic drift with each other.

(a) Loschbour

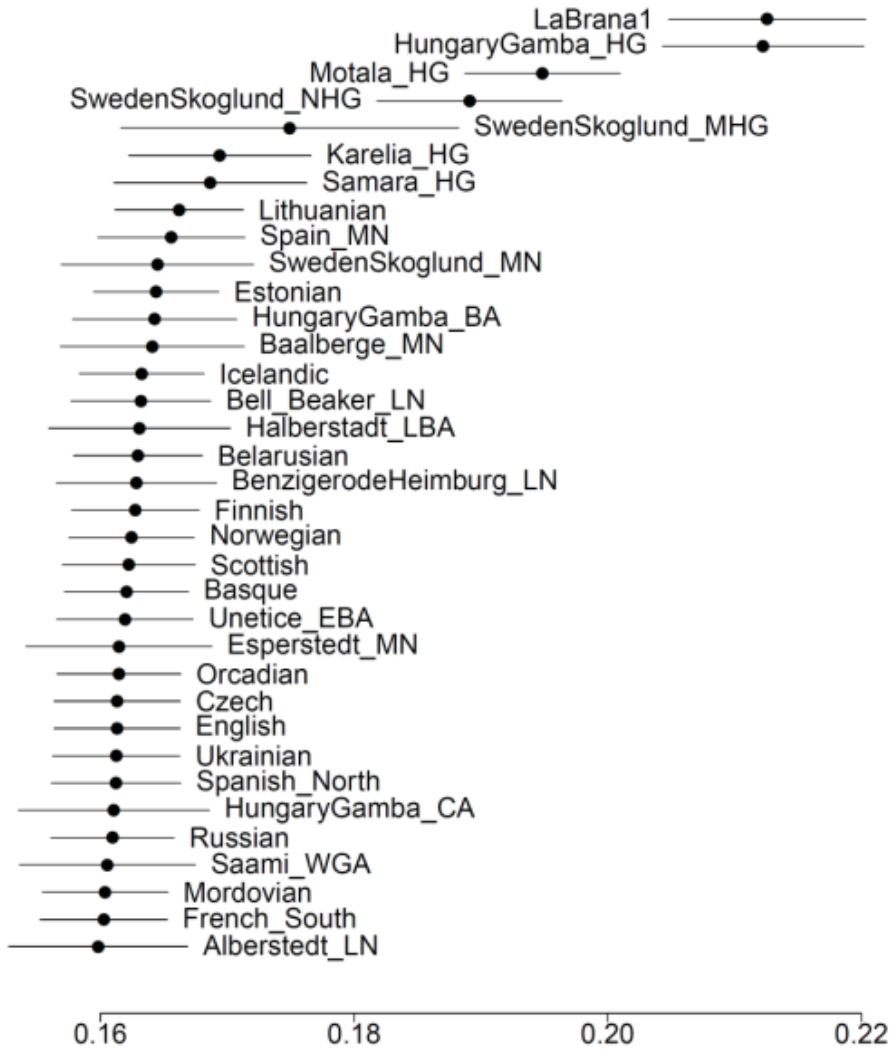

(b) LaBrana1

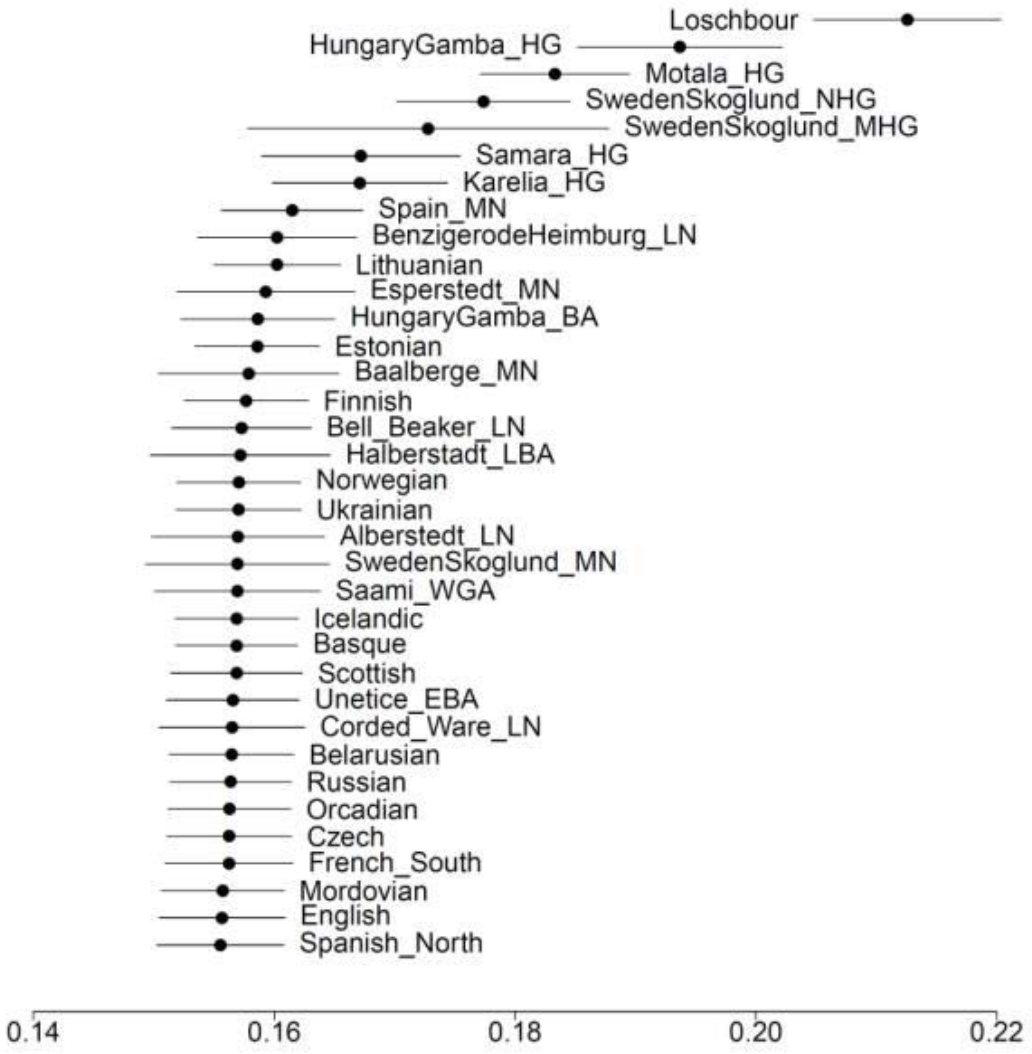

(c) HungaryGamba_HG

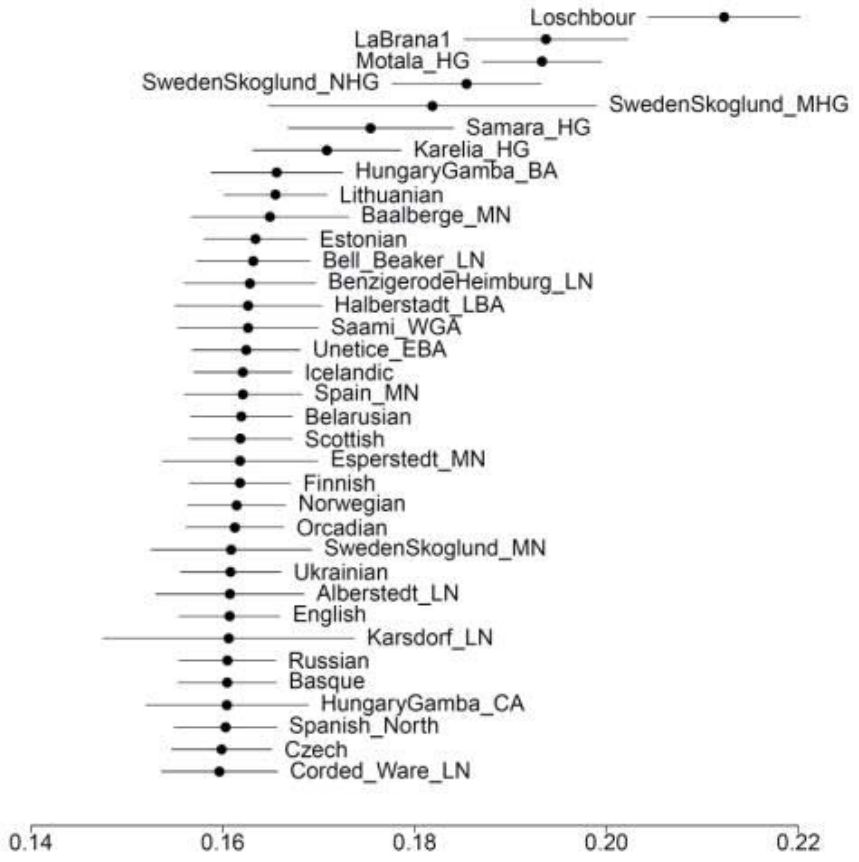

**Figure S7.4 [Ancient North Eurasians and Eastern European hunter-gatherers]: Populations that share most genetic drift with a population *X* according to the statistic $f_3$(Dinka; *X*, *Y*). The estimated value and ±3 standard errors are indicated.** The MA1 Upper Paleolithic shares more genetic drift with the two eastern European hunter-gatherers Karelia_HG and Samara_HG; these two EHG samples share more drift with each other.

(a) MA1

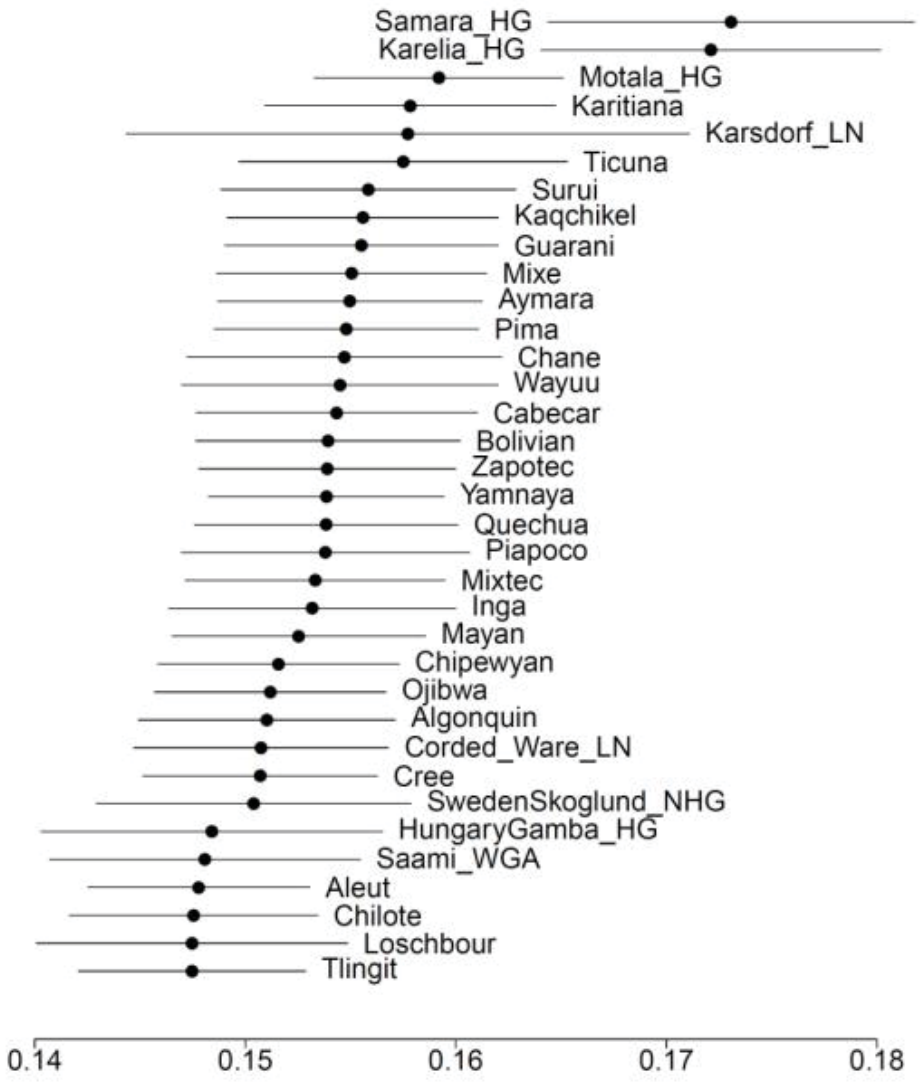

(b) Karelia_HG

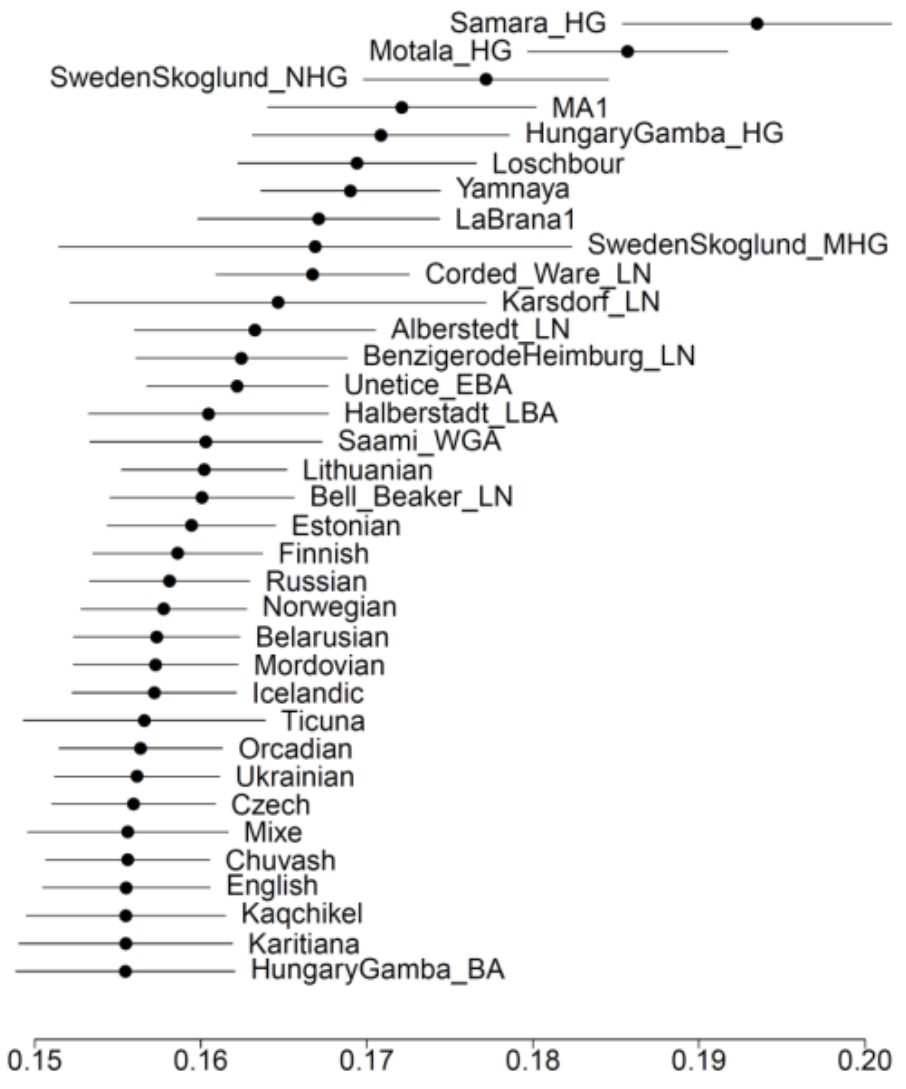

(c) Samara_HG

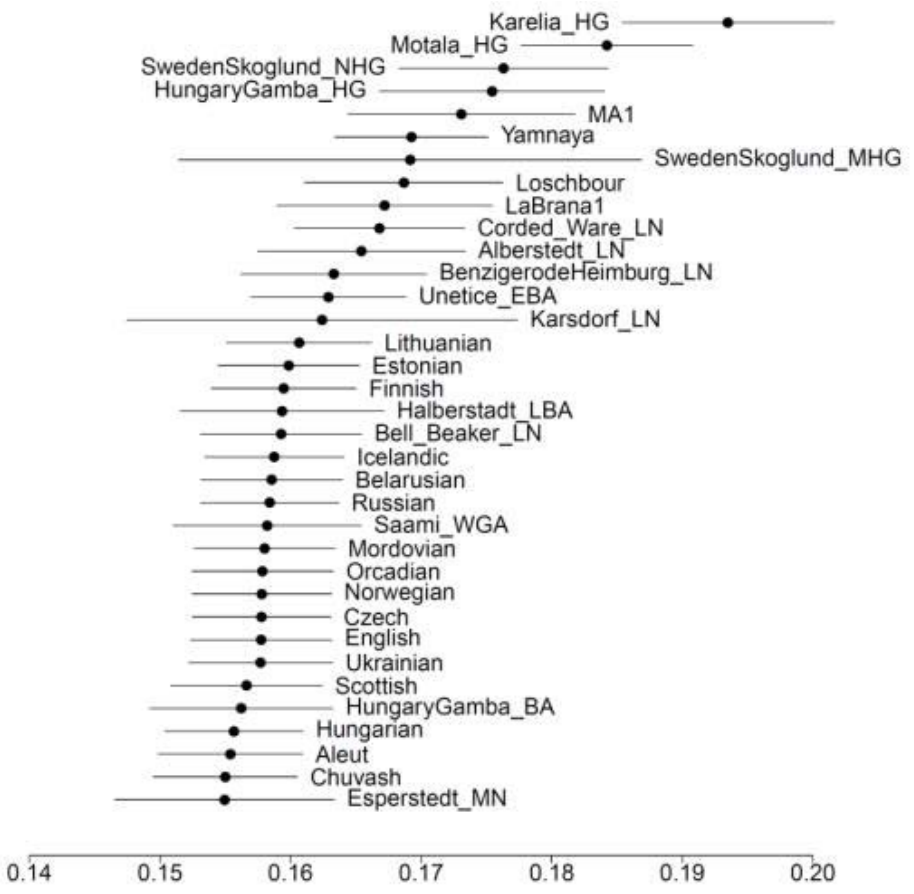

**Figure S7.5 [Early Neolithic Europeans]: Populations that share most genetic drift with a population X according to the statistic $f_3$(Dinka; X, Y). The estimated value and ±3 standard errors are indicated.** Early Neolithic Europeans share more genetic drift either with other EN populations, or the MN populations that succeed them.

(a) LBK_EN

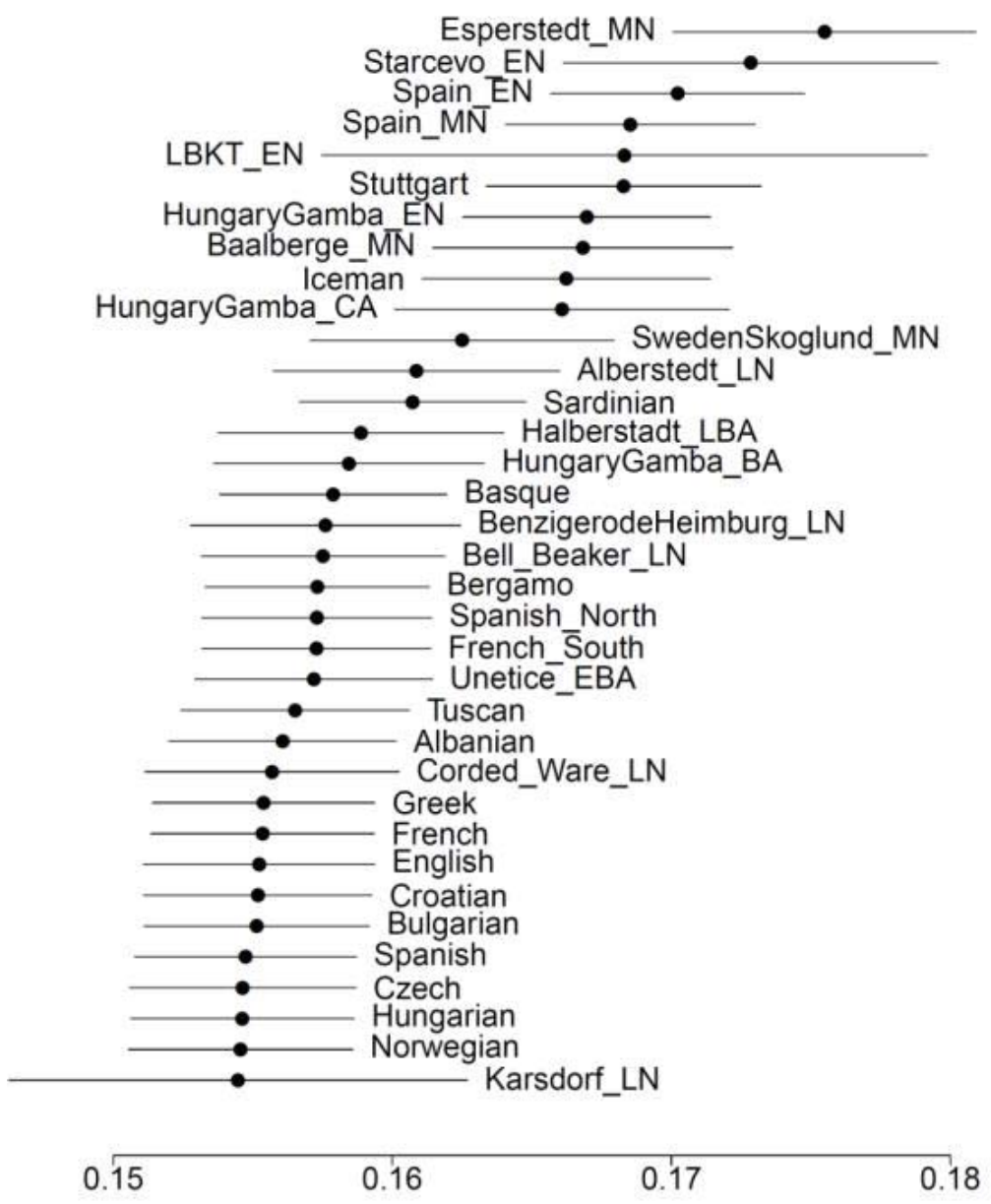

(b) LBKT_EN

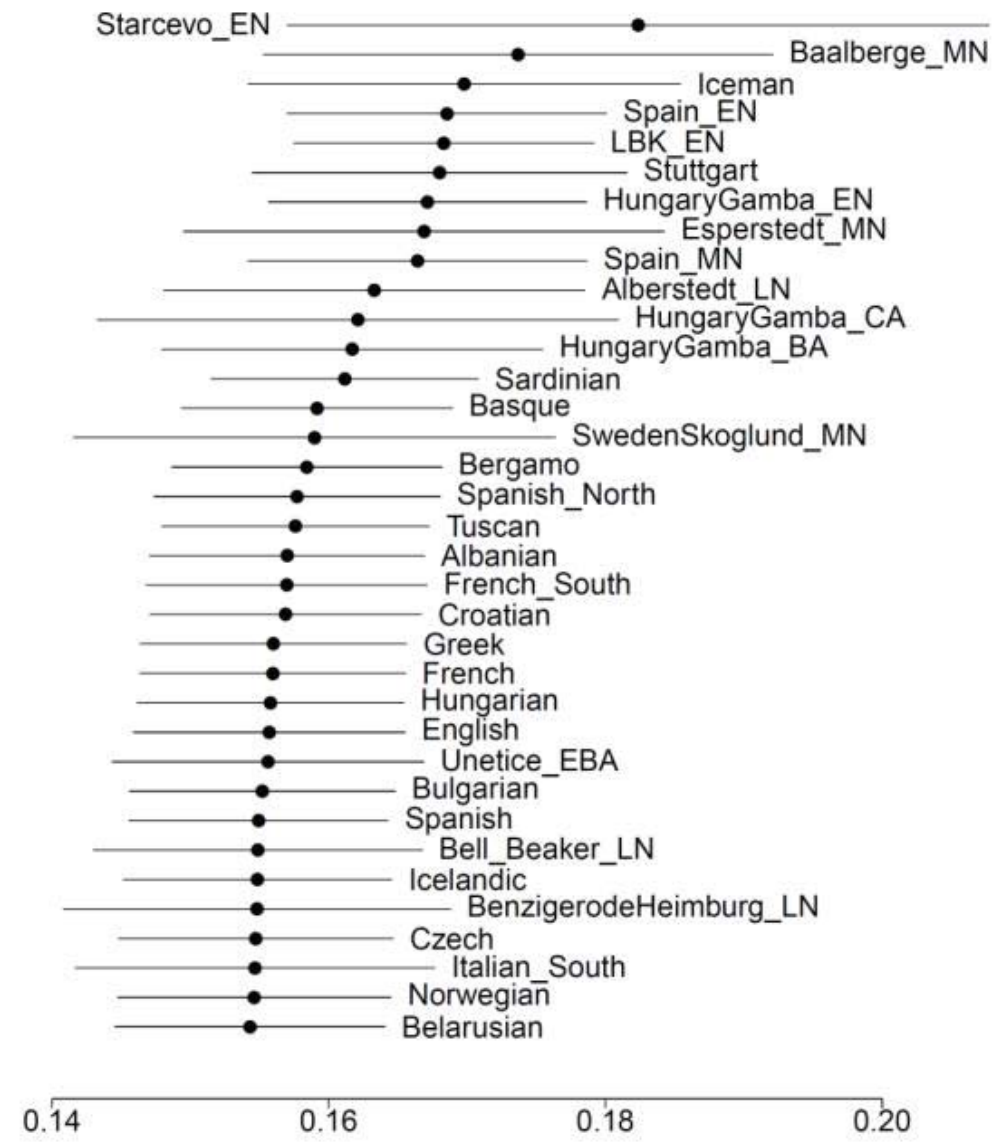

(c) Starcevo_EN

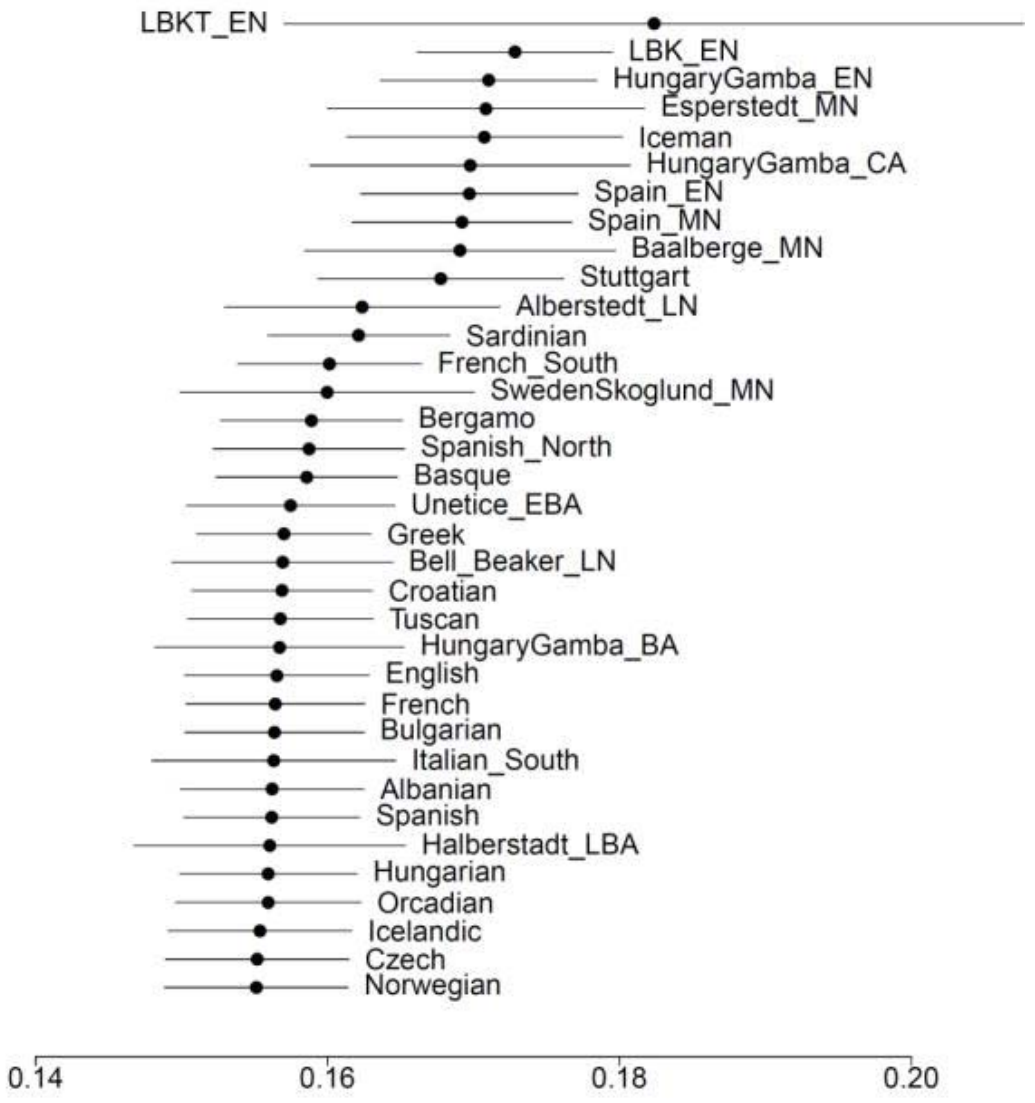

(d) Spain_EN

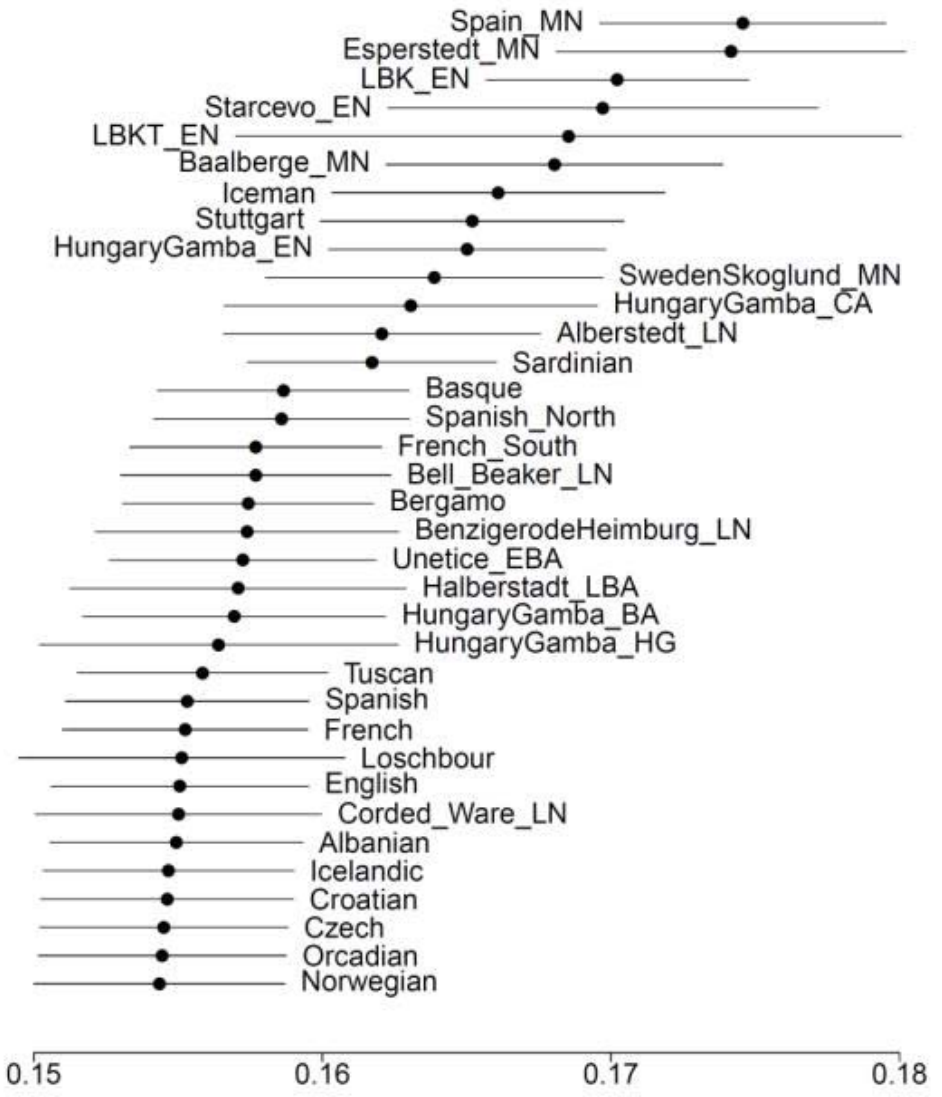

**Figure S7.6 [Middle Neolithic Europeans]: Populations that share most genetic drift with a population *X* according to the statistic $f_3$(Dinka; *X*, *Y*). The estimated value and ±3 standard errors are indicated.** Middle Neolithic Europeans share more genetic drift either with other MN populations, or the EN populations that precede them.

(a) Esperstedt_MN

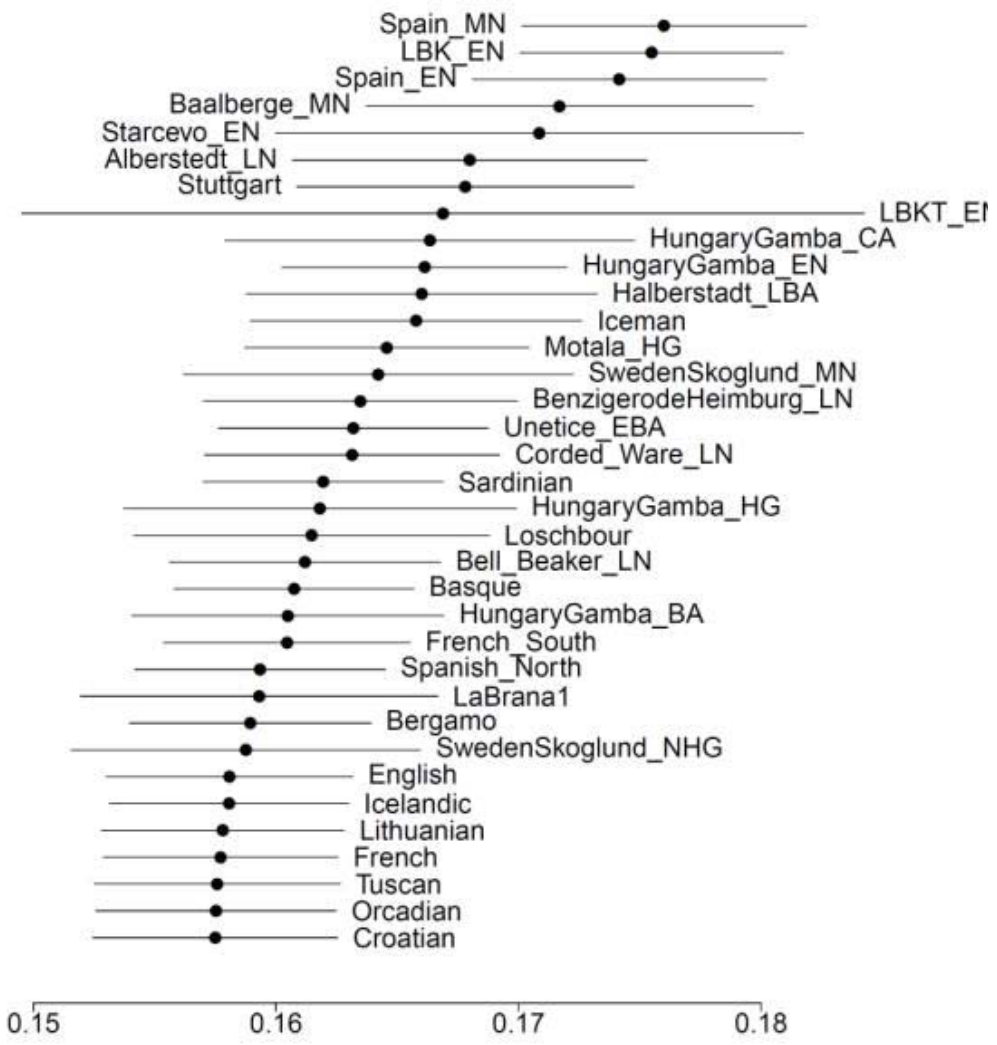

(b) HungaryGamba_CA

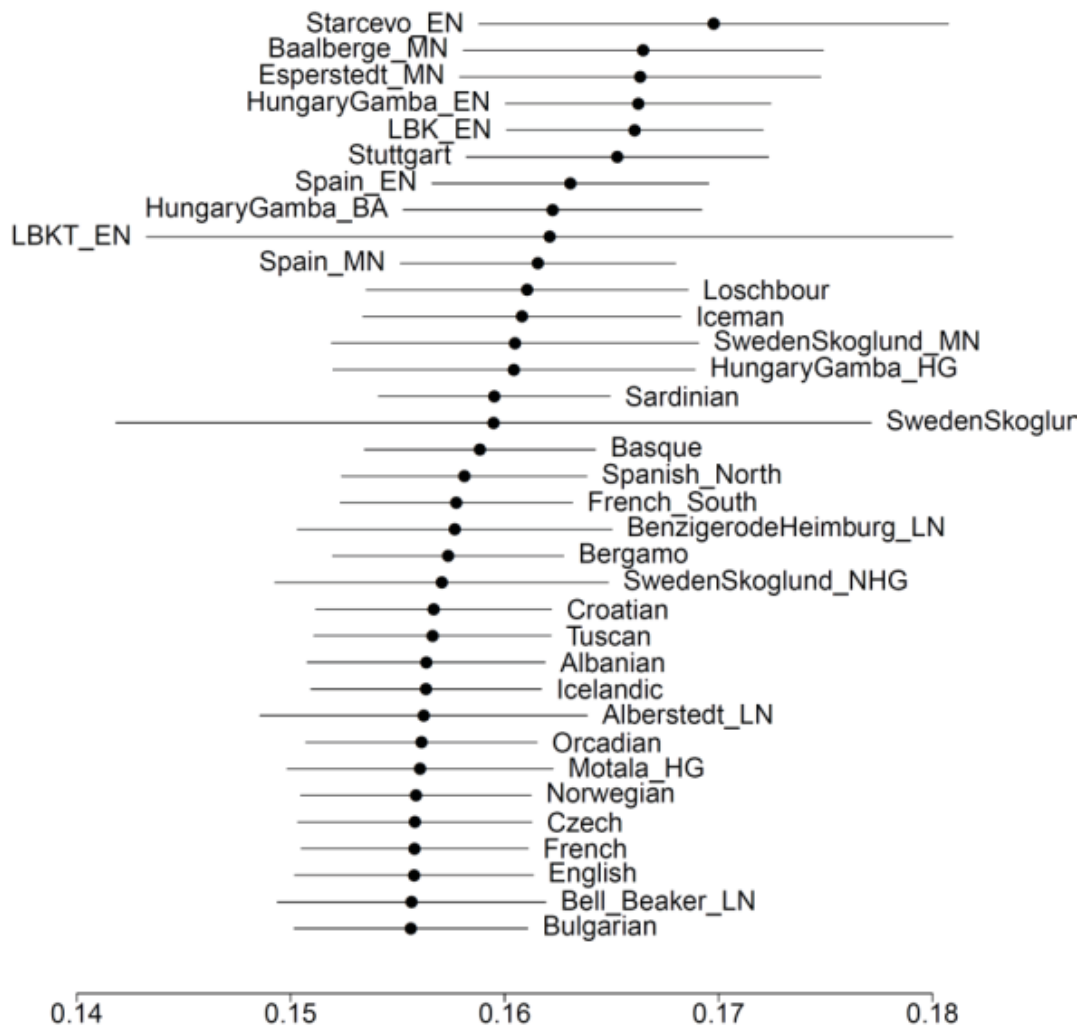

(c) Spain_MN

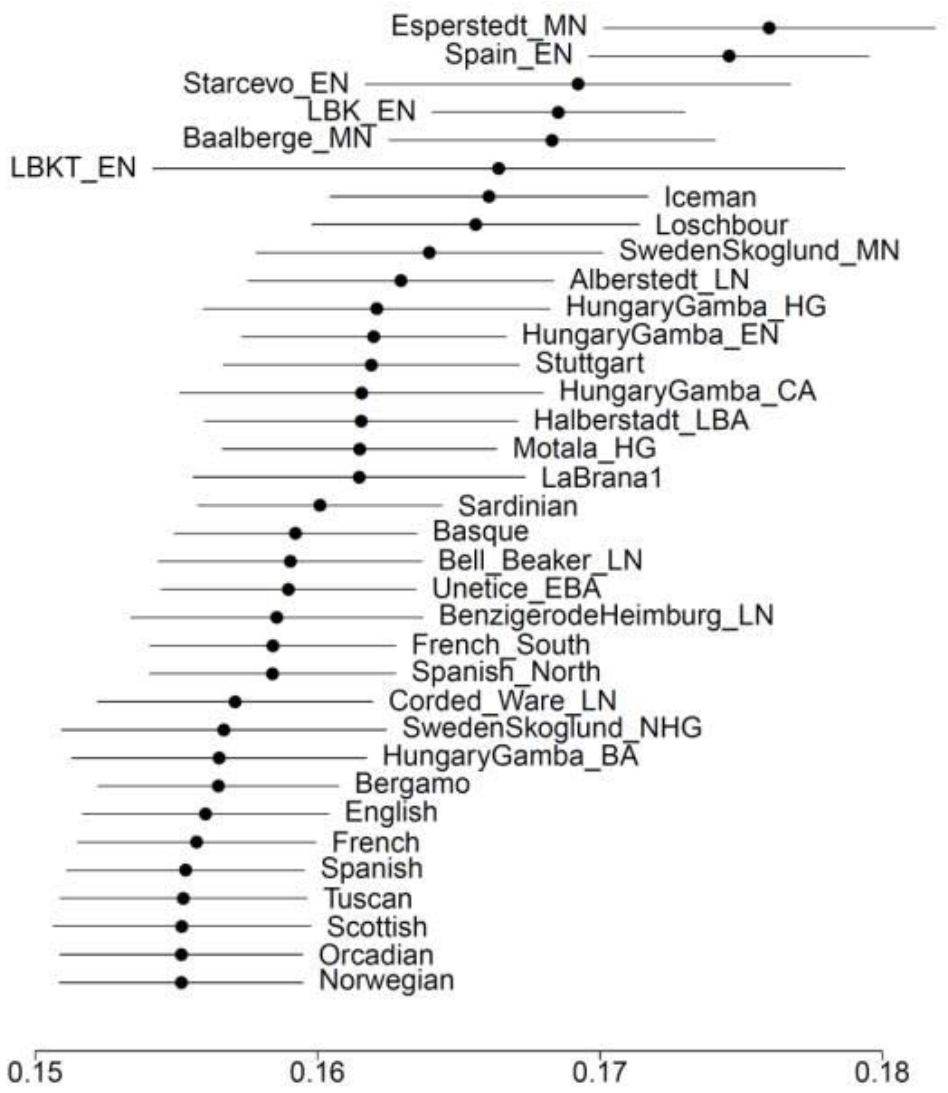

(d) SwedenSkoglund_MN

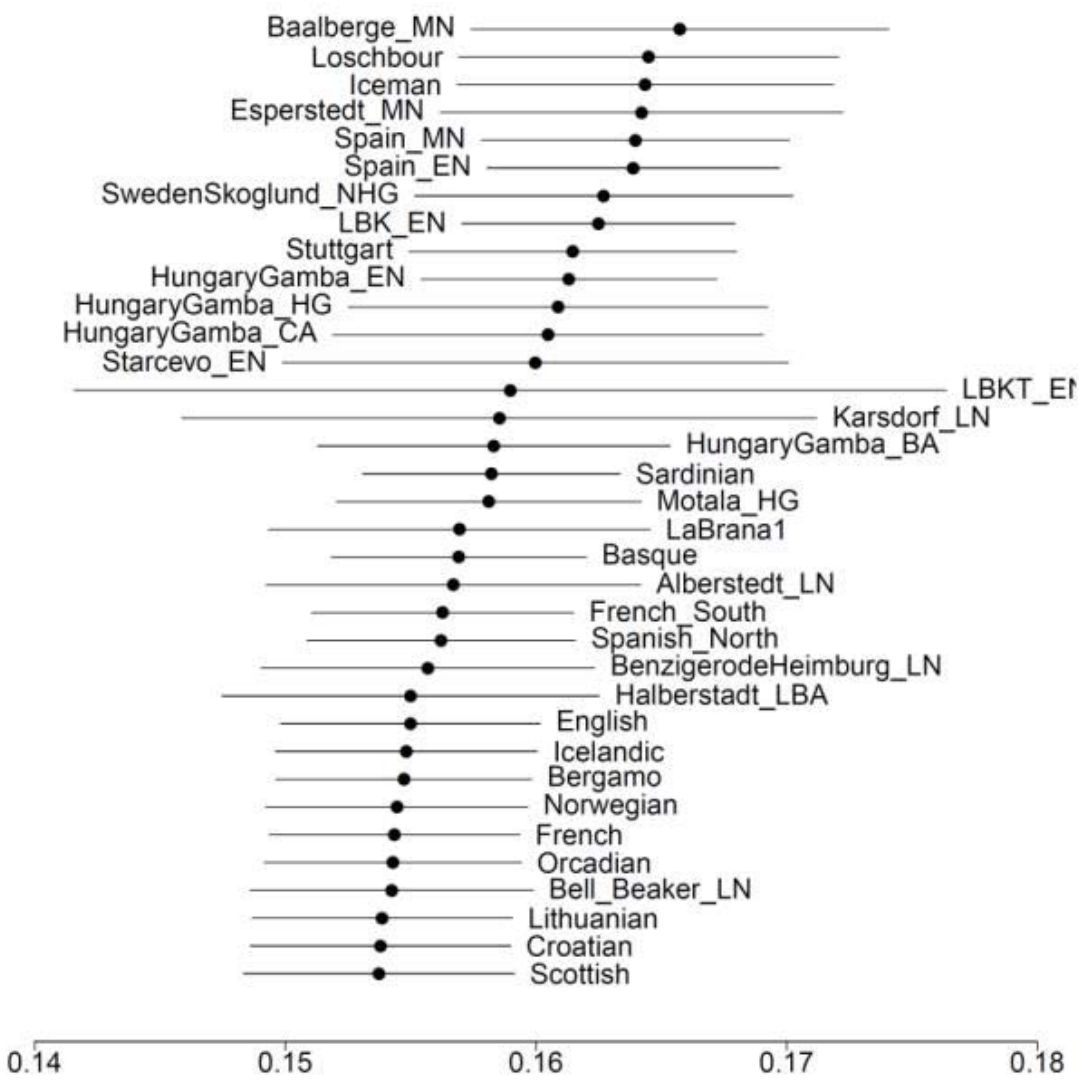

**Figure S7.7 [Late Neolithic / Bronze Age Europeans]: Populations that share most genetic drift with a population *X* according to $f_3$(Dinka; *X*, *Y*). The estimated value and ±3 standard errors are indicated.** These populations share most drift with hunter-gatherers, other LN/BA populations, Yamnaya, or Esperstedt_MN, the youngest sample from the Middle Neolithic of Germany.

(a) Corded_Ware_LN

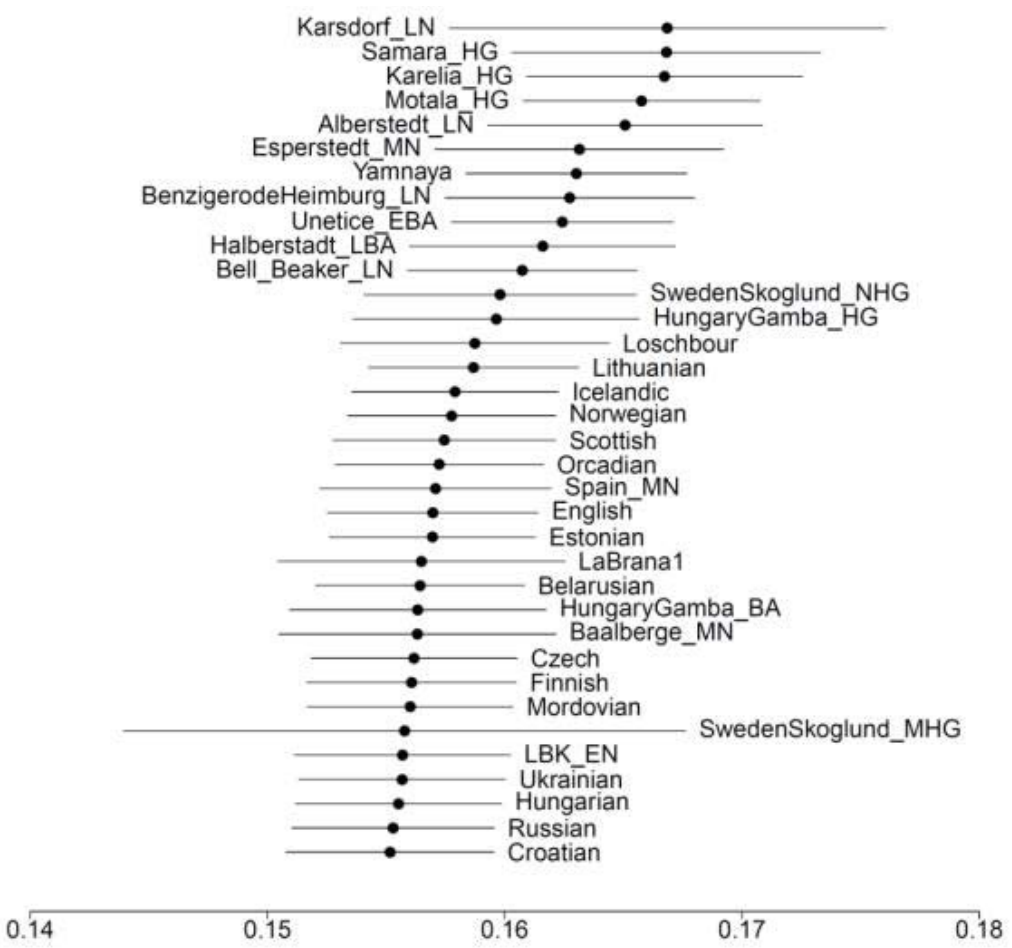

(b) Bell_Beaker_LN

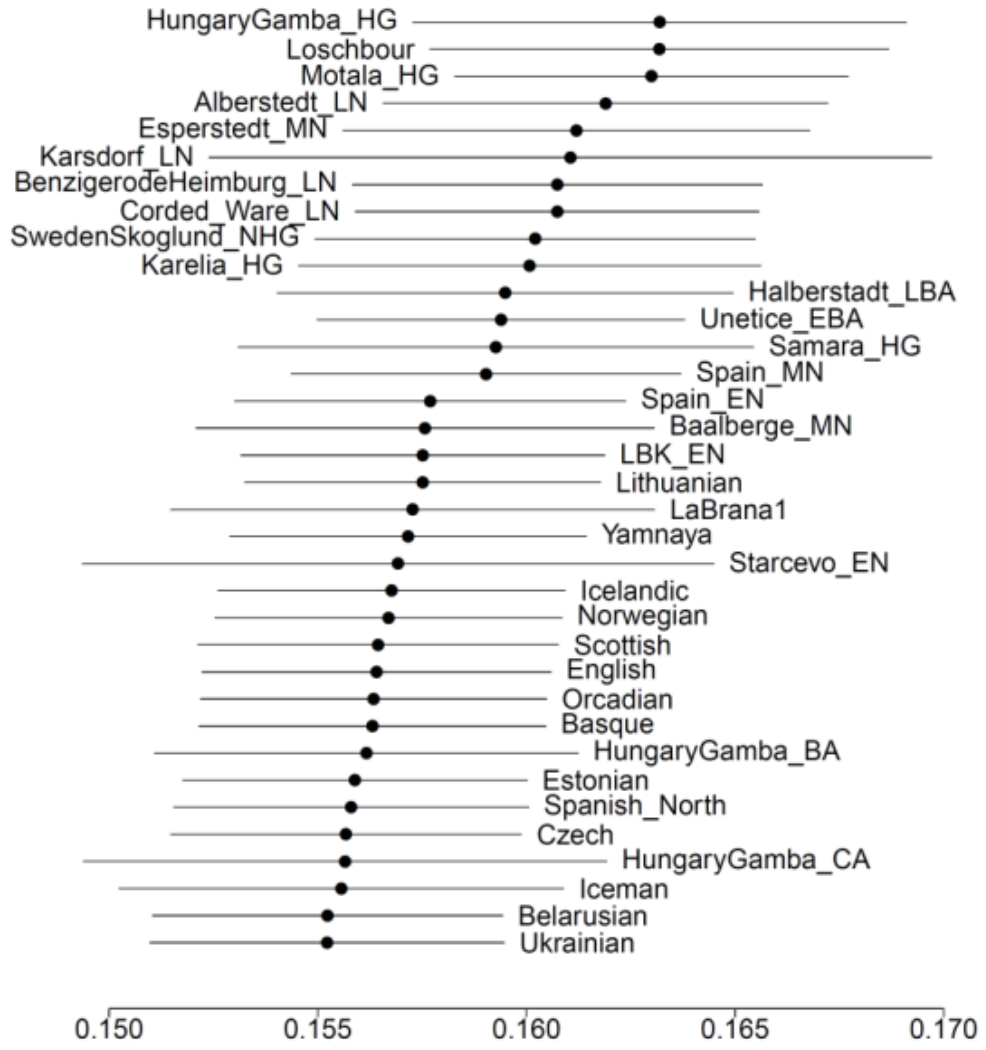

(c) Unetice_EBA

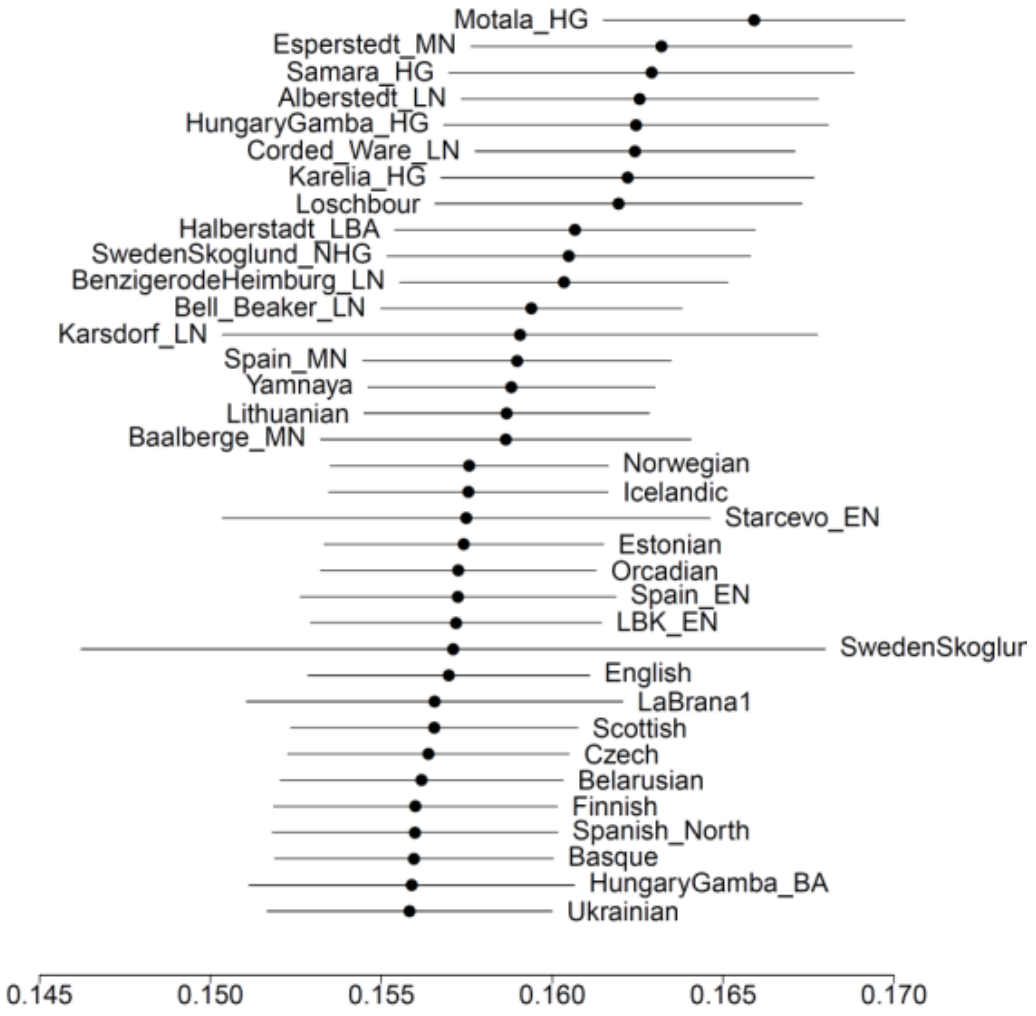

(d) Halberstadt_LBA

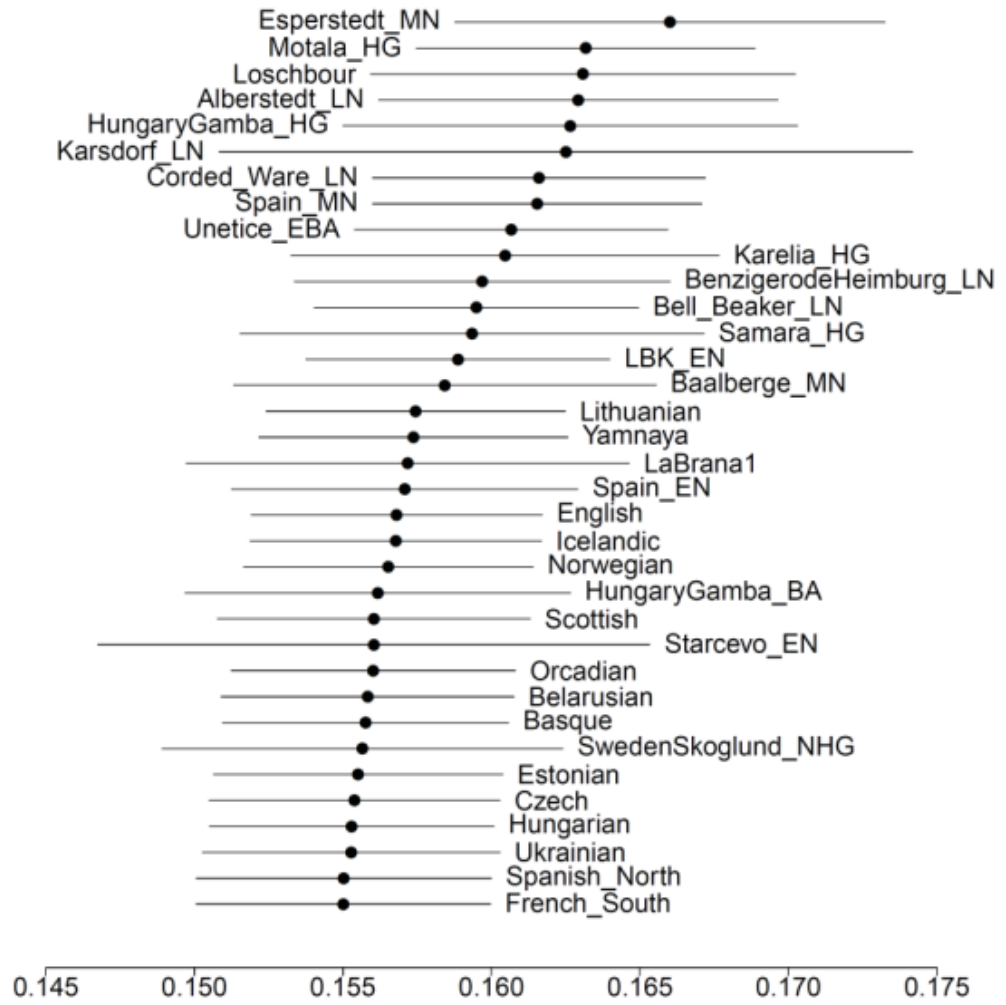

**Western European hunter-gatherers (WHG) in Luxembourg, Spain, and Hungary**

In a previous study[13], Western European hunter-gatherers (WHG) were defined by the Loschbour Mesolithic hunter-gatherer from Luxembourg, while noting that the LaBrana1 Mesolithic hunter-gatherer from Spain was closely related to it, a result that is again shown in Fig. S7.3. An unknown hunter-gatherer (UHG) was proposed to have contributed WHG-related ancestry into the early European farmers of Germany, thus postulating the existence of WHG-related populations somewhere in the route from the Near East to central Europe. The recent publication of KO1, an early Neolithic individual from Hungary[12] (named HungaryGamba_HG here, Fig. S7.3p) showed that such an individual with hunter-gatherer ancestry did in fact live in central Europe during the early Neolithic period.

To confirm that this was indeed related to Loschbour, we used the statistic $f_4$(Loschbour, HungaryGamba_HG; *Test*, Chimp), iterating *Test* over all other ancient or modern populations. All but one of these statistics were not significant (|Z|<3), confirming the impression from PCA (Fig. 2a) and outgroup $f_3$-statistics (above), that HungaryGamba_HG is related to WHG. The significant statistic was $f_4$(Loschbour, HungaryGamba_HG; LaBrana1, Chimp) = 0.00429 (Z=5.6) which suggests some structure within the WHG population, as LaBrana1 shares more alleles with Loschbour than with HungaryGamba_HG. This statistic continues to be significant when limiting to transversions (Z=3.9).

**Eastern European hunter-gatherers (EHG) in Russia**

We studied individuals from Russia to identify potential sources of "Ancient North Eurasian" ancestry in later Europeans, as the best representative of that ancestry, the Mal'ta (MA1) specimen from Upper Paleolithic Siberia, is both geographically and temporally remote from Neolithic and present-day Europeans. The two hunter-gatherers, Samara_HG and Karelia_HG are from far-apart locations in European Russia: Samara_HG is on the Volga, on the steppe, and Karelia_HG in Lake Onega, far to the north. Despite this fact, these two individuals appear to be mutually share the most drift, and MA1 shares more drift with them than with any other populations (Fig. S7. 4).

We confirmed these observations directly by taking the statistic $f_4$(Karelia_HG, Samara_HG; *Test*, Chimp) which is not significant for all modern/ancient *Test* populations (|Z|<1.9), indicating that these two hunter-gatherers did in fact form a clade with respect to all other humans. Moreover, the statistic $f_4$(Karelia_HG, *Test*; MA1, Chimp) is significantly positive (|Z|>4.7 on all SNPs, and |Z|>3.4 on transversions) for all *Test* populations except three (*Test*=Samara_HG, |Z|=0.3, consistent with Samara_HG and Karelia_HG forming a clade, and *Test*=Karsdorf_LN or SwedenSkoglund_MHG where <43,000 sites are available for co-analysis in all three ancient populations involved in the statistic). We conclude that MA1 shares more alleles with Karelia_HG than with other ancient/modern human populations; it was previously shown that MA1 shares more alleles with Native Americans[7] than any other modern humans, but our results show that it shares even more alleles with EHG than with Native Americans, e.g., $f_4$(Karelia_HG, Karitiana; MA1, Chimp) = 0.00388 (Z=5.3).

However, MA1 and EHG do not form a clade with respect to other humans, as, e.g., $f_4$(Karelia_HG, MA1; Loschbour, Chimp) = 0.00625 (Z=7.4). In SI8 we show that MA1, EHG, and WHG cannot be related to each other by a simple tree and at least one of them must be admixed. While the direction of gene flow cannot be resolved uniquely, the above statistic clearly shows that there is some common genetic drift shared by all "European hunter-gatherers" (both WHG and EHG) at the exclusion of MA1. This is consistent with the PCA (Fig. 2a) which shows all European hunter-gatherers occupying the left end of PC1.

**Scandinavian Mesolithic and Neolithic hunter-gatherers**

Co-analyzing our Motala_HG sample of Mesolithic hunter-gatherers with the StoraForvar11 (SwedenSkoglund_MHG) and later Pitted Ware (PWC) Neolithic hunter-gatherers from Sweden[8] (SwedenSkogldun_NHG), it is evident that 8-5 thousand years ago, the hunter-gatherers of Sweden were fairly similar in the context of West Eurasian populations (PCA, Fig. 2a). Indeed, when we study the statistic $f_4$(Motala_HG, SwedenSkoglund_MHG *or* SwedenSkoglund_NHG; *Test*, Chimp), its Z-score is always Z>-2.3 for all other ancient or modern populations, suggesting that the hunter-gatherers of ref. 8 have no other discernible ancestry than the Motala_HG of Mesolithic individuals from Sweden. The SwedenSkoglund_NHG sample is <3,000 cal BCE old, but seems (within the limits of our resolution) not to have acquired any admixture from other populations that was not already present in Motala_HG ~2 thousand years earlier.

It was previously shown that both Motala_HG[13] and SwedenSkoglund_NHG[8] could be modeled as mixtures of WHG and MA1. We estimate $f_4$(Motala_HG, Loschbour; Karelia_HG, Chimp) = 0.0041 (Z=7.6), and $f_4$(SwedenSkoglund_NHG, Loschbour; Karelia_HG, Chimp) = 0.00198 (Z= 2.9), which suggests that Eastern European hunter-gatherers share more alleles with Scandinavian hunter-gatherers (SHG) than with the WHG. In SI9 we show that SHG can be modeled as mixtures of EHG and WHG.

**Table S7.2: Relationship of Holocene hunter-gatherers to Upper Paleolithic ones.**

| ***$HG_1$*** | ***$HG_2$*** | **Upper Paleolithic** | **$f_4$(*$HG_1$*, *$HG_2$*; *Upper Paleolithic*, Chimp)** | **Z** |
|---|---|---|---|---|
| Karelia_HG | Motala_HG | Kostenki14 | -0.00075 | -1.3 |
| Loschbour | Motala_HG | Kostenki14 | 0.00158 | 2.9 |
| Loschbour | Karelia_HG | Kostenki14 | 0.00247 | 3.4 |
| Karelia_HG | Ust_Ishim | Kostenki14 | 0.00661 | 7.7 |
| Motala_HG | Ust_Ishim | Kostenki14 | 0.00731 | 10.0 |
| Loschbour | Ust_Ishim | Kostenki14 | 0.00897 | 11.0 |
| Karelia_HG | Kostenki14 | Ust_Ishim | -0.00044 | -0.6 |
| Motala_HG | Kostenki14 | Ust_Ishim | -0.00026 | -0.4 |
| Karelia_HG | Motala_HG | Ust_Ishim | -0.00016 | -0.3 |
| Loschbour | Kostenki14 | Ust_Ishim | 0.00037 | 0.5 |
| Loschbour | Karelia_HG | Ust_Ishim | 0.00064 | 0.9 |
| Loschbour | Motala_HG | Ust_Ishim | 0.00052 | 1.1 |

**Ust Ishim and Kostenki14**

To test whether the three main groups of Holocene hunter-gatherers (WHG, EHG, SHG) differed in their relationship to Upper Paleolithic Eurasians like Kostenki14[10] and Ust Ishim[11], we tested statistics of the form $f_4$(*$HG_1$*, *$HG_2$*; *Upper Paleolithic*, Chimp), which we report in Table S7.2. Holocene hunter-gatherers (and Kostenki14) appear to be equally related to Ust_Ishim, consistent with this individual being an outgroup to later populations[11]. Kostenki14 shares more alleles with Holocene European hunter-gatherers than with Ust_Ishim, consistent with it being an early European[10]. There is marginal signal (Z=3.4), confirmed in transversions only (Z=3.0) that Kostenki14 shares more alleles with Loschbour than with Karelia_HG. The signal also exists when replacing Karelia_HG with Samara_HG, although weaker (Z=2.7 on all SNPs and Z=2.5 on transversions only), given the lower number of SNPs available for analysis for that sample. This statistic casts doubt on the long-term continuity of the hunter-gatherers of eastern Europe, which must be further investigated with additional hunter-gatherers from different times in that region.

**Early Neolithic Europeans were descended from a common ancestral population**

We study statistics of the form $f_4$(LBK_EN, *Test*; *Other EN*, Chimp), varying *Other EN* to be any Early Neolithic population from the set: Starcevo_EN, LBKT_EN, Spain_EN, HungaryGamba_EN, Stuttgart, and *Test* to be any modern or ancient (but not Early Neolithic) population. We use LBK_EN as a baseline for early farmers, as this is the population with the highest sample size (N=12).

We find only two significantly negative statistics, $f_4$(LBK_EN, Spain_MN; Spain_EN, Chimp) = -0.00133 (Z=-4.2), and $f_4$(LBK_EN, Esperstedt_MN; Spain_EN, Chimp) = -0.00173 (Z=-3.6). The first can be interpreted as Spain_MN having some drift in common with Spain_EN (but not LBK_EN), indicating continuity of Neolithic populations in Spain. The second is surprising, as it indicates some gene flow from Spanish related populations into Germany. With these two exceptions, all other statistics are not significant (|Z|<3) or significantly positive (Z>3), suggesting that all Early Neolithic Europeans had ancestry in common with the LBK_EN and are thus descended from a common ancestral population.

We also studied statistics of the form $f_4$($EN_1$, $EN_2$; *Test*, Chimp) which test whether a pair of Early Neolithic populations ($EN_1$, $EN_2$) form a clade with respect to a *Test* population, varying *Test* to be any other ancient (but not Early Neolithic) or modern population. We list significant statistics (|Z|>3) in Table S7.3. We note that in our capture data (which have been identically processed which is likely to reduce systematic biases due to differences in error processes), the only significant statistic is $f_4$(LBK_EN, Spain_EN; Loschbour, Chimp) < 0.

**Variable WHG-related ancestry of Early and Middle Neolithic Europeans**

Early European farmers had WHG-related ancestry[13] and it was observed that a Funnel Beaker culture farmer from Sweden (SwedenSkoglund_MN in our analysis) had more hunter-gatherer ancestry than the approximately contemporaneous (~5,000 years ago) Tyrolean Iceman[8]. We were able to assess the extent of hunter-gatherer admixture in early European farmers with the statistic $f_4$(LBK_EN, *European Farmer*; Loschbour, Chimp), which measures whether Loschbour (the best sample from the WHG population) shares more alleles with LBK_EN or with another *European Farmer* population. We chose LBK_EN as the comparative baseline as it is the population with the most (N=12) individuals. This statistic is plotted in Fig. S7. 8, showing a striking pattern: Loschbour shares more alleles with all European farmer groups younger than 6,000 BCE, from Germany, Spain, Sweden, Italy, and Hungary, than with the early Neolithic LBK_EN sample from Germany. Thus, the early Neolithic population of central Europe in Germany and Hungary (LBK_EN as well as HungaryGamba_EN, Stuttgart, LBKT_EN, and Starcevo_EN) had less hunter-gatherer ancestry than the people that followed them across Europe. We can document the resurgence at a local level for the countries of Germany, Hungary and Spain, in which we have Neolithic samples from different periods (Table S7.4).

**Table S7.3: Significant statistics of the form $f_4(EN_1, EN_2; Test, \text{Chimp})$**

| $EN_1$ | $EN_2$ | *Test* | $f_4(EN_1, EN_2; Test, \text{Chimp})$ | Z |
|---|---|---|---|---|
| LBK_EN | HungaryGamba_EN | Alberstedt_LN | 0.00170 | 5.4 |
| LBK_EN | HungaryGamba_EN | Algerian | 0.00071 | 3.0 |
| LBK_EN | HungaryGamba_EN | Bell_Beaker_LN | 0.00121 | 4.4 |
| LBK_EN | HungaryGamba_EN | BenzigerodeHeimburg_LN | 0.00143 | 4.6 |
| LBK_EN | HungaryGamba_EN | Canary_Islanders | 0.00081 | 3.1 |
| LBK_EN | HungaryGamba_EN | Corded_Ware_LN | 0.00164 | 5.9 |
| LBK_EN | HungaryGamba_EN | Halberstadt_LBA | 0.00212 | 6.3 |
| LBK_EN | HungaryGamba_EN | Karelia_HG | 0.00102 | 3.1 |
| LBK_EN | HungaryGamba_EN | Mbuti | 0.00063 | 3.5 |
| LBK_EN | HungaryGamba_EN | Motala_HG | 0.00117 | 4.3 |
| LBK_EN | HungaryGamba_EN | Tunisian | 0.00075 | 3.3 |
| LBK_EN | HungaryGamba_EN | Tuscan | 0.00073 | 3.0 |
| LBK_EN | HungaryGamba_EN | Unetice_EBA | 0.00138 | 5.3 |
| LBK_EN | Spain_EN | Loschbour | -0.00108 | -3.0 |
| LBK_EN | Stuttgart | Alberstedt_LN | 0.00177 | 3.8 |
| LBK_EN | Stuttgart | BenzigerodeHeimburg_LN | 0.00174 | 4.3 |
| LBK_EN | Stuttgart | Halberstadt_LBA | 0.00175 | 3.8 |
| LBK_EN | Stuttgart | Karelia_HG | 0.00145 | 3.1 |
| LBK_EN | Stuttgart | Unetice_EBA | 0.00130 | 3.8 |
| Spain_EN | HungaryGamba_EN | Alberstedt_LN | 0.00214 | 5.3 |
| Spain_EN | HungaryGamba_EN | Algerian | 0.00090 | 3.2 |
| Spain_EN | HungaryGamba_EN | Basque | 0.00093 | 3.3 |
| Spain_EN | HungaryGamba_EN | Bell_Beaker_LN | 0.00142 | 4.3 |
| Spain_EN | HungaryGamba_EN | BenzigerodeHeimburg_LN | 0.00159 | 4.3 |
| Spain_EN | HungaryGamba_EN | Canary_Islanders | 0.00108 | 3.4 |
| Spain_EN | HungaryGamba_EN | Corded_Ware_LN | 0.00163 | 4.9 |
| Spain_EN | HungaryGamba_EN | Halberstadt_LBA | 0.00178 | 4.6 |
| Spain_EN | HungaryGamba_EN | Karelia_HG | 0.00134 | 3.4 |
| Spain_EN | HungaryGamba_EN | LaBrana1 | 0.00134 | 3.3 |
| Spain_EN | HungaryGamba_EN | Motala_HG | 0.00152 | 4.6 |
| Spain_EN | HungaryGamba_EN | Saharawi | 0.00093 | 3.2 |
| Spain_EN | HungaryGamba_EN | Sardinian | 0.00109 | 3.8 |
| Spain_EN | HungaryGamba_EN | Spanish | 0.00090 | 3.2 |
| Spain_EN | HungaryGamba_EN | Unetice_EBA | 0.00149 | 4.8 |
| Spain_EN | Stuttgart | Alberstedt_LN | 0.00224 | 4.4 |
| Spain_EN | Stuttgart | BenzigerodeHeimburg_LN | 0.00190 | 4.2 |
| Spain_EN | Stuttgart | Karelia_HG | 0.00173 | 3.4 |
| Spain_EN | Stuttgart | LaBrana1 | 0.00203 | 3.9 |
| Spain_EN | Stuttgart | Motala_HG | 0.00143 | 3.4 |
| Spain_EN | Stuttgart | Unetice_EBA | 0.00142 | 3.6 |

**Table S7.4: Resurgence of WHG ancestry in three European countries during the Middle Neolithic.** (The sample from Hungary is from the Copper Age, but shows the same pattern of resurgence as the earlier Middle Neolithic samples from Germany and Spain).

| *EN* | *MN* | $f_4(EN, MN; \text{Loschbour, Chimp})$ | Z |
|---|---|---|---|
| LBK_EN | Baalberge_MN | -0.00376 | -6.3 |
| LBK_EN | Esperstedt_MN | -0.00343 | -6.1 |
| Spain_EN | Spain_MN | -0.00275 | -6.4 |
| HungaryGamba_EN | HungaryGamba_CA | -0.00206 | -3.3 |

**Figure S7.8: Western European hunter-gatherers like Loschbour shared more alleles with Middle Neolithic European farmers, documenting variable amounts of hunter-gatherer ancestry in Neolithic Europe. The estimated value of the statistic $f_4$(LBK_EN, *European Farmer*; Loschbour, Chimp) and ±3 standard errors are shown.**

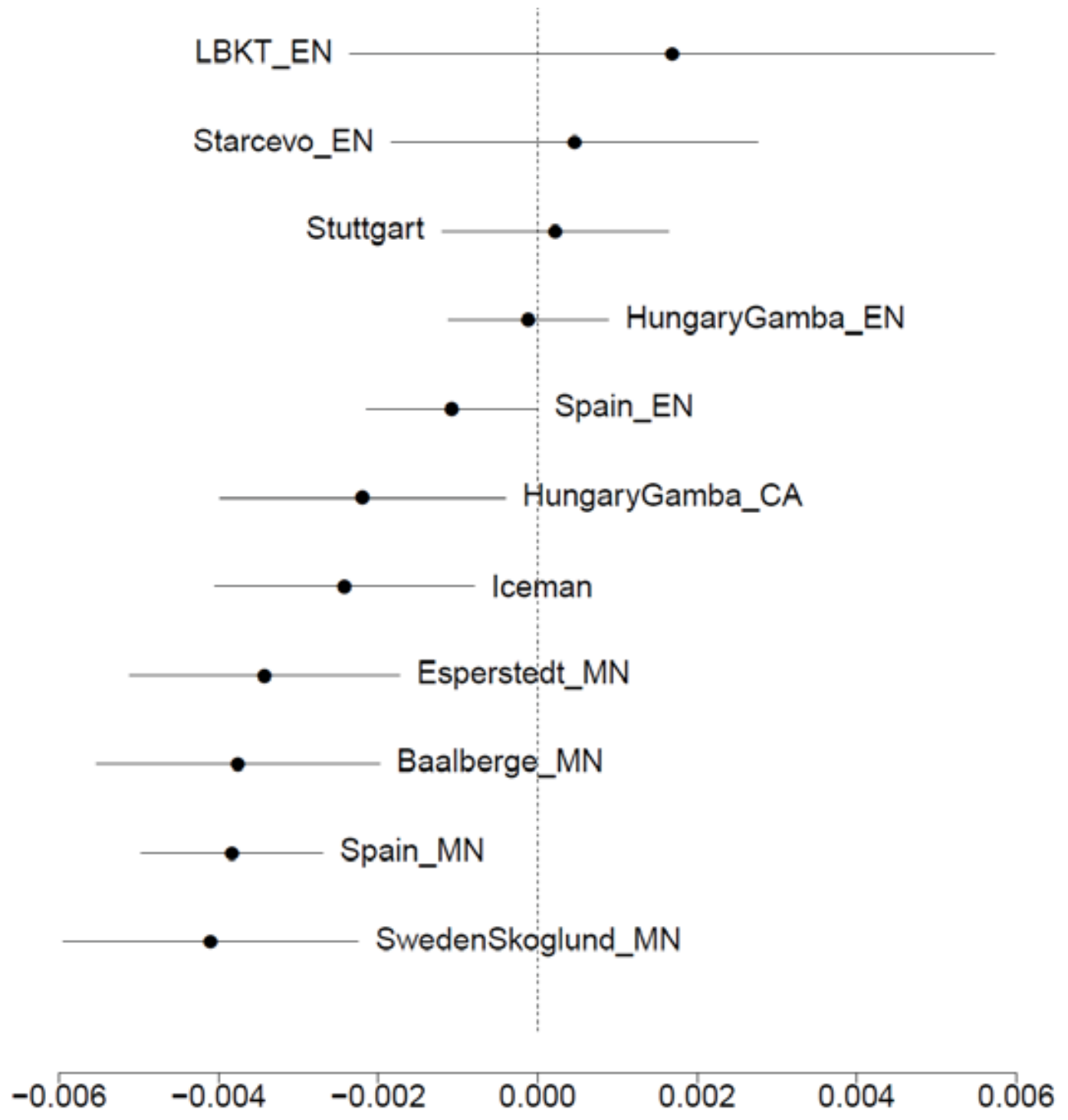

We also see (Fig. S7.8) that Loschbour shares more alleles (|Z|>4) with both the Tyrolean Iceman and SwedenSkoglund_MN, two samples from ~5,000 years ago, than with LBK_EN. We cannot speak of "WHG resurgence" in these two cases, as there are no earlier farmer samples from these two Sweden and Italy. Loschbour does not (|Z|<1.4) share more alleles with SwedenSkoglund_MN than it does with Baalberge_MN and Esperstedt_MN, Middle Neolithic samples from Germany, and neither does Motala_HG (|Z|<1.8). Thus, there is no statistical evidence for the occurrence of additional hunter-gatherer ancestry in farmers from Scandinavia above that which existed in contemporaneous farmers of central Europe. As for the Tyrolean Iceman, it is necessary to collect additional earlier samples from southern Europe to see if its elevated (relative to LBK_EN) affinity to WHG is due to a resurgence of such ancestry, or reflects higher WHG ancestry in that region since the early Neolithic. We do see some evidence that even early Neolithic Europeans varied in their WHG ancestry (Fig. S7.8) as $f_4$(LBK_EN, Spain_EN; Loschbour, Chimp) = -0.00108 (Z=-3.0), a marginal signal which, however, persists when limiting to transversions (Z=-3.5), which suggests that while early farmers in Germany and Spain were closely related, they may have had different levels of WHG-related admixture.

**Source of WHG resurgence**

The fact that the resurgence of hunter-gatherer ancestry occurred in several European countries raises the question of its source. It is possible that pockets of hunter-gatherers[14] existed in Europe long after the arrival of first farmers; their later admixture with farmer communities would account for the observed phenomenon. Alternatively, there may have existed larger areas of Europe where populations with substantial hunter-gatherer ancestry persisted and gene flow from these refuge areas increases the amount of this ancestry in the European farmers. We tested statistics of the form $f_4(EN, MN; WHG_1, WHG_2)$ for the (*EN*, *MN*) population pairs of Table S7.4 and the three WHG samples. However, all of these were not significant (|Z|<2.4), and thus, at present we cannot identify the source of WHG-related resurgence in any of the available genomes of WHG individuals. However, we can confidently exclude either the SHG or the EHG as sources of this phenomenon, as $f_4$(EN, MN; WHG, EHG) = -0.00108 (Z=-6.2), and $f_4$(EN, MN; WHG, SHG) = -0.00057 (Z=-4.5), which shows that the WHG are a better source of this ancestry in Middle Neolithic Europeans than the hunter-gatherers of Sweden or Russia.

**Dilution of EHG ancestry in the Yamnaya**

The Yamnaya pastoralists from the Samara district are not descended only from the EHG that precede them in eastern Europe. To show this, we estimate the value of the statistic $f_4$(Karelia_HG, Yamnaya; *Test*, Chimp) for different modern/ancient *Test* populations. In Table S7.5, we show the populations that maximize/minimize this statistic. European hunter-gatherers, MA1, and Native Americans share more alleles with Karelia_HG than with Yamnaya, and Near Eastern populations share more alleles with Yamnaya than Karelia_HG.

**Table S7.5: Dilution of EHG in the Yamnaya: $f_4$(Karelia_HG, Yamnaya; *Test*, Chimp)**

| *Test* | Statistic | Z | *Test* | Statistic | Z |
|---|---|---|---|---|---|
| Iraqi_Jew | -0.00176 | -4.2 | Samara_HG | 0.00594 | 8.9 |
| Armenian | -0.00161 | -3.9 | Motala_HG | 0.00532 | 11.3 |
| Abkhasian | -0.00157 | -3.7 | SwedenSkoglund_NHG | 0.00476 | 8.2 |
| Georgian | -0.00151 | -3.6 | MA1 | 0.00436 | 6.7 |
| Yemenite_Jew | -0.00149 | -3.6 | SwedenSkoglund_MHG | 0.00414 | 2.9 |
| Saudi | -0.00137 | -3.3 | Mixe | 0.00401 | 8.3 |
| Iranian_Jew | -0.00131 | -3.2 | Wayuu | 0.00401 | 7.3 |
| Druze | -0.00131 | -3.2 | Piapoco | 0.00396 | 7.8 |
| Tunisian_Jew | -0.00130 | -3.1 | Cabecar | 0.00390 | 7.6 |
| Lebanese | -0.00125 | -3.1 | Aymara | 0.00388 | 8.1 |

It seems that an unknown population, related to present-day Near Eastern populations, entered eastern Europe at some time between the EHG and the formation of the Yamnaya. Nonetheless, the EHG shared more alleles with the Yamnaya than with nearly all Early, Middle, Late Neolithic and Bronze Age populations from Europe outside Russia (Table S7.6).

**Arrival of Eastern European hunter-gatherer ancestry during the Late Neolithic**

Both PCA (Fig. 2a) and ADMIXTURE analysis (Fig. 2b) show that the shift that occurred in Europe between the Middle and Late Neolithic was qualitatively different from the WHG resurgence. We can show this formally by studying the statistic $f_4$(LBK_EN, *Test*; Loschbour, Karelia_HG) which tests whether allele frequency differences between the early LBK_EN farmers and a *Test* population are correlated to differences between eastern and western European hunter-gatherers. Fig. S7.9 shows the value of this statistic for different MN/LN/BA populations. For Middle Neolithic Europeans (bottom half), the statistic is negative, corresponding to the WHG-related resurgence, but for Late Neolithic/Bronze Age ones (and maximally for the Yamnaya steppe population) the statistic is positive, demonstrating the arrival of eastern populations related to the Yamnaya and to the EHG during this period.

**Table S7.6: Karelia_HG shares more alleles with Yamnaya than with most EN/MN/BA Europeans outside Russia**

| ***Test*** | $f_4$(***Test*, Yamnaya, Karelia_HG, Chimp)** | **Z** |
|---|---|---|
| LBKT_EN | -0.00811 | -5.5 |
| Stuttgart | -0.00704 | -13.0 |
| HungaryGamba_EN | -0.00669 | -16.0 |
| Starcevo_EN | -0.00639 | -7.0 |
| HungaryGamba_CA | -0.00607 | -9.0 |
| LBK_EN | -0.00565 | -14.7 |
| Spain_EN | -0.00538 | -12.1 |
| Baalberge_MN | -0.00470 | -7.1 |
| Iceman | -0.00466 | -8.0 |
| SwedenSkoglund_MN | -0.00440 | -6.7 |
| Spain_MN | -0.00361 | -8.1 |
| HungaryGamba_BA | -0.00328 | -6.4 |
| Esperstedt_MN | -0.00282 | -4.3 |
| Halberstadt_LBA | -0.00186 | -3.2 |
| Bell_Beaker_LN | -0.00181 | -4.7 |
| BenzigerodeHeimburg_LN | -0.00176 | -3.6 |
| Unetice_EBA | -0.00161 | -4.5 |
| Alberstedt_LN | -0.00152 | -2.6 |
| Karsdorf_LN | -0.00060 | -0.6 |
| Corded_Ware_LN | -0.00001 | 0.0 |

**Figure S7.9: The statistic $f_4$(LBK_EN, *Test*; Loschbour, Karelia_HG) demonstrates the "eastern" shift of Late Neolithic/Bronze Age Europeans. The estimated value of the statistic and ±3 standard errors are shown.**

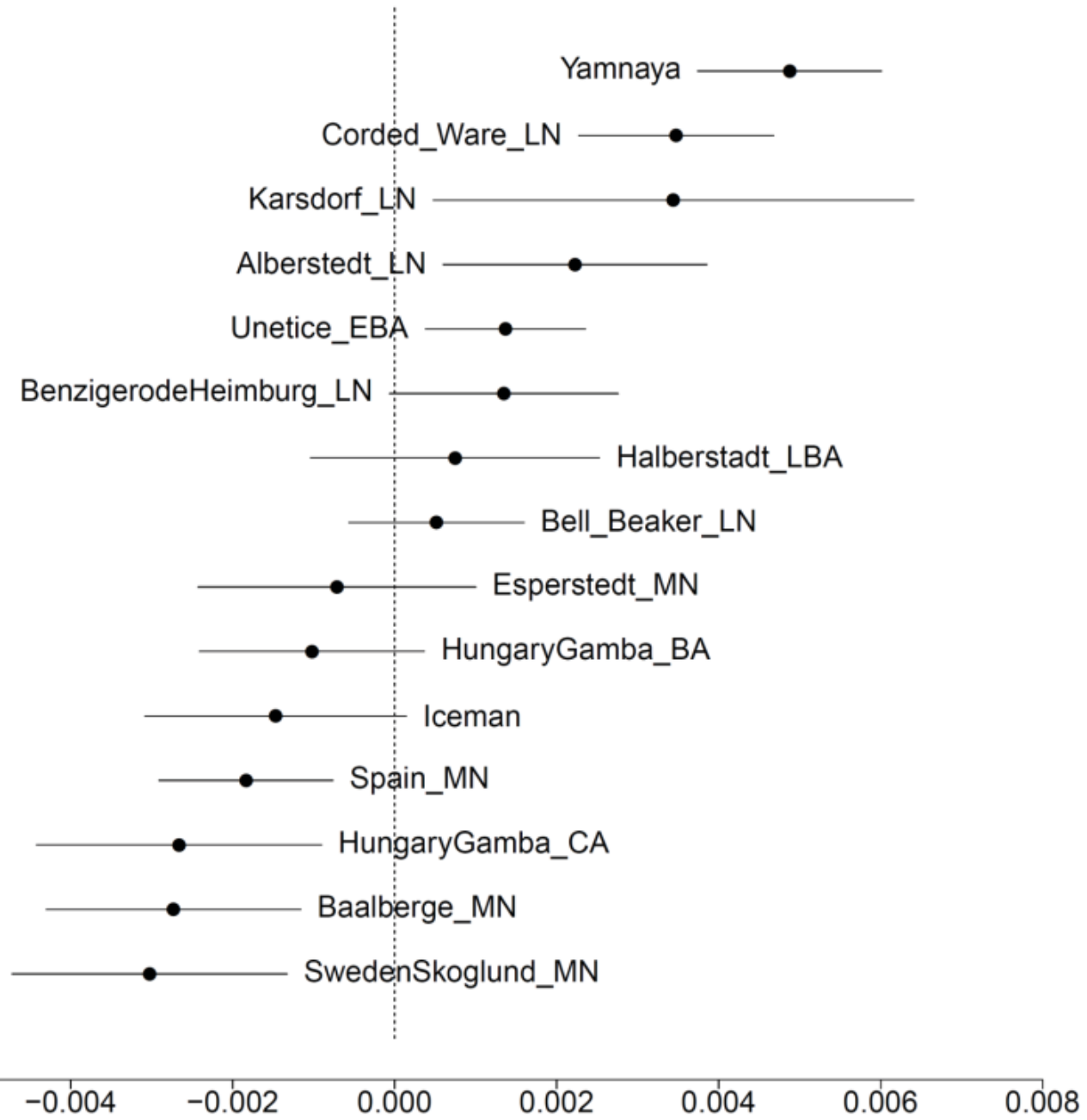

We can also show that the change between the Middle and Late Neolithic is related to eastern Europe by the statistics $f_4$(MN, LNBA; EHG, WHG) = -0.0032 (Z=-16.4) , and $f_4$(MN, LNBA; EHG, SHG) = -0.00189 (Z=-11.2), which show that allele frequency changes of populations between these two periods show an affinity to the EHG in the LNBA.

**The Corded Ware are intermediate between the Yamnaya and the Middle Neolithic**

To determine that the Corded Ware (the earliest Late Neolithic population from Germany, and the one which maximally expresses the "eastern" shift in Fig. S7. 9) differ from Middle Neolithic Europeans and the Yamnaya that temporally precede them in central and eastern Europe respectively, we study the statistics $f_4$(Corded_Ware, Esperstedt_MN, *Test*, Chimp), and $f_4$(Corded_Ware, Yamnaya; *Test*, Chimp) (Fig. S7.10). The first of these statistics shows that in relation to the Middle Neolithic population of Germany, the Corded Ware share more alleles with eastern populations, including EHG, MA1, Yamnaya, and Native Americans. The second shows that in relation to the Yamnaya, the Corded Ware share more alleles with Middle and Early Neolithic European populations. Thus, these statistics paint a similar picture as the PCA analysis (Fig. 2a) which places the Corded Ware in an intermediate position between the Yamnaya and Middle Neolithic Europeans.

**Figure S7.10: The Corded Ware are intermediate between the Yamnaya and the Middle Neolithic Europeans as assessed by the statistics $f_4$(Corded_Ware, Esperstedt_MN, *Test*, Chimp) (panel a), and $f_4$(Corded_Ware, Yamnaya; *Test*, Chimp) (panel b). The estimated value of the statistic and ±3 standard errors are shown.**

(a) (b)

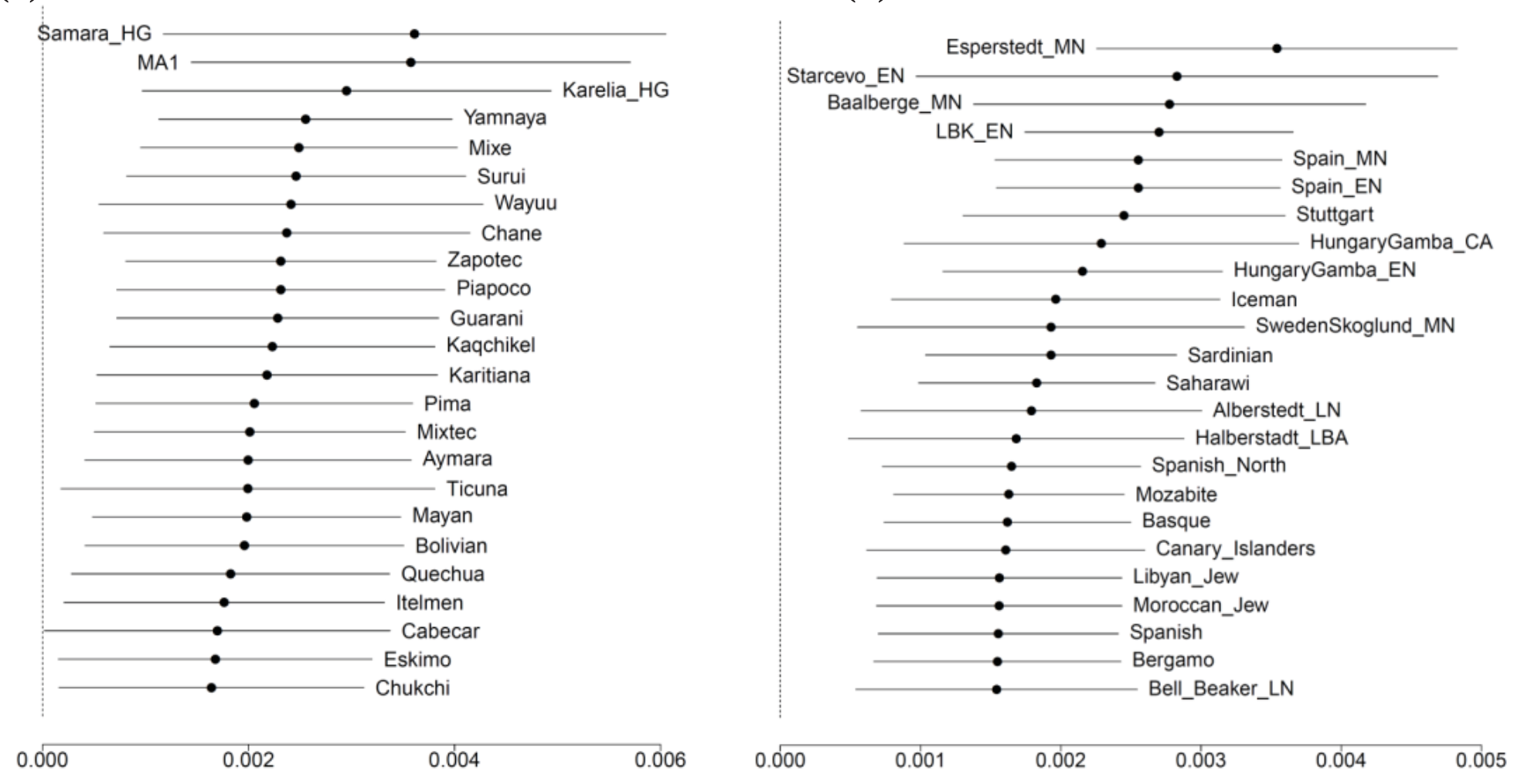

**Differences between Late Neolithic/Bronze Age groups**

While the Late Neolithic/Bronze Age groups appear to differ from the Middle Neolithic inhabitants of Europe in the same way (having more ancestry related to the Eastern European hunter-gatherers, MA1, and Yamnaya), they are not completely homogeneous. We studied differences between them using the statistic $f_4$(*X*, *Y*; *Test*, Chimp), allowing (*X*, *Y*) to be a pair of LN/BA populations, and *Test* to be Loschbour, Karelia_HG, Yamnaya, and LBK_EN; by doing so, we wanted to examine if these populations differed from each other with respect to WHG, EHG, Yamnaya, and EN populations of Europe. The significant statistics ($|Z|>3$) are shown in Table S7.7 and show that the Corded_Ware_LN had more ancestry related to eastern groups Karelia_HG and Yamnaya than any of the other LN/BA Europeans. It appears that the "eastern" ancestry introduced into Central Europe via the Corded Ware was later diluted in this region. This finding is inconsistent with a process of continuous migration from the east during the LN/BA period, as that would imply that temporally later populations would share more alleles with eastern populations as a result of having experienced more gene flow.

**Table S7.7: Significant (|Z|>3) statistics of the form $f_4$(*X*, *Y*; *Test*, Chimp) for (*X*, *Y*) a pair of LN/BA populations, and *Test* either Loschbour, Karelia_HG, Yamnaya, or LBK_EN.**

| *X* | *Y* | *Test* | Z | X |
|---|---|---|---|---|
| Alberstedt_LN | HungaryGamba_BA | Yamnaya | 0.00161 | 3.1 |
| Bell_Beaker_LN | HungaryGamba_BA | Yamnaya | 0.00121 | 3.0 |
| Corded_Ware_LN | Bell_Beaker_LN | Karelia_HG | 0.00182 | 4.4 |
| Corded_Ware_LN | Unetice_EBA | Karelia_HG | 0.00162 | 4.0 |
| Corded_Ware_LN | Halberstadt_LBA | Karelia_HG | 0.00187 | 3.2 |
| Corded_Ware_LN | HungaryGamba_BA | Karelia_HG | 0.00327 | 6.2 |
| Corded_Ware_LN | Bell_Beaker_LN | Yamnaya | 0.00162 | 5.1 |
| Corded_Ware_LN | BenzigerodeHeimburg_LN | Yamnaya | 0.00144 | 3.6 |
| Corded_Ware_LN | Unetice_EBA | Yamnaya | 0.00151 | 4.9 |
| Corded_Ware_LN | Halberstadt_LBA | Yamnaya | 0.00165 | 3.7 |
| Corded_Ware_LN | HungaryGamba_BA | Yamnaya | 0.00286 | 7.3 |
| Unetice_EBA | HungaryGamba_BA | Karelia_HG | 0.00161 | 3.2 |
| Unetice_EBA | HungaryGamba_BA | Yamnaya | 0.00129 | 3.4 |

**Differences between the Corded Ware and present-day Europeans**

We also studied differences between the Corded Ware and present-day Europeans using statistics of the form $f_4$(*European*, Corded_Ware_LN; *Other*, Chimp), with *Other* chosen from the list: LBK_EN, Loschbour, Karelia_HG, Yamnaya. These statistics are plotted in Fig. S7. 11, and show that both the EHG and the Yamnaya share more alleles with the Corded Ware than with any present-day European population. This is expected in the case of southern Europeans (as the Corded Ware horizon was a central/northern European phenomenon, and one might not expect present-day southern Europeans to form a clade with the Corded Ware population), but we find that it is also true for all present-day northern Europeans as well. This suggests that the ancestry introduced into Europe from the steppe during the Late Neolithic was later diluted, a process that had already begun during the Late Neolithic period itself (Table S7.7). This dilution may have involved the pre-existing farming population of Europe, but in parts of Europe may have included populations with substantial hunter-gatherer ancestry, as indicated by the fact that the statistic $f_4$(*European*, Corded_Ware_LN; Loschbour, Chimp) is significantly positive for some European populations such as Lithuanians, Estonians, and Icelanders (Fig S7.11c).

**Evidence of admixture with $f_3$-statistics**

The $f_4$-statistics examined so far do not resolve the direction of gene flow unambiguously. For example, negative $f_4$(*EN*, *MN*; *WHG*, Chimp) could indicate gene flow into *MN* from *WHG* or into *WHG* from *MN*. Likewise, negative $f_4$(*MN*, *LNBA*; *EHG*, Chimp) may be interpreted as either gene flow into *LNBA* from *EHG* or into *EHG* from *LNBA*. Our interpretation favoring Middle Neolithic WHG "resurgence" and Late Neolithic EHG "arrival" is based in part on the chronological order of the samples, as it seems unlikely that the MN and LN/BA populations of Europe would represent lineages that admixed into the earlier WHG and EHG than vice versa. However, we can also prove that our interpretation is correct by examining statistics of the form $f_3$(*Test*; *X*, *Y*) , using the *inbreed: YES* parameter of the qp3Pop program of ADMIXTOOLS[1]. Post-admixture genetic drift may mask a negative signal of admixture[1] and application of $f_3$-statistics on pseudo-haploid data must account for the pseudo-inbreeding induced by the complete homozygosity of the *Test* population individuals in a manner analogous to the real inbreeding present in some modern human populations[2]. Significantly negative statistics demonstrating the three main events proposed in our paper (WHG resurgence during the MN, EHG dilution in the Yamnaya, and MN/Yamnaya admixture in the formation of LN/BA populations) are listed in Table S7. 8. These statistics can only be estimated when *Test* has at least 2 individuals.

**Figure S7. 11: Present-day Europeans do not form a clade with the Corded Ware.**

(a) $f_4$(*European*, Corded_Ware_LN; Yamnaya, Chimp) (b) $f_4$(*European*, Corded_Ware_LN; Karelia_HG, Chimp)

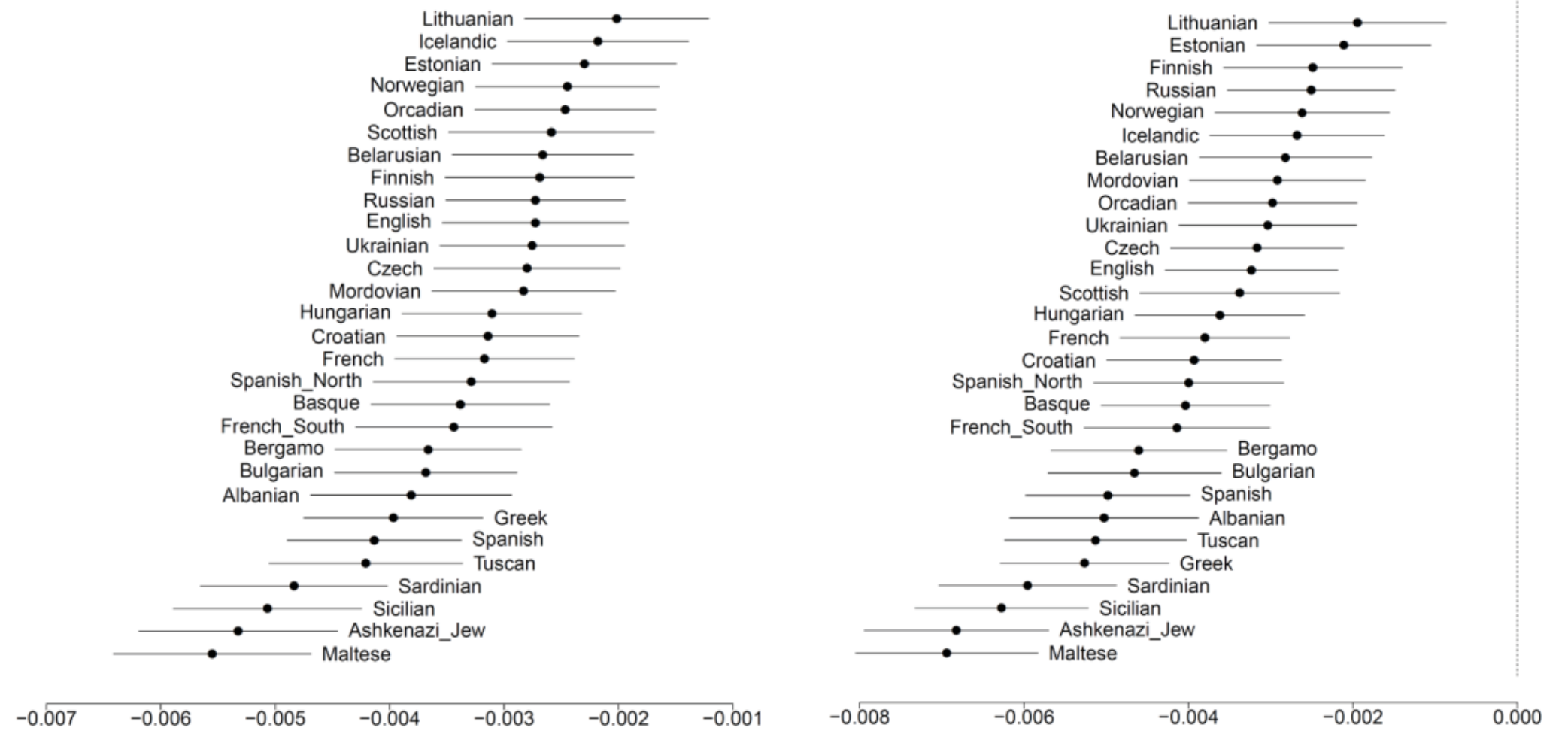

(c) $f_4$(*European*, Corded_Ware_LN; Loschbour, Chimp) (d) $f_4$(*European*, Corded_Ware_LN; LBK_EN, Chimp)

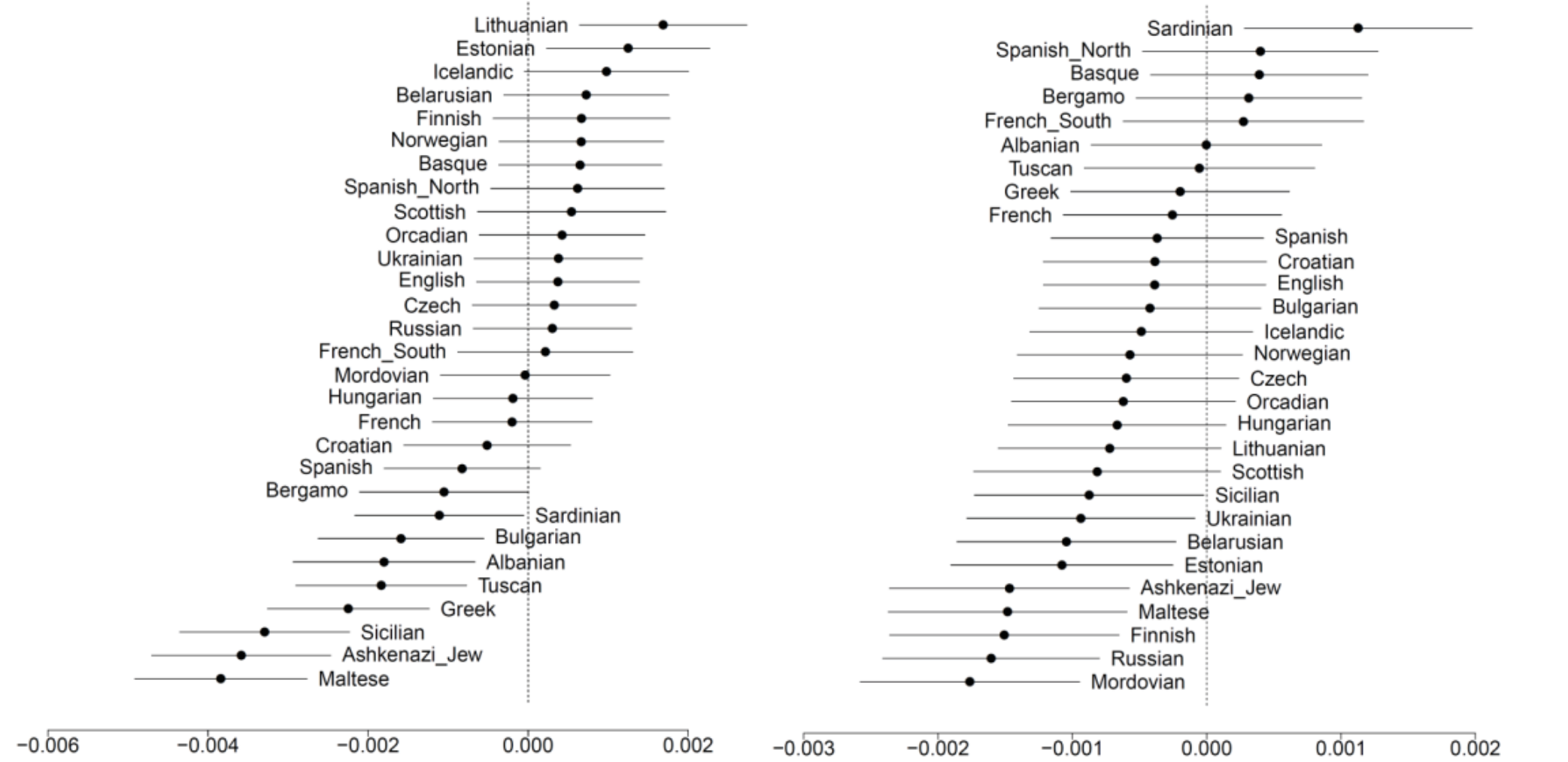

## Conclusions

We can discern three different groups of hunter-gatherers who lived in Europe before the arrival of the first farmers: western European hunter-gatherers (WHG) in Spain, Luxembourg, and Hungary; eastern European hunter-gatherers (EHG) in Russia, and Scandinavian hunter-gatherers (SHG) in Sweden. We can show that the early farmers of Europe descended from a common ancestral population. However, by the Middle Neolithic period they had experienced a resurgence of ancestry related to the WHG. In the Russian steppe, where farming did not get established, the Yamnaya pastoralists emerged as a mixture of the EHG and a Near Eastern population. Thus, while hunter-gatherer ancestry was increasing in western Europe, it was decreasing in eastern Europe. The two regions of Europe came into contact during the Late Neolithic period, when, beginning with the Corded Ware, migrants from the east, related to the EHG or the Yamnaya, arrived in central Europe. This does not seem to have been a process of slow infiltration of eastern populations over a prolonged period of time, as the earliest population (Corded Ware) was more closely related to the eastern groups than all the later ones. Other Late Neolithic and Bronze Age populations, as well as all present-day Europeans, have less ancestry related to these eastern migrants than the Corded Ware population did. In SI9 we quantify the magnitude of these population transformations by estimating mixture proportions in the affected populations.

**Table S7.8: Confirmation of main inferences using $f_3$-statistics.**

| ***Test*** | ***X*** | ***Y*** | **$f_3$(*Test*; *X*, *Y*)** | ***Z*** |
|---|---|---|---|---|
| Spain_MN | Spain_EN | Loschbour | -0.01001 | -5.2 |
| Baalberge_MN | LBK_EN | Loschbour | -0.01129 | -2.8 |
| Yamnaya | Karelia_HG | Armenian | -0.00765 | -6.3 |
| Corded_Ware_LN | Esperstedt_MN | Yamnaya | -0.00791 | -4.0 |
| Bell_Beaker_LN | Esperstedt_MN | Yamnaya | -0.00815 | -5.0 |
| BenzigerodeHeimburg_LN | Esperstedt_MN | Yamnaya | -0.00829 | -3.2 |
| Unetice_EBA | Esperstedt_MN | Yamnaya | -0.01055 | -7.2 |
| Corded_Ware_LN | Baalberge_MN | Yamnaya | -0.00394 | -1.8 |
| Bell_Beaker_LN | Baalberge_MN | Yamnaya | -0.00689 | -4.2 |
| BenzigerodeHeimburg_LN | Baalberge_MN | Yamnaya | -0.00747 | -3.0 |
| Unetice_EBA | Baalberge_MN | Yamnaya | -0.01139 | -8.0 |
| HungaryGamba_BA | Yamnaya | HungaryGamba_CA | -0.01088 | -3.3 |

# Supplementary Information 8

## Phylogenetic relationships of ancient Eurasians

Iosif Lazaridis*, Nick Patterson, and David Reich

* To whom correspondence should be addressed (lazaridis@genetics.med.harvard.edu)

### Overview

In this section we model the phylogenetic relationships of Eurasian hunter-gatherers using the ADMIXTUREGRAPH[1,2] software that allows one to propose an admixture graph model of history (a tree augmented with admixture edges), fit model parameters (genetic drift edges and admixture proportions), and report the difference between estimated and modeled *f*-statistics as the number of standard errors (Z-score) using a block jackknife[1,3].

The goals of this note are two-fold:

(1) *We show that the model we previously developed for the deep history of Eurasian populations[4] fits the new data reported in this study.*
We do not simply add the newly available samples to the previous model and test if they fit. Instead, we start afresh, exploring the space of possible models more fully by (i) exhaustively iterating tree topologies for a set of populations given a specified number of admixture events, and (ii) when adding a new population to an existing admixture graph, iterating over all possible placements of this population as either a simple branch or a 2-way mixture. By doing so, we mitigate against the subjectivity of specifying an admixture graph "by hand", or relying on an approach such as TreeMix[5] that does not search the space of models exhaustively and may overlook topologies that fit the data.

(2) *We show that several very different models of population relationships can fit the data once we add Eastern Hunter Gatherers to the modeling.*
When we add Karelia_HG (representing EHG), we are able to obtain fits to the data when we allow an additional admixture event. However, several very different models of history can be made to fit and are equally parsimonious in the sense of specifying the same number of admixture events. Because we do not want our inferences about mixture to be biased by arbitrary choices about a phylogenetic model relating all the populations, we developed new techniques, described in detail in SI9 and SI10, which are able to infer mixture proportions without precisely modeling phylogenetic relationships.

Point #2 is the key motivation for this note. While we could in theory have tried alternative model-fitting approaches like TreeMix[5], we did not attempt to do so because these fit the same underlying data as our ADMIXTUREGRAPH analysis. We know from this note that several very different models equally parsimoniously fit the data. Thus, even if TreeMix found additional models that fit, it would not change our conclusion that the whole exercise of model building based on allele frequency correlation statistics does not produce a unique solution. In other words, the main point of the model-building in this note is to motivate the development of methods, shown in SI9 and SI10, that work robustly regardless of the true underlying model.

**Details**

We begin by fitting the basic model[4] to the new data, substituting the Stuttgart Early European Farmer (EEF), which belonged to the Linearbandkeramik archaeological culture with the LBK_EN group (n=12). This model is fitted successfully (Fig. S8.1), with no *f*-statistics differing between estimated and modeled values by |Z|>3.

**Figure S8.1: Basic model from ref.4 applied to data of this study.**

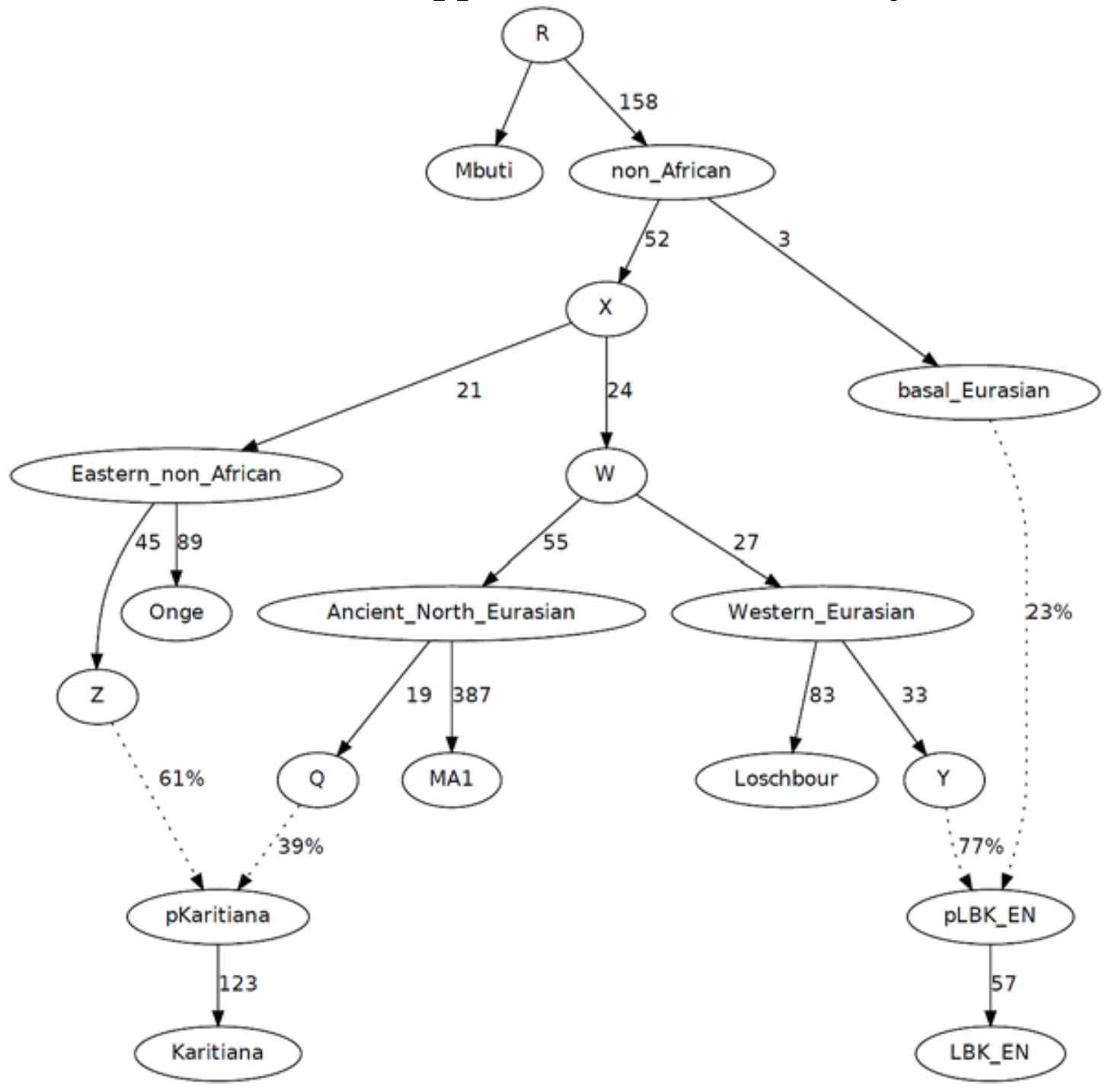

In ref.4, the amount of “Basal Eurasian” admixture in the Stuttgart farmer was also estimated using the $f_4$-ratio $f_4$(Stuttgart, Loschbour; Onge, MA1) / $f_4$(Mbuti, MA1; Onge, Loschbour) as 44±10%. When we estimate this parameter using LBK_EN instead of Stuttgart and obtain an estimate of 40.0±9.3%. Some basal Eurasian admixture into LBK_EN is necessary, as shown by the fact that when we force 0% “Basal Eurasian” admixture into LBK_EN, the model does not fit, with the worst discrepancy being for the statistic $f_3$(LBK_EN; Onge, MA1) whose fitted value is lower than the estimated one by Z=11.3. A model without Basal Eurasian admixture into LBK_EN seriously underestimates the shared genetic drift between MA1 and Onge; by introducing “Basal Eurasian” ancestry into LBK_EN the fact that eastern non-Africans share more alleles with European hunter-gatherers and MA1 is taken into account[4].

**Fitting Ust-Ishim**

Existence of “Basal Eurasian” admixture into Europeans was further supported by a study of the Ust’-Ishim[6] Upper Paleolithic Siberian (~45 thousand years ago). Ust’-Ishim was inferred to occupy a basal position to eastern non-African and European hunter-gatherer populations but not to modern Europeans, which was interpreted as due to admixture from a population occupying an even more basal position to Europeans, as proposed in ref. 4. We attempted to fit Ust-Ishim anywhere on the graph of Fig. S8.1, but all phylogenetic placements failed (|Z|>3.8) except the placement of Ust-Ishim on the non_African→X edge which is shared by all Eurasians other than Basal Eurasians (Fig. S8.2).

**Figure S8.2: Ust-Ishim is basal to all Eurasians except "Basal Eurasians" in the sense of ref.4**

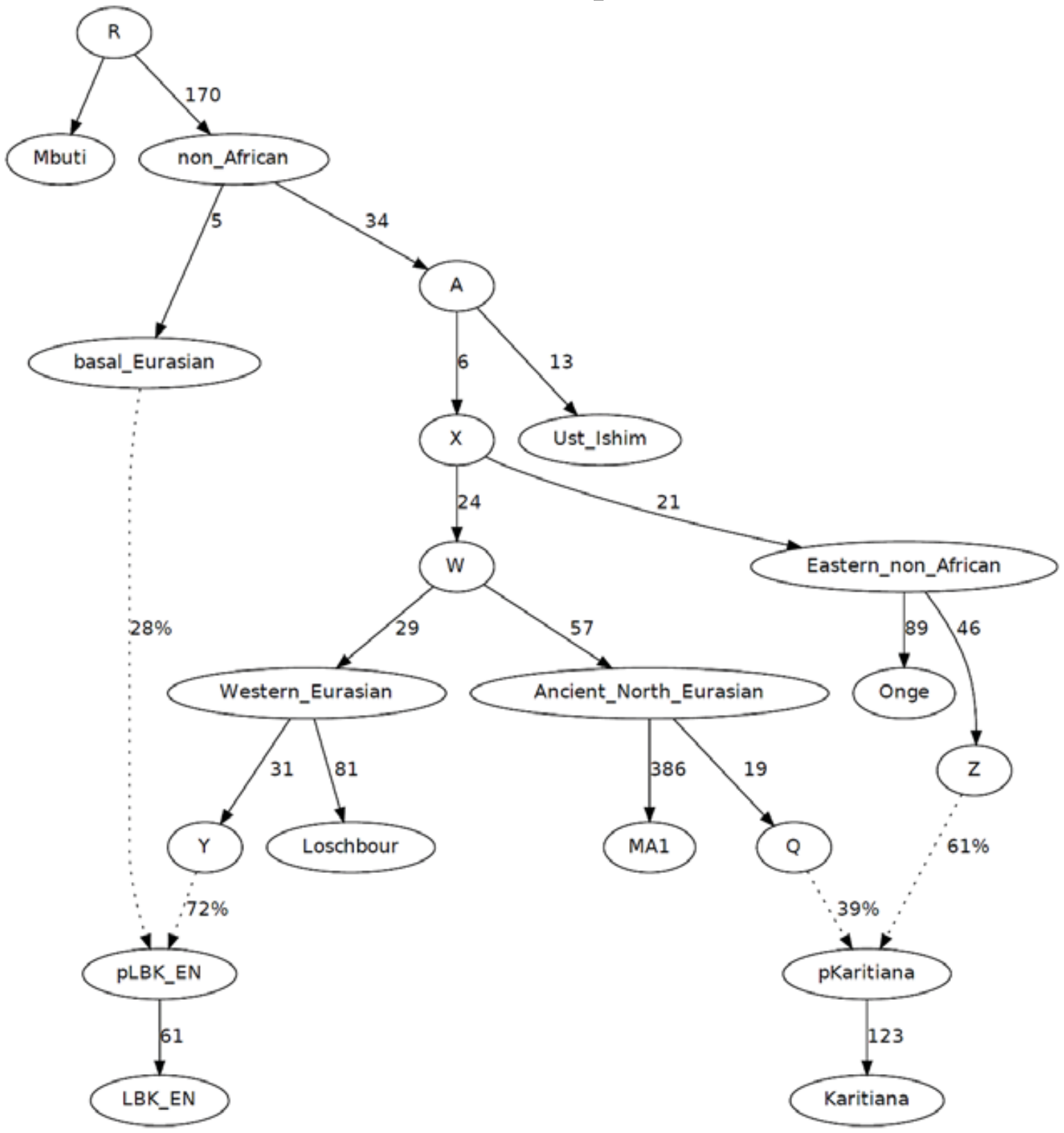

Thus, the split between "Basal" and other Eurasians must have taken place no later than ~45,000 years ago (the age of the Ust-Ishim individual).

**Fitting Kostenki14**

A study of the Kostenki14 Upper Paleolithic individual from Russia[7] (K14) proposed that this individual already had Basal Eurasian ancestry ~36,000 years ago, almost 30,000 years before the appearance of this type of ancestry in Europe with the early farmers[4]. This suggestion was based on three lines of evidence (Fig. 2 of ref. [7]):

1. D(Mbuti, East Asia; Mesolithic European Hunter Gatherer, K14) < 0. We confirm this asymmetry in our data, as D(Mbuti, Han; Loschbour, Kostenki14) = -0.0177 (Z=-3.7).
2. D(Mbuti, East Asia; Early European Farmer, K14) = 0. We confirm this symmetry in our data as D(Mbuti, Han; LBK_EN, Kostenki14) = 0.0034 (Z=0.8).
3. Presence of "Middle Eastern" admixture in ADMIXTURE analysis of this individual. We also confirm this point in our ADMIXTURE analysis (SI6). However, we note that interpretation of this finding is not straightforward, as old samples such as Ust_Ishim and Kostenki14 may predate the differentiation of populations, and the fact that they are seen as admixed from multiple sources in ADMIXTURE analysis (SI6), does not indicate that they are mixtures of these much later populations.

Arguments #1, 2 do not suffice to draw the conclusion that Kostenki14 had the same Basal Eurasian ancestry as Neolithic European farmers. To illustrate this point, we use the model of Fig. S8.2 as a baseline. First, we attempt to fit Kostenki14 into this model as either a simple branch or a 2-way mixture.

No phylogenetic placement of Kostenki14 as a simple branch fits the data (|Z|>4.6). However, we could fit Kostenki14 as a 2-way mixture (Fig. S8.3-5). All successful models agree that Kostenki14 shares genetic drift with Loschbour and LBK_EN to the exclusion of non-European populations and is "European" in this sense. However, the model of Fig. S8.3 suggests it has Basal Eurasian ancestry that split off *before* the split of this type of ancestry in LBK_EN, while those of Fig. S8.4-5 are equivalent and suggest that its Basal Eurasian ancestry is the same as that found in LBK_EN and there is a trifurcation of the Basal Eurasian ancestry for the three lineages {LBK_EN, Kostenki14, main Eurasian ancestry node "A"}.

The hypothesis of Basal Eurasian ancestry in Kostenki14 needs to be further tested, as the negative D(Mbuti, Han; Loschbour, Kostenki14) statistic could also reflect gene flow between Han↔Loschbour; this is *a priori* plausible, as these populations are much younger than Kostenki14 and may share intra-Eurasian genetic drift that Kostenki14 lacks because of its age. The possibility of later gene flow between Europeans and eastern non-Africans must be further tested with additional ancient samples from Upper Paleolithic Europe and Asia.

**Figure S8.3 Kostenki14 as a 2-way mixture with basal Eurasian admixture *before* that of the early farmers**

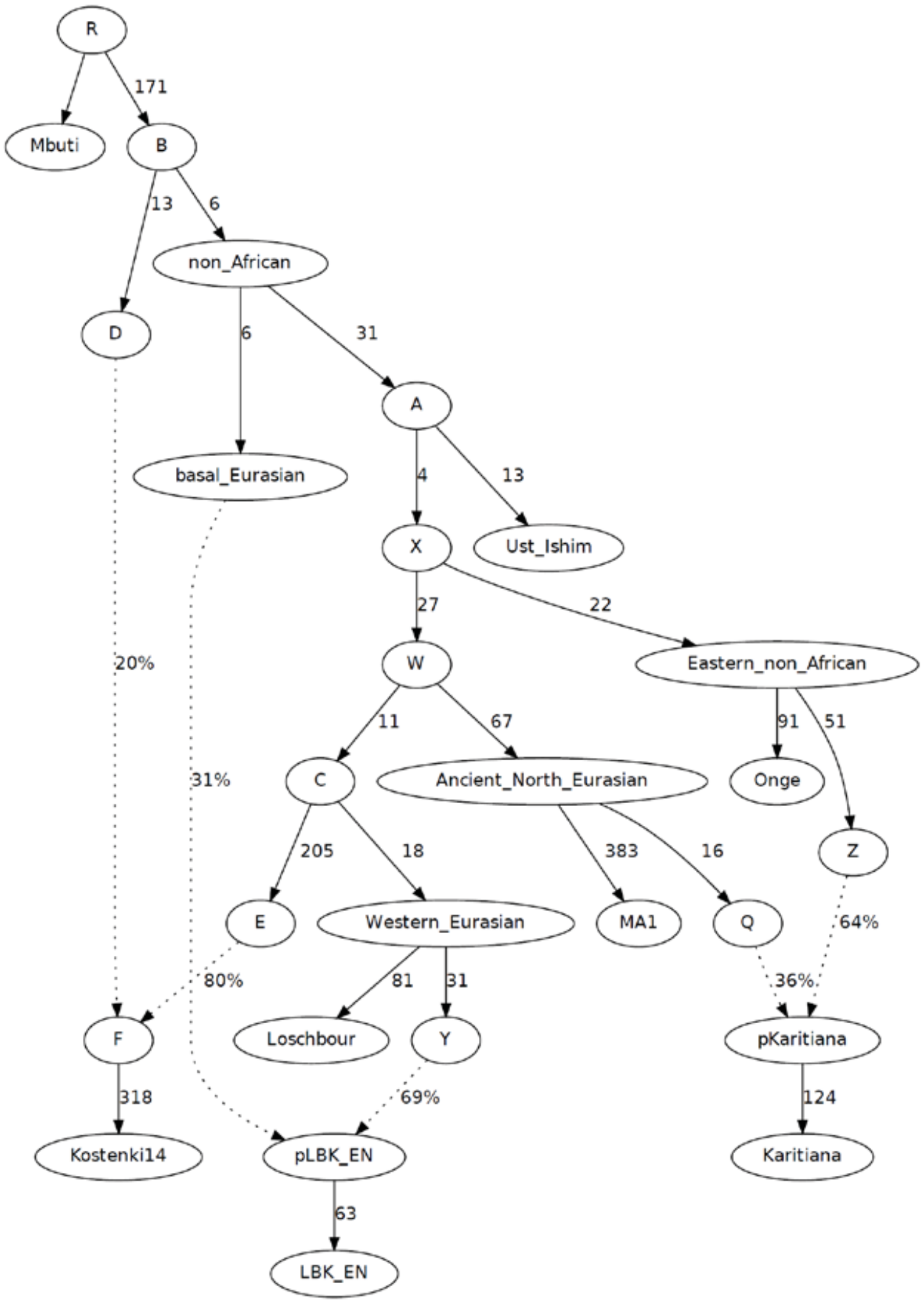

**Figure S8.4 Kostenki14 as a 2-way mixture with basal Eurasian admixture like that of the first farmers (however, topologically equivalent to Fig. S8.5 due to 0 drift in non_African→C branch)**

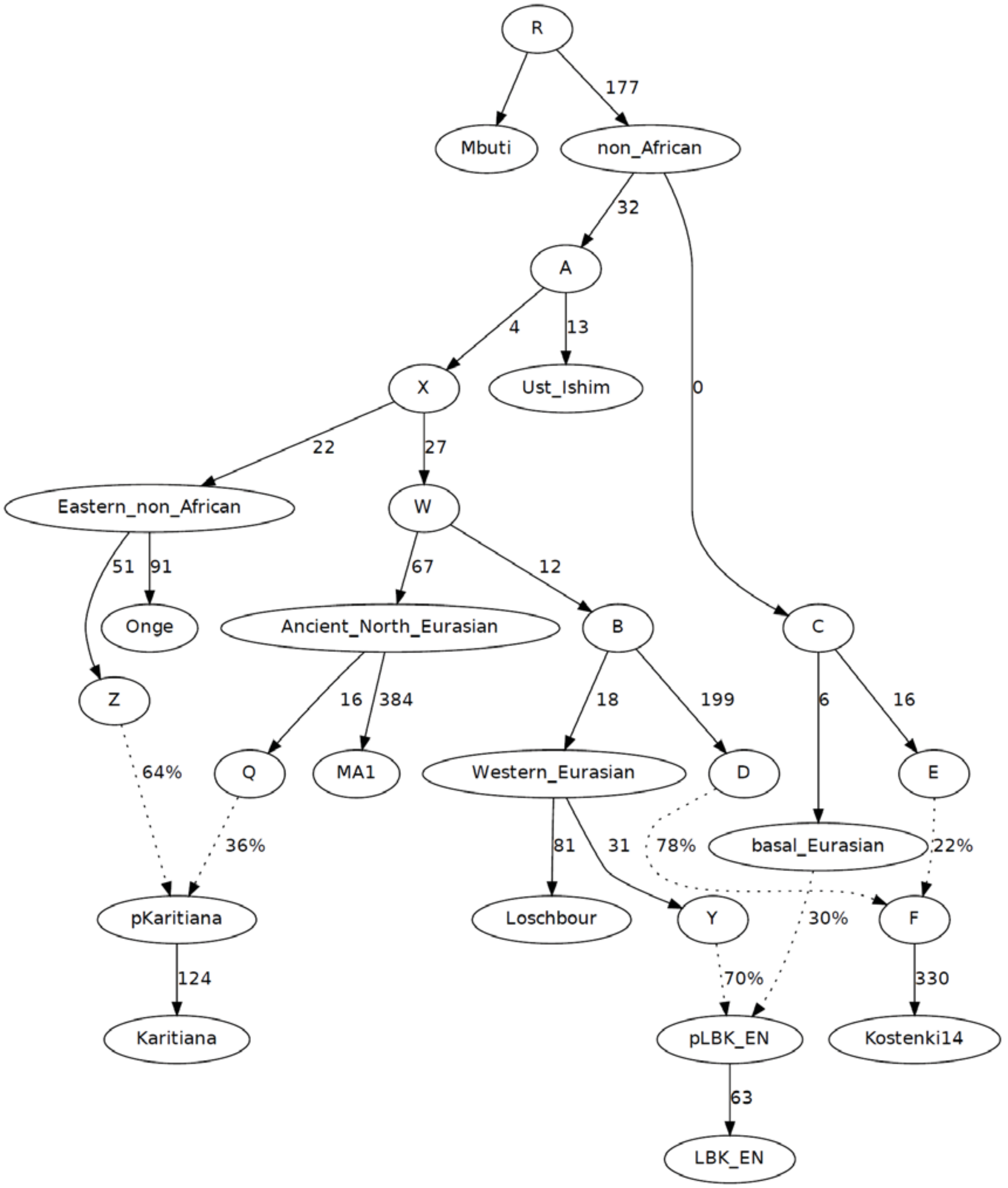

**Figure S8.5 Kostenki14 as a 2-way mixture with basal Eurasian admixture after that of the first farmers (but topologically equivalent to Fig. S8.4 due to 0 drift in non_African→C branch).**

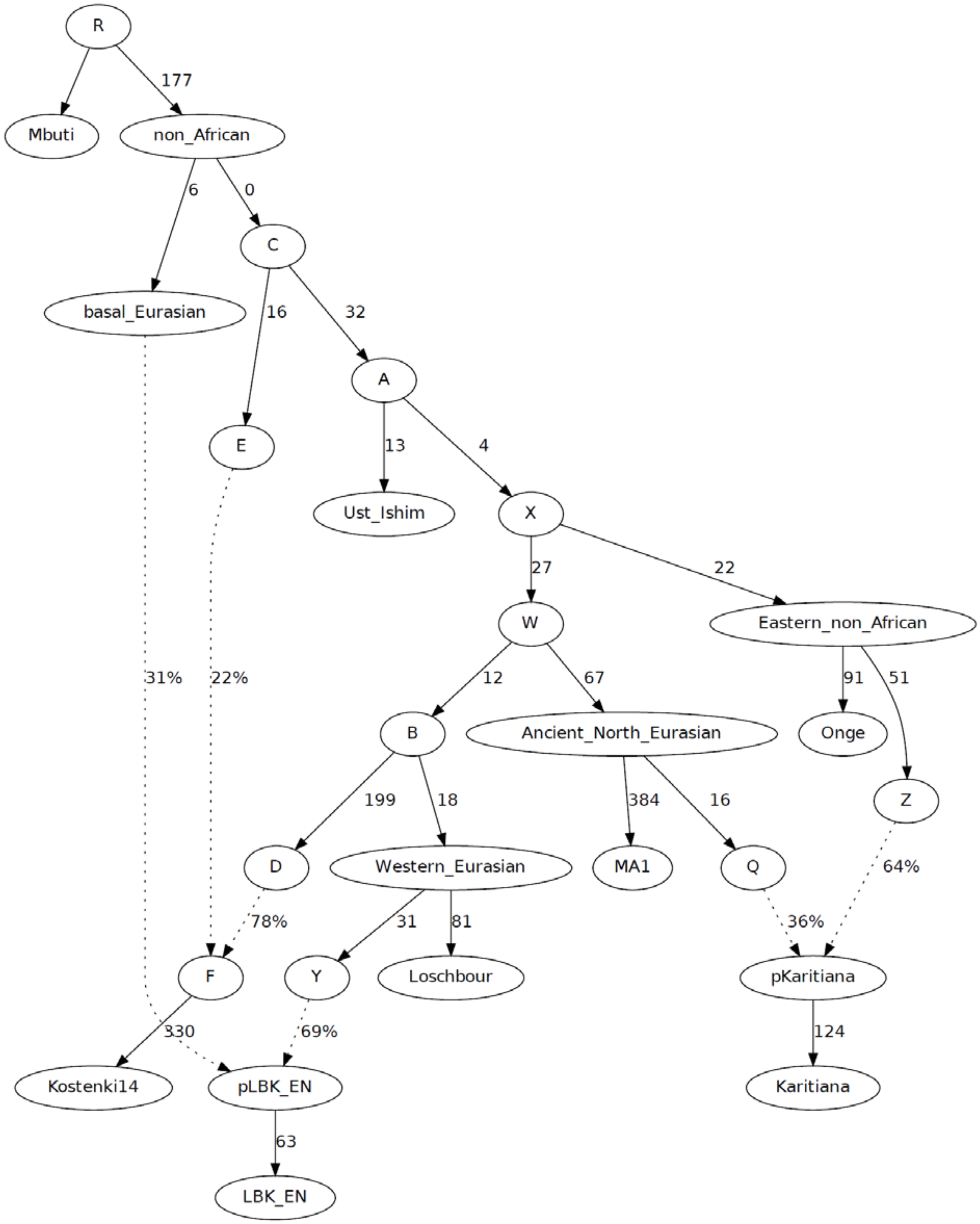

**Figure S8.6: Karelia_HG as 2-way mixture of MA1 and EHG (the model fails marginally with |Z|=3.1)**

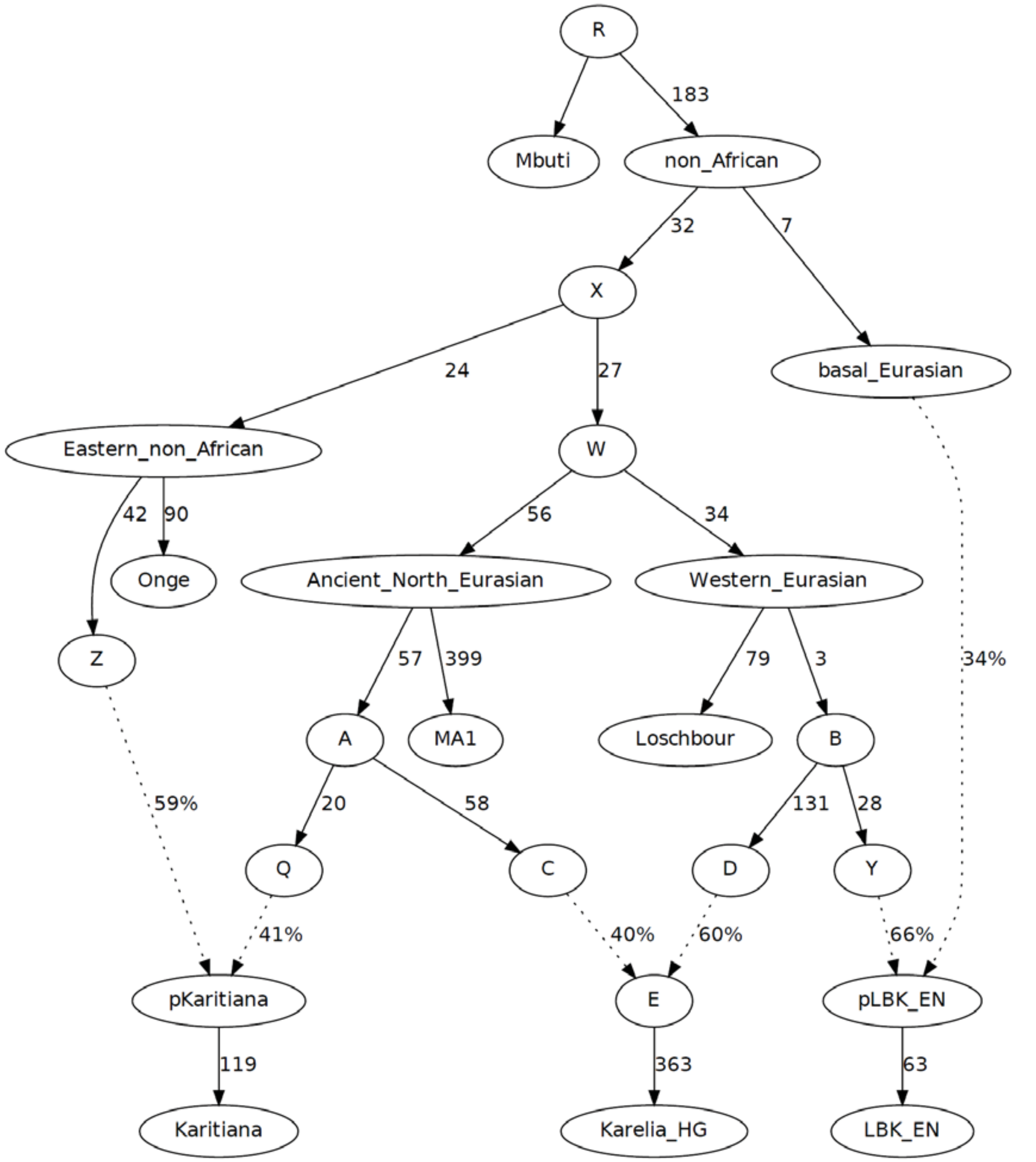

**Eastern European hunter-gatherers (EHG) as a WHG-ANE mixture**

A population that had not been previously studied is the Eastern European hunter-gatherers (EHG) from European Russia, so we attempt to fit Karelia_HG (the best sample from the EHG) to the model of Fig. S8.1. No models could accommodate Karelia_HG as a simple (not admixed) population (|Z|>9), or a 2-way mixture (|Z|>3). However, three models failed only marginally (3.1<|Z|<3.7) and we show one of them in Fig. S8.6. The model of Fig. S8.6 derives the Western Eurasian-related ancestry in Karelia_HG from a node D related to the population that admixed into the LBK_EN. The other two models (not shown) differ only slightly, by deriving the admixture into Karelia_HG either from the Loschbour branch, or a node that is basal to the Western_Eurasian node. Given the number of *f*-statistics computed, a largest |Z|-score of 3.1-3.7 does not provide very strong rejection of the model

All three of these 2-way mixture models arrive at a similar inference of 38-40% ANE and 60-62% WHG ancestry in Karelia-HG, which is higher than the ~20% ANE ancestry inferred for Motala12 in a previous study[4], consistent with the fact that EHG are the population sharing more alleles with "Ancient North Eurasians" (SI 7) than any other. This is also consistent with a previous hypothesis[4] of a "Hunter" population with 60-80% WHG/(WHG+ANE) ancestry that combined with early European farmers (EEF) to produce (in different proportions) present-day Europeans. The estimate 60% (Fig. S8.6) is at the low end of the inferred 60-80% range proposed in this previous study[4]. However, the present study reveals the Early European Farmers experienced a "resurgence" of WHG-related ancestry during the Middle Neolithic (SI 9), and thus, the WHG-related ancestry in present-day Europeans could come both from that event (that is, a Middle Neolithic population with a WHG/(WHG+ANE) ratio of 100%), and an eastern population bearing EHG-related ancestry (with a WHG/(WHG+ANE) ratio of 60%). The sum of these two processes may explain the 60-80% range of this ratio of the previous study[4].

A different detail of Fig. S8.6 is that the ANE ancestry in Karelia_HG is derived from the branch of "Ancient_North_Eurasian" that goes into the Karitiana Native Americans, rather than the MA1 branch. This is plausible, given the fact that modern humans arrived in the Americas after the time of MA1 (~24,000 years ago), giving added opportunity for more genetic drift to accumulate in the "Ancient North Eurasian" population; this could be shared by the EHG hunter-gatherers of European Russia from the Holocene period.

**Eastern European hunter-gatherers as an unadmixed population**

Fig. S8.6 shows Karelia_HG as a 2-way mixture of ANE and WHG, but this may be due to the fact that the model of Fig. S8.1 was developed before the EHG were sampled. We do not want to reify ANE and WHG simply on the basis of their having been described earlier in the literature, so we also explored models in which EHG are not admixed. This was driven, in part, by our observation that the statistic $f_4$(Karelia_HG, MA1; Karitiana, Chimp) = -0.00014 (Z=-0.2) which appears to be show a perfect symmetry in the relationship of (Karelia_HG, MA1) with Native Americans. Fig. S8.6 accounts for this "symmetry" between MA1 and Karelia_HG with respect to Native Americans by proposing a balancing of two processes: first, Karelia_HG share more alleles with Native Americans due to sharing additional common genetic drift (and thus Native Americans should share more alleles with Karelia_HG than with MA1), but, second, Karelia_HG has WHG-related ancestry which dilutes this affinity to Native Americans. However, this requires that the additional common drift shared by Karelia_HG and Native Americans to be nearly perfectly balanced with the dilution due to WHG ancestry, resulting in the observed symmetry, which is not very parsimonious.

To study the relationship between ANE, EHG, and WHG without necessarily accepting the model of Fig. S8.1, we attempted to fit MA1, Karelia_HG, and Loschbour as a simple tree using Mbuti as an outgroup. There are 3 possible arrangements of this kind (depending on whether MA1, Karelia_HG, or Loschbour is chosen as an outgroup to the two others), but none fit successfully (|Z|>6), suggesting that at least one of the 3 populations is admixed. We have already explored the possibility that MA1 and Loschbour are unadmixed (Fig. S8.1) which leads to the inference that EHG are admixed (Fig. S8.6), so we only examine the two remaining possibilities: (i) that MA1 is admixed and Loschbour, Karelia_HG are unadmixed, and (ii) that Loschbour is admixed and MA1, Karelia_HG are unadmixed. These two models fit successfully, and are shown in Fig. S8.7.

**Figure S8.7: Alternative phylogenies that fit the data in which (a) MA1 is admixed, or (b) Loschbour is admixed.**

(a) (b)

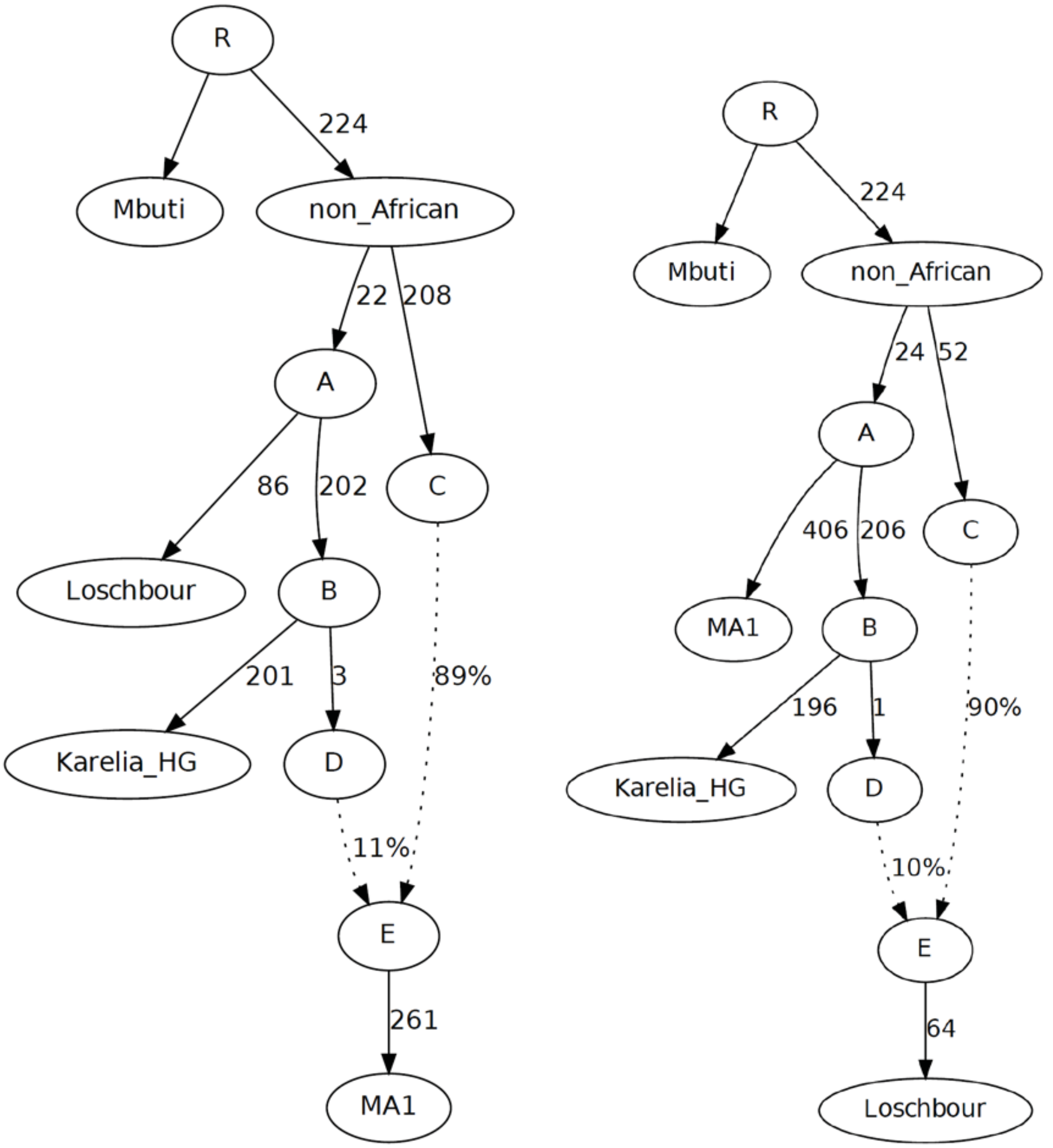

We next added new populations to the models of Fig. S8.7. First, we added the Onge[8], as an isolated eastern non-African population that could constrain these models (Fig. S8.8), and obtained good fits.

Next, we added the Karitiana Native Americans, to account for the different relationship between Native Americans to west Eurasian groups, as they share more alleles with MA1 and EHG than with Loschbour. The Karitiana cannot be added as an unadmixed population to the models of Fig. S8.5 ($|Z|>8$). They can, however, be modeled as a 2-way mixture, with successful ($|Z|<3$) models shown in Fig. S8.9. Thus, for every model we analyzed, our results support ref. [9] that Native Americans are anciently admixed.

**Figure S8.8: Adding Onge as an eastern non-African constraint to the models of Fig. S8.4**

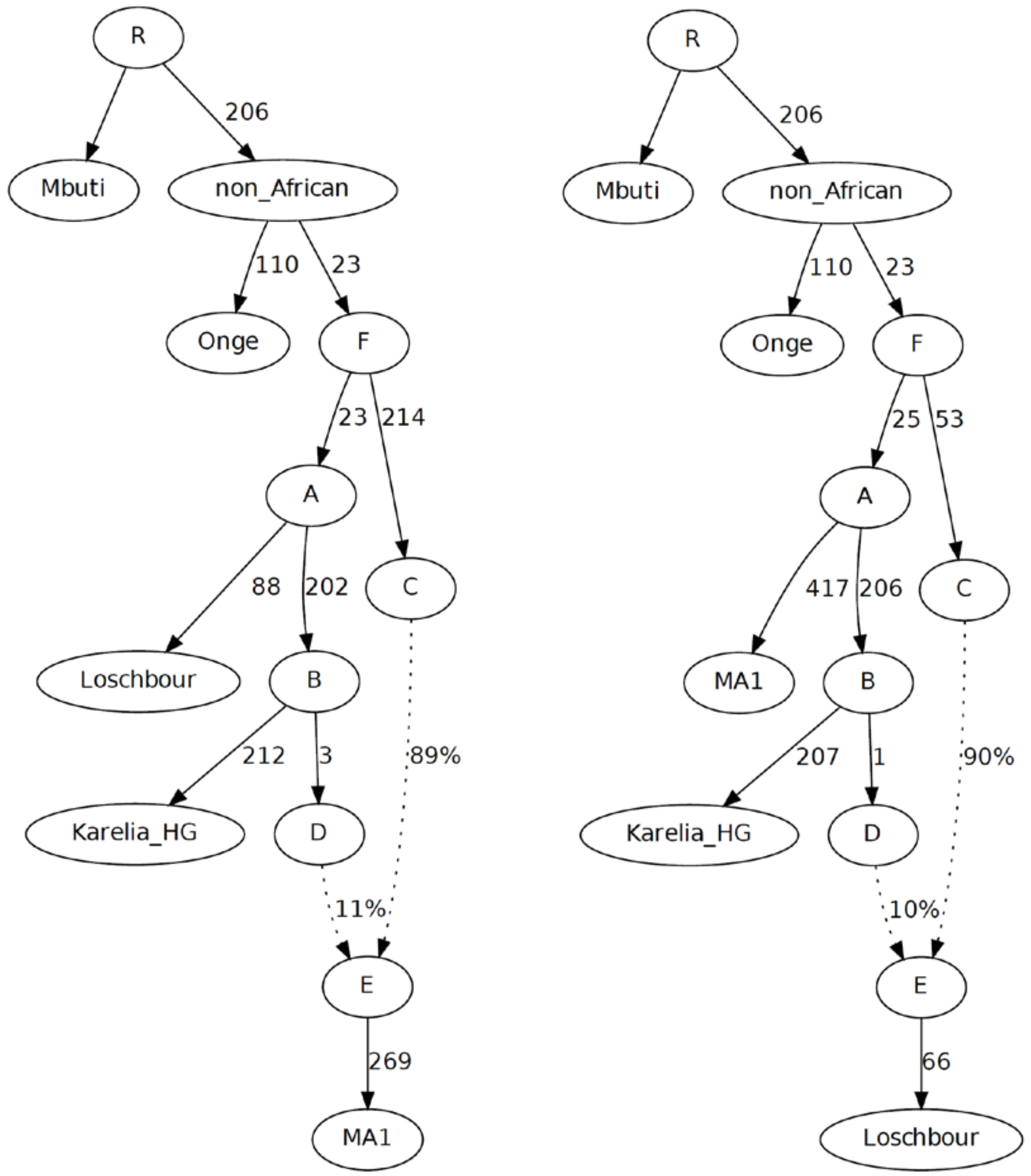

In Fig. S8.9a (when MA1 is admixed), Native Americans shown to be a mixture of a population J related to the Onge and a different population K related to MA1 and Karelia_HG. Note that this model accounts for the symmetry in the relationship of Native Americans to MA1 and EHG in a different way than that of Fig. S8.6. In Fig. S8.9a, Native American ancestry is more closely related to MA1 than to EHG, however, it is MA1 that derives part of its ancestry from an early node C which "dilutes" its affinity to Native Americans, while Karelia_HG shares the genetic drift on the path F→A→B fully.

Fig. S8.9b and Fig. S8.9c present two different solutions (when Loschbour is admixed). Both (b) and (c) agree with (a) in deriving ancestry of the Karitiana from both eastern non-Africans and populations related to west Eurasians[4,9,10]. Unlike the models of Fig. S8.9a and Fig. S8.6, however, they do not resolve the symmetric relationship of Native Americans to MA1 and Karelia_HG by postulating that Native Americans share additional common genetic drift with MA1 or Karelia_HG, but this ancestry is diluted in MA1 and Karelia_HG, resulting in the observed symmetry. Rather, Fig

S8.9b and Fig. S8.9c propose a trifurcation of Ancestral North Eurasians, whereby the ancestry that goes into Native Americans is equally related to both MA1 and Karelia_HG, both of which are unadmixed. (Note the 0 length edges A→G and G→A in the models of Fig. S8.6b and Fig. S8.6c, reflecting this trifurcation).

**Figure S8.9: Adding Karitiana as a constraint for ancient North Eurasian-related populations**

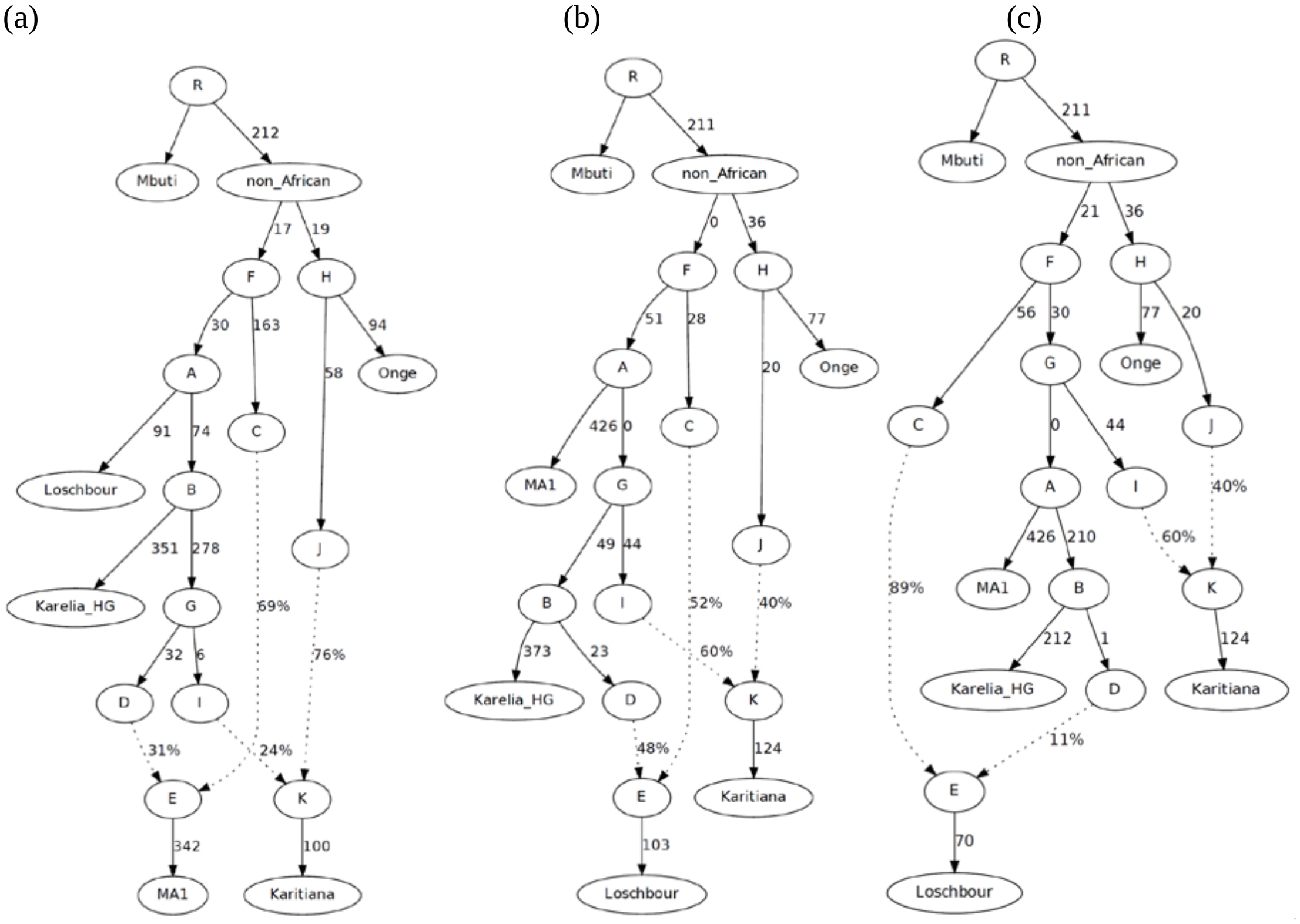

We finally added LBK_EN to the population models of Fig. S8.9. This could only be fit as a 2-way mixture that included a "Basal Eurasian" component, with the successful models shown in Fig. S8.10-11. The remainder of the ancestry of LBK_EN is inferred to be a sister group of Loschbour (Figs. S8.10, 11b, 12b) or a sister group of a component of Loschbour (Figs. S8.11a, 12a).

This exploration of possible phylogenies has revealed that seven models with three admixture events are consistent (|Z|<3) or marginally inconsistent (3<|Z|<4) with the studied set of populations (Fig. S8.6, 10, 11, 12). The models of Fig. S8.6 have an advantage in postulating that EHG are a mixture of populations related to MA1 and Loschbour, which are actual individuals, while those of Figs S8.10, 11, 12 propose that MA1 or Loschbour are admixed populations, although at present there are no actual individuals that represent some of the admixing populations. Nonetheless, all these models agree on several points: the dual origin of Native Americans[9], the existence of Basal Eurasian ancestry in Early Neolithic Europeans[4], the fact that MA1 shares more alleles with Karelia_HG than with Loschbour, but MA1 and Karelia_HG are symmetrically related to Native Americans, and, finally, the fact that the three group of Eurasian hunter-gatherers (EHG, WHG, and ANE) cannot be related to each other by a simple tree, and at least one of them must be admixed.

Uncertainty about the modeling of ancient Europeans may be a consequence of two factors.

First, it is possible that these populations were related in complex ways with gene flow between them. We show that at least one admixture event is required to jointly fit (EHG, WHG, and ANE). Models with more admixture events or bidirectional gene flow are more complex and thus less parsimonious; however, it is possible that they better capture the actual history of these populations.

Second, the sparse sampling of pre-Neolithic Europeans makes inferences about their relationships problematic. With the exception of Holocene hunter-gatherers from Sweden where multiple individuals are now known (ref. [4,10] and this study), a total of nine individuals of likely hunter-gatherer ancestry have autosomal data that is amenable to analysis (Samara_HG and Karelia_HG in this study, Loschbour[4], LaBrana[11], KO1 (HungaryGamba_HG)[12], Kostenki14[7], MA1[9], Ust-Ishim[6], and Tianyuan[13]). Tens of thousands of years since the earliest settlement of Europe by anatomically modern humans[14] and until the Holocene remain to be filled with data points that will doubtless inform our understanding of the distant past of European populations.

In the current paper, we do not address these issues, except to show, as we did above, the most salient points shared by some successful models, and to reject others. Fortunately, for the much shorter period since the Neolithic transition, there is an abundance of data from different parts of Europe and different points in time, allowing us to reconstruct some key transitions without explicitly modeling the deep phylogenetic relationships of Eurasian populations (SI9).

**Figure S8.10: Adding LBK_EN to the Fig. S8.9a model (MA1 admixed; this weakly fails |Z|=3.2)**

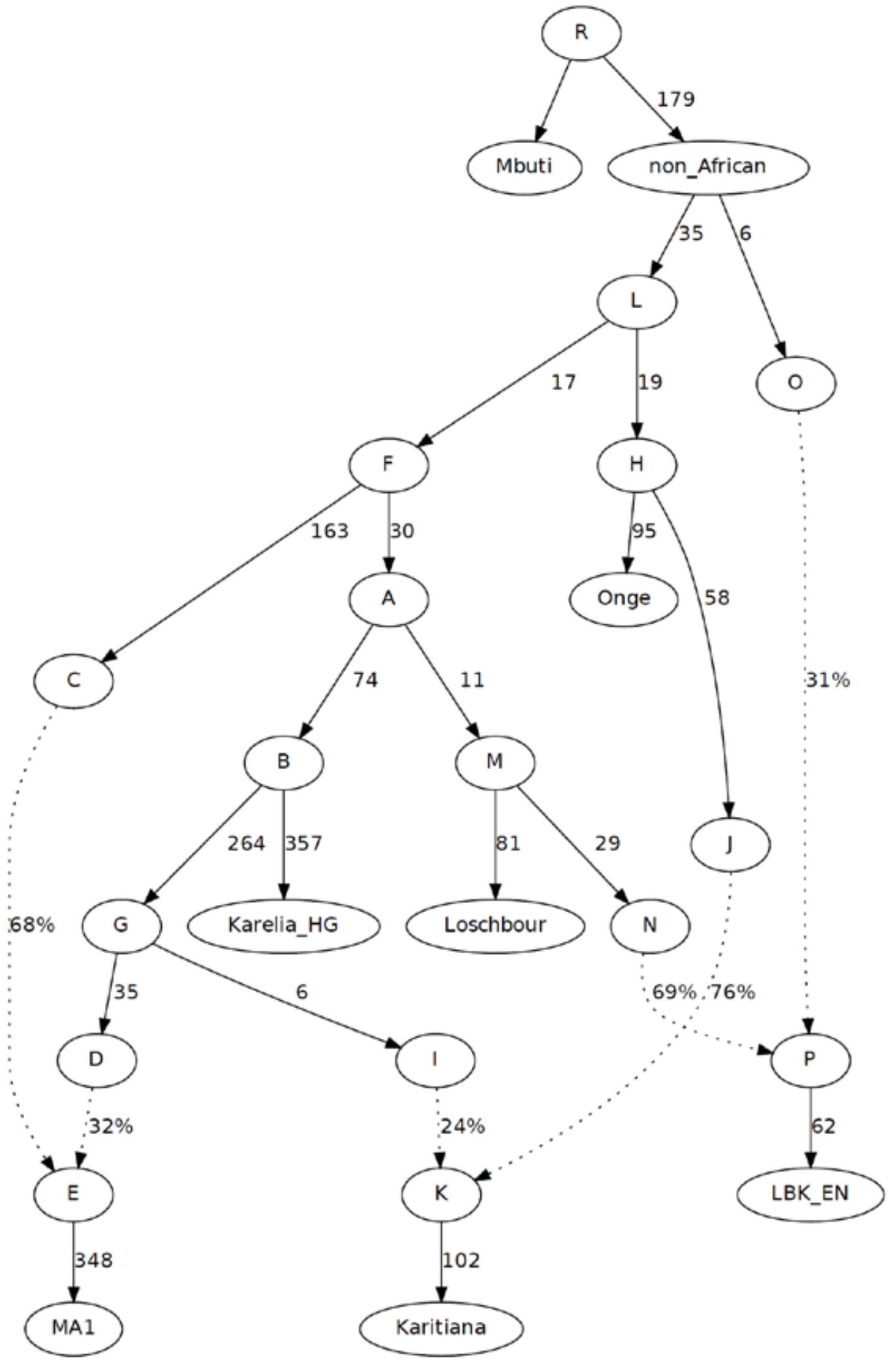

**Figure S8.11: Adding LBK_EN to the models of Fig. S8.9b (Loschbour is admixed)**

(a)

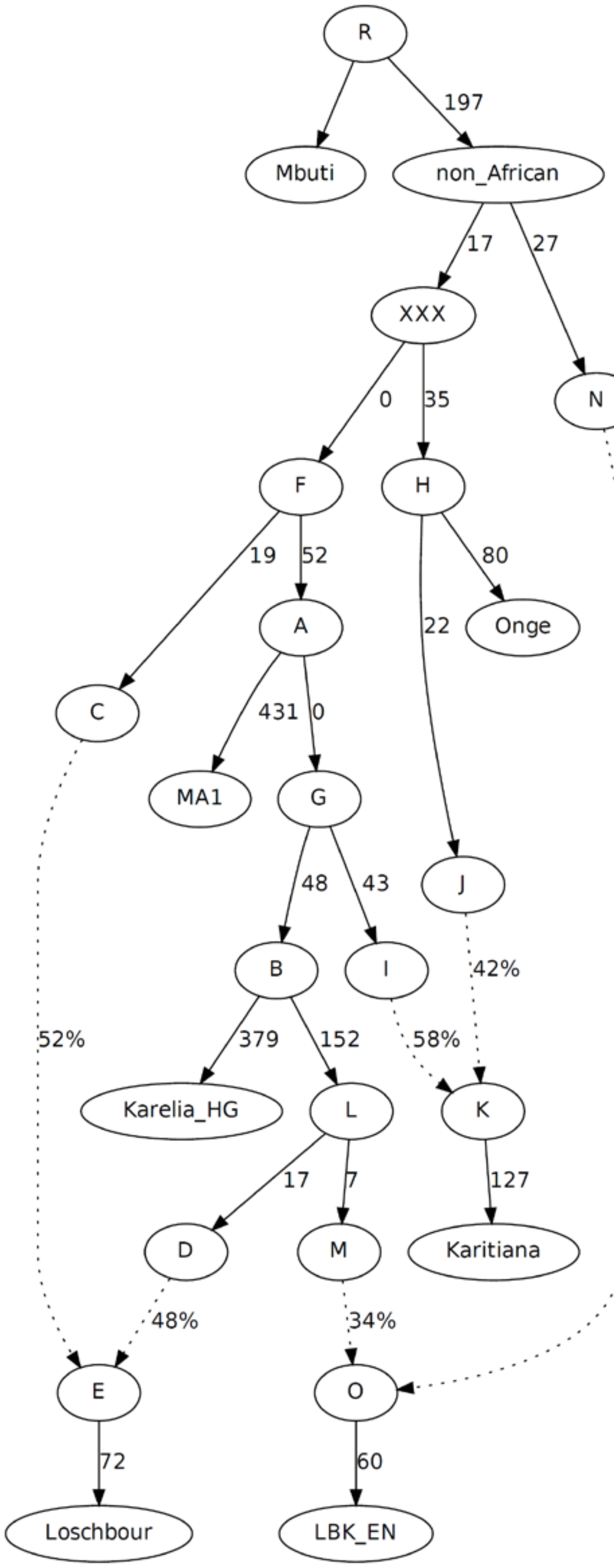

(b)

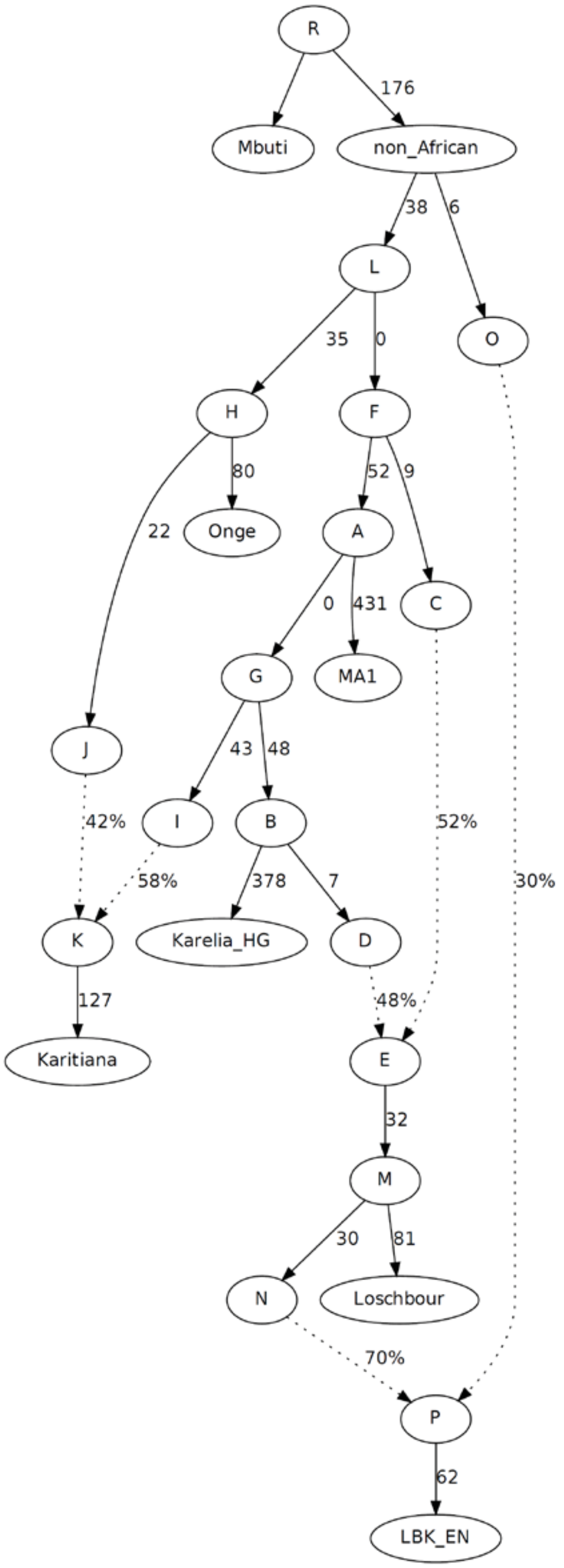

**Figure S8.12: Adding LBK_EN to the models of Fig. S8.9c (Loschbour is admixed)**

(a) (b)

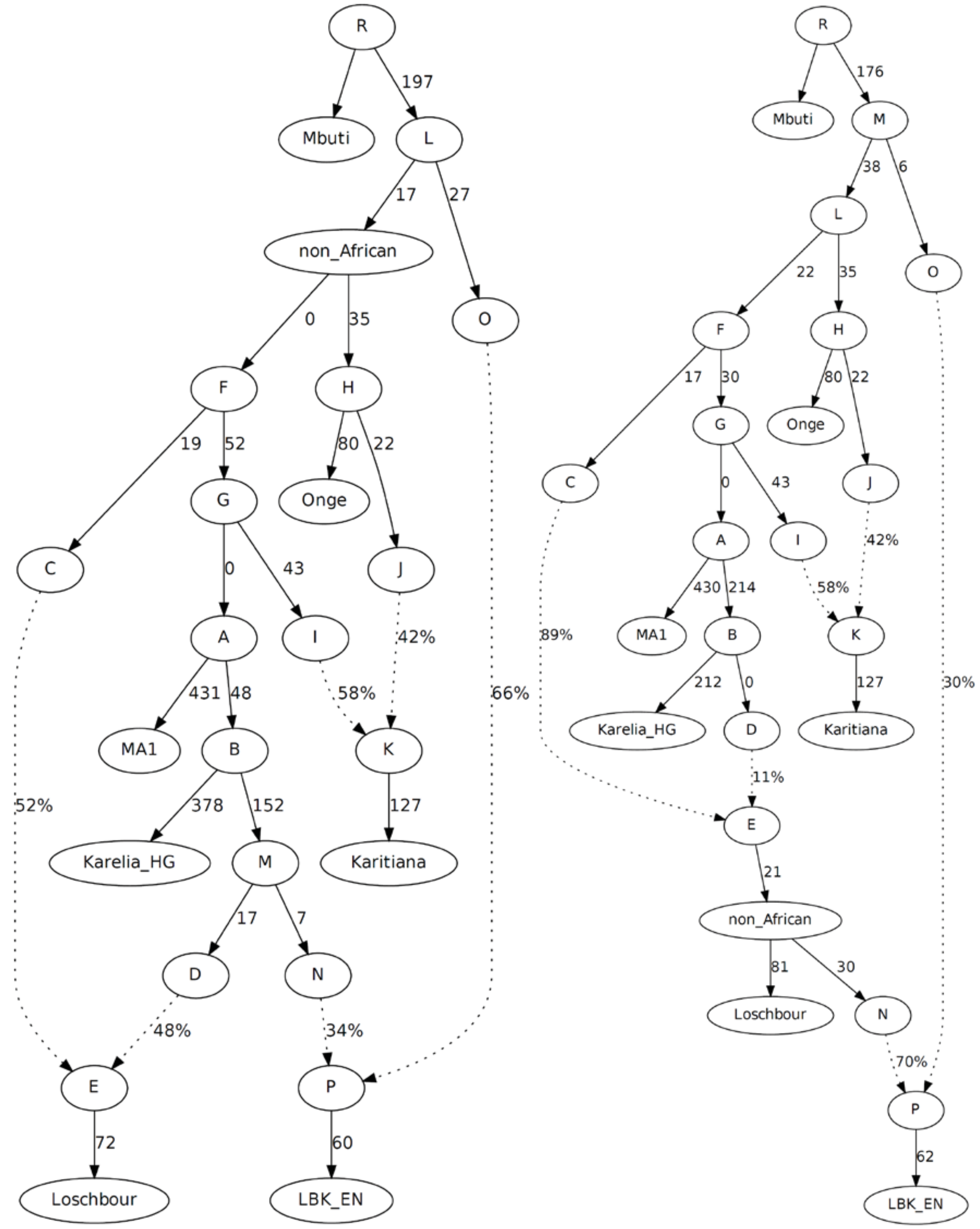

# Supplementary Information 9
## Inference of admixture proportions without detailed phylogenetic modeling

Iosif Lazaridis*, Nick Patterson, and David Reich

* To whom correspondence should be addressed (lazaridis@genetics.med.harvard.edu)

In this section, we develop a method that can be used (i) to identify reference populations that may have contributed ancestry to a *Test* population, and (ii) to estimate mixture proportions from these reference populations for *Test*. The method uses the intuition that the reference populations are not identically related to a panel of "focal" or "outgroup" populations, but share different amounts of genetic drift with them as a result of their deep evolutionary history (which is, however, not explicitly modeled). These "outgroups" must be devoid of recent gene flow with either the *Test* or the candidate reference population, as such gene flow introduces additional common genetic drift. One way to identify them is to pick a varied set of world populations that are (i) geographically remote from the area under study, and (ii) do not show evidence of admixture from that area using an algorithm such as ADMIXTURE[1] that can identify recently admixed populations and individuals. In practice, we will use the following set $\mathcal{O}$ of 15 previously identified[2] outgroups for West Eurasia:

**"World Foci 15" set of outgroups $\mathcal{O}$:** Ami, Biaka, Bougainville, Chukchi, Eskimo, Han, Ju_hoan_North, Karitiana, Kharia, Mbuti, Onge, Papuan, She, Ulchi, Yoruba

**Figure S9.1: If *Test* and *Ref*$_1$ form a clade with respect to the outgroups, then statistics involving either one of them and the outgroups are identical.**

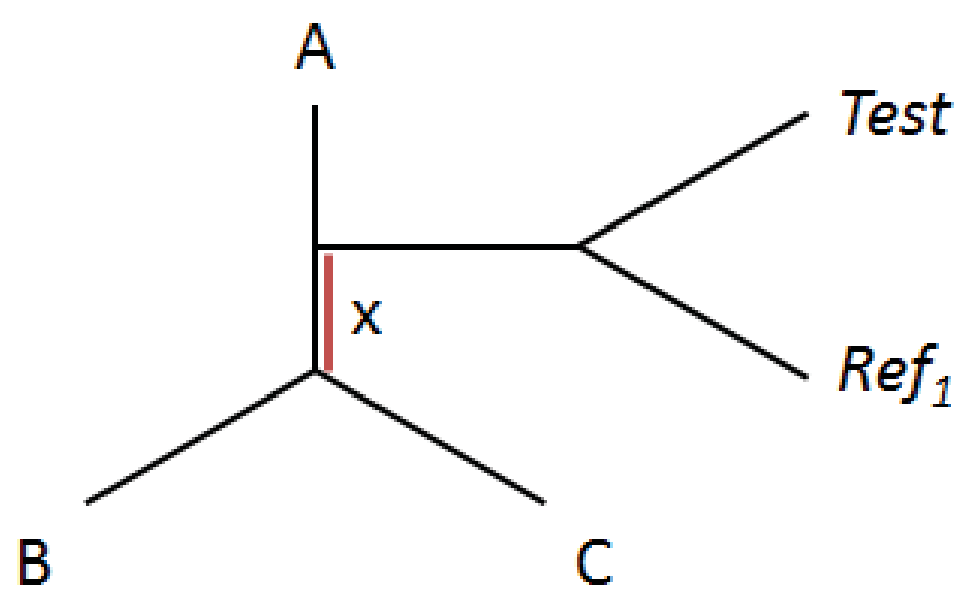

| | X=Test | X=Ref$_1$ |
|---|---|---|
| $f_4(X, A; B, C)$ | 0 | 0 |
| $f_4(X, B; A, C)$ | x | x |
| $f_4(X, C; A, B)$ | x | x |

**Admixture proportion estimation with non-negative least squares**

If *Test* forms a clade with a reference population $Ref_1$ since their divergence from a triple of outgroups *A*, *B*, *C* (Fig. S9.1) we expect that $f_4(Test, A; B, C) \approx f_4(Ref_1, A; B, C)$, $f_4(Test, B; A, C) \approx f_4(Ref_1, B; A, C)$, $f_4(Test, C; A, B) \approx f_4(Ref_1, C; A, B)$. However, if *Test* also has ancestry related to another population $Ref_2$ then it will be the case that $f_4(Test, A; B, C) \approx \alpha f_4(Ref_1, A; B, C) + (1\text{-}\alpha) f_4(Ref_2, A; B, C)$ where $\alpha$ is the fraction of ancestry from $Ref_1$ and 1-$\alpha$ from $Ref_2$ (Fig. S9.2). More generally, if *Test* has ancestry related to *N* different populations, then we can write:

$$f_4(Test; A; B, C) \approx \sum_{i=1}^{N} \alpha_i f_4(Ref_i; A; B, C) \qquad \text{(S9.1)}$$

with $\sum_{i=1}^{N} \alpha_i = 1$ and $\alpha_i \geq 0, i = 1, \ldots, N$. Given a set of *n* outgroups, there are $n\binom{n-1}{2}$ arrangements of the outgroups, as each of them can take the position of *A* in the above expression, and all pairs of the remaining *n-1* ones could take the position of *B*, *C*. For the set of 15 outgroups there are 1,365 different arrangements, and hence 1,365 equations of the above form. We can then write $t \approx R\alpha^{\mathrm{T}}$ where $t$ is a column vector of 1,365 $f_4$-statistics involving the *Test* population, $R$ is a 1,365-by-$N$ matrix of $f_4$-statistics (each column for one of the $N$ references) and $\alpha$ is a column vector of $N$ mixture proportions (to be estimated). This problem can be solved for $\alpha$ by least squares, minimizing $\|t - R\alpha^{\mathrm{T}}\|_2^2$ subject to the constraint that $\alpha$ has non-negative elements and $\|\alpha\|_1 = 1$. We use the implementation of least squares in the *lsqlin* function of Matlab (http://www.mathworks.com/help/optim/ug/lsqlin.html) which estimates the vector $\hat{\alpha}$ of the mixture proportions and also reports *resnorm* = $\|t - R\hat{\alpha}^{\mathrm{T}}\|_2^2$, the squared 2-norm of the residuals.

**Figure S9.2: *Test* is a mixture related to two reference populations *Ref*$_1$ and *Ref*$_2$.** Three possible topologies are shown: (a) *Ref*$_1$ and *Ref*$_2$ are symmetrically related to *A*, *B*, *C*. (b) *Ref*$_1$ and *Ref*$_2$ are differentially related to one outgroup *A*, and symmetrically related to the other two *B*, *C*. (c) *Ref*$_1$ and *Ref*$_2$ are related to two of the outgroups *A*, *C* and symmetrically related to the third *B*. In case (a) all $f_4$-statistics involving *Test*, *Ref*$_1$, *Ref*$_2$ are identical and cannot be used to estimate mixture proportions. In cases (b, c) $f_4$-statistics involving *Test* are a weighted average of those involving *Ref*$_1$ and *Ref*$_2$.

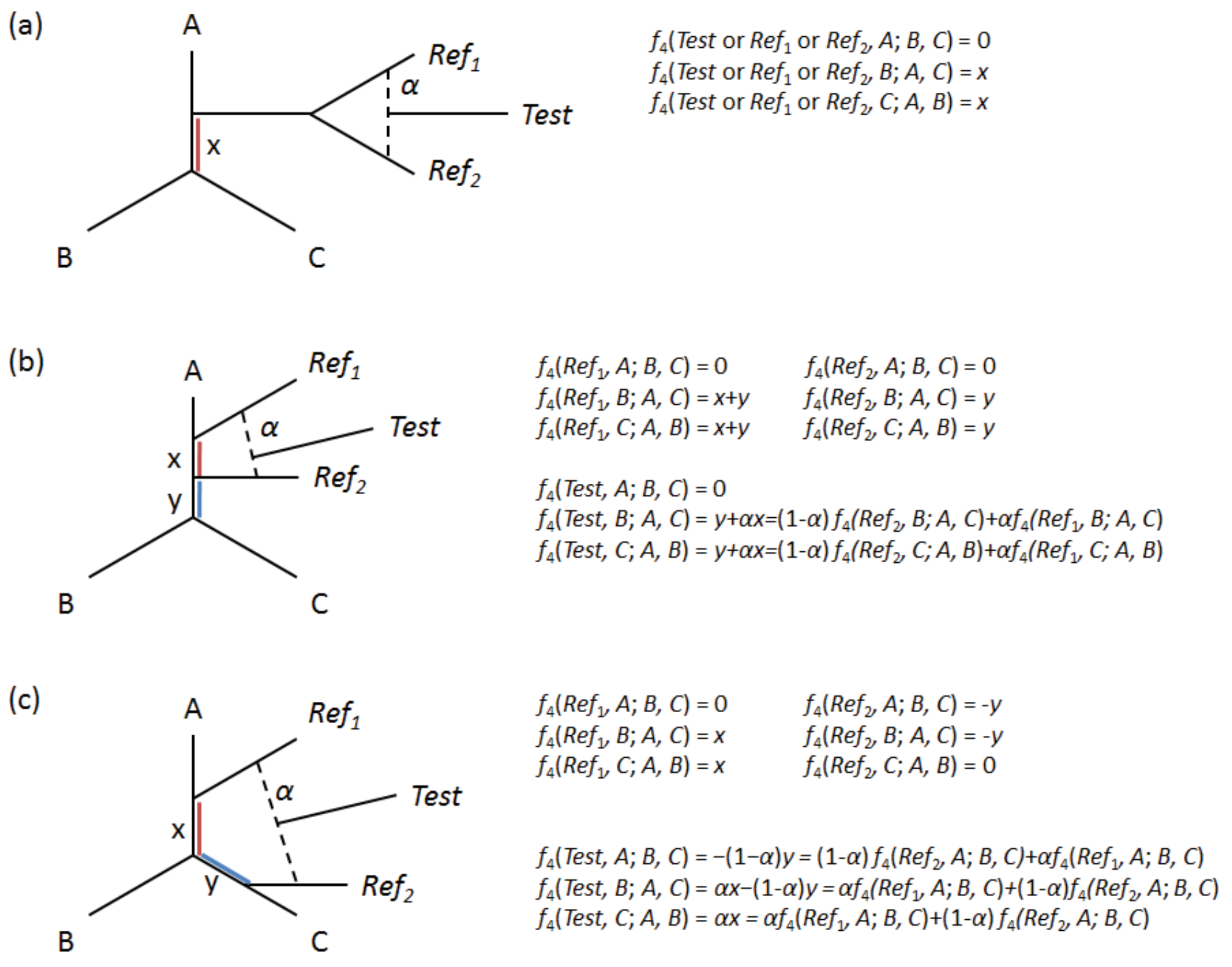

The advantage of this method is that the relationship of *Ref*$_1$, *Ref*$_2$, and *A*, *B*, *C* need not be explicitly modeled. Each of them may have a history of admixture in its evolutionary past that is arbitrarily complex, but for each portion of its ancestry one of the topological relationships of Fig. S9.2 will hold and thus the $f_4$-statistics involving *Test* will be the weighted average of the reference populations.

**Figure S9.3: LBK_EN and Spain_EN appear to be symmetrically related to the outgroups, but there is some variation between LBK_EN and Karelia_HG which may be used to infer admixture estimates for populations having ancestry from sources related to them.**

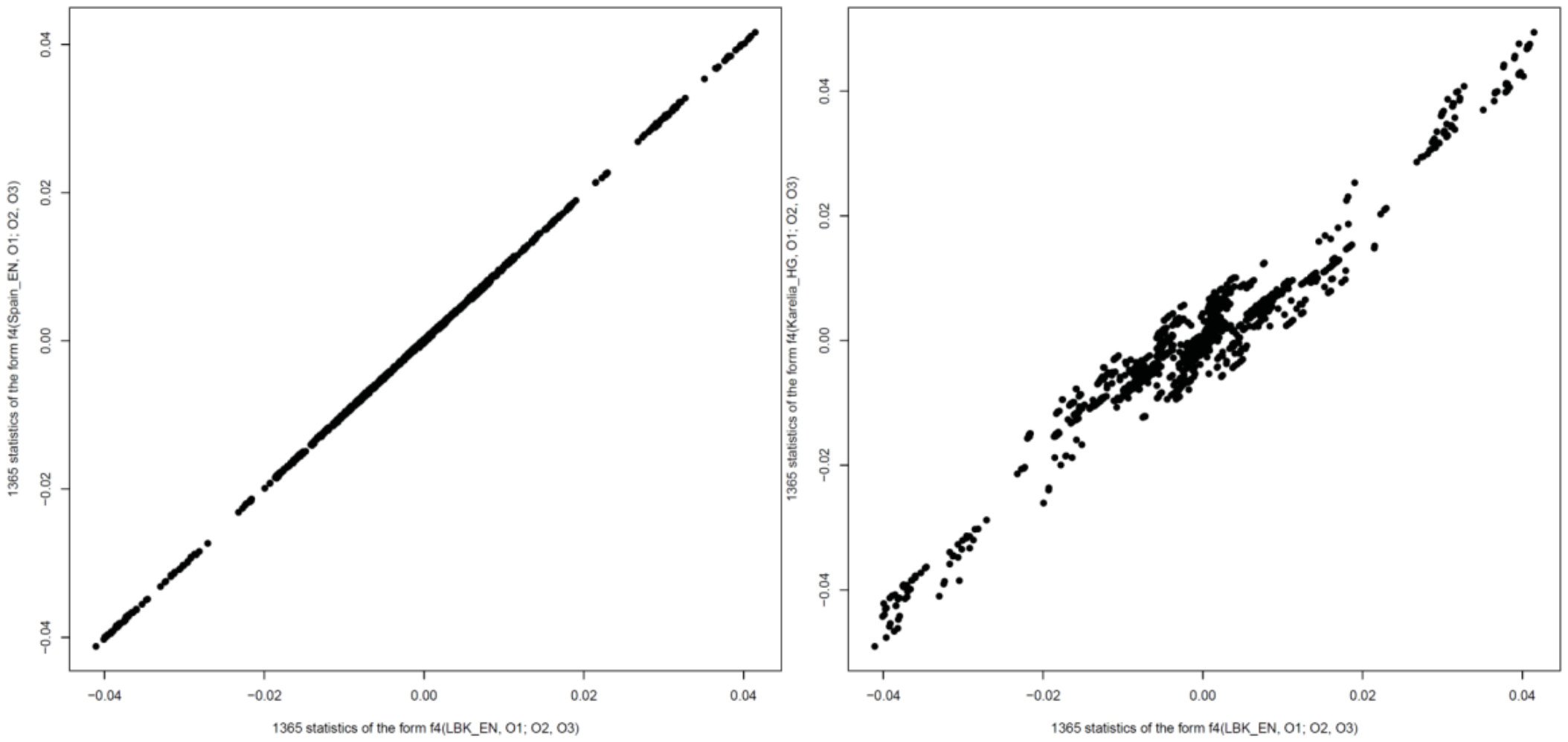

This method works if the reference populations are not all identically related to the outgroups, but does not work if they are identically related to the outgroups. Concretely, if the outgroups are archaic humans and primates, or alternatively Loschbour and LaBrana[2], they are likely to be identically related (from the perspective of shared genetic drift) to the references populations. Thus there will be no leverage to discern whether test samples are more closely related to $Ref_1$ or $Ref_2$.

It is useful to plot $f_4(Ref_1, A; B, C)$ and $f_4(Ref_2, A; B, C)$ against each other (Fig. S9.3). Two closely related populations (e.g., LBK_EN and Spain_EN) have a virtually identical relationship to the outgroups, while two other populations (e.g., LBK_EN and Karelia_HG) have a visibly differentiated relationship. This difference is not huge, as LBK_EN and Karelia_HG are West Eurasian populations that are mostly symmetrically related to the outgroups. However, it is sufficient to infer ancestry proportions for populations of mixed LBK_EN and Karelia_HG-related ancestry, as we will show.

We do not assume that particular reference populations contributed ancestry to a *Test* population. We also do not assume that a *Test* population is admixed or that it is a mixture of a particular number *N* of reference populations. Our approach is as follows: for each *Test* population the set of reference populations $\mathcal{R}$ is a set of temporally preceding ancient populations. For the Middle Neolithic populations these include:

**“preMN”:** Karelia_HG, Motala_HG, LBK_EN, Spain_EN, HungaryGamba_EN, Yamnaya, Loschbour

For the Late Neolithic/Bronze Age populations they include:

**“preLN/BA”:** Karelia_HG, Motala_HG, LBK_EN, Spain_EN, Baalberge_MN, Esperstedt_MN, Spain_MN, HungaryGamba_EN, Yamnaya, Loschbour

This list includes the best representative (higher number of SNPs) of WHG and EHG (Loschbour and Karelia_HG), the best representative of early farmers from Spain, Germany, and Hungary (Spain_EN,

LBK_EN, and HungaryGamba_EN), and Middle Neolithic farmers (Spain_MN, Baalberge_MN, Esperstedt_MN), and Yamnaya.

Given a *Test* population we vary *N*=1, 2, 3. If *N*=1 then we attempt to make *Test* a sister group of a preceding population; $N=2$ a 2-way mixture, etc. We test all $\binom{|\mathcal{R}|}{N}$ combinations of reference populations, fitting $\alpha$ for each one and recording its *resnorm*. We record the residuals:

$$Residual(Test; A, B, C) = \sum_{i=1}^{N} \hat{a}_i f_4(Ref_i, A; B, C) - f_4(Test, A; B, C) \qquad \text{(S9.2)}$$

We also record the Z-score. If our estimate of a statistic is $f_4(Test, A; B, C) = m \pm e$ then, the Z-score is $\frac{1}{e} Residual(Test; A, B, C)$. The maximum |Z| provides an additional measure of how well the mixture model explains all the observations.

**Middle Neolithic Europeans**

We have seen (SI7) that Middle Neolithic Europeans are a mixture of Early Neolithic Europeans and western European hunter-gatherers. Our analysis of the three MN populations seems to confirm this (Figs. S9.4-6). For *N*=1, the best fitting population is an Early Neolithic one, however a marked lowering of *resnorm* is seen for $N=2$ with the best model suggesting an EN+WHG mixture, and virtually no change for *N*=3. This is consistent with our hypothesis of Middle Neolithic WHG-related resurgence in Europe. It is important to note that the chosen set of outgroups have no power to discriminate between different Early Neolithic populations (as was clear from Fig. S9.3). So it is not surprising that Spain_EN+Loschbour and LBK_EN+Loschbour models provide a virtually identical fit for the different MN populations. However, EN and WHG are clearly differentially related to the outgroups, which makes it possible to correctly infer that MN is a mixture of EN+WHG.

**Late Neolithic/Bronze Age Europeans**

We have seen (SI7) that Late Neolithic Europeans in central Europe (Germany) are not a continuation of the earlier farmers of Europe but harbor ancestry from an "eastern" population (EHG or Yamnaya). Our method seems to confirm this (Figs. S9.7-10). For none of the LN/BA populations is ancestry from an EN/MN farmer group the best *N*=1 model, and in all populations a marked improvement is observed between *N*=1 and *N*=2, but not between *N*=2 and *N=3*. The *N*=2 models shown in panel (b) of Figs. S9.7-10 always show the best fit when a "western" and "eastern" population is paired. For the Corded_Ware_LN, the best models involve European farmers and Yamnaya in proportions of approximately ~1/5 and ~4/5 or European farmers and Karelia_HG in proportions of ~2/3 and ~1/3. Lower "eastern" contributions are inferred for the other LN/BA populations, consistent with the results of SI7 that demonstrate that EHG and Yamnaya share more alleles with Corded_Ware_LN than with other LN/BA populations. Once again, the method cannot distinguish between the closely related populations such as LBK_EN and Spain_EN, however, the key point is that it arrives at similar estimates of turnover regardless of which population is chosen as a reference.

**Figure S9.4: Modeling Esperstedt_MN as a mixture of earlier populations.** (a) We present in the left column a histogram of raw $f_4$-statistic residuals and on the right Z-scores for the best-fitting (lowest *resnorm*) model at each *N*. In the bottom row, we show on the left how *resnorm* and the maximum |Z| score change for different *N*. (b) *resnorm* of different *N*=2 models.

(a)

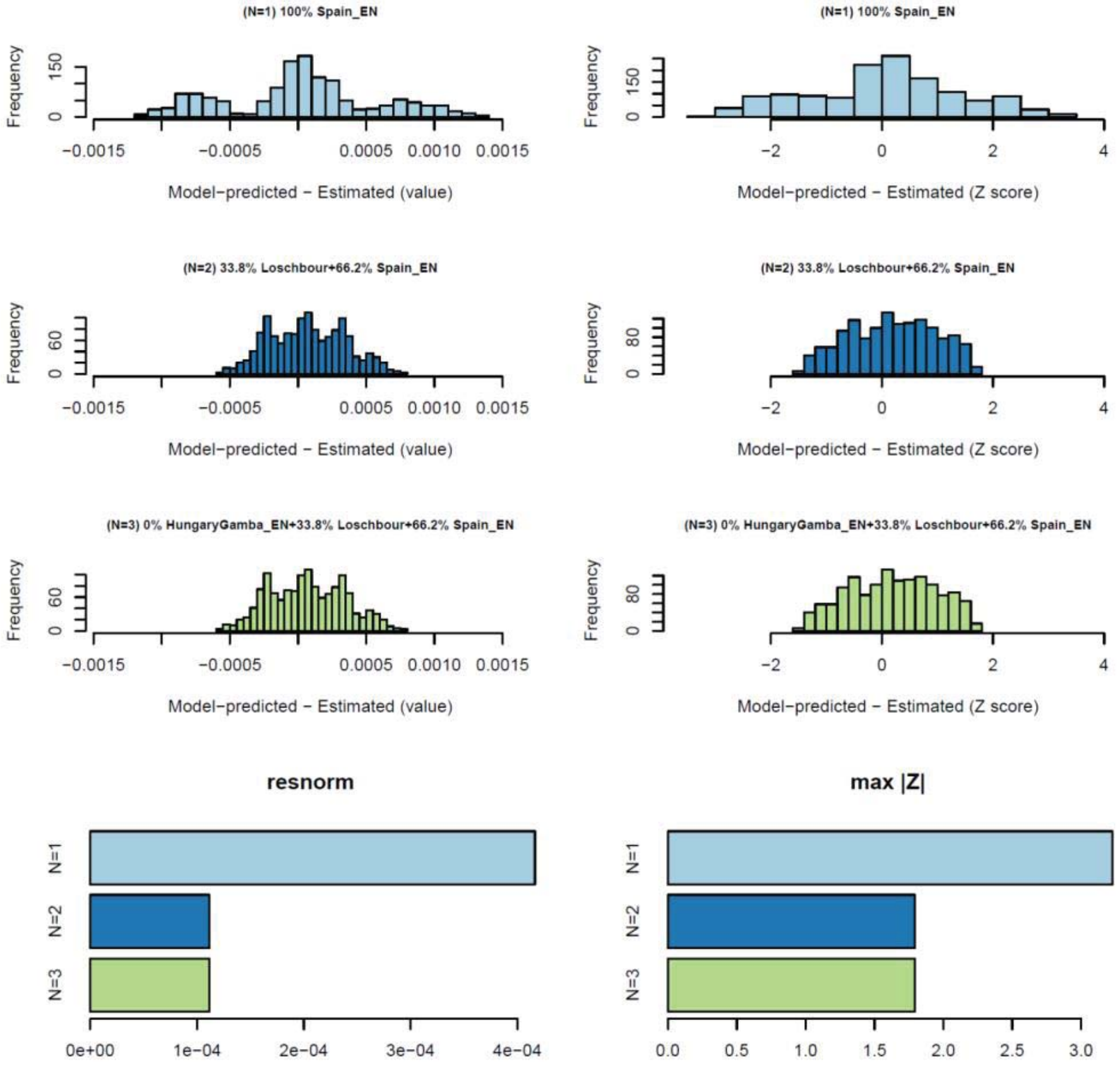

(b)

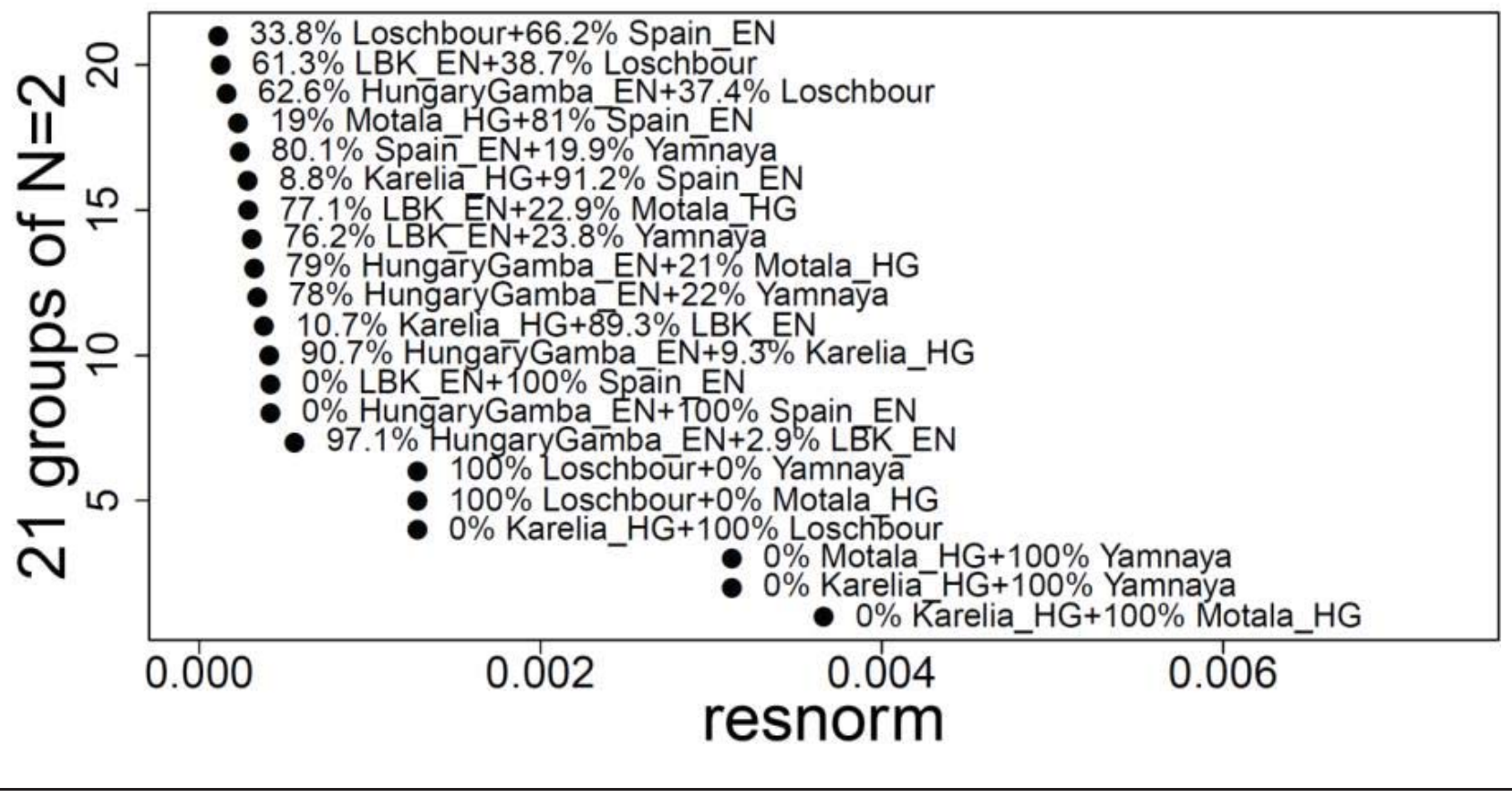

**Figure S9.5: Modeling Baalberge_MN as a mixture of earlier populations.** (a) We present in the left column a histogram of raw $f_4$-statistic residuals and on the right Z-scores for the best-fitting (lowest *resnorm*) model at each $N$. In the bottom row, we show on the left how *resnorm* and the maximum |Z| score change for different $N$. (b) *resnorm* of different $N$=2 models.

(a)

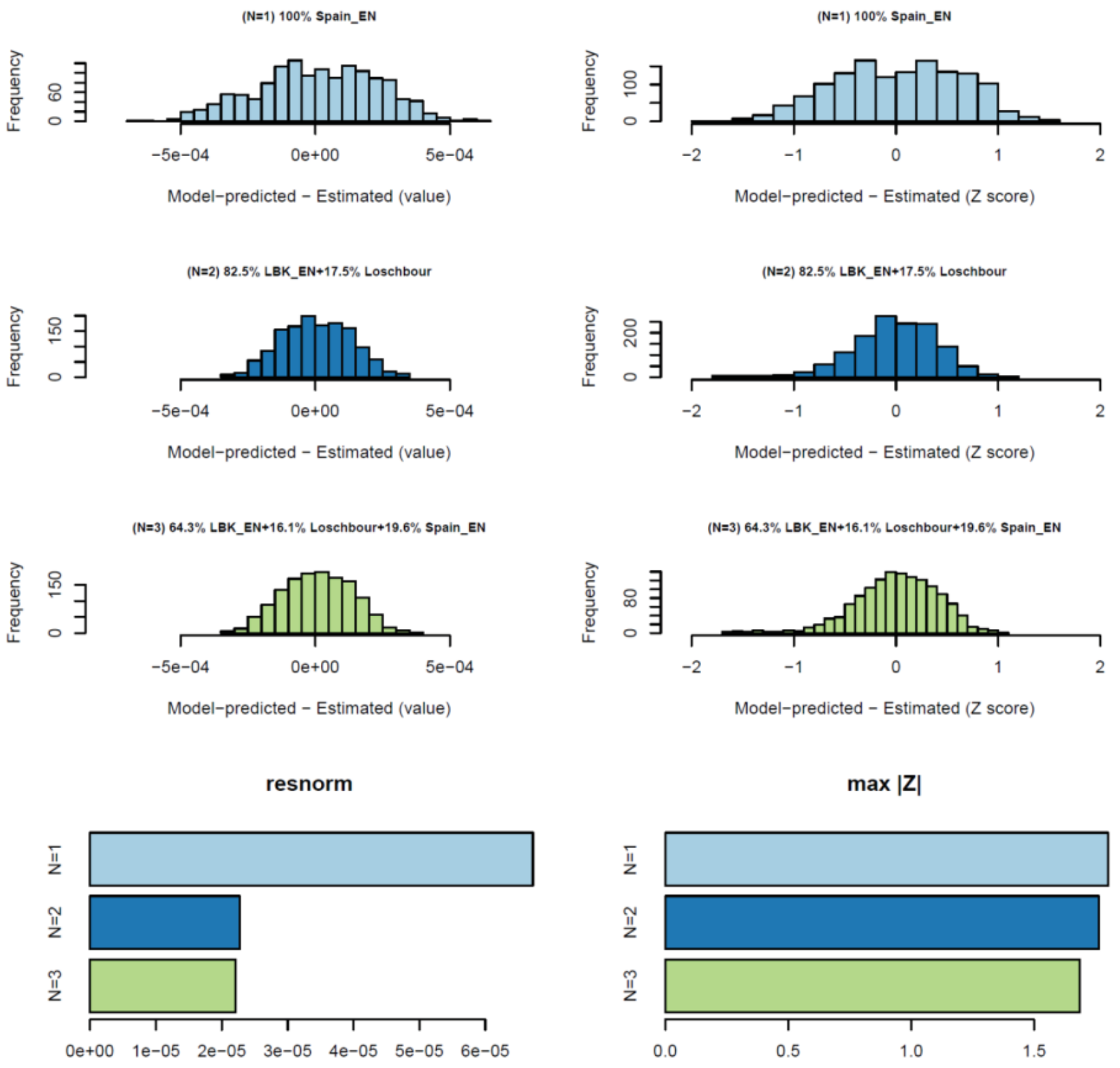

(b)

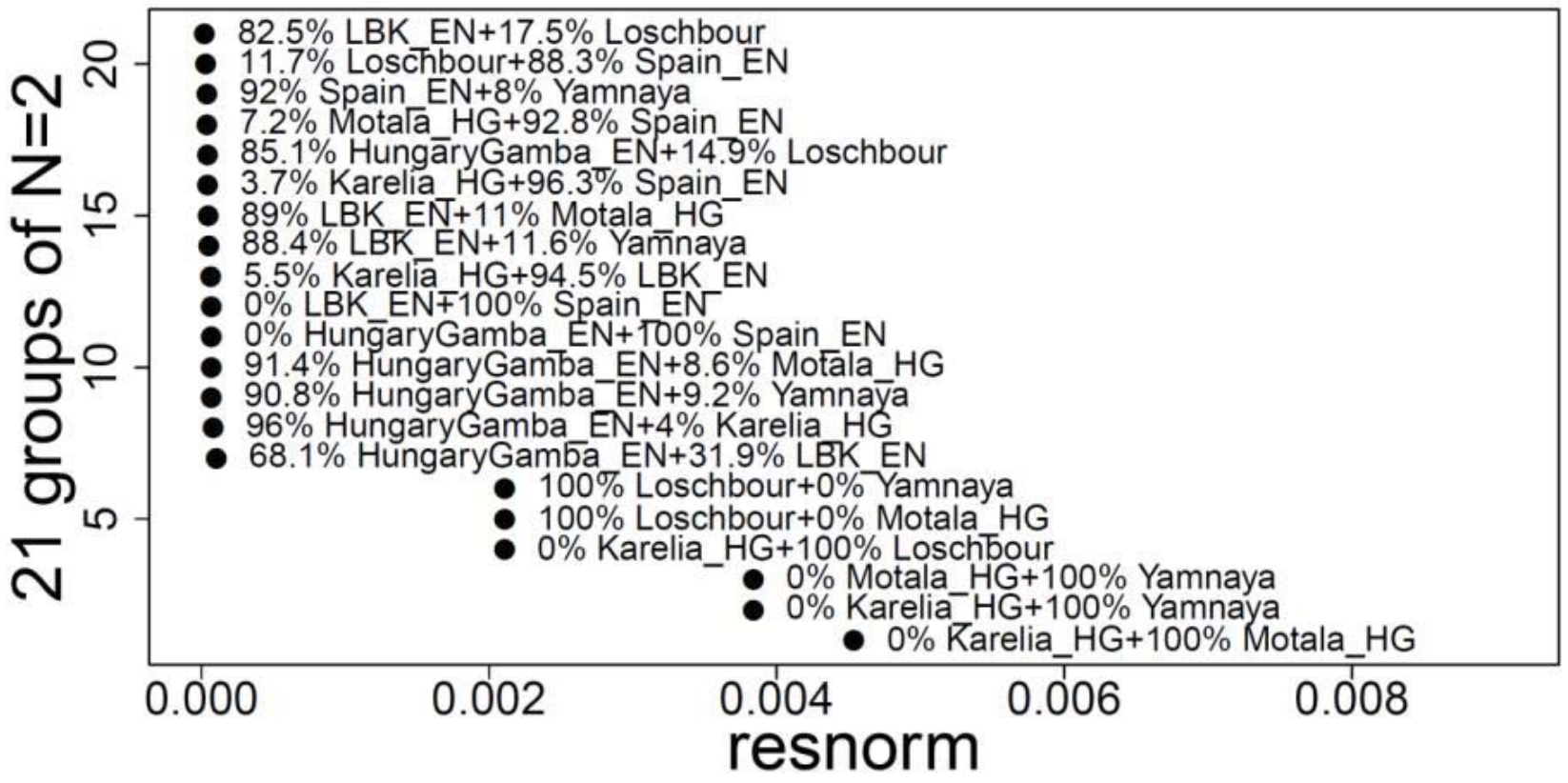

**Figure S9.6: Modeling Spain_MN as a mixture of earlier populations.** (a) We present in the left column a histogram of raw $f_4$-statistic residuals and on the right Z-scores for the best-fitting (lowest *resnorm*) model at each *N*. In the bottom row, we show on the left how *resnorm* and the maximum |Z| score change for different *N*. (b) *resnorm* of different *N*=2 models.

(a)

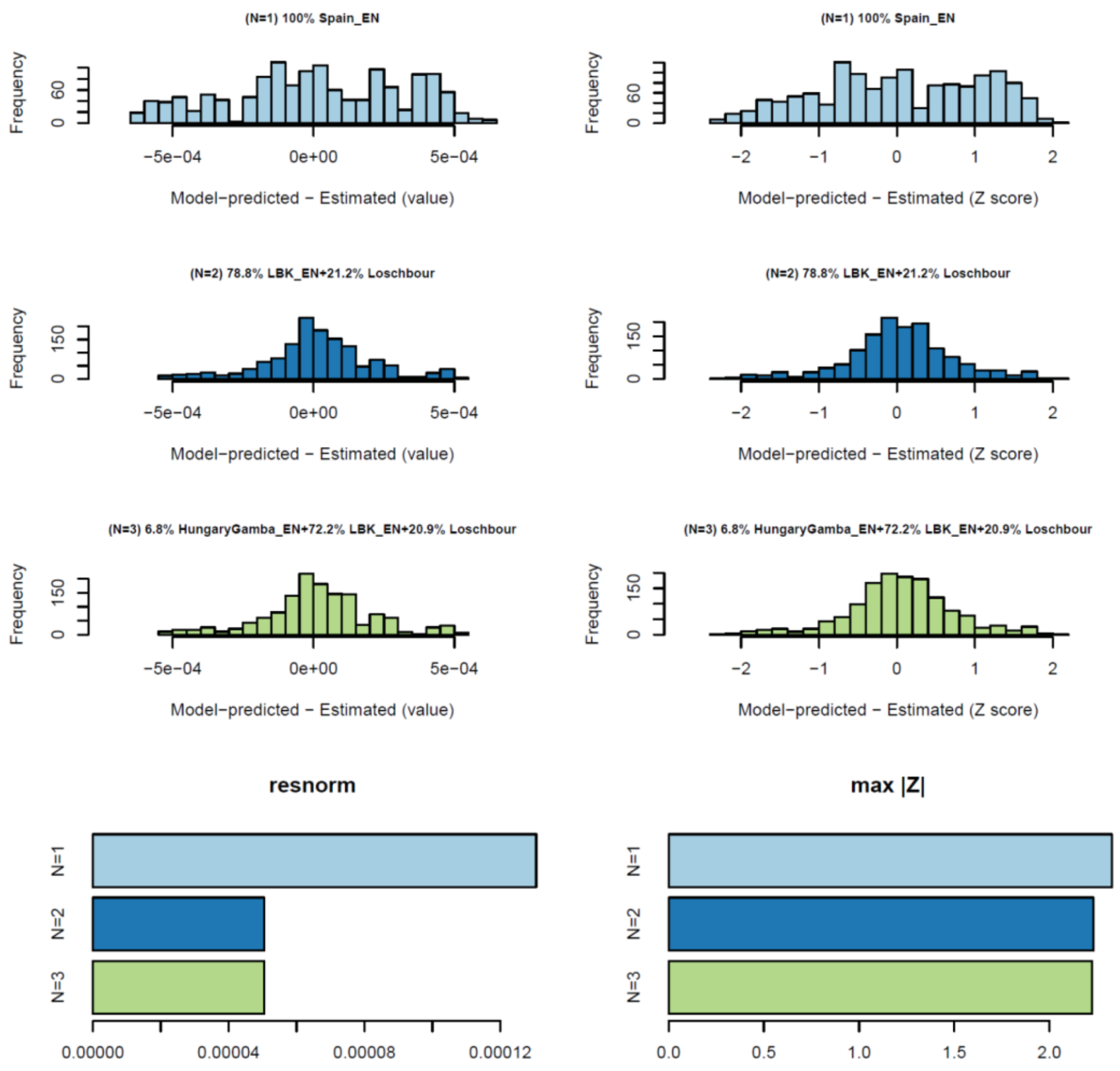

(b)

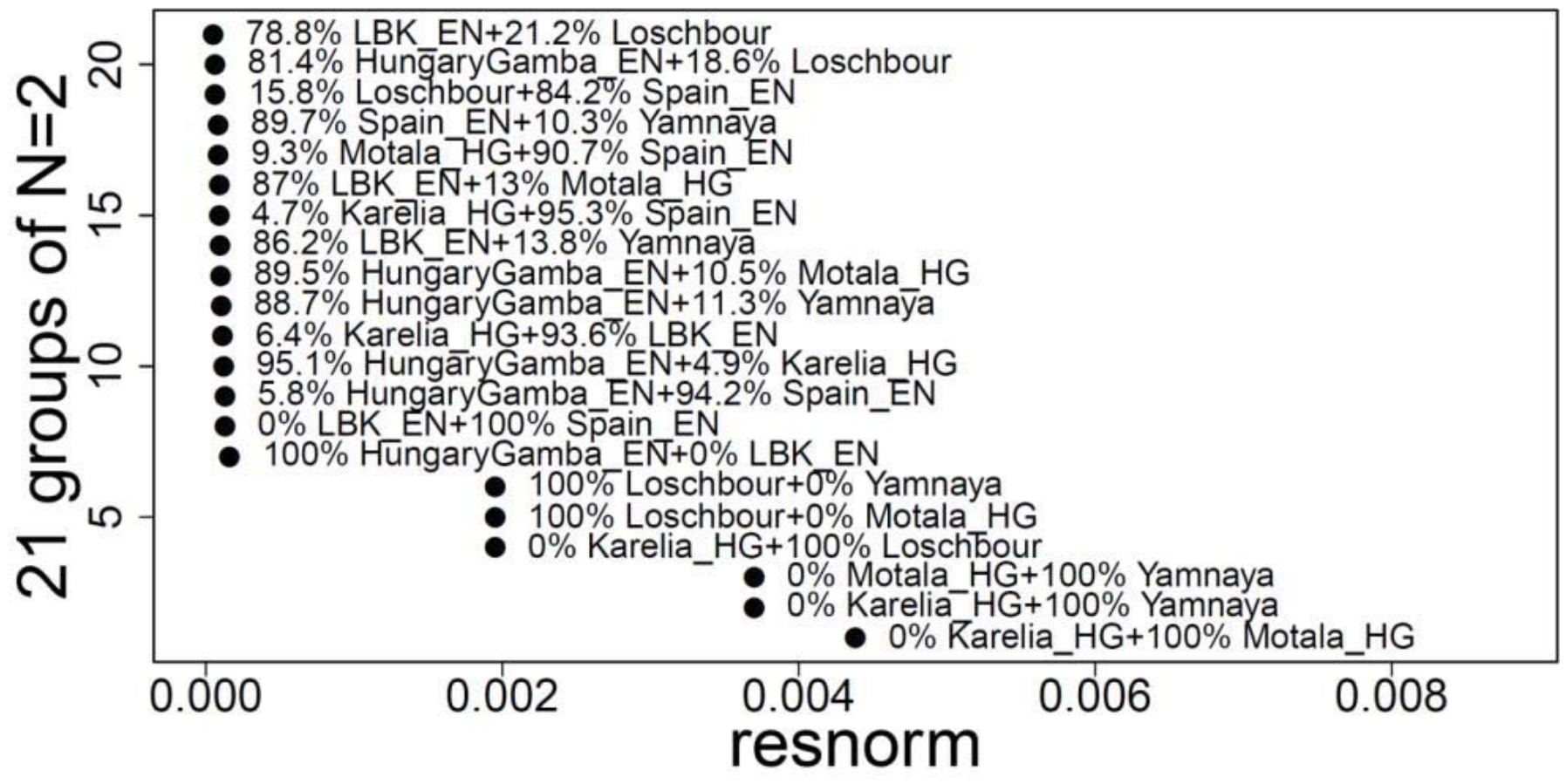

**Figure S9.7: Modeling Corded_Ware_LN as a mixture of earlier populations.** (a) We present in the left column a histogram of raw $f_4$-statistic residuals and on the right Z-scores for the best-fitting (lowest *resnorm*) model at each *N*. In the bottom row, we show on the left how *resnorm* and the maximum |Z| score change for different *N*. (b) *resnorm* of different *N*=2 models.

(a) (b)

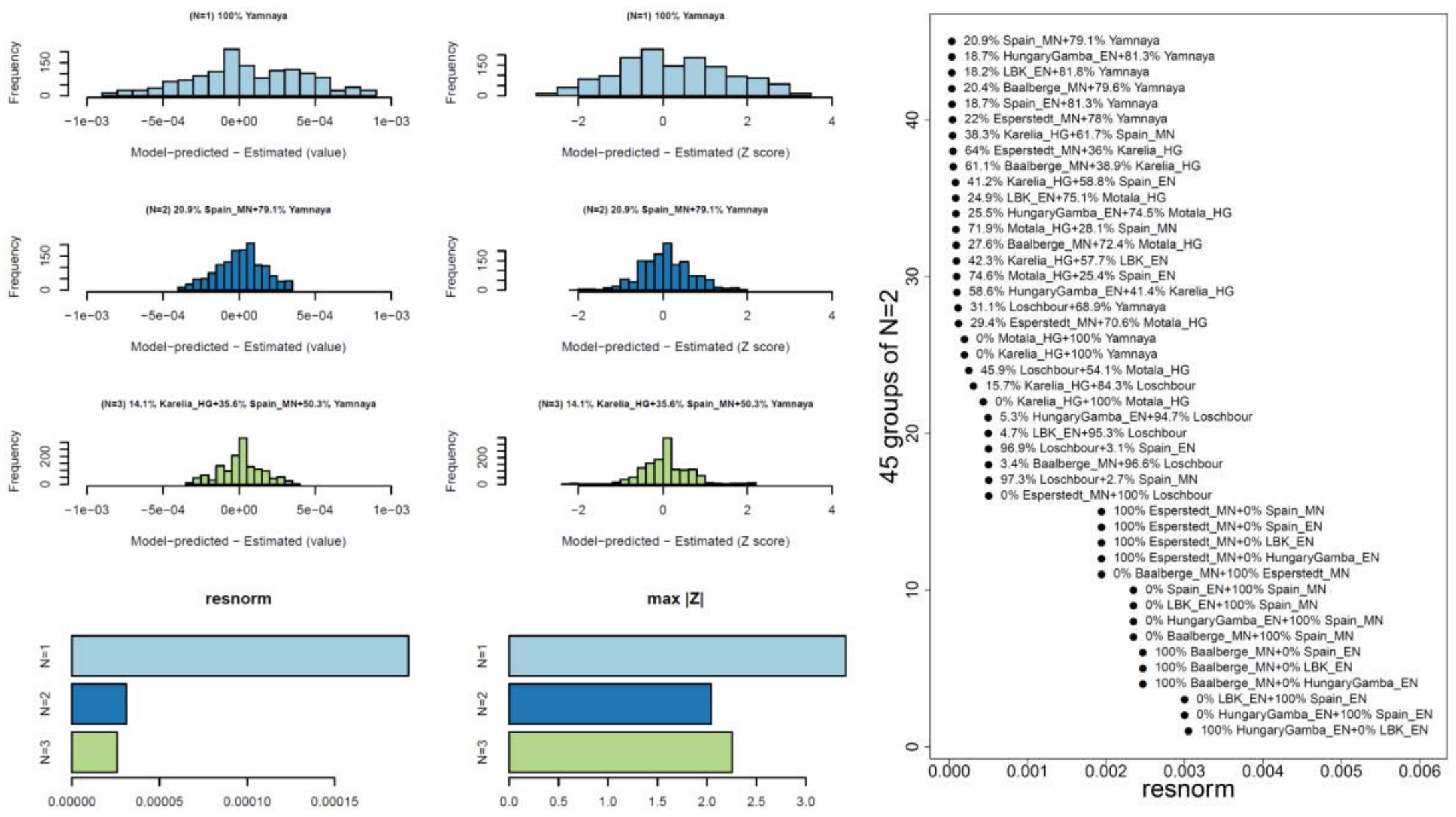

**Figure S9.8: Modeling Bell_Beaker_LN as a mixture of earlier populations.** (a) We present in the left column a histogram of raw $f_4$-statistic residuals and on the right Z-scores for the best-fitting (lowest *resnorm*) model at each *N*. In the bottom row, we show on the left how *resnorm* and the maximum |Z| score change for different *N*. (b) *resnorm* of different *N*=2 models.

(a) (b)

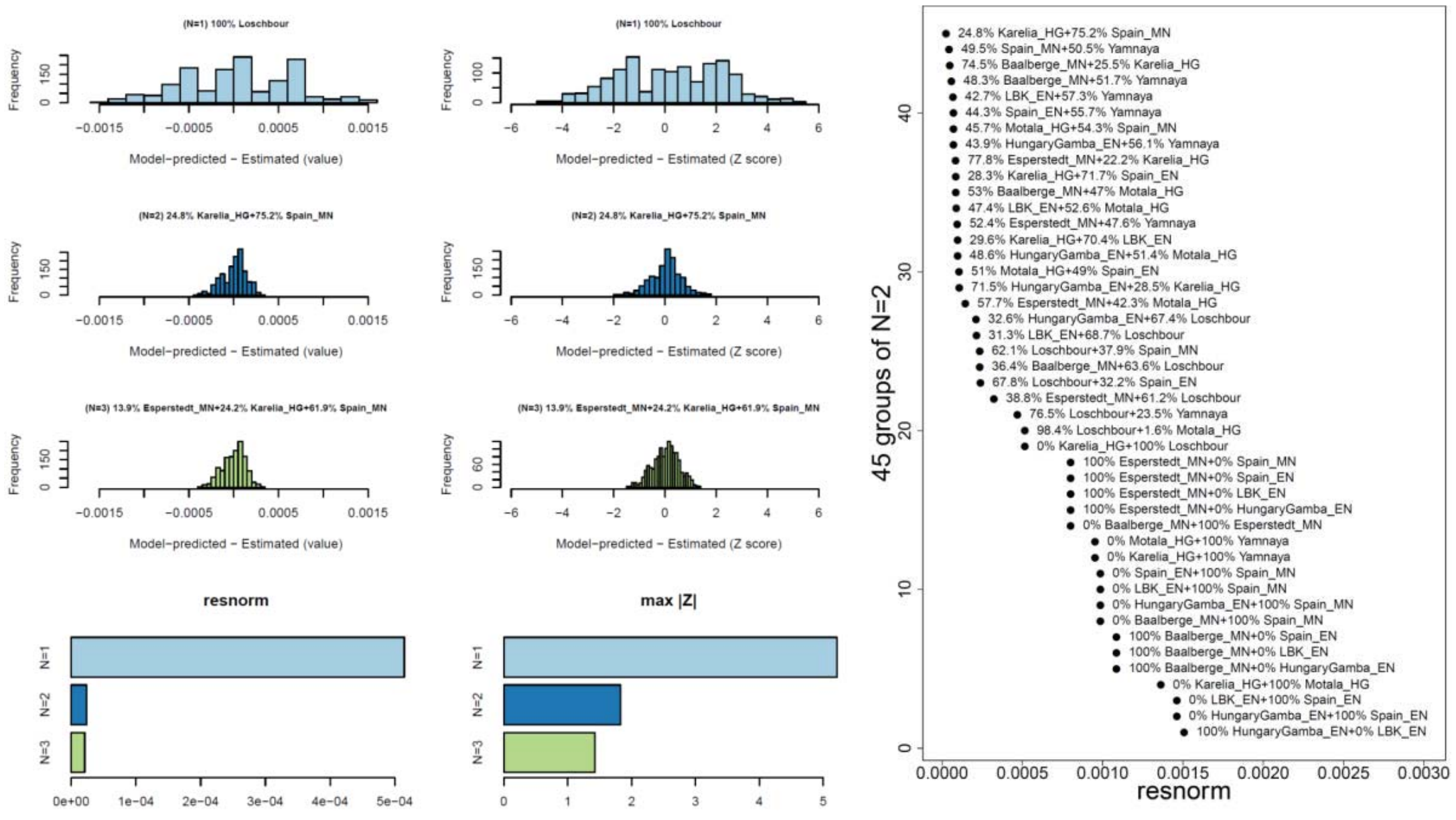

**Figure S9.9: Modeling BenzigerodeHeimburg_LN as a mixture of earlier populations.** (a) We present in the left column a histogram of raw $f_4$-statistic residuals and on the right Z-scores for the best-fitting (lowest *resnorm*) model at each *N*. In the bottom row, we show on the left how *resnorm* and the maximum |Z| score change for different *N*. (b) *resnorm* of different *N*=2 models.

(a) (b)

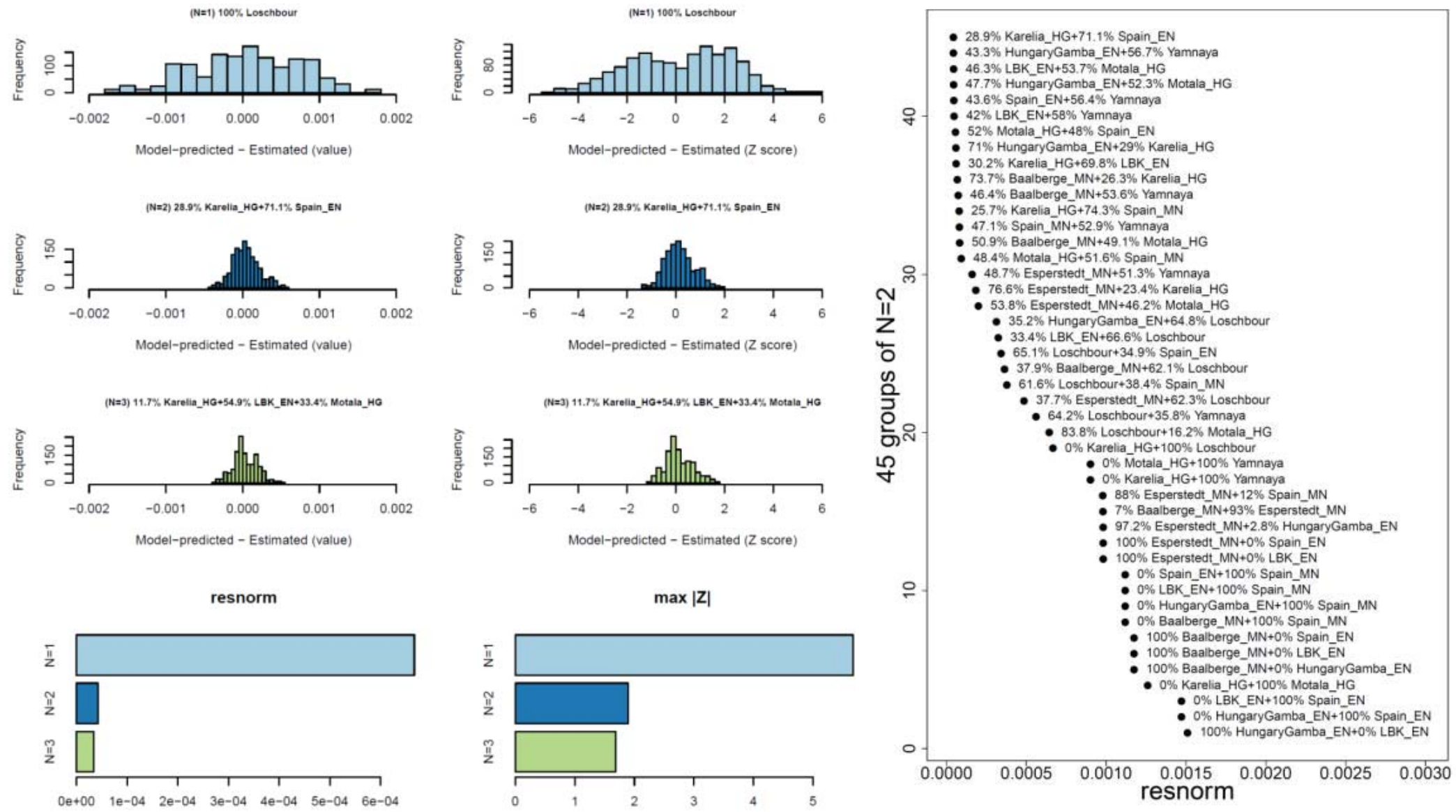

**Figure S9.10: Modeling Unetice_EBA as a mixture of pre-Late Neolithic populations.** (a) We present in the left column a histogram of raw $f_4$-statistic residuals and on the right Z-scores for the best-fitting (lowest *resnorm*) model at each *N*. In the bottom row, we show on the left how *resnorm* and the maximum |Z| score change for different *N*. (b) *resnorm* of different *N*=2 models.

(a) (b)

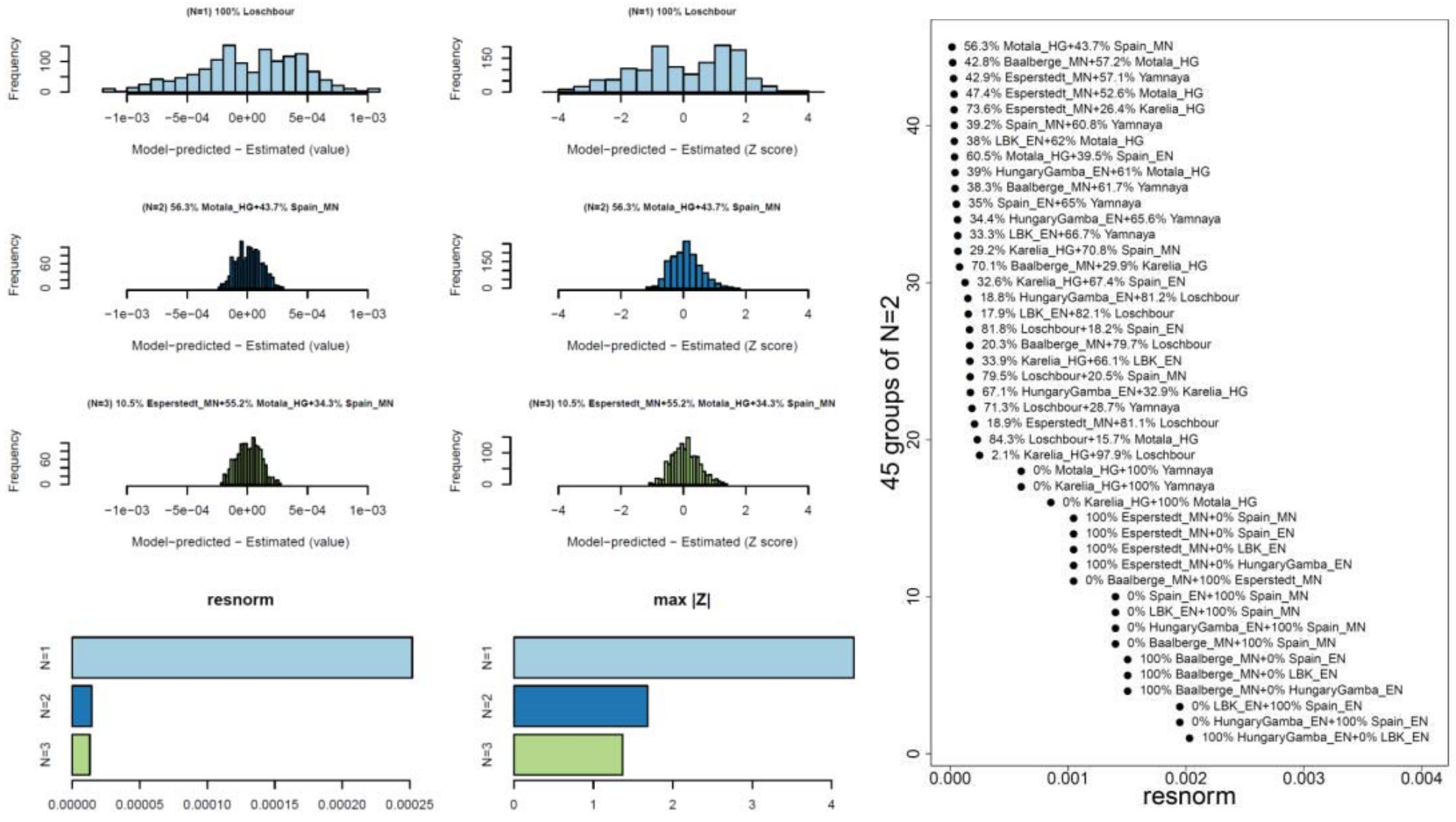

**Figure S9.11: U-shaped curve of *resnorm* and max|Z| for Corded_Ware_LN as a variable 2-way mixture of LBK_EN and Karelia_HG or Yamnaya.**

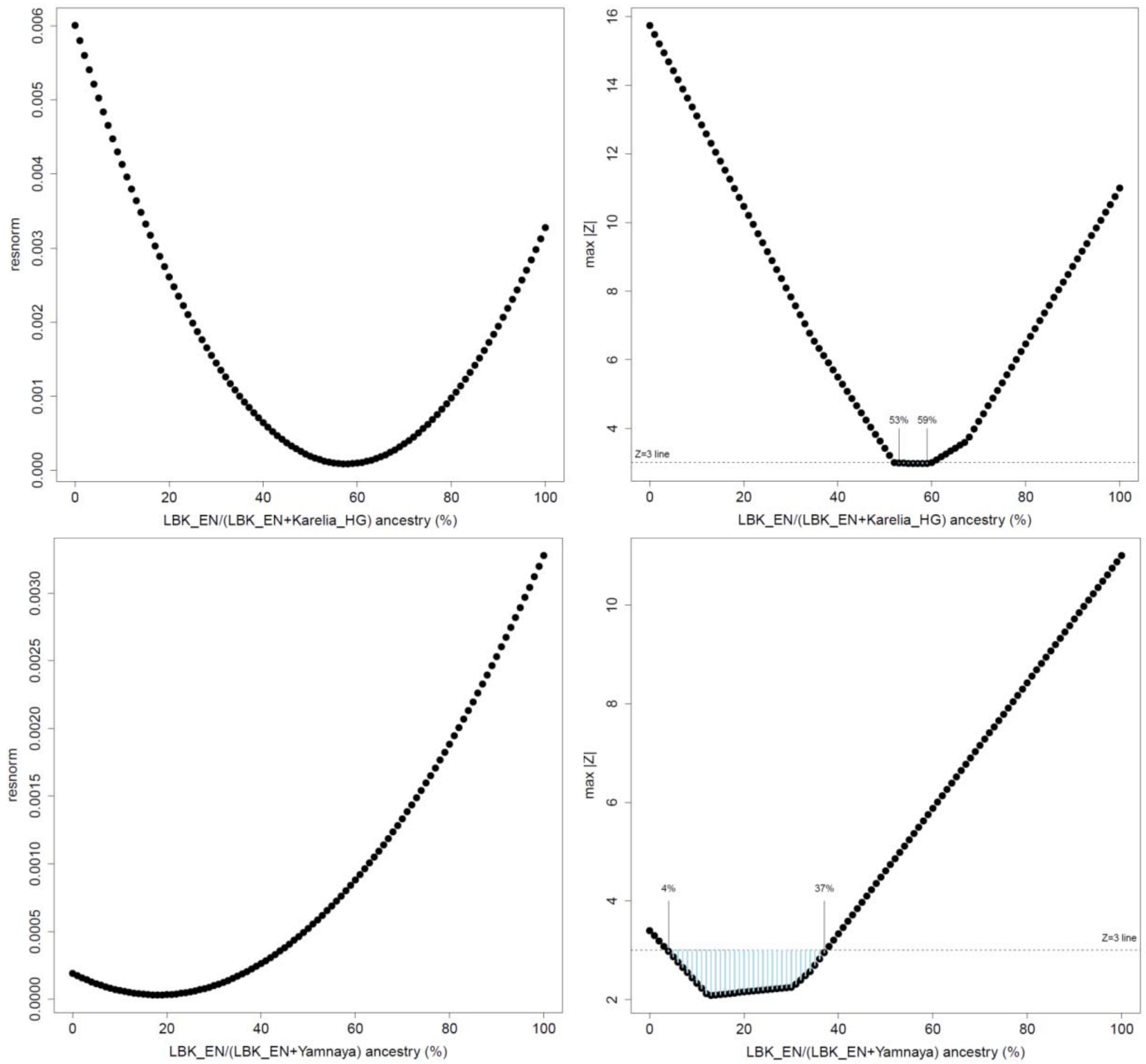

**U-shaped curve for an admixed population**

A useful way of visualizing that a *Test* population is indeed formed by admixture between two others $Ref_1$ and $Ref_2$ is to show the variation in *resnorm* and max |Z| as we model *Test* as a variable mixture of ($x$, 1-$x$) fraction of ancestry from $Ref_1$ and $Ref_2$, sliding $x$ between 0 and 1. This provides a useful confirmation about the existence of admixture. We do this for Corded_Ware_LN in Fig. S9.11 as a mixture of LBK_EN and Yamnaya or Karelia_HG (similar curves exist for the other pairings of Fig. S9.7b . The optimal proportion corresponds well to the 18.2% LBK_EN and 81.8% Yamnaya derived with our least squares optimizer (Fig. S9.7b). The jagged appearance of the max |Z| plot is due to the fact that it is dominated by the single worst behaving Z-score at each value of $x$, whereas the *resnorm* is an overall measure of fit that takes account all 1,365 $f_4$-statistics.

**Regression-based estimation of admixture proportions**

For $N$=2 it is also possible to detect the presence of admixture and estimate its parameters using simple linear regression. Detection of admixture in a *Test* population can be done using the statistic $f_3$(*Test*; $Ref_1$, $Ref_2$) which we have used in SI7 to prove that admixture took place in the Middle and Late Neolithic populations under study. This statistic makes use of the fact that allele frequencies in *Test* are intermediate between those of $Ref_1$ and $Ref_2$ and thus allele frequency differences $Test$-$Ref_1$ and $Test$-$Ref_2$ are negatively correlated. We have seen (Fig. S9.2) that statistics of the form $f_4(Test, A; B, C)$ are intermediate between $f_4(Ref_1, A; B, C)$ and $f_4(Ref_2, A; B, C)$ when *Test* is formed by admixture between populations related to $Ref_1$ and $Ref_2$ and $Ref_1$ and $Ref_2$ are asymmetrically related to the outgroups. If that is true, then the difference $f_4(Test, A; B, C)$- $f_4(Ref_1, A; B, C)$ will be anti-correlated with $f_4(Test, A; B, C)$- $f_4(Ref_2, A; B, C)$. These differences are equal to $f_4(Test, Ref_1; B, C)$ and $f_4(Test, Ref_2; B, C)$. We plot them in Fig. S9.12 for *Test*=Corded_Ware_LN, $Ref_1$=Karelia_HG or Yamnaya and $Ref_1$=Esperstedt_MN or Baalberge_MN or LBK_EN showing that they do in fact show a pattern of negative correlation with Early/Middle Neolithic populations from Germany that precede them. Such patterns of negative correlation also exist when using the other Neolithic populations as references for the Corded Ware, which are genetically very similar to those from Germany, and one should bear in mind that the admixture detected here need not have taken place in Germany itself, but might involve other Neolithic populations between Germany and eastern Europe.

The pattern observed in Fig. S9.12 is more "crisp" when using Karelia_HG as an eastern reference, but this does not necessarily indicate that Karelia_HG is the admixing population. As the proportion of the eastern population increases the line becomes more vertical (notice the reduced range of the statistic on the x-axis when $Ref_1$=Yamnaya in all the right-side plots of Fig. S9.12), and at the limit (for a hypothetical population *X* with 100% descent from the eastern population), all variation in the statistics of the horizontal axis would be noise, as the expected value on the horizontal axis would be zero. Another interesting case is when the two references form a clade with each other relative to the *Test* population. We show an example of this in Fig. S9.13 where we model Corded_Ware_LN as a mixture of LBK_EN and Spain_EN. A line with positive slope ≈1 is observed, also seen in Fig. S9.3 above. A positive correlation indicates that Corded_Ware_LN differs from Spain_EN and LBK_EN *in the same manner*, indicating that it is *not* a mixture of these two populations.

**Figure S9.12: Evidence for admixture in the Corded_Ware_LN between a Neolithic central European population from Germany (LBK_EN, Baalberge_MN, or Esperstedt_MN) and an eastern European population (Karelia_HG or Yamnaya).** A linear regression through (0,0) is shown.

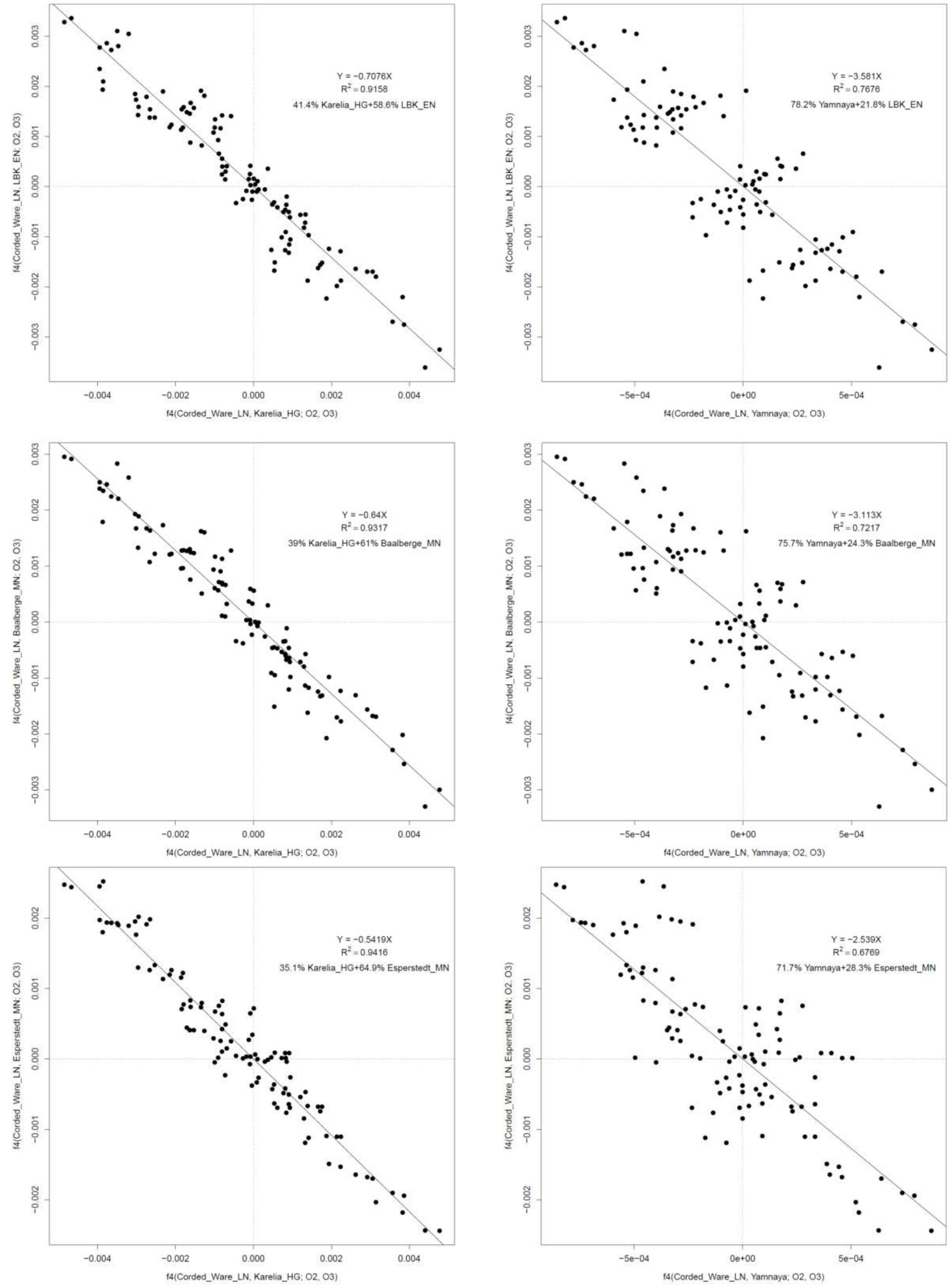

**Figure S9.13: When the two references are a clade with respect to the *Test* (as Spain_EN and LBK_EN are with respect to Corded_Ware_LN), a line with positive slope is observed.**

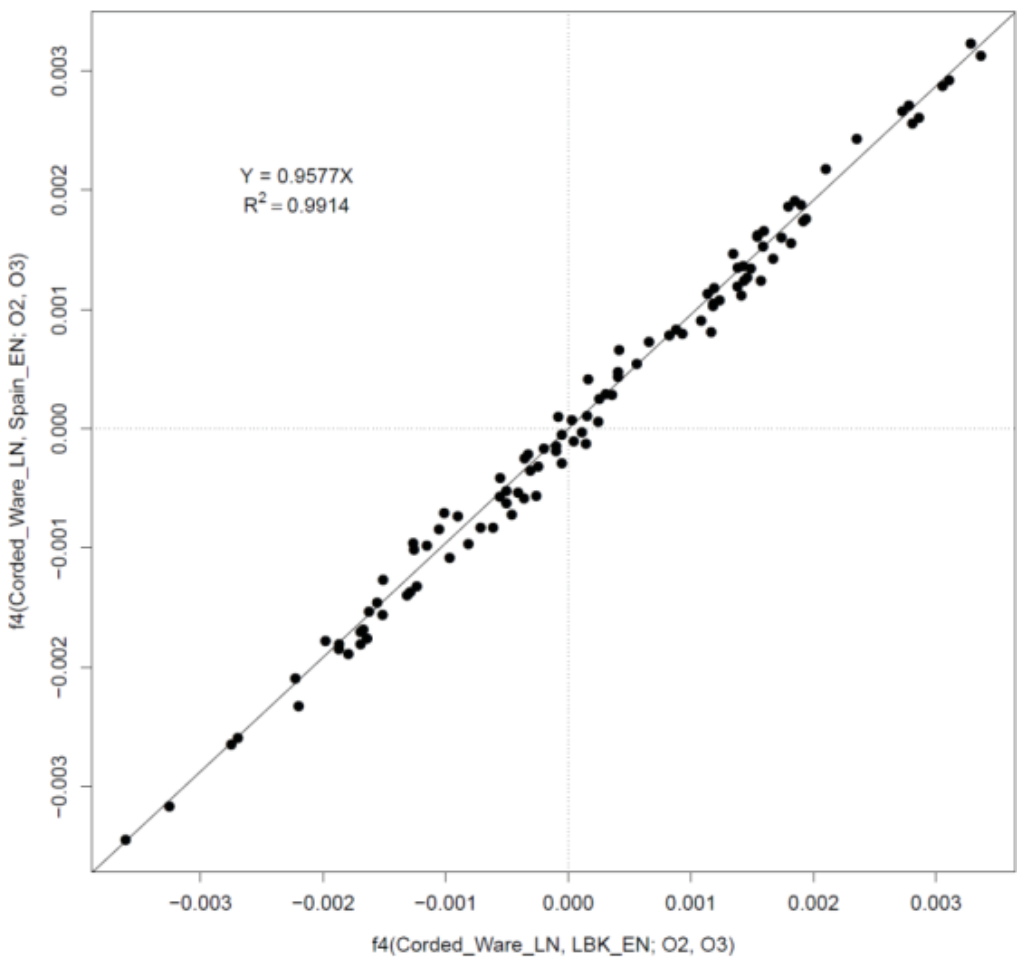

Plotting the $f_4$(*Test*, $Ref_1$; *B*, *C*) and $f_4$(*Test*, $Ref_2$; *B*, *C*) statistics to detect admixture has an advantage over the use of the $f_3$(*Test*; $Ref_1$, $Ref_2$) statistic in that these statistics are not affected by post-admixture drift in the admixed population, but rather rely on allele frequency correlations deep in the phylogeny. An example of this power is furnished by the case of Motala_HG which does not have a negative $f_3$-statistic using Loschbour and Karelia_HG as reference populations. When we test statistics of the form $f_3$(Motala_HG; *X*, *Y*), iterating over all pairs (*X*, *Y*) of ancient or modern populations, we obtain the lowest statistic $f_3$(Motala_HG; Loschbour, Karelia_HG) = 0.00404, (Z=1.8), which is, however positive. Additional evidence for admixture in Motala_HG is provided by the statistics $f_4$(Motala_HG, Loschbour, Karelia_HG, Chimp) = 0.0041 (Z=7.6) and $f_4$(Karelia_HG, Motala_HG; Loschbour, Chimp) = -0.00696 (Z=-11.5), which show that Karelia_HG shares more alleles with Motala_HG than with Loschbour and Loschbour shares more alleles with Motala_HG than with Karelia, which is expected if Motala_HG is a mixture of these two populations. When we plot $f_4$(Motala_HG, Loschbour; *B*, *C*) and $f_4$(Motala_HG, Karelia_HG; *B*, *C*) statistics (Fig. S9.14a, b), we observe a clear pattern of negative correlation, which shows that allele frequencies of Motala_HG are intermediate between those of Loschbour and Karelia_HG once we remove the post-admixture genetic drift in Motala_HG which is uncorrelated to allele frequencies between the outgroups *B*, *C*.

**Figure S9.14: Motala_HG is a mixture of (a) Loschbour and Karelia_HG or (b) LaBrana1 and Samara_HG.** By plotting these statistics we recover a signal of admixture that is not shown by the $f_3$(Motala_HG; Loschbour, Karelia_HG) statistic due to post admixture genetic drift in Motala_HG ancestors. However, **Karelia_HG (c) and Samara_HG (d) do not appear to be a mixture of Loschbour and MA1.** The lack of a negative correlation does not disprove the presence of such admixture, but it provides an additional motivation for considering the alternative models in SI8.

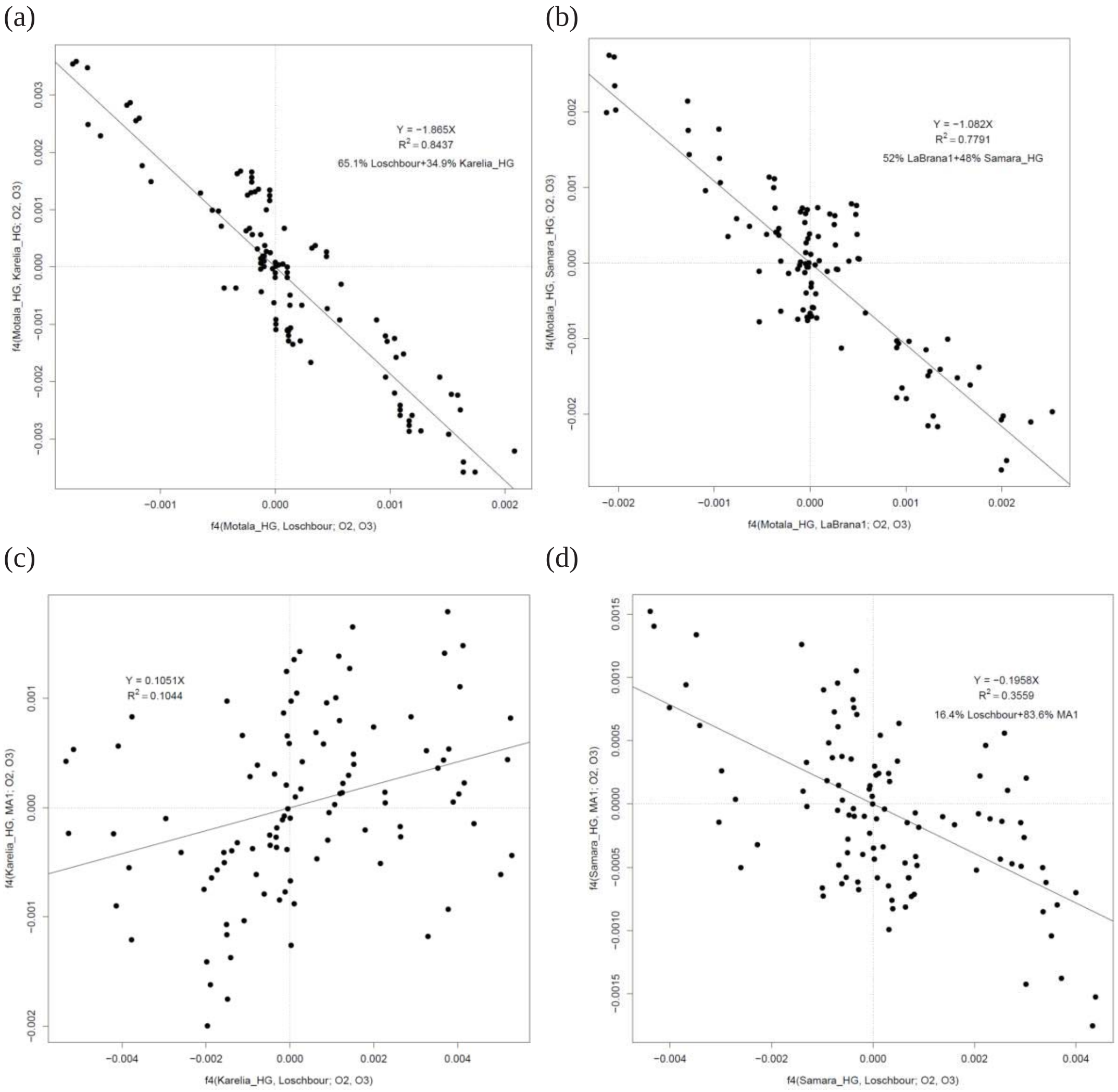

We also explored the possibility that the eastern European hunter-gatherers like Karelia_HG and Samara_HG were admixed, as we observed that they are intermediate between Loschbour and MA1 as measured by the statistics $f_4$(Karelia_HG, MA1; Loschbour, Chimp) = 0.00625 (Z=7.4) and $f_4$(Karelia_HG, Loschbour; MA1, Chimp) = 0.00601 (Z=7.4), and it can be modeled as a mixture of these two elements (SI8). However, when we use the regression method, we obtain a slightly positive correlation for Karelia_HG and a slightly negative one for Samara_HG (Fig. S9.14c, d), which do not appear convincing signals of admixture as in the case of Motala_HG. Note, however, that the lack of a negative correlation does not, in itself, indicate lack of admixture, as it is possible that allele frequency differences between the reference populations (Loschbour and MA1) may not be sufficiently correlated with those of the outgroups, which is a necessary condition for the preservation of this signal of admixture.

It is also possible to extract admixture proportions (for the case $N$=2) from a regression like the one in Fig. S9.12 and Fig. S9.14, and we indicate estimated mixture proportions in these two Figures. Remember that $f_4(Test, A; B, C) \approx \alpha f_4(Ref_1, A; B, C)+(1-\alpha)f_4(Ref_2, A; B, C)$. This can be re-written as $(\alpha+1-\alpha)f_4(Test, A; B, C) \approx \alpha f_4(Ref_1, A; B, C)+(1-\alpha)f_4(Ref_2, A; B, C)$ or $\alpha(f_4(Test, A; B, C)- f_4(Ref_1, A; B, C)) \approx (1-\alpha)(f_4(Ref_2, A; B, C)- f_4(Test, A; B, C))$, or, $\alpha f_4(Test, Ref_1; B, C)\approx (\alpha-1)(f_4(Test, Ref_2; B, C))$, or, $f_4(Test, Ref_2; B, C) \approx \frac{a}{a-1} f_4(Test, Ref_1; B, C)$. The coefficient $\frac{a}{a-1}$ can be fitted using a large number of outgroups (as in Figs. S9.12 and S9.14), yielding an estimate for $\alpha$.

For a fixed pair of outgroups ($B$, $C$):

$$\hat{a} = \left(1 - \frac{f_4(Test, Ref_1; B, C)}{f_4(Test, Ref_2; B, C)}\right)^{-1} \quad \text{(S9.3)}$$

So, an estimate can be derived analytically (similar to an $f_4$-ratio[3,4]). However, by using a larger number of outgroups, the mixture proportion can be estimated from the regression equation which makes use of the correlation between $f_4$-statistics involving a large number of outgroups, and also obviates the need of choosing particular outgroup populations when a large number of them are available for analysis.

More generally, we view this admixture estimation procedure as a substantial generalization of $f_4$-ratio estimation[3,4]. Unlike $f_4$-ratio estimation, the method can work even without a formal model of the historical relationships of all the analyzed populations. All that is required is that the *Test* population be related to the outgroups via ancestral populations that form clades with the $Ref_1$ and $Ref_2$ populations (the reference populations themselves may be complicated and poorly understood mixtures of the outgroups).

**Interpretation of negative correlation**

We have so far treated the 15 world outgroups as a “magic set” whose members are unequally related to ancient reference populations and can thus help detect admixture (and estimate its proportions) in a *Test* population. To gain understanding on why this method works, we examined the particular statistics plotted in Fig. S9.12 and noticed that the pattern of linear anti-correlation was most strongly affected by Native American and North Asian populations. We re-plot the data points of Fig. S9.13b in Fig. S9.15, color-coding five clusters that were inferred in unsupervised manner with a model-based clustering algorithm (mclust[5,6]).

We examined the most extreme “orange” and “purple” clusters of Fig. S9.15 which contribute to the inference of admixture, and list the relevant statistics in Table S9.1. The statistics involving Yamnaya and LBK_EN have consistently opposing signs (thus Corded_Ware_LN is intermediate between these populations), and involve comparisons with the Native American Karitiana and the Chukchi and Eskimo populations from northeast Siberia. The inclusion of these populations (who have “Ancient North Eurasian” ancestry[2]) can discriminate between LBK_EN and Yamnaya, as Yamnaya share most drift with them, LBK_EN least, and Corded_Ware_LN intermediate.

**Figure S9.15:** Admixture in the Corded_Ware_LN between Esperstedt_MN and Yamnaya reference populations can be detected due to the statistics color-coded "orange" "blue" "green" and "purple", as inferred by mclust[5]. The cluster of "red square" statistics has a mean close to (0,0) and consists of outgroup population pairs to which the two references are approximately equally related. The line passing through the means of the "purple" and "orange" clusters has intercept ~0. It is these statistics that primarily contribute to admixture inference.

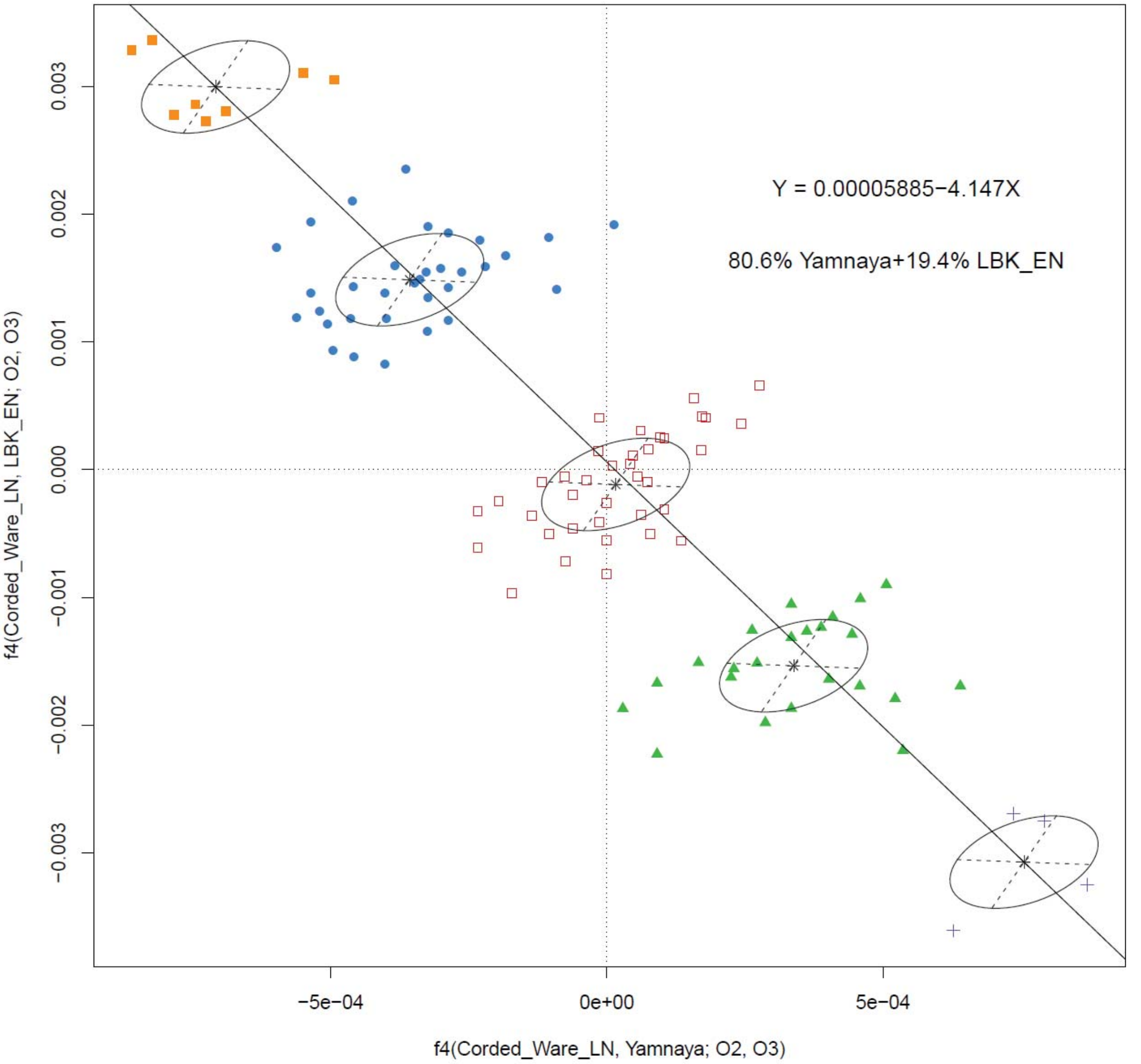

**Table S9.1: Corded_Ware_LN is intermediate between Yamnaya and LBK_EN in sharing alleles with “Siberian/Native American” populations**

| $O_2$ | $O_3$ | $f_4$(Corded_Ware_LN, Yamnaya; $O_2$, $O_3$) | Z | $f_4$(Corded_Ware_LN, LBK_EN; $O_2$, $O_3$) | Z |
|---|---|---|---|---|---|
| Biaka | Chukchi | 0.00074 | 3.3 | -0.00270 | -11.7 |
| Biaka | Eskimo | 0.00079 | 3.5 | -0.00275 | -11.4 |
| Biaka | Karitiana | 0.00087 | 3.3 | -0.00325 | -11.8 |
| Chukchi | Ju_hoan_North | -0.00049 | -2.1 | 0.00305 | 12.7 |
| Chukchi | Mbuti | -0.00069 | -2.9 | 0.00281 | 11.4 |
| Chukchi | Yoruba | -0.00073 | -3.4 | 0.00273 | 12.5 |
| Eskimo | Ju_hoan_North | -0.00055 | -2.3 | 0.00311 | 12.3 |
| Eskimo | Mbuti | -0.00075 | -3.0 | 0.00286 | 11.2 |
| Eskimo | Yoruba | -0.00078 | -3.6 | 0.00278 | 12.2 |
| Ju_hoan_North | Karitiana | 0.00063 | 2.2 | -0.00361 | -12.7 |
| Karitiana | Mbuti | -0.00082 | -2.9 | 0.00336 | 11.7 |
| Karitiana | Yoruba | -0.00086 | -3.3 | 0.00328 | 12.5 |

**Robustness to outgroup removal**

Estimated admixture proportions rely on the set of 15 outgroups, so a potential source of error is the choice of these outgroups. If estimated admixture proportions are sensitive to this choice then this would reduce our confidence in them, as they would be driven by (little understood) details in the history of the outgroups in relation to West Eurasians, rather than the admixture processes in the history of the tested groups. To investigate the resilience of our method to the outgroup populations, we estimated mixture proportions by dropping each of the outgroups in turn. Results are summarized in Table S9.2 and indicate that the method is not sensitive to the particular set of outgroups.

**Figure S9.16: Adding EN or MN populations to the set of outgroups reduces the amount of Esperstedt_MN related ancestry (“red”) and increases the amount of Karelia_HG related ancestry (“blue”) inferred for LN/BA populations.**

(a) Corded_Ware_LN

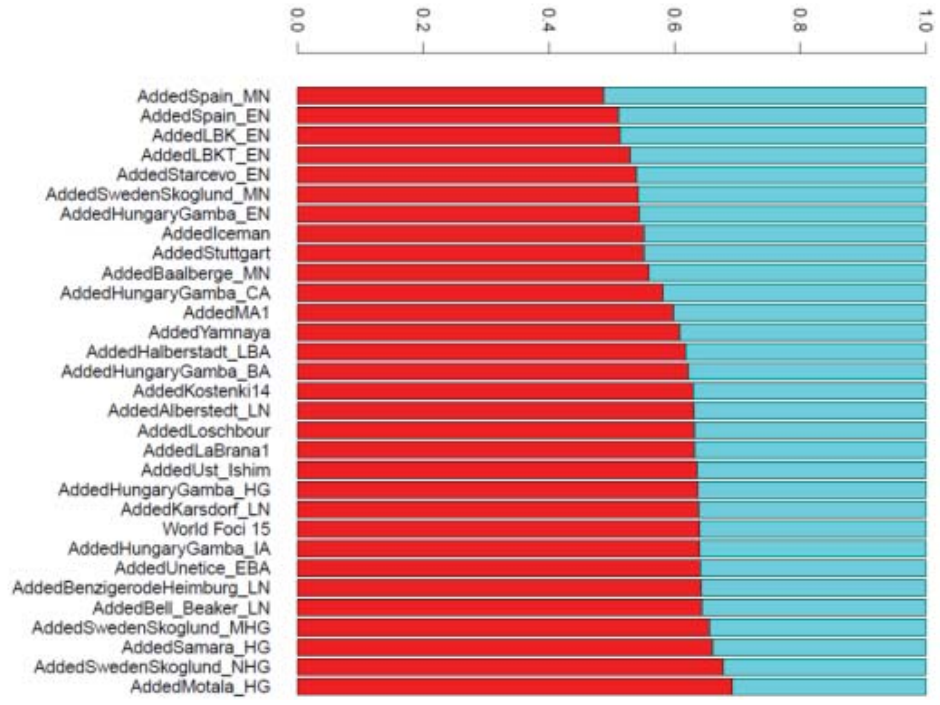

(b) Bell_Beaker_LN

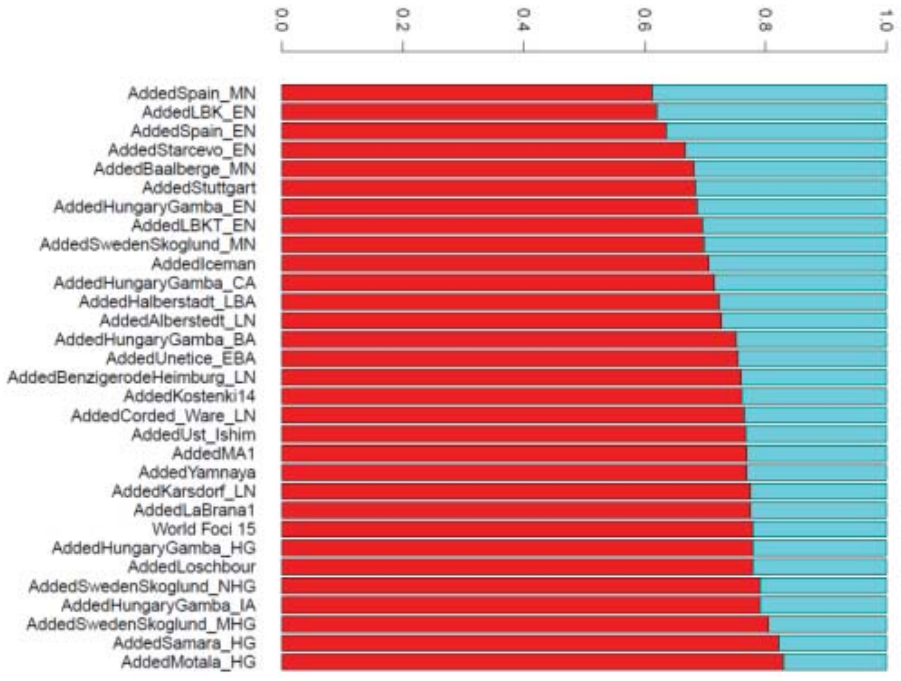

(a) BenzigerodeHeimburg_LN

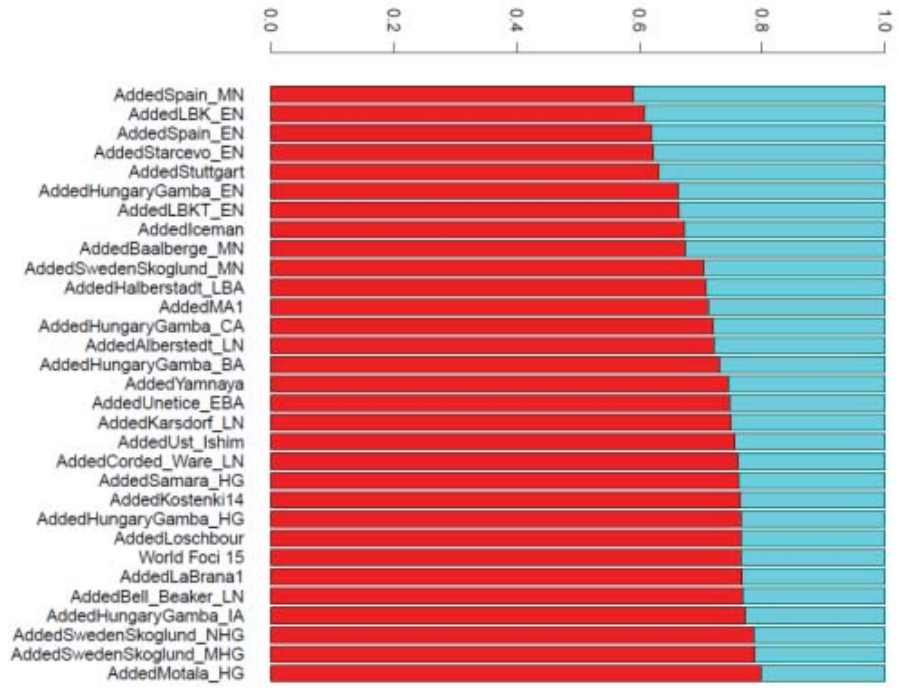

(b) Unetice_EBA

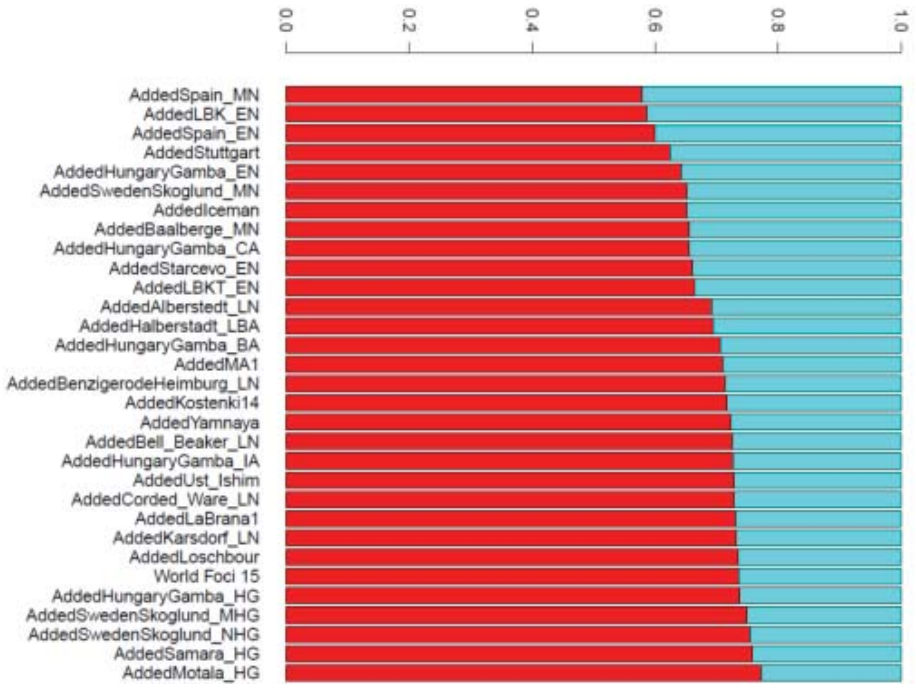

**Table S9.2: Estimated admixture proportions are robust to the choice of outgroups.** (These proportions are estimated using the non-negative least squares method). The estimate for the best $N$=2 pair of reference populations when using all 15 outgroups is given in the top row; in the bottom row the maximum difference between this estimate and the estimate when one of the 15 outgroups is dropped is given. This is always ≤4%, suggesting that inferences are not driven by any single outgroup.

| | **Baalberge_MN** | | **Esperstedt_MN** | | **Spain_MN** | | **Karsdorf_LN** | | **Corded_Ware_LN** | | **Bell_Beaker_LN** | | **Alberstedt_LN** | | **BenzigerodeHeimburg_LN** | | **Unetice_EBA** | | **Halberstadt_LBA** | |
|---|---|---|---|---|---|---|---|---|---|---|---|---|---|---|---|---|---|---|---|---|
| | LBK_EN | Loschbour | Loschbour | Spain_EN | LBK_EN | Loschbour | Motala_HG | Spain_MN | Spain_MN | Yamnaya | Karelia_HG | Spain_MN | Karelia_HG | Spain_MN | Karelia_HG | Spain_EN | Motala_HG | Spain_MN | Karelia_HG | LBK_EN |
| **WorldFoci15** | **0.825** | **0.175** | **0.338** | **0.662** | **0.788** | **0.212** | **0.816** | **0.184** | **0.209** | **0.791** | **0.248** | **0.752** | **0.250** | **0.750** | **0.289** | **0.711** | **0.563** | **0.437** | **0.297** | **0.703** |
| Ami | 0.825 | 0.175 | 0.333 | 0.667 | 0.790 | 0.210 | 0.814 | 0.186 | 0.208 | 0.792 | 0.247 | 0.753 | 0.250 | 0.750 | 0.289 | 0.711 | 0.563 | 0.437 | 0.297 | 0.703 |
| Biaka | 0.824 | 0.176 | 0.335 | 0.665 | 0.783 | 0.217 | 0.814 | 0.186 | 0.203 | 0.797 | 0.245 | 0.755 | 0.246 | 0.754 | 0.288 | 0.712 | 0.557 | 0.443 | 0.299 | 0.701 |
| Bougainville | 0.823 | 0.177 | 0.343 | 0.657 | 0.784 | 0.216 | 0.817 | 0.183 | 0.209 | 0.791 | 0.248 | 0.752 | 0.250 | 0.750 | 0.289 | 0.711 | 0.562 | 0.438 | 0.294 | 0.706 |
| Chukchi | 0.827 | 0.173 | 0.349 | 0.651 | 0.795 | 0.205 | 0.824 | 0.176 | 0.203 | 0.797 | 0.253 | 0.747 | 0.253 | 0.747 | 0.289 | 0.711 | 0.566 | 0.434 | 0.293 | 0.707 |
| Eskimo | 0.819 | 0.181 | 0.346 | 0.654 | 0.788 | 0.212 | 0.811 | 0.189 | 0.200 | 0.800 | 0.249 | 0.751 | 0.263 | 0.737 | 0.293 | 0.707 | 0.571 | 0.429 | 0.295 | 0.705 |
| Han | 0.825 | 0.175 | 0.335 | 0.665 | 0.791 | 0.209 | 0.817 | 0.183 | 0.208 | 0.792 | 0.248 | 0.752 | 0.250 | 0.750 | 0.289 | 0.711 | 0.563 | 0.437 | 0.297 | 0.703 |
| Ju_hoan_North | 0.827 | 0.173 | 0.298 | 0.702 | 0.788 | 0.212 | 0.815 | 0.185 | 0.224 | 0.776 | 0.241 | 0.759 | 0.247 | 0.753 | 0.284 | 0.716 | 0.552 | 0.448 | 0.294 | 0.706 |
| Karitiana | 0.814 | 0.186 | 0.374 | 0.626 | 0.752 | 0.248 | 0.829 | 0.171 | 0.218 | 0.782 | 0.251 | 0.749 | 0.234 | 0.766 | 0.287 | 0.713 | 0.562 | 0.438 | 0.301 | 0.699 |
| Kharia | 0.825 | 0.175 | 0.336 | 0.664 | 0.787 | 0.213 | 0.818 | 0.182 | 0.208 | 0.792 | 0.248 | 0.752 | 0.254 | 0.746 | 0.289 | 0.711 | 0.564 | 0.436 | 0.298 | 0.702 |
| Mbuti | 0.843 | 0.157 | 0.342 | 0.658 | 0.810 | 0.190 | 0.826 | 0.174 | 0.204 | 0.796 | 0.249 | 0.751 | 0.256 | 0.744 | 0.285 | 0.715 | 0.567 | 0.433 | 0.297 | 0.703 |
| Onge | 0.832 | 0.168 | 0.327 | 0.673 | 0.790 | 0.210 | 0.809 | 0.191 | 0.210 | 0.790 | 0.248 | 0.752 | 0.248 | 0.752 | 0.286 | 0.714 | 0.562 | 0.438 | 0.300 | 0.700 |
| Papuan | 0.828 | 0.172 | 0.341 | 0.659 | 0.790 | 0.210 | 0.821 | 0.179 | 0.213 | 0.787 | 0.245 | 0.755 | 0.248 | 0.752 | 0.293 | 0.707 | 0.562 | 0.438 | 0.294 | 0.706 |
| She | 0.825 | 0.175 | 0.331 | 0.669 | 0.791 | 0.209 | 0.817 | 0.183 | 0.208 | 0.792 | 0.247 | 0.753 | 0.250 | 0.750 | 0.289 | 0.711 | 0.563 | 0.437 | 0.298 | 0.702 |
| Ulchi | 0.818 | 0.182 | 0.340 | 0.660 | 0.792 | 0.208 | 0.815 | 0.185 | 0.211 | 0.789 | 0.247 | 0.753 | 0.249 | 0.751 | 0.287 | 0.713 | 0.563 | 0.437 | 0.297 | 0.703 |
| Yoruba | 0.825 | 0.175 | 0.345 | 0.655 | 0.788 | 0.212 | 0.796 | 0.204 | 0.202 | 0.798 | 0.249 | 0.751 | 0.250 | 0.750 | 0.289 | 0.711 | 0.564 | 0.436 | 0.306 | 0.694 |
| **Max. diff** | **0.018** | **0.018** | **0.040** | **0.040** | **0.036** | **0.036** | **0.020** | **0.020** | **0.016** | **0.016** | **0.006** | **0.006** | **0.016** | **0.016** | **0.004** | **0.004** | **0.011** | **0.011** | **0.009** | **0.009** |

**Robustness to outgroup addition**

We can also add more populations to the set of 15 outgroups and see whether our conclusions are affected. The 15 outgroups are a "conservative" choice as they lack (both due to geography and as empirically assessed with ADMIXTURE) close ties to Europe. However, this comes at the cost of potentially having little or no relevance to the different relatedness of European populations. We have seen that this is not the case, as $f_4$-statistics relating them to ancient European populations make it possible to recover the presence of admixture and to estimate its proportions. However, by attempting to include outgroups that are genetically closer to the populations under study (and which thus share more genetic drift), we can increase our ability to perform these tasks and gain new insight.

We add ancient populations to the set of outgroups and re-estimate admixture proportions using the new set of 15+1 outgroups. An interesting pattern occurs when we attempt to fit Corded_Ware_LN as a mixture of Esperstedt_MN (which is the youngest Middle Neolithic sample from Germany predating the appearance of the Corded Ware) and Karelia_HG (Fig. S9.16). When any of the EN/MN European populations is added to the set of outgroups, the fraction of Esperstedt_MN ancestry inferred for Corded_Ware_LN is reduced. The same pattern is also observed for other LN/BA populations. This hints that European farmer populations do not function as outgroups, violating in some way our assumptions.

To better understand how European farmer populations influence admixture proportions when added as outgroups, we plot statistics of the form $f_4$(Corded_Ware_LN, Karelia_HG; $O_2$, $O_3$) and $f_4$(Corded_Ware_LN, Esperstedt_MN; $O_2$, $O_3$) in Fig. S9.17. In each panel of Fig. S9.17 we add a different European farmer population (LBK_EN, Spain_EN, HungaryGamba_EN, and Iceman), representing different locations in Europe. All present a similar pattern: statistics of the form $f_4$(Corded_Ware_LN, Esperstedt_MN; $O_2$, *European Farmer*) do not fall on the same line as those with the "World Foci 15" outgroups, but appear to be transposed by a constant positive value. This is unexpected if Corded_Ware_LN shares alleles with *European Farmer* only via a population like Esperstedt_MN. While the regression of "World Foci 15" outgroups passes clearly through the origin, in accordance with the modeling prediction that $f_4$(*Test*, Esperstedt_MN; $O_2$, $O_3$) $\approx \frac{\alpha}{\alpha-1}$ $f_4$(*Test*, Karelia_HG; $O_2$, $O_3$) derived above, the statistics $f_4$(*Test*, Karelia_HG; $O_2$, *European Farmer*) and $f_4$(*Test*, Esperstedt_MN; $O_2$, *European Farmer*) clearly do not have this behavior.

Fig. S9.18 suggests an explanation for these findings. We empirically observe that $f_4$(*Test*, Karelia_HG; $O_2$, *European Farmer*) = -(1-$\alpha$)$z$ and $f_4$(*Test*, Esperstedt_MN; $O_2$, *European Farmer*) =$x$+$\alpha z$. Thus:

$$\frac{f_4(Test, \text{Esperstedt_MN}; O_2, European\ Farmer)}{f_4(Test, \text{Karelia_HG}; O_2, European\ Farmer)} = \frac{x+\alpha z}{-(1-\alpha)z} = \frac{\alpha}{\alpha-1} + \frac{x}{-(1-\alpha)z}$$
(S9.4)

Multiplying by $f_4$(*Test*, Karelia_HG; $O_2$, *European Farmer*), and using the fact that $f_4$(*Test*, Karelia_HG; $O_2$, *European Farmer*) = -(1-$\alpha$)$z$:

$$f_4(Test, \text{Esperstedt_MN}; O_2, European\ Farmer) = \frac{\alpha}{\alpha-1} f_4(Test, \text{Karelia_HG}; O_2, European\ Farmer) + x \qquad \text{(S9.5)}$$

This is the functional form observed in Fig. S9.17; the positive offset *x* is seen for all European farmer populations when added as outgroups, and this suggests that the admixing population *Q* (Fig. S9.18) split before the differentiation of European farmers from each other. Note also that the positive offset would appear if at least some of the ancestry of the *Test* population came from *Q* (for which *European Farmer* is not a "real outgroup") and some of it came from a population closely related to Esperstedt_MN (for which *European Farmer* is a real outgroup).

**Figure S9.17: Adding early European farmers to the set of outgroups (Esperstedt_MN+Karelia_HG mix).** In blue are statistics involving only "World Foci 15" outgroups. In red are statistics where $O_3$ is the early European farmer population. The two fitted lines differ by a positive vertical offset.

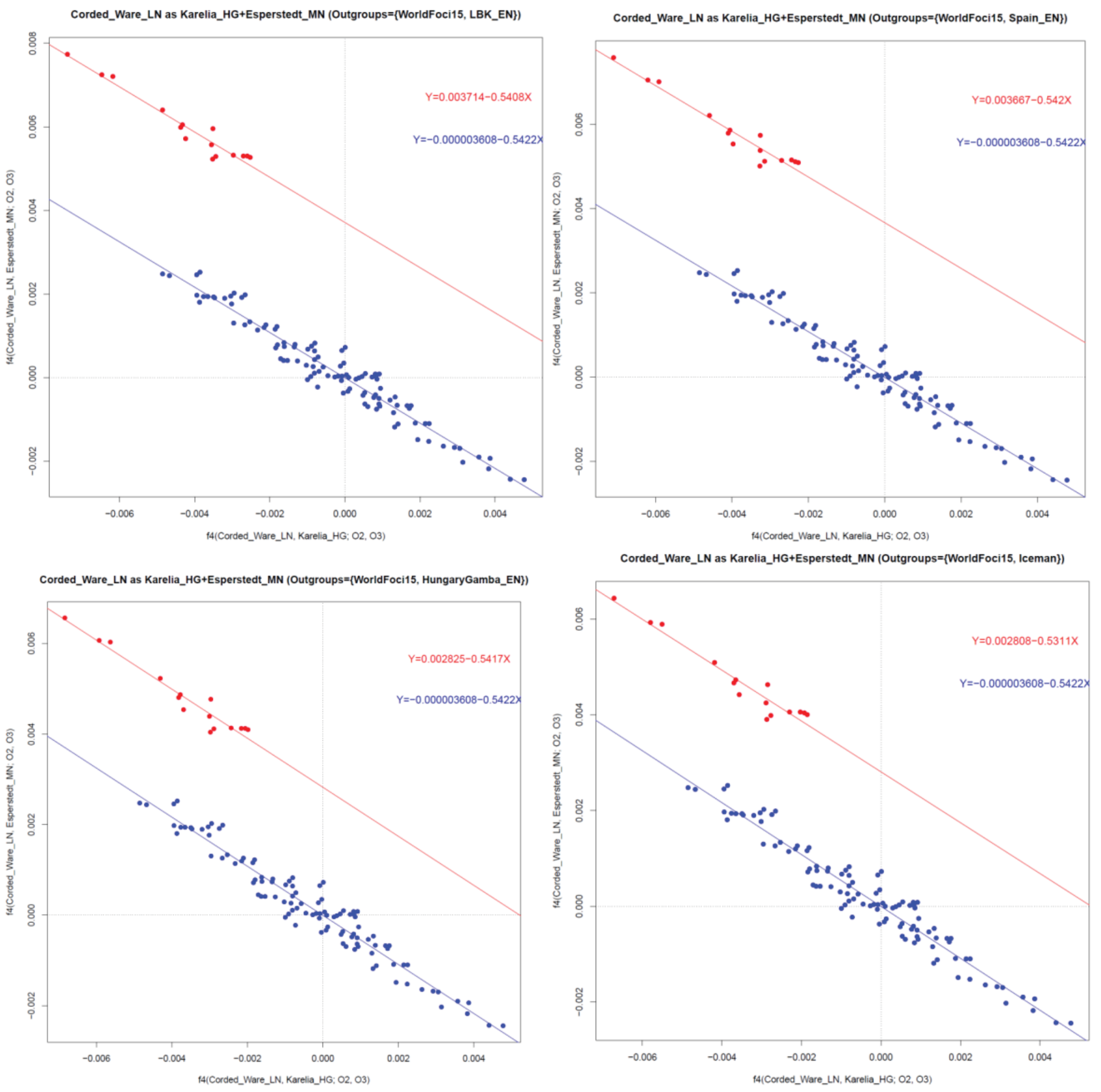

**Figure S9.18: When a *European Farmer* population is treated as an outgroup, but in fact has ancestry that is as closely related to the true admixing population *Q* as population Esperstedt_MN that we use as a reference, the assumptions of our negative regression analysis are violated.** In this simplified model, the *European Farmer* population shares genetic drift *x* with Esperstedt_MN after the split of the true admixing population *Q*, which results in an offset of the scatterplot by an amount *x* for $f_4$-statistics involving *European Farmer*. In truth, the history is more complicated (elsewhere in this paper, we show that both *European Farmer* and Esperstedt_MN have WHG-related ancestry in different proportions). However, we hypothesize that the patterns in Figure S9.17 are driven by the fact that a substantial amount of the ancestry in *European Farmer* and Esperstedt conforms to these relationships.

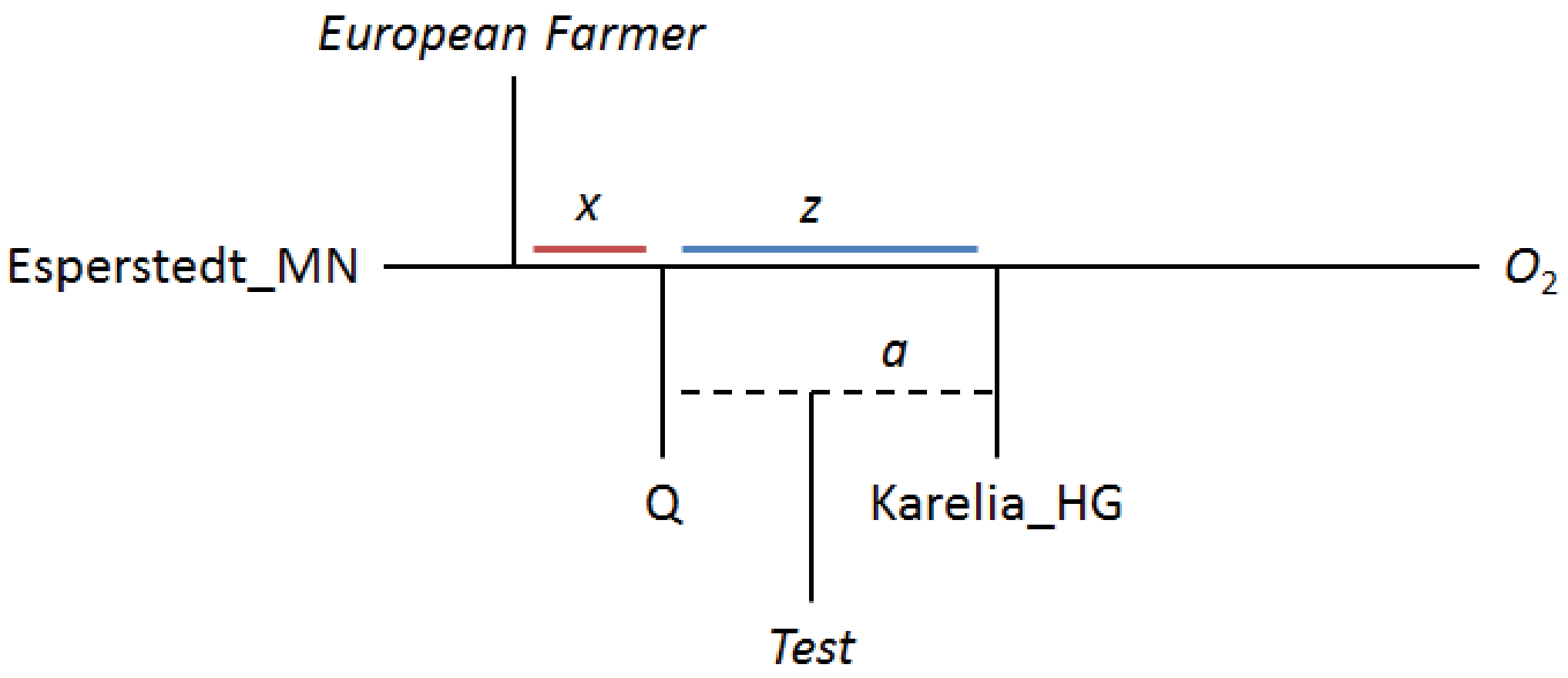

In Fig. S9.19 we show the same statistics as Fig. S9.17, but using Yamnaya as an eastern reference population. The substantial offset seen in Fig. S9.17 is not evident when using Yamnaya. There is evidence (SI7) that Yamnaya has diluted EHG (Karelia_HG-related) ancestry from a source related to present-day Near Eastern populations. If some of the farmer-related ancestry in the Corded Ware was not of local European origin but came with the Yamnaya, ultimately from the Near East, then this ancestry would represent a farmer-related population that was basal to European farmers and thus occupy a phylogenetic position similar to that of *Q* in Fig. S9.18.

**Figure S9.19: Adding early European farmers to the set of outgroups (Esperstedt_MN+Yamnaya mix).** Unlike Fig. S9.17, the red statistics involving a European farmer as $O_3$ line up fairly well with the blue "World Foci 15" outgroups.

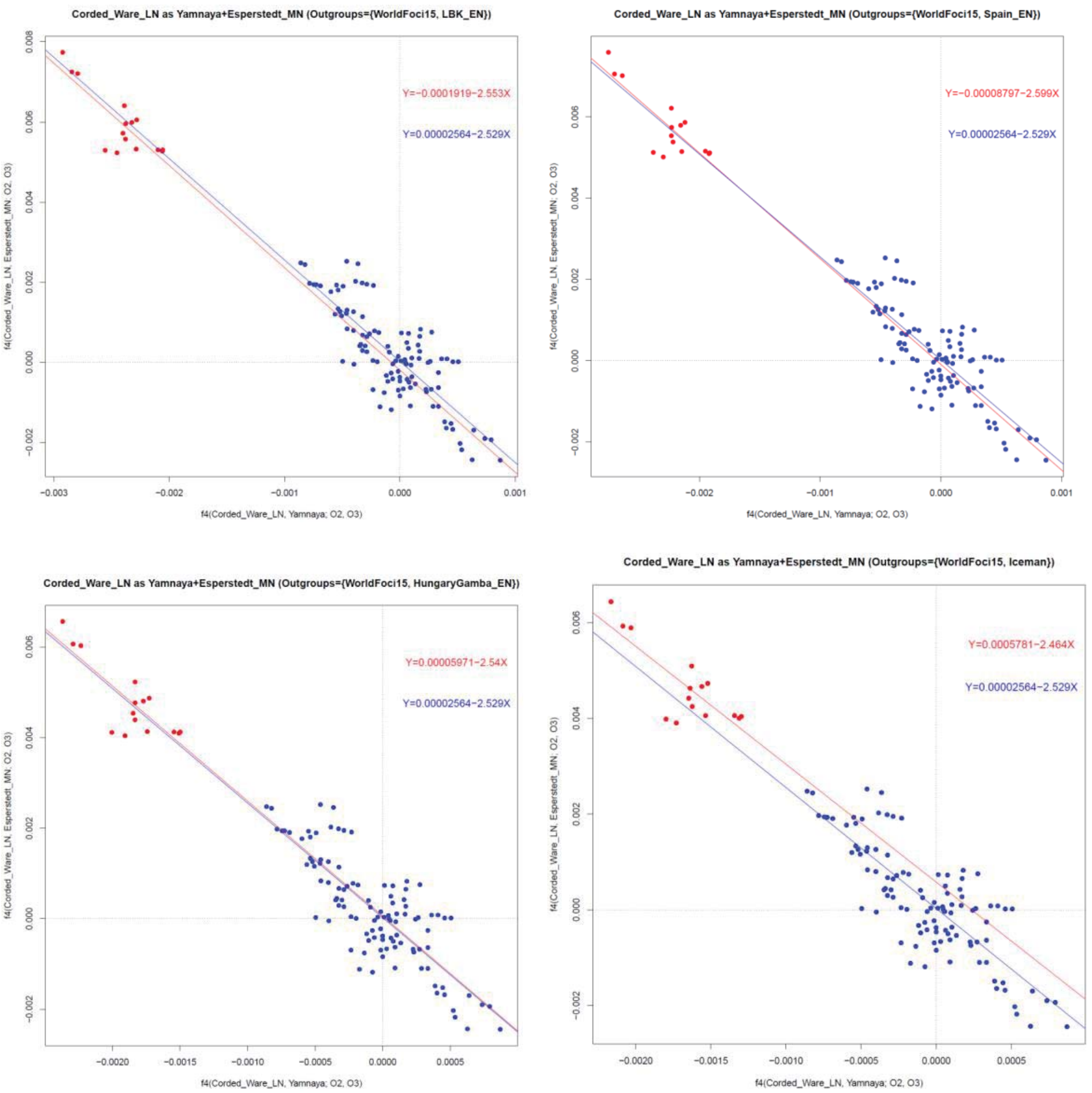

Ancient genomes from the Caucasus, the Near East, and Central Asia might reveal the existence of Neolithic populations there that may be involved in the ancestry of ancient steppe populations and central Europeans. The existence of such ancestry in the Yamnaya is visualized in Fig. S9.20, where this population is plotted as a mixture of Karelia_HG and either Iraqi_Jew or Armenian, showing a clear pattern of negative correlation.

**Figure S9.20: The Yamnaya show a pattern of negative correlation when using Karelia_HG and either Iraqi_Jew or Armenian as references.** This is consistent with it having a component of ancestry related to the Caucasus/Near East, suggesting that ancient DNA work in these regions may reveal a good surrogate for this type of ancestry.

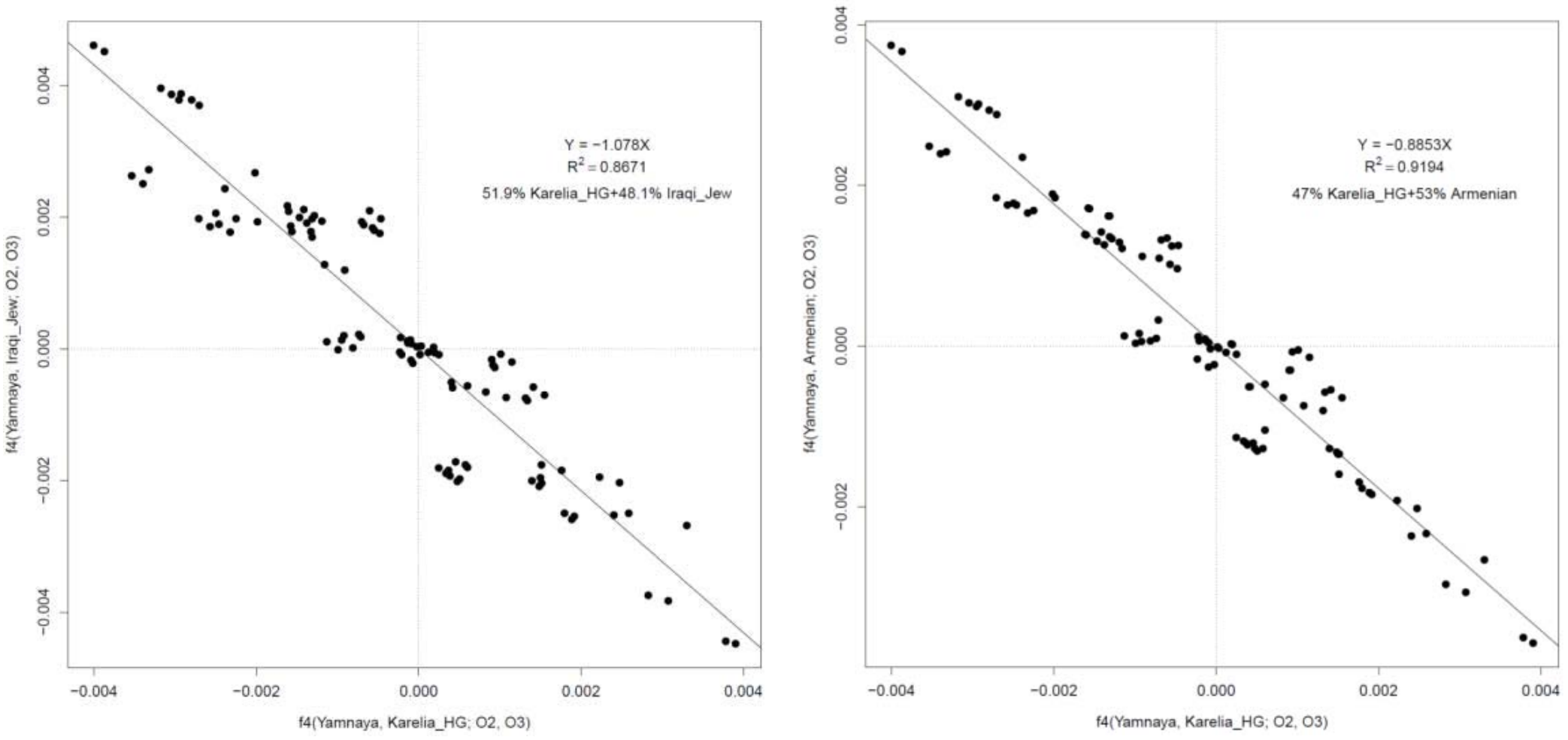

**Figure S9.21: Adding different outgroups in position $O_3$ of the statistics $f_4$(Corded_Ware, *Eastern*; $O_2$, $O_3$) vs. $f_4$(Corded_Ware, Esperstedt_MN; $O_2$, $O_3$). (a) Eastern=Karelia_HG, (b) Eastern=Yamnaya**

(a) (b)

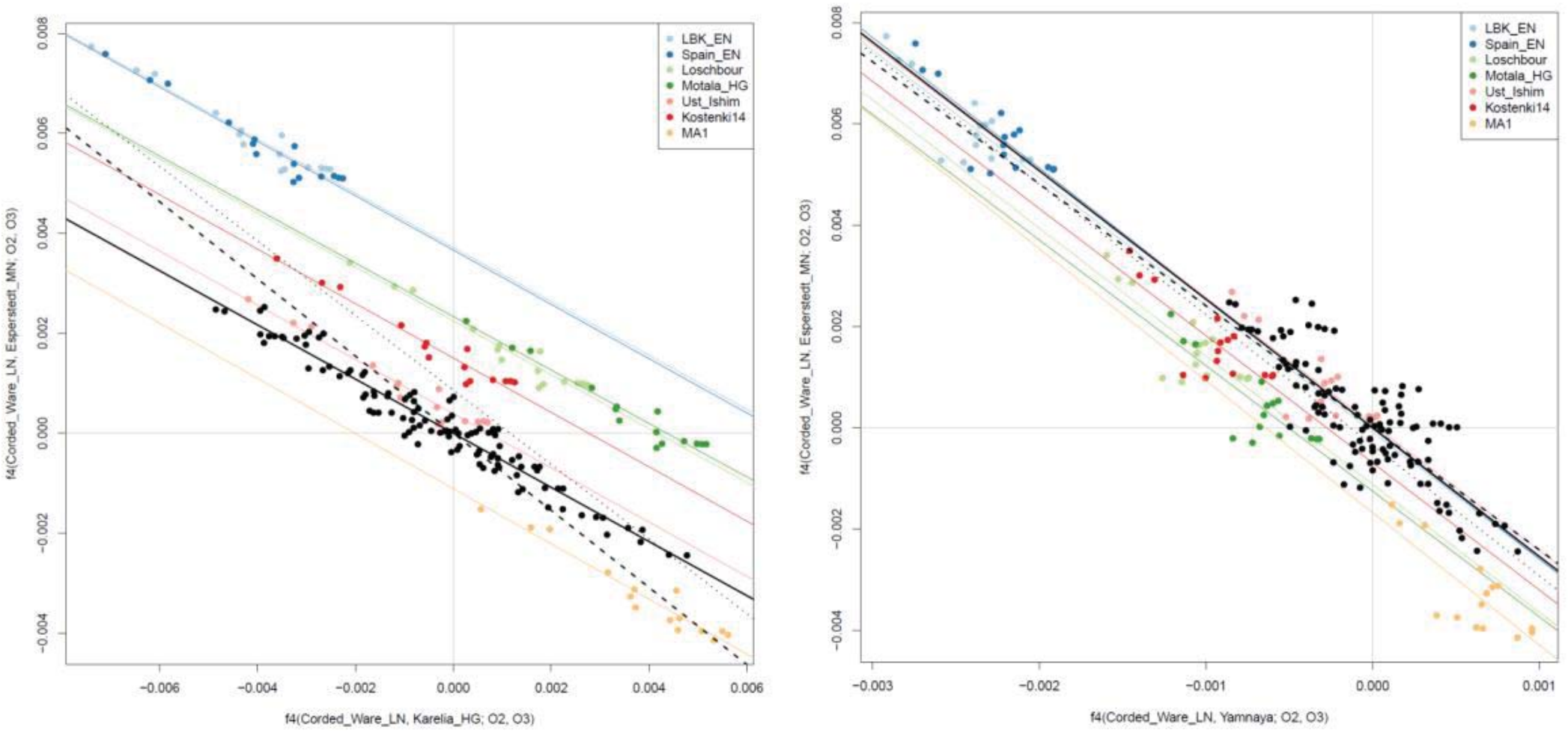

In Fig. S9.21 we plot statistics of the form $f_4$(Corded_Ware, *Eastern*; $O_2$, $O_3$) vs. $f_4$(Corded_Ware, Esperstedt_MN; $O_2$, $O_3$), color coding different ancient European outgroups for $O_3$. We plot a number of different regression equations on this plot: the solid black line corresponds to the regression using only the world outgroups (which is the same as in Fig. S9.12). The dashed black line is the regression through (0,0) using all plotted data points, and the dotted black line is the regression using all data points but with a non-zero intercept. The color-coded lines are regressions for the subsets of color-coded points representing the different choices for $O_3$. Fig. S9.21 visually explains why the addition of early farmers as outgroups results in a change of ancestry estimates (Fig. S9.16): the inclusion of early farmers in the set of outgroups results in a regression (dashed line) with a different slope to account for the "blue" statistics involving European farmers in the top left quadrant of Fig. S9.16a).

Fig. S9.16a also shows that statistics involving European hunter-gatherers (Kostenki14, Loschbour, and Motala_HG) are also transposed by a positive amount along the vertical axis corresponding to the statistic $f_4$(Corded_Ware_LN, Esperstedt_MN; *O*2, *European hunter-gatherer*). This reinforces the idea of shared drift between Esperstedt_MN and *European hunter-gatherer* that is not shared by a component in the ancestry of Corded_Ware_LN. Thus, Corded_Ware_LN has ancestry from a component that is basal to both European farmers (who have the maximum positive offset) and European hunter-gatherers. A slight negative offset is seen in Fig. S9.16b. for the statistics involving European hunter-gatherers and world outgroups. This hints that the Yamnaya population, while visibly a much better ancestral source for the Corded Ware, may in fact not be the exact admixing population. This also agrees with the results of *N*=3 discussed below and presented in Extended Data Fig. 3, which estimates that the Corded Ware can be modeled as 29.1% Esperstedt, 9.4% Samara_HG, and 61.5% Yamnaya, which suggests that the population of eastern migrants had a slightly higher proportion of EHG ancestry in its makeup than the Yamnaya sample from Samara. Such a conclusion might also be drawn from the $f_4$-statistic presented in SI7 (Table S7.6) that shows $f_4$(Corded_Ware_LN, Yamnaya; Karelia, Chimp) = -0.00001 (Z=0.0). If Corded_Ware_LN was a simple mixture of a population related to our Yamnaya sample and of Neolithic Europeans, this statistic should be negative. However, if Corded_Ware_LN is descended from a population that has a higher proportion of EHG ancestry than the Yamnaya population, then the dilution of EHG ancestry due to European Neolithic admixture (which would cause the statistic to be negative), would be counterbalanced by its increase due to this higher EHG ancestry (which would cause it to be positive). It is quite possible that the variable mixtures of EHG and farmer populations existed in the European steppe, and our Yamnaya population represents only a point in a continuum of such mixtures.

**Figure S9.22: Adding multiple ancient outgroups.** The regression using (Esperstedt_MN+Yamnaya) is virtually identical when using World Foci 15 + Ancient outgroups and when using only World Foci 15 (blue dots). However, the two lines are noticeably different for the (Esperstedt_MN+Karelia_HG) model.

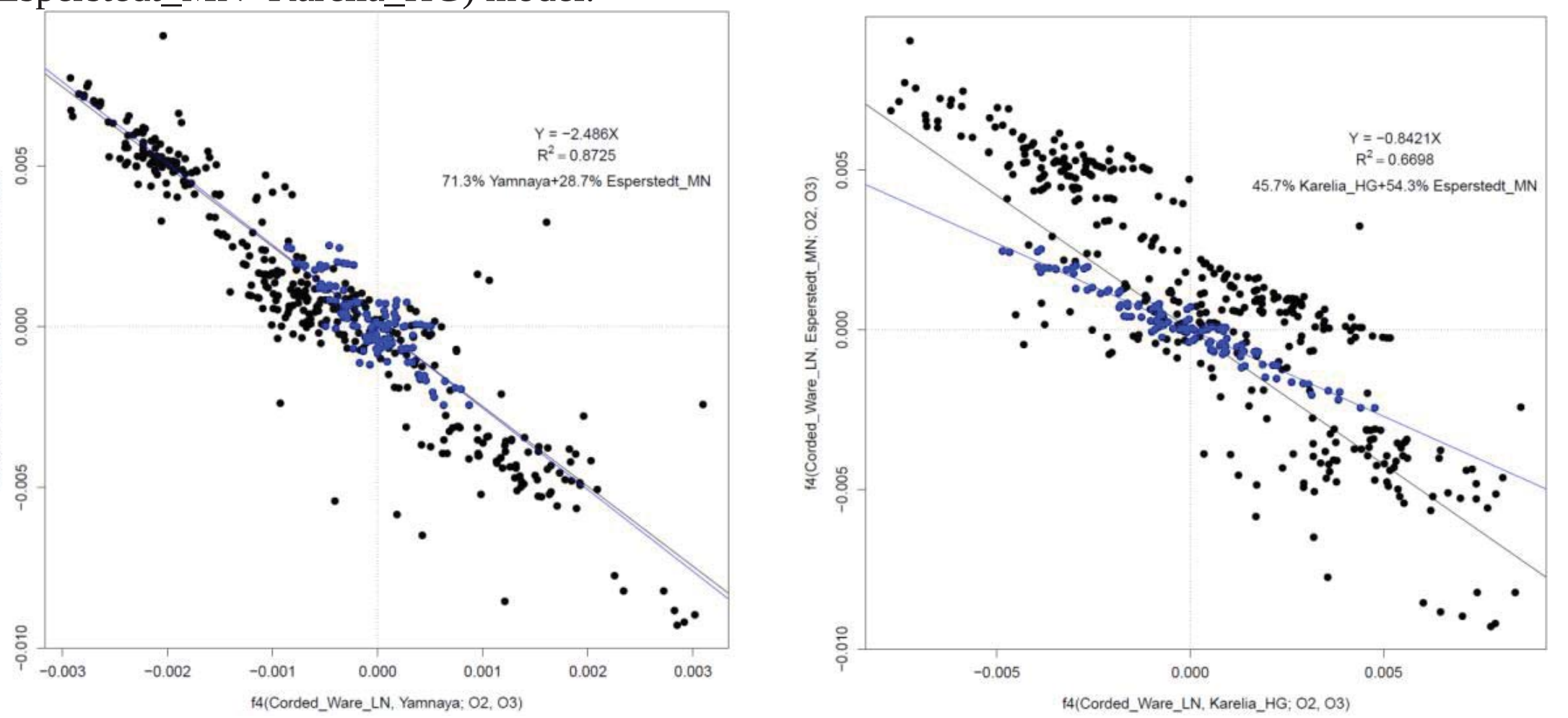

Finally, we included all ancient populations predating the Corded Ware in the set of outgroups (excluding the Samara_HG, Karelia_HG, Yamnaya, Esperstedt_MN, Baalberge_MN that were possible reference populations for the Corded Ware):

**"World Foci 15 + Ancients":** Ami, Biaka, Bougainville, Chukchi, Eskimo, Han, Ju_hoan_North, Karitiana, Kharia, Mbuti, Onge, Papuan, She, Ulchi, Yoruba, Starcevo_EN, LBKT_EN, LBK_EN, Spain_EN, HungaryGamba_EN, Stuttgart, Spain_MN, Loschbour, LaBrana1, HungaryGamba_HG, Motala_HG, SwedenSkoglund_MHG, SwedenSkoglund_NHG, MA1, Kostenki14, Ust_Ishim

We estimate mixture proportions using the regression method in Fig. S9.22. The regression plot is visibly similar when using only "World Foci 15" or "World Foci 15 +Ancients" when Yamnaya is the eastern reference population, but visibly different when Karelia_HG is the eastern reference.

In Extended Data Fig. 3 we also estimate mixture proportions for the Corded_Ware_LN using the non-negative least squares optimizer using "World Foci 15 + Ancients" as outgroups. The two worst performing $N$=2 models (Extended Data Fig. 3), involve either descent from only the Middle Neolithic populations *or* the Eastern European hunter-gatherers. The two best performing ones involve descent from the Yamnaya and either of the Middle Neolithic populations of Germany in proportions ~3/4 and ~1/4 respectively. The estimated mixture proportions are very similar to those derived using a related method in SI10.

**Proximate ancestral populations for present-day Europeans**

We also attempted to estimate ancestry proportions for present-day European populations in terms of the ancient European populations. We recognize that some European populations have East Eurasian, Near Eastern, or African ancestry that may not be modeled by the ancient European populations included in our study (from Iberia and Central/Northern/Eastern Europe). It is important to remember that inference of particular ancestral populations does not imply direction of migration for present-day European populations. For example, Sardinians often appear similar to ancient Neolithic farmers from continental Europe[2,7-9], but this does not imply migration from continental Europe into Sardinia. Instead, the finding that Cardial Ware-related Neolithic Iberians were similar to those from central Europe suggests that early farmers that followed a Mediterranean route into Europe may have also settled Sardinia. Similarly, there are no ancient samples from southeastern Europe in our study, yet groups from the Balkans can be modeled[2] as a 3-way mixture of the same ancestral populations that contributed to central, western, and northern Europe[2]. The fact that many European groups today can be well-described by mixtures of populations represented by ancient individuals from a small number of European locations suggests that the sampled ancient individuals are representative of widely distributed populations.

In Fig. S9.23 we show the best-fitting (lowest *resnorm*) models for European populations using the "World Foci 15" outgroups and reference populations by period: WHG, EHG, SHG, EN, MN, LN/BA, Yamnaya. Varying $N$=1, 2, 3 we observe some interesting patterns. At $N$=1, present-day Europeans from central/northern Europe prefer the LN/BA population as a reference, while those from southern Europe prefer the EN/MN ones. This is not surprising given the fact that LN/BA Europeans are temporally closer to the present and belonged to archaeological cultures (such as Corded Ware or Bell Beaker) with wide geographical ranges. There is also substantial variation in the goodness of fit for different populations, hinting that many of them are not well-modeled with $N$=1 (and in agreement with our findings in SI7 that indicate discontinuity between LN/BA populations and present-day ones).

**Figure S9.23: Identifying proximate ancestral populations for present-day Europeans**. (a, b, c): Best reference populations for present-day Europeans with *N*=1, 2, 3 admixing populations. (d): Change in *resnorm* as *N* changes.

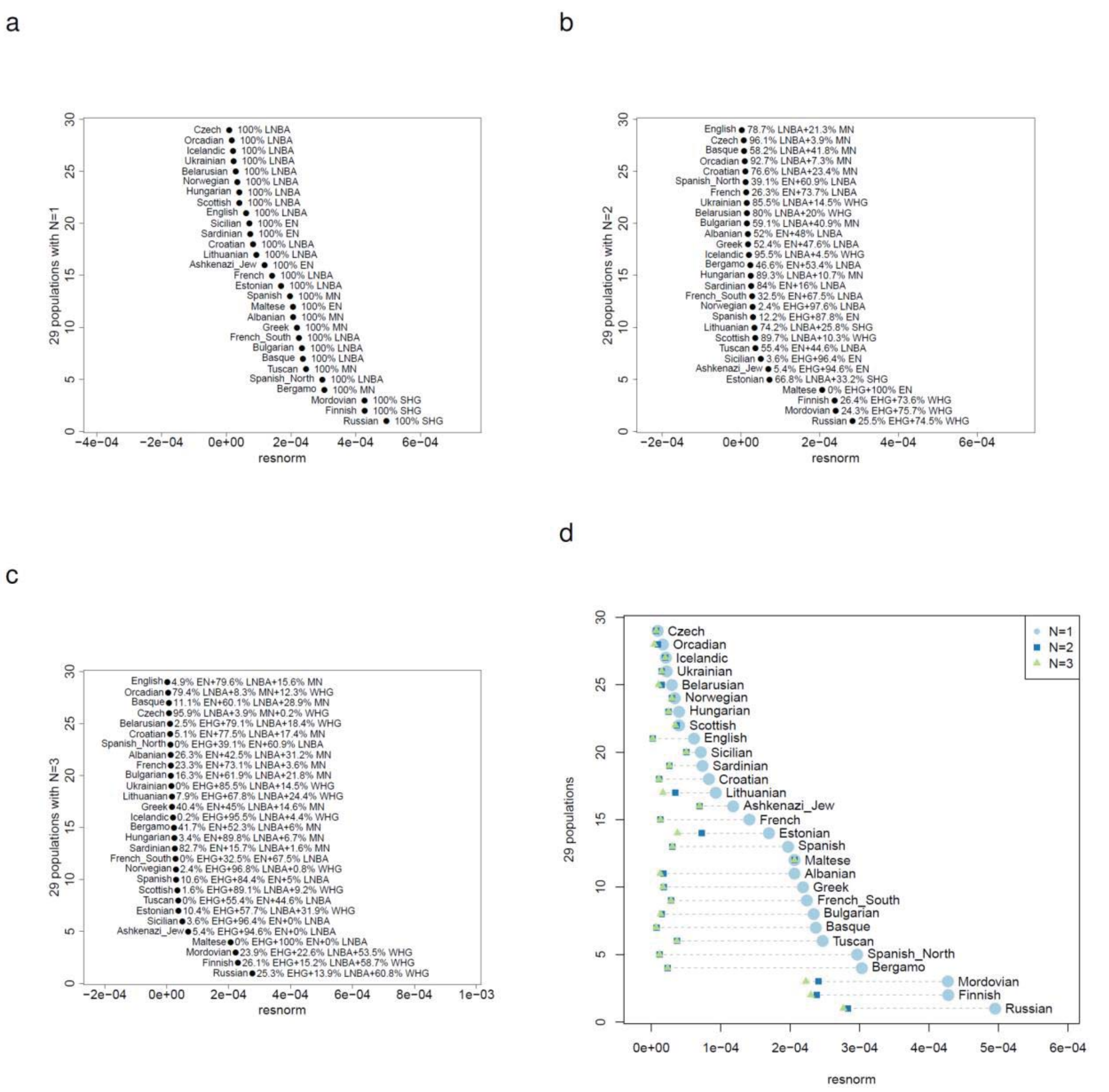

At *N*=2 most populations appear to be a mixture of two different elements which are EN/MN and LN/BA. There are interesting exceptions such as the Lithuanians (who have LN/BA plus SHG, consistent with our finding that they have more hunter-gatherer ancestry than Late Neolithic Europeans, SI7). For most other northern European populations it seems that the best *N*=2 model involves admixture between LN/BA and MN, consistent with our observations about the dilution of the ancestry brought into Late Neolithic Europe in modern populations (SI7). Conversely, modern southern Europeans (who generally fit worse than northern ones for *N*=1) appear to be a mixture of EN/MN and LN/BA. This does not necessarily imply that LN/BA populations of central northern Europe admixed into the populations of southern Europe as the LN/BA central Europeans may be the best proxy in our dataset for the ANE-bearing populations that affected the entire European continent[2]. Unlike central Europe where the appearance of ANE-related populations can be dated to the Late Neolithic by the results of our paper, we do not know when such populations arrived elsewhere in Europe. Did they arrive at the same time, as part of a general phenomenon affecting the entire European continent? Alternatively, did they arrive later, perhaps related to migrations from north to south that overlap the period of recorded history[10,11]?

*N*=3 seems to provide very little benefit in the fit of the mixture model, indicating that three proximate sources are not needed for most European populations. This finding is not in contradiction with the three source population model for Europeans.[2] Instead, the finding shows that these source populations contributed to many present-day European populations through the intermediary of as few as two proximate populations (in different proportions), which we are able to model as clade with samples for which we have ancient DNA data.

Fig. S9.24d shows visually how model fit improves as *N* increases: significant *resnorm* at *N*=1 is greatly reduced at *N*=2 but there is virtually no change at *N*=3. Some populations, including previously identified Jewish, Mediterranean, and Northeastern European outliers[2], differ markedly from most Europeans even at *N*=3, suggesting that they possess ancestry from additional sources not included in the set of references.

**Proportions of EN, WHG, and EHG/Yamnaya ancestry in ancient and modern Europeans**

The previous analysis used MN and LN/BA Europeans as reference populations for modern ones, but it is also useful to treat them as test populations and estimate their ancestral proportions together with those of modern Europeans. Fig. S9.24 shows that when WHG admixture is added to EN, residuals for most European populations are reduced, consistent with most Europeans not being descended from EN farmers of Europe. Four populations indicate no change in residuals: Sardinians are the population that is closest to early European farmers[2,7-9,12] with an estimated ~90% descent from them (ref. 2 and Fig. S9.23b), while Maltese, Ashkenazi Jews, and Sicilians may have Near Eastern admixture not mediated via early European farmers[2].

Fig. S9.25 shows that when EHG/Yamnaya admixture is added on top of the EN+WHG one, residuals continue to improve for European populations. Three noticeable exceptions are Sardinians (who may have little if any "Ancient North Eurasian" admixture[2]), MN Europeans (which is not surprising as they precede the arrival of EHG/Yamnaya-related populations into central Europe), and Maltese (an island European population that may also have been less affected by this type of ancestry). For all other European populations (including the LN/BA and modern Europeans across the continent), the addition of EHG/Yamnaya admixture visibly improves the residuals. This indicates that ancient eastern European populations eventually contributed ancestry to most European populations. In central Europe this occurred as early as the Late Neolithic. Future ancient DNA research is likely to show whether the rest of Europe was also affected at that time or whether this type of ancestry spread earlier or later to different parts of Europe.

Three northeastern European populations (Finns, northwestern Russians, and Mordovians) continue to have noticeably higher residuals at *N*=3 in Fig. S9.25. We have previously determined that this is due to more recent gene flow from Siberia[2], perhaps related to the spread of Finno-Ugric languages. Unfortunately, we do not have access to an ancient DNA sample that might be representative of this population; clearly Eastern European hunter-gatherers like Karelia_HG who lived in northeastern Europe are not good representatives of this population (as their effect is Europe-wide and not limited to northeastern Europe and as they fail to fully bring northeastern European residuals in line with those of other Europeans). We therefore tried using Nganasan as a new reference population. The Nganasan are a north Siberian Samoyedic-speaking population with little evidence of West Eurasian admixture in ADMIXTURE K=3 (SI6) analysis and with a high frequency[13] (92.1%) of Y-haplogroup N, which may be associated with the arrival of a separate stream of Siberian influence into parts of Europe. Fig. S9.26 shows that an $N$=4 model with the addition of the Nganasan to either the EN+WHG+EHG or the EN+WHG+Yamnaya model visibly improves the residuals for the northeastern European outlier populations, but has little effect on most others.

**Figure S9.24: EN+WHG admixture improves residuals across most European populations.** (a) *resnorm* for $N$=1 EN model. (b) *resnorm* for $N$=2 EN+WHG model. (c) change in *resnorm* between $N$=1 and $N$=2.

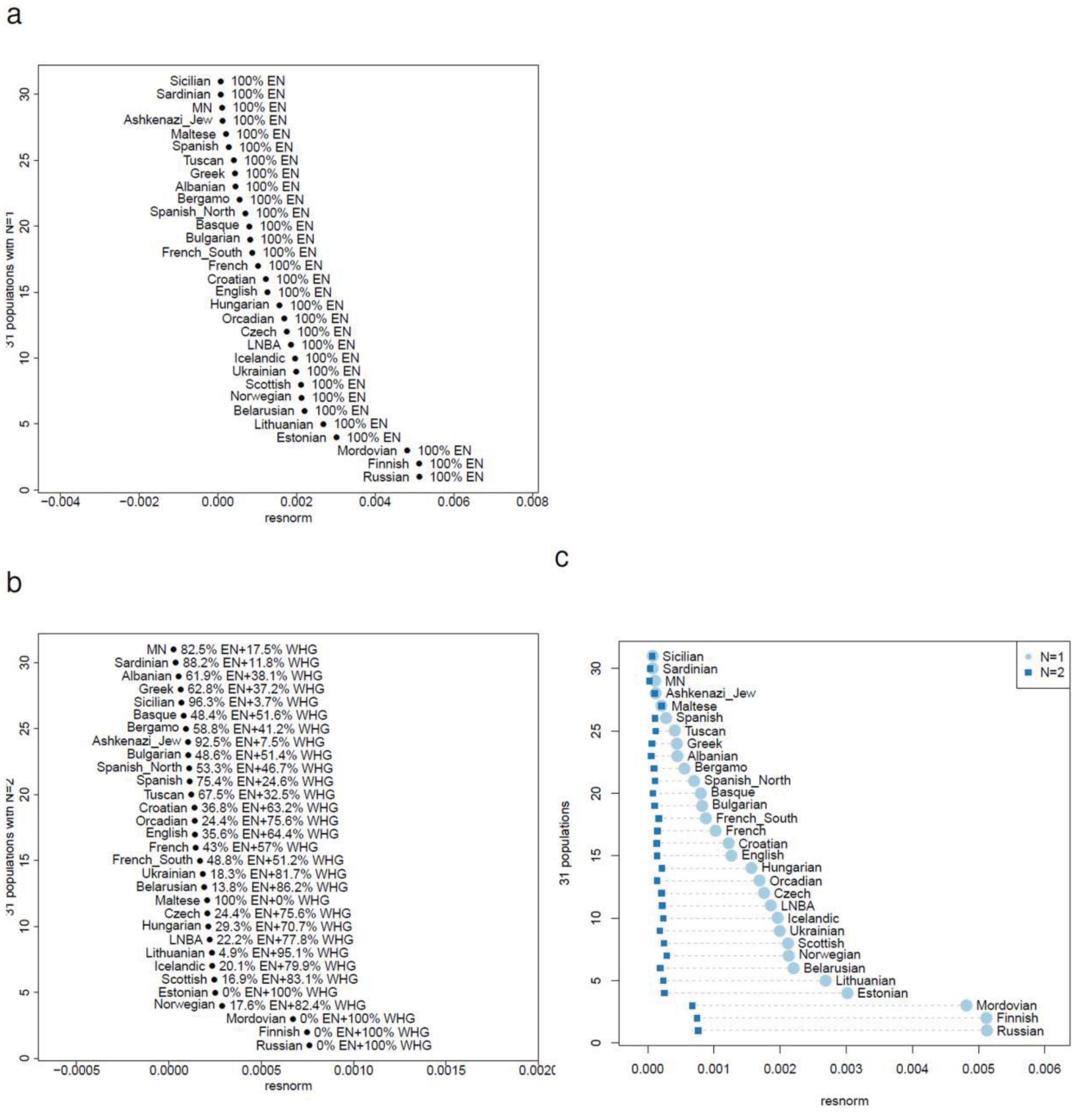

**Figure S9.25: EHG or Yamnaya admixture improves residuals across most European populations.** (a) *resnorm* for *N*=3 EN+WHG+EHG model. (b) change in *resnorm* between *N*=2 and *N*=3 EN+WHG+EHG model. (c) *resnorm* for *N*=3 EN+WHG+Yamnaya model. (d) change in *resnorm* between *N*=2 and *N*=3 EN+WHG+Yamnaya model.

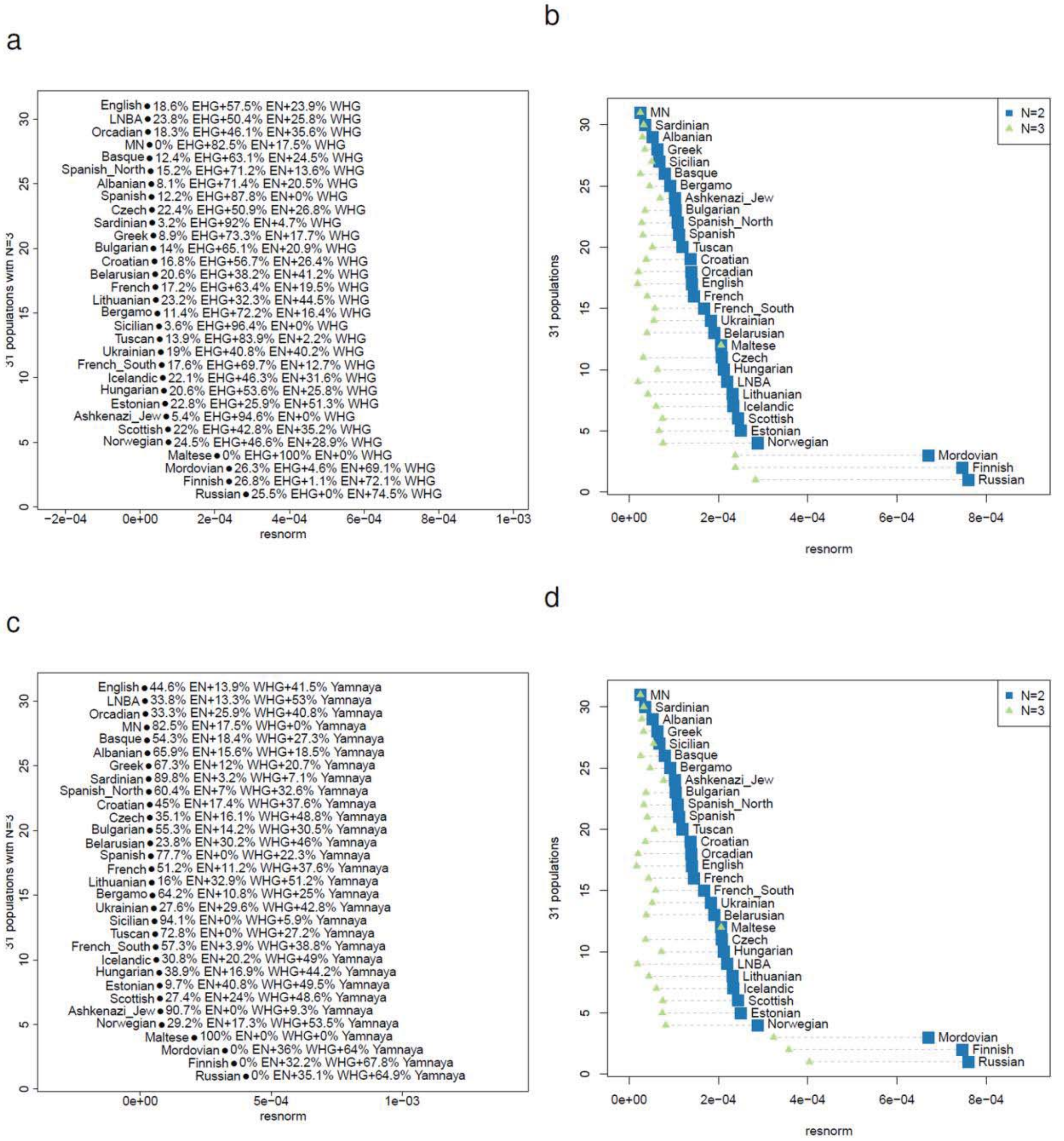

**Figure S9.26: Adding Nganasan as a 4th reference population improves residuals for northeastern European populations.** The effect is strongest for Finns, Russians, and Mordovians, followed by Hungarians and Estonians. A slighter effect is seen in a few other populations, especially from eastern Europe; low levels of recent East Eurasian admixture in eastern Europe have been proposed using haplotype sharing[10].

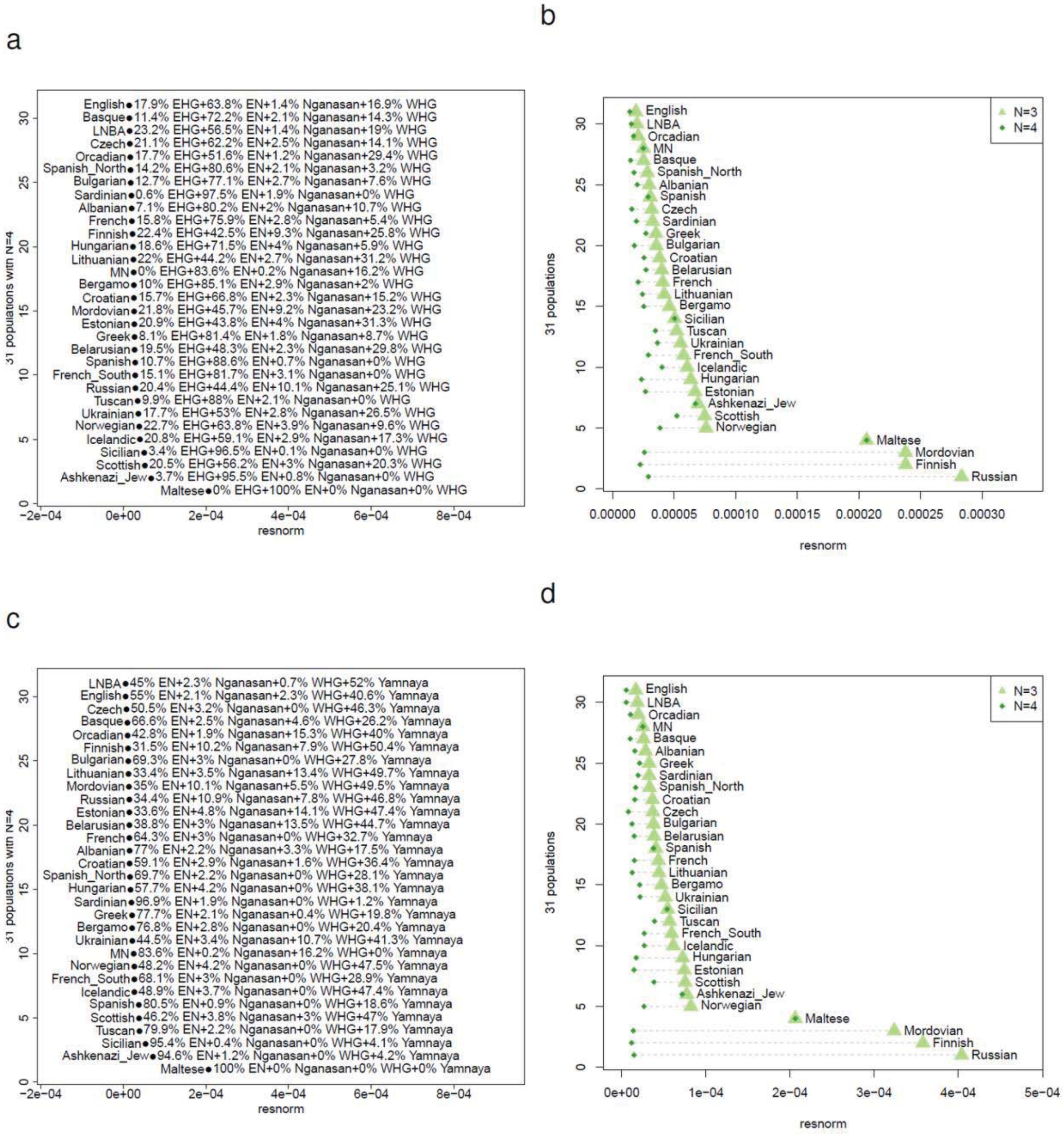

The three populations with the highest *resnorm* at *N*=4 in Fig. S9.26 are Maltese, Sicilians, and Ashkenazi Jews. We thus added the BedouinB as a 5$^{th}$ reference population, as there is evidence that these populations may have an excess of Near Eastern ancestry not mediated via the early European farmers. In Fig. S9.27 we show that this addition visibly improves the residuals for the three outlier European populations (and to a lesser extent the Spanish), consistent with this evidence. As the number of reference populations increases it may become increasingly difficult to accurately estimate ancestry proportions, especially when using modern reference populations (who may have their own history of admixture) and a particular set of outgroups (that may not be sufficient to distinguish between closely related reference populations).

In choosing to add Nganasan and BedouinB as 4$^{th}$ and 5$^{th}$ reference populations we used prior knowledge of the additional mixtures that may have taken place in Europe. It is also possible to identify the sources of such mixtures without a priori knowledge. To show that our choice of a Siberian and Near Eastern population could have been discovered in an unsupervised manner, we added all modern and ancient populations to the EN+WHG+EHG model and identified (for each outlier population) the additional references that minimize its *resnorm*. These are shown in Table S9.3 and indeed the two groups of outliers fit better when either Near Eastern or contemporary north Asian populations are added as a 4$^{th}$ ancestral population.

**Summary**

In this section we devised methods for detecting admixture and estimating its proportions without detailed phylogenetic modeling. Our assumptions were that a *Test* population is formed as a mixture of an arbitrary number of other *reference* populations that are differentially related to a set of *outgroup* populations. This approach allowed us to study admixture in ancient Europe without using a tree or admixture graph model. We were able to show that Middle Neolithic Europeans had experienced <30% resurgence of western European hunter-gatherer ancestry while the Corded Ware (the earliest of the Late Neolithic populations) had experienced >1/3 gene flow from a population related to eastern European hunter-gatherers. Furthermore, we showed that the Yamnaya population from the early Bronze Age in the Russian steppe may have been related to the population that admixed into Late Neolithic central Europeans, as it was not of purely eastern European hunter-gatherer descent but had additional ancestry of a farmer-related population that split off before the differentiation of European farmers, perhaps from the Caucasus or Near East, although more work is needed to identify the source of this population. Taking the Yamnaya as a source of the eastern admixing population, the level of population replacement between the Middle Neolithic and the Corded Ware becomes ~3/4, suggesting a massive influence of eastern populations in the formation of the Late Neolithic in central Europe. This influence is evident in other Late Neolithic populations (at a lower level), as well as in present-day Europeans, who differ from the Late Neolithic central Europeans by having additional farmer or west European hunter-gatherer ancestry. The robustness of our inference is supported by the fact that by adding a different set of outgroups from ancient Europe, the ~3/4 estimate was little affected. Further research of populations from central Europe, Eastern Europe, and the Caucasus may reveal the formation of steppe groups that succeeded the eastern European hunter-gatherers and how they affected (and were affected by) the populations surrounding them. The fact that sizeable admixture accompanied major transitions in European prehistory in at least three locations (Spain, Germany, and European Russia) suggests that the formation of present-day European populations was effected by sizeable local population replacement. Archaeological models of the past must accommodate the possibility of migration and admixture at a massive scale long after the Neolithic transition in Europe.

**Figure S9.27: Adding BedouinB as a 5th reference population improves residuals for Maltese, Sicilians, Ashkenazi Jews, and Spanish.**

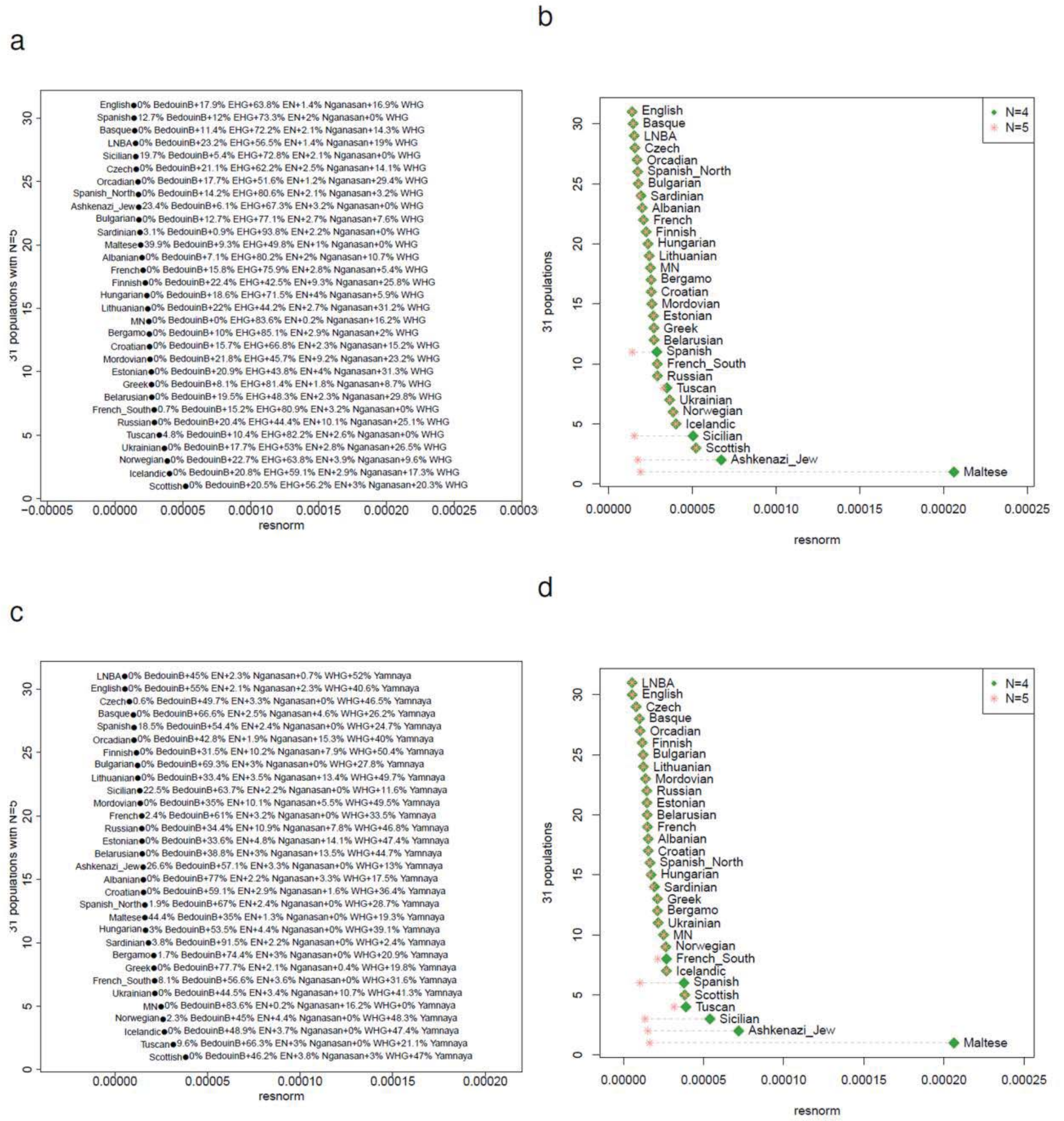

**Table S9.3: Populations that improve *resnorm* for European outlier populations when added to a model of EN/WHG/EHG admixture as a 4th ancestral population.** We show the top 20 populations that reduce *resnorm* the most when added to the mixture model. For Maltese, Sicilians, and Ashkenazi Jews these populations tend to be from the Middle East and North Africa. For Finns, Russians, and Mordovians they tend to be populations from Central Asia, Siberia, and Turkic/Uralic populations from Western Eurasia.

| **Maltese** | | **Sicilian** | | **Ashkenazi_Jew** | | **Finnish** | | **Russian** | | **Mordovian** | |
|---|---|---|---|---|---|---|---|---|---|---|---|
| $4^{th}$ anc. pop. | *resnorm* | $4^{th}$ anc. pop. | *resnorm* | $4^{th}$ anc. pop. | *resnorm* | $4^{th}$ anc. pop. | *resnorm* | $4^{th}$ anc. pop. | *resnorm* | $4^{th}$ anc. pop. | *resnorm* |
| Moroccan_Jew | 0.000006 | Turkish_Jew | 0.000006 | Cypriot | 0.000006 | Chuvash | 0.000013 | Chuvash | 0.000016 | Turkmen | 0.000011 |
| Lebanese | 0.000010 | Cypriot | 0.000006 | Iraqi_Jew | 0.000018 | Mansi | 0.000020 | Turkmen | 0.000020 | Uzbek | 0.000012 |
| Syrian | 0.000010 | Moroccan_Jew | 0.000009 | Turkish_Jew | 0.000018 | Uzbek | 0.000020 | Uzbek | 0.000021 | Nogai | 0.000012 |
| Tunisian_Jew | 0.000011 | Druze | 0.000012 | Moroccan_Jew | 0.000020 | Even | 0.000021 | Nogai | 0.000022 | Chuvash | 0.000014 |
| Saudi | 0.000014 | Iraqi_Jew | 0.000012 | Druze | 0.000025 | Selkup | 0.000021 | Altaian | 0.000025 | Altaian | 0.000017 |
| Turkish_Jew | 0.000016 | Syrian | 0.000014 | Lebanese | 0.000027 | Dolgan | 0.000021 | Mansi | 0.000025 | Tubalar | 0.000017 |
| Libyan_Jew | 0.000017 | Lebanese | 0.000016 | Syrian | 0.000028 | Turkmen | 0.000022 | Tubalar | 0.000025 | Hazara | 0.000017 |
| Jordanian | 0.000017 | Tunisian_Jew | 0.000019 | Tunisian_Jew | 0.000037 | Nogai | 0.000022 | Selkup | 0.000026 | Kyrgyz | 0.000019 |
| Palestinian | 0.000019 | Saudi | 0.000021 | Saudi | 0.000039 | Nganasan | 0.000022 | Even | 0.000026 | Tuvinian | 0.000019 |
| Druze | 0.000022 | Jordanian | 0.000022 | Iranian_Jew | 0.000039 | Yakut | 0.000022 | Yakut | 0.000026 | Uygur | 0.000019 |
| Yemenite_Jew | 0.000022 | Libyan_Jew | 0.000025 | Jordanian | 0.000039 | Altaian | 0.000023 | Dolgan | 0.000026 | Yakut | 0.000019 |
| BedouinB | 0.000022 | Palestinian | 0.000026 | Palestinian | 0.000044 | Tubalar | 0.000023 | Tuvinian | 0.000027 | Oroqen | 0.000020 |
| BedouinA | 0.000023 | BedouinB | 0.000027 | Libyan_Jew | 0.000044 | Tuvinian | 0.000024 | Kyrgyz | 0.000029 | Kalmyk | 0.000020 |
| Tunisian | 0.000024 | Yemenite_Jew | 0.000028 | Yemenite_Jew | 0.000045 | Yukagir | 0.000025 | Hazara | 0.000029 | Even | 0.000020 |
| Mozabite | 0.000024 | Tunisian | 0.000028 | BedouinB | 0.000045 | Oroqen | 0.000026 | Nganasan | 0.000029 | Dolgan | 0.000021 |
| Algerian | 0.000025 | Mozabite | 0.000028 | BedouinA | 0.000046 | Kyrgyz | 0.000026 | Oroqen | 0.000029 | Selkup | 0.000021 |
| Egyptian | 0.000025 | BedouinA | 0.000028 | Tunisian | 0.000046 | Kalmyk | 0.000027 | Yukagir | 0.000030 | Hezhen | 0.000021 |
| Saharawi | 0.000026 | Egyptian | 0.000028 | Egyptian | 0.000046 | Hazara | 0.000029 | Kalmyk | 0.000030 | Daur | 0.000021 |
| Yemen | 0.000028 | Algerian | 0.000029 | Mozabite | 0.000047 | Hezhen | 0.000030 | Uygur | 0.000032 | Mansi | 0.000022 |
| Esan | 0.000030 | Saharawi | 0.000029 | Algerian | 0.000047 | Aleut | 0.000031 | Hezhen | 0.000032 | Xibo | 0.000023 |

# Supplementary Information 10

## Number of streams of migration into ancient Europe

Nick Patterson*, Iosif Lazaridis*, David Reich

* To whom correspondence should be addressed
(nickp@broadinstitute.org or lazaridis@genetics.med.harvard.edu)

### Number of streams of migration

Take a set of *left* populations $U$ and a set of *right* populations $V$ and consider the matrix

$$X(u, v) = F_4(u_0, u; v_0, v)$$

where $u_0$, $v_0$ are some fixed populations of $U$ and $V$, and $u$, $v$ range over all choices of populations of $U$, $V$. We can assume that u $u \neq u_0$, and $v \neq v_0$, so that the matrix $X$ is $(a - 1) \times (b - 1)$, where $a$, $b$ is the cardinality of $U$, $V$ respectively. We have shown[1] that if $X$ had rank $r$ and there had been $n$ waves of migration from $V$ to $U$ with no back-migration from $U$ to $V$ then:

$$r+1 \leq n$$

We describe our computational strategy in a little more detail. We compute $\hat{X}$ an estimate of $X$ so that in the notation of ref. 2:

$$\hat{X}(u, v) = f_4(u_0, u; v_0, v)$$

Thus:

$$E(X) = \hat{X}$$

where expectations are calculated from the true (and unknown) phylogeny. We can use the block jackknife[3] to compute $Q$ an estimate of the error covariance of $X$. To test if $\hat{X}$ has rank $r$ we write:

$$\hat{X} = A.B + E$$

where A is $(a\text{-}1)\times r$ and $B$ is $r\times(b\text{-}1)$ and $E$ is a matrix of residuals. The (log) likelihood for $(A, B)$ and implicitly $r$ is:

$$\mathcal{L}(A, B) = -\frac{1}{2}\sum_{i,j,k,l} Q^{-1}_{i,j;k,l}\, E_{i,j}E_{k,l}$$

where the residual matrix $E$ is defined by:

$$E = \hat{X} - A.B$$

For each rank $r$ we set $A$, $B$ initially by an SVD analysis of $X$, and then iterate, minimizing $\mathcal{L}$ with respect to $A$, $B$ in turn. For fixed $A$, $\mathcal{L}(A, B)$ is quadratic in $B$ and can be minimized by solving linear equations, and similarly if we fix $B$. Since $A$, $B$ only enter into the likelihood through a matrix product:

$$A.B = (A.Z).(Z^{-1}.B)$$

for any non-singular $r \times r$ matrix $Z$. Thus, the number of degrees of freedom is:

$$d(r) = ((a-1) + (b-1))r - r^2 = r(a+b-(r+2))$$

As a check, if $r$ is the maximal rank $Min(a\text{-}1, b\text{-}1)$, then $d(r)=(a\text{-}1)(b\text{-}1)$ which is obviously correct. This is the *saturated* model, where we fit the data perfectly. We compute statistics with a likelihood ratio test (LRT), obtaining statistics that are asymptotically $\chi^2$ distributed. The $\chi^2$ statistics here, using the LRT, are computed using a fixed covariance $Q$. It would be more correct to re-estimate $Q$ simultaneously with $A$, $B$. This would greatly increase complexity, without adding much precision.

**Finding mixture coefficients**

Let $T$ be a target population, and $S = \{s_1, s_2, \ldots, s_n\}$ a set of source populations. In the easiest case to consider, in which $T$ is an admixture of populations of $S$, we can write symbolically:

$$T = \sum_{i=1}^{n} w_i s_i$$

where $w_i$ is the admixture coefficient (ancestry proportion) from the $i^{th}$ source population. It then follows that for any populations $o_1$, $o_2$:

$$\sum_i w_i F_4(T, s_i, o_1, o_2) = F_4(T, T; o_1, o_2) = 0$$

A little thought shows that this is true even if the populations $s_i$ are descendants of the true source populations, provided that there has been no gene flow between the most recent ancestor of $T$, $S$ on the one hand and the most recent ancestor of $o_1$, $o_2$ on the other. (We previously used $f_3$ statistics to derive mixing coefficients[4]. The methods there require samples of the actual source and admixed populations, but do not require outgroup populations as we do here with the $o_i$.)

Thus, if $T$ is admixed, as above, pick a set of outgroup populations $O$, and

1. Check, setting left populations L = $S$, and right populations $O$ that the matrix $X$ has full rank $n$-1.
2. Check, again setting L = $\{T, S\}$ that there is no strong evidence that the rank of $X$ increases with the addition of $T$.

We now will take $T$ as the base population of L = $\{T, S\}$, which simplifies the algebra. We calculate matrices $A$, $B$ as above, with the rank set to $n$-1 (corank 1). Thus, the recovered $A$ is of dimension $n \times (n\text{-}1)$. We have $n$ source populations so L has $n$+1 populations after including $T$. It then follows that estimates $w = (w_1, w_2, \ldots, w_n)$ of the admixture weights can be found by solving the equations:

$$\boldsymbol{w}.\boldsymbol{A} = \boldsymbol{0}$$
$$\sum_{i=1}^{n} w_i = 1$$

We can use the block jackknife to compute a covariance matrix for the errors. Formally, we should re-estimate $Q$, the estimated covariance of $X$ as we delete blocks in the jackknife. This is not done at present, as it would add complexity and seems unlikely to make a material difference.

**Application to ancient Europe**

We use this methodology to study some of the proposed admixture events discussed in the paper. We begin by using the same 15 World Outgroups as in SI9 (this is the right population list *V*).

> **World Foci 15 outgroups:** Ami, Biaka, Bougainville, Chukchi, Eskimo, Han, Ju_hoan_North, Karitiana, Kharia, Mbuti, Onge, Papuan, She, Ulchi, Yoruba

In the following discussion we will follow the idea described above of first testing a pair of source populations (left list *U*) to ensure that they have rank=1 with respect to the outgroups, which shows that they are descended from two independent streams of ancestry. We will then add a third population *Test*, making *U*={*U*, *Test*}, showing that it does not increase the rank, and can thus be modelled as a mixture of the initial pair.

**Western and Eastern European hunter-gatherers not a clade with respect to world outgroups**

We first test the pair (Loschbour, Karelia_HG). Model with rank=0 is rejected (p=1.6e-43), thus the WHG and EHG are descended from at least two ancestral populations.

**Scandinavian hunter-gatherers not a third ancestral population independent of WHG and EHG**

We next add Scandinavian hunter-gatherers to (Loschbour, Karelia_HG) (Table S10.1). Rank=0 is excluded, but rank=1 is not. Two ancestral populations suffice for Scandinavian hunter-gatherers.

**Table S10.1: Testing (*SHG*, Loschbour, Karelia_HG)**

| ***SHG* population added** | **p for rank=0** | **p for rank=1** |
|---|---:|---:|
| Motala_HG | 8.93E-44 | 7.09E-01 |
| SwedenSkoglund_MHG | 4.26E-13 | 2.15E-01 |
| SwedenSkoglund_NHG | 7.49E-44 | 4.93E-01 |

**Early Neolithic Europeans and Western European hunter-gatherers (EN and WHG) are not a clade with respect to world outgroups**

We first show that the pair of an early Neolithic population *EN* and Loschbour are descended from two streams of migration (Table S10.2). We add all Early Neolithic populations in turn, and in each case rank=0 is rejected. Thus, EN and WHG are descended from at least two ancestral populations.

**Table S10.2: Testing (*EN*, Loschbour)**

| ***EN* population added** | **p for rank=0** |
|---|---:|
| Starcevo_EN | 9.49E-09 |
| LBKT_EN | 4.16E-04 |
| LBK_EN | 7.17E-15 |
| Spain_EN | 1.84E-11 |
| HungaryGamba_EN | 6.24E-13 |
| Stuttgart | 9.73E-10 |

**Middle Neolithic Europeans are not a 3$^{rd}$ ancestral population independent of WHG and EN**

We next add a Middle Neolithic European population *MN* to (Loschbour, LBK_EN) (Table S10.3). With the exception of the Iceman, most Middle Neolithic Europeans appear to be well-modeled as a mixture of Loschbour and LBK_EN.

**Table S10.3: Testing (*MN*, LBK_EN, Loschbour)**

| ***MN* population added** | **p for rank=0** | **p for rank=1** |
|---|---|---|
| Baalberge_MN | 1.92E-11 | 7.59E-01 |
| Esperstedt_MN | 4.57E-15 | 4.28E-01 |
| Spain_MN | 2.18E-14 | 1.52E-01 |
| HungaryGamba_CA | 4.49E-12 | 2.44E-01 |
| SwedenSkoglund_MN | 4.85E-13 | 1.46E-01 |
| Iceman | 1.34E-15 | 1.87E-04 |

**Yamnaya and Eastern European hunter-gatherers not a clade with respect to world outgroups**

We test the pair (Yamnaya, Karelia_HG). The model with rank=0 is rejected (p= 2.5e-28). Thus the Yamnaya and EHG are descended from at least two ancestral populations. This is related to the process of dilution of EHG ancestry discussed in SI7, SI9.

**Yamnaya can be modeled as a mixture of Armenians and Karelia_HG**

To understand the source of the dilution of EHG ancestry in the Yamnaya, we test the triple (Yamnaya, Karelia_HG, Armenian). We do not strongly reject rank=1 for (Yamnaya, Karelia_HG, Armenian) (p=0.0365). The p-value is low, which may be driven by complexity in the ancestry of present-day Armenians. We may not currently have a good surrogate for the population that diluted the EHG to form the Yamnaya, although the analysis of *f*-statistics (SI7) suggests a source related to Near Eastern populations, and the Yamnaya show evidence of such admixture (SI7, SI9).

**Table S10.4: Testing (*MN*, *Eastern*)**

| ***Eastern*** | ***MN*** | **p for rank=0** |
|---|---|---|
| Yamnaya | Baalberge_MN | 4.95E-29 |
| Karelia_HG | Baalberge_MN | 4.57E-68 |
| Yamnaya | Esperstedt_MN | 5.31E-21 |
| Karelia_HG | Esperstedt_MN | 7.82E-56 |
| Yamnaya | Spain_MN | 4.58E-61 |
| Karelia_HG | Spain_MN | 3.68E-103 |
| Yamnaya | HungaryGamba_CA | 4.55E-18 |
| Karelia_HG | HungaryGamba_CA | 4.66E-52 |
| Yamnaya | SwedenSkoglund_MN | 1.60E-35 |
| Karelia_HG | SwedenSkoglund_MN | 4.26E-76 |
| Yamnaya | Iceman | 3.29E-36 |
| Karelia_HG | Iceman | 1.91E-78 |

**Middle Neolithic Europeans and Yamnaya or Karelia_HG not a clade versus world outgroups**

We next test whether a Middle Neolithic European population *MN* is a clade with the populations of eastern Europe (Yamnaya or Karelia_HG). Rank=0 is universally rejected (Table S10.4).

**The Corded Ware are modeled as a mixture of Middle Neolithic and Eastern Europeans**

We have determined that eastern European populations (Karelia_HG or Yamnaya) and Middle Neolithic populations from the rest of Europe are descended from two ancestral populations. Next, we add the Corded Ware. In Table S10.5 we list the p-values for models that involve Esperstedt_MN, Baalberge_MN, i.e., the populations that precede the Corded Ware in Germany. The p-value for rank=1 is ≥.179, thus rank=1 is not excluded, and the Corded Ware can be modeled as a mixture of an eastern population (Yamnaya or Karelia_HG) and a Middle Neolithic population from Germany (Baalberge_MN or Esperstedt_MN).

**Table S10.5: The Corded Ware can be modeled as mixtures of Eastern European and Middle Neolithic central Europeans**

| ***Eastern population*** | ***MN population*** | **p for rank=0** | **p for rank=1** |
|---|---|---|---|
| Yamnaya | Baalberge_MN | 6.24E-29 | 2.38E-01 |
| Karelia_HG | Baalberge_MN | 5.59E-68 | 1.79E-01 |
| Yamnaya | Esperstedt_MN | 2.38E-22 | 4.21E-01 |
| Karelia_HG | Esperstedt_MN | 8.34E-59 | 7.07E-01 |

**Modeling the Corded Ware as a mixture of Middle Neolithic Europeans and Yamnaya is robust to choice of outgroups, while modeling it as a mixture of Middle Neolithic Europeans and Eastern European hunter-gatherers is not**

Table S10.5 shows that the Corded Ware can be modeled as mixtures of Middle Neolithic Europeans and either Yamnaya or Karelia_HG. To test the robustness of this inference, we added Eurasian hunter-gatherers and Early Neolithic farmers as additional outgroups. The full set of outgroups is:

> **8 Ancient Eurasian Outgroups:** HungaryGamba_EN, Kostenki14, LBK_EN, Loschbour, MA1, Motala_HG, Spain_EN, Ust_Ishim

This set includes all Upper Paleolithic Eurasians, representatives of early European farmers from Germany, Spain, and Hungary, as well as Western and Scandinavian hunter-gatherers. We add each of these 8 populations to the set of 15 World Outgroups, thus resulting in 8 different sets of 16 outgroups. In Table S10.6 we show the p-value for rank=1 for these different sets.

Rank=1 for models involving Karelia_HG and a Middle Neolithic population can be excluded with $p<7.64E-05$ for at least one added outgroup. Rank=1 involving Yamnaya and Baalberge_MN can be very weakly excluded ($p=.0234$, when adding Motala_HG as an outgroup). However, rank=1 for the pairing (Yamnaya, Esperstedt_MN) cannot be excluded ($p\geq0.272$ for all outgroups added), suggesting that Corded Ware can be modeled as a simple 2-way mixture of these two groups.

To further test the choice of (Yamnaya, Esperstedt_MN), we modeled all Late Neolithic / Bronze Age populations as such mixtures and show the p-value for rank=1 in Table S10.7. All such p-values are $\geq0.017$ for the Late Neolithic / Bronze Age populations of Germany, suggesting that Late Neolithic/Bronze Age populations in general can be reasonably well-modelled as such mixtures.

**Table S10.6: p-values for rank=1 of *U* {Corded_Ware_LN, *Eastern* population, *MN* population} using outgroups WorldFoci15 plus an ancient Eurasian indicated in the column names.**

| *Eastern* population | *MN* population | Ust_Ishim | Kostenki14 | MA1 | LBK_EN | Spain_EN | Hungary Gamba_EN | Loschbour | Motala_HG |
|---|---|---|---|---|---|---|---|---|---|
| Karelia_HG | Baalberge_MN | 1.98E-01 | 3.91E-01 | 5.51E-01 | 3.30E-03 | 6.32E-03 | 7.64E-05 | 8.53E-05 | 1.80E-02 |
| Karelia_HG | Esperstedt_MN | 7.15E-01 | 7.94E-01 | 2.88E-01 | 3.70E-05 | 4.04E-04 | 4.96E-03 | 3.24E-02 | 5.07E-05 |
| Yamnaya | Baalberge_MN | 2.84E-01 | 1.93E-01 | 5.68E-01 | 2.18E-01 | 2.35E-01 | 2.99E-01 | 2.57E-01 | 2.34E-02 |
| Yamnaya | Esperstedt_MN | 4.86E-01 | 6.44E-01 | 2.72E-01 | 4.83E-01 | 5.13E-01 | 4.79E-01 | 4.25E-01 | 3.99E-01 |

**Table S10.6: p-values for rank=1 of *U* {*Late Neolithic / Bronze Age*, Yamnaya, Esperstedt_MN} using outgroups WorldFoci15 plus an ancient Eurasian indicated in the column names.**

| *Late Neolithic / Bronze Age population* | Ust_Ishim | Kostenki14 | MA1 | LBK_EN | Spain_EN | Hungary Gamba_EN | Loschbour | Motala_HG |
|---|---|---|---|---|---|---|---|---|
| Karsdorf_LN | 8.98E-01 | 8.40E-01 | 7.98E-01 | 9.02E-01 | 8.70E-01 | 9.00E-01 | 9.12E-01 | 8.67E-01 |
| Corded_Ware_LN | 4.86E-01 | 6.44E-01 | 2.72E-01 | 4.83E-01 | 5.13E-01 | 4.79E-01 | 4.25E-01 | 3.99E-01 |
| Bell_Beaker_LN | 2.70E-01 | 1.28E-01 | 4.98E-01 | 1.67E-02 | 4.35E-02 | 1.54E-01 | 1.93E-01 | 2.10E-01 |
| Alberstedt_LN | 2.34E-01 | 2.71E-01 | 3.87E-01 | 1.65E-01 | 2.65E-01 | 2.02E-01 | 1.97E-01 | 5.74E-02 |
| BenzigerodeHeimburg_LN | 8.07E-02 | 8.19E-02 | 1.56E-01 | 3.32E-02 | 4.90E-02 | 5.65E-02 | 5.46E-02 | 7.22E-02 |
| Unetice_EBA | 3.57E-01 | 1.95E-01 | 6.88E-01 | 2.48E-01 | 2.25E-01 | 3.56E-01 | 1.49E-01 | 1.74E-01 |
| Halberstadt_LBA | 1.34E-01 | 4.84E-02 | 2.69E-01 | 1.45E-01 | 1.42E-01 | 1.40E-01 | 3.27E-02 | 1.44E-01 |
| HungaryGamba_BA | 8.66E-01 | 7.79E-01 | 9.49E-01 | 8.43E-03 | 3.66E-03 | 6.02E-01 | 8.13E-01 | 2.32E-01 |

We estimate mixture proportions for all Late Neolithic / Bronze Age of Germany populations in Table S10.8, showing estimates when using 15 world outgroups, 8 ancient Eurasians as outgroups, and 23 outgroups (combining both of the above).

**Table S10.8: Yamnaya mixture proportions in Late Neolithic / Bronze Age Germans**

| *LN/BA population* | WorldFoci15 Outgroups | | | 8 Ancient Outgroups | | | All 23 Outgroups | | |
|---|---|---|---|---|---|---|---|---|---|
| Halberstadt_LBA | 0.556 | +/- | 0.108 | 0.488 | +/- | 0.041 | 0.520 | +/- | 0.032 |
| Alberstedt_LN | 0.422 | +/- | 0.098 | 0.568 | +/- | 0.040 | 0.531 | +/- | 0.034 |
| Bell_Beaker_LN | 0.387 | +/- | 0.071 | 0.534 | +/- | 0.030 | 0.555 | +/- | 0.023 |
| BenzigerodeHeimburg_LN | 0.472 | +/- | 0.097 | 0.650 | +/- | 0.032 | 0.618 | +/- | 0.026 |
| Unetice_EBA | 0.543 | +/- | 0.062 | 0.632 | +/- | 0.024 | 0.625 | +/- | 0.019 |
| Corded_Ware_LN | 0.764 | +/- | 0.066 | 0.743 | +/- | 0.027 | 0.731 | +/- | 0.022 |
| Karsdorf_LN | 0.823 | +/- | 0.128 | 0.821 | +/- | 0.081 | 0.745 | +/- | 0.064 |

The standard errors are highest with the WorldFoci15 set of outgroups, as one might expect given that these populations share little genetic drift with those of Europe. They are smaller when using Ancient Eurasians as outgroups, and smaller still when combining both sets. The standard errors are also larger for populations that consist of single individuals (such as Halberstadt_LBA or Karsdorf_N), as one might expect due to the limited data.

Overall, there is reasonable agreement between the inferences using the two disjoint sets of outgroups, which increases our confidence that our inference that Late Neolithic / Bronze Age populations were formed by this type of mixture between local Middle Neolithic and migrant Yamnaya-related ancestors. All estimates agree that a sizeable portion of the ancestry of LN/BA populations is related to the Yamnaya, and quantify this portion to ~3/4 for the Corded Ware, in agreement with the inference of SI9 with a related method and the results of PCA and ADMIXTURE analysis (Fig. 2) and $F_{ST}$ analysis (Extended Data Table 3) that also reveal the close relationship between the Corded Ware of Late Neolithic Germany with the Yamnaya population of the Samara district sampled ~2,600km to the east.

# Supplementary Information 11

## Relevance of ancient DNA to the problem of Indo-European language dispersals

Iosif Lazaridis, Wolfgang Haak, Nick Patterson, David Anthony and David Reich*

* To whom correspondence should be addressed (reich@genetics.med.harvard.edu)

**Main hypotheses of Indo-European language dispersals**

Indo-European languages are widely spoken from the Atlantic Ocean to the Indian subcontinent[1], and, through more recent migrations, have expanded into much of the rest of the world. The set of living Indo-European languages includes, among others, English, Greek, Russian, Italian, Farsi, and Hindi, and is complemented by a number of extinct languages like Hittite and later Phrygian in Anatolia, Tocharian in the Xinjiang province of China[2,3], and the languages of the Scythians of the Eurasian steppe. With the advent of the written word and modern means of travel and communication, it has become possible to acquire a new language by cultural transmission alone. However, in the past, languages must have spread through direct contact[4] and, at least to some extent, with migrants, and thus the study of ancient DNA, which can trace the migration of people, is relevant to evaluating models of language dispersal.

Ancient DNA can roll back the clock and reveal earlier stages in the formation of human populations, just as comparative linguistics can reveal earlier stages in the formation of human languages. Most experts agree that Proto-Indo-European, a reconstructed ancestral language to living and historically attested Indo-European languages, must have once been spoken in a geographically circumscribed and relatively small part of Eurasia[1,5-9]. Many theories have been proposed since the discovery of the Indo-European language family in the eighteenth century as to where that homeland was and how the language was transmitted. Such theories invariably postulate at least some migrations of people. These hypotheses can be tested with the methods of genetics applied to ancient individuals.

The two leading theories of Indo-European language dispersal are:

1. The "Steppe hypothesis"[8,10-14], which derives Indo-European languages from the people buried under kurgans on the Pontic-Caspian steppe in modern Ukraine and southern Russia.

2. The "Anatolian hypothesis"[9,15], which suggests that early farmers who migrated from Anatolia to Europe during the Early Neolithic brought Indo-European languages with them.

Other hypotheses have also been formulated and we will discuss the following two where relevant:

3. The "Balkan hypothesis"[6] that resembles the Anatolian hypothesis in linking the migration of Indo-Europeans to farmers, but with the difference that Indo-European languages spread from southeastern Europe at a later period.

4. The "Armenian plateau hypothesis"[7,16] which resembles the Steppe hypothesis in postulating a role for the steppe in the dispersal of languages into Europe, but places the homeland of Proto-Indo-European speakers south of the Caucasus.

A key argument in favor of the Anatolian hypothesis has been that the migration of early farmers from Anatolia ~8,000-7,000 years ago was a major transformation that affected most of Europe. A massive population turnover is the most widely accepted and easiest explanation for the introduction of a new language across a large geographic area. There are of course other ways by which languages can spread, such as elite dominance[7], whereby a politically dominant minority transmits its language to the resident majority (e.g. Hungarian). Another argument for the Anatolian hypothesis is the inferred topology of the Indo-European family tree, in which Anatolian languages like Hittite occupy a basal position, which could plausibly be explained by an Anatolian homeland and migrations from Anatolia to Europe through the Balkans[17]. Finally, the recent application of phylogenetic methods adopted from biology to the estimation of language divergence times[18,19] has been used to argue in favor of an earlier breakup of Proto-Indo-European into daughter languages.

The Steppe hypothesis dates the breakup of Proto-Indo-European to several millennia later: to 4,000-3,000 BCE. One of the prominent arguments in favor of the Steppe hypothesis is that reconstructions of the ancestral Proto-Indo-European language have suggested a shared vocabulary associated with wheeled vehicles, with cognates in Indo-European languages ranging from Europe to India, although not all linguists agree that it is possible to make meaningful inferences about the material culture of an ancient people based on descendant languages[20]. If this reconstruction is right, Proto-Indo-European could not have been the language spread by the first farmers moving into Europe, as wheels and wheeled vehicles were only invented thousands of years after the spread of farming. While the earliest finds of wheels are not from the steppe[21], the intensification of the use of wheeled vehicles and domestic horses is linked to the spread of steppe pastoralism. Another attractive feature of the Steppe hypothesis is that the steppe geography could account for the evidence of loan words between Proto-Indo-European and Uralic languages to the north, and Kartvelian languages to the south[5].

**Hypotheses of Indo-European origins in light of the new genetic data presented in this paper**

Genetic data is a valuable source of information that is useful for evaluating competing hypotheses of Indo-European language dispersals, to the extent that theories of such dispersals invoke migratory movements to explain their extensive distribution. Such hypotheses can be tested both with archaeology (which tests for the spread of material culture, which can spread not only by migration but also through trade or an exchange of ideas), and with genetics (which can test directly whether the movement of people accompanied perceived changes in the material record).

Past genetic data from ancient DNA has confirmed one of the major predictions of the Anatolian hypothesis – the migration of early farmers from the Near East (inclusive of Anatolia) to Europe – using both mitochondrial DNA[22-24] and whole genome analysis[25,26]. The results of our study are consistent with these findings, and also extend them by showing that not only the early farmers of central Europe (Germany and Hungary) and Scandinavia, but also of Iberia were descended from a common stock. In this sense, ancient DNA is consistent with migrations following the predictions of the Anatolian hypothesis, and indeed our ancient DNA results match the scenario outlined by Bellwood for the initial dispersal of farming into Europe remarkably well[27]. This is also true for the Balkan hypothesis, as geographically, southeastern Europe is a plausible place where early farmers could have diverged into an inland Danubian route toward central Europe, and a Mediterranean route toward Iberia. The evidence of a relatively homogeneous population of early European farmers with substantial Near Eastern ancestry[25] is indeed a reasonable candidate for the spread of a single language family across Europe.

Our new genetic data are important in showing that a second major migration from the steppe into Europe occurred at the end of the Neolithic period (between 5000-4500 years ago). Moreover, we have demonstrated that these migrants accounted for at least ~3/4 of the ancestry of the Corded Ware people of Germany, and much of the ancestry of other Late Neolithic / Bronze Age populations of Germany and present-day northern Europeans (Fig. 3, SI9, SI10). Thus, the main argument in favor of the Anatolian hypothesis (that major language change requires major migration) can now also be applied to the Steppe hypothesis. While we cannot go back in time to learn what languages the migrants spoke, it seems more likely than not that the Corded Ware people we sampled spoke the languages of the people who contributed the great majority of their ancestry (Yamnaya), rather than the local languages of the people who preceded them. Thus, our results increase the plausibility that the Corded Ware people and those genetically similar groups who followed them in central Europe spoke a steppe-derived Indo-European language. More generally, our results level the playing field between the two leading hypotheses of Indo-European origins, as we now know that both the Early Neolithic and the Late Neolithic were associated with major migrations.

While our results do not settle the debate about the location of the proto-Indo European homeland, they increase the plausibility of some hypotheses and decrease the plausibility of others as follows:

1. The Steppe hypothesis gains in plausibility by our discovery of a migration during the Late Neolithic from the steppe into central Europe. This migration was predicted by some proponents of the Steppe hypothesis and we have now shown (definitively) that it occurred. We also note that our results help to differentiate between variants of the steppe hypothesis: we do not find evidence of an influence of steppe migrants earlier than the Corded Ware, although we cannot rule out the possibility that such evidence might be found with larger sample sizes and more sampling locations in central Europe. However, we can definitely reject that the breakup of Indo-European occurred as late as 4000 years ago[28], as by ~4500 years ago the migration into Europe had already taken place. Moreover, this migration clearly resulted in a large population turnover, meaning that the Steppe hypothesis does not require elite dominance[9] to have transmitted Indo-European languages into Europe. Instead, our results show that the languages could have been introduced simply by strength of numbers: via major migration in which both sexes participated (SI2, SI4)

2. The Anatolian hypothesis becomes less plausible as an explanation for the origin of all Indo-European languages in Europe, as it can no longer claim to correspond to the only major population transformation in European prehistory, and it must also account for the language of the steppe migrants. However, the Anatolian hypothesis cannot be ruled out entirely by our data, as it is possible that it still accounts for some of the major branches of the Indo-European language family in Europe, especially the branches of the south where the proportion of steppe ancestry today is smaller than in central and northern Europe (Figure 3).

3. The Balkan hypothesis faces similar difficulties as the Anatolian. If the early farmers of southeastern Europe were genetically similar to their descendants in central and western Europe, a spread of Indo-European speaking migrants from the Balkans to the rest of Europe would simply introduce another layer of "Early Neolithic" genes similar to those present elsewhere in Europe, but would not account for the migration from the steppe and its associated language. Furthermore if the steppe immigrants spoke Indo-European languages, these languages are unlikely to have been acquired by migration from Europe, as our Yamnaya samples show no sign of a major component of ancestry derived from European Early or Middle Neolithic farmers (Fig. 2).

4. The Armenian plateau hypothesis gains in plausibility by the fact that we have discovered evidence of admixture in the ancestry of Yamnaya steppe pastoralists, including gene flow from a population of Near Eastern ancestry for which Armenians today appear to be a reasonable surrogate (SI4, SI7, SI9). However, the question of what languages were spoken by the "Eastern European hunter-gatherers" and the southern, Armenian-like, ancestral population remains open. Examining ancient DNA from the Caucasus and Near East may be able to provide further insight about the dynamics of the interaction between these regions and the steppe. Our results show that southern populations diluted the ancestry of populations from the steppe, but also that ancestry related to Ancient North Eurasians forms a major ancestral component of the populations of the present-day Caucasus[25]. Thus, both south-north and north-south genetic influence across the Caucasus is plausible.

**Pitfalls in using genetic data to make inferences about language spread**

The study of Indo-European origins and language dispersals has been controversial, in part because of the history of misuse of the concept of the Proto-Indo-European homeland for ideological reasons[29].

In the early 20$^{th}$ century, Gustav Kossinna proposed the idea of 'settlement archaeology': that a material culture identified by archaeology, specifically the Corded Ware, might correspond to a genetically well-defined people and homogeneous language group, specifically the Proto-Indo-Europeans. Our data directly contradict Kossinna's theories in showing that the Corded Ware are not a locally derived central European population but instead are to a significant degree descended from eastern migrants. V. Gordon Childe[12], following linguistic arguments by Otto Schrader[14], proposed a migration from the steppe into Europe, which seems, in view of the results of our study, to have been closer to the mark. However, following the Second World War, and especially in the 1960s and 1970s, archaeologists responded to the history of misuse of archaeology by rejecting sweeping migrations and the settlement archaeology framework altogether[30], and suggesting that in practice, it would not ever be possible to show that archaeological, linguistic, and genetic groupings overlap[31,32]. This climate in the archaeological community has made it challenging to propose migration as an explanation for similarities or differences in material culture across time and space. Although migration is today accepted more widely by archaeologists[33] than it was 30 years ago[11], it is usually discussed in connection with demographic-growth models linked to the expansion of agriculture[27], while migrations linked to the evolution of new socio-political structures among long-established food-producing populations are less understood, less recognizable and often viewed with skepticism.

Genetic data is important to this debate as it changes the equation, bringing to bear a new type of data that can directly speak to whether or not migration occurred. This type of fact could never be clearly established before the advent of ancient DNA, except by use of stable isotope analysis which only works to detect migration if the studied samples are from the first generation of migrants. While migration from the steppe had been proposed on the basis of human skeletal morphological data by Childe and others, a more recent study[34], which dealt with precisely the problem of relationships between the Corded Ware and steppe Kurgan (Pit Grave) groups, concluded that "The local [Corded Ware] groups of the core area (Central Germany, Czechoslovakia, Poland) form a very homogeneous block, issued from the local "Old Europe" substratum and persisting until Aunjetitz (Unetice) at least. This block show[s] no biological affinities to the Ukrainian Kurgan populations. There is no evidence for physical presence of Kurgan tribes in this area." Similarly, on the basis of archaeology it has been claimed by a key proponent of the steppe hypothesis[8] that "there is no real case for an expansion of Yamnaya invaders across the North European plain, producing the Corded Ware horizon", highlighting how controversial the idea of migration has become. With ancient DNA, one can establish whether or not migrations occurred, and tie migrations to well-defined archaeological cultures by sequencing DNA from radiocarbon dated skeletons buried with diagnostic grave goods. Thus, it is possible, using ancient DNA, to evaluate directly whether a particular material culture could have spread through migration or whether cultural transmission occurred. Our paper has now made an overwhelming case for just such a migration, showing the power of genetics to contribute meaningfully to debates about past cultural interactions.

An important caveat to using ancient DNA to make arguments about the origins of languages is that prior to the invention of writing, we have no way to directly tie ancient cultures to a language. Nevertheless, by establishing that major migrations or exchanges of genes occurred, we identify movements of people that would have been plausible vectors for the spread of languages, and we can establish some periods in time as the most plausible ones for language spread. Thus, genetic data can change the balance of probabilities among competing hypotheses as we outline above.

Although in this study we have focused on the genetic findings, our data are also interesting from the point of view of archaeological methodology. Specifically, our findings challenge the idea of a limited role of migration in human population history[11] by providing unambiguous evidence of two major episodes of migration and population turnover in Europe. By documenting not only that these major migrations occurred, but that they were both followed to a degree by the genetic resurgence of the local populations (SI7, Fig. 3), we hope that our study will help to spur new debate on the interactions between migrants and indigenous peoples long after the occurrence of migration.

**Prospects for further genetic insight into Indo-European language origins**

The ultimate question of the Proto-Indo-European homeland is unresolved by our data.

One important future direction for genetic research into Indo-European origins is to obtain ancient DNA data from India, Iran, northwestern China, and intervening regions to test hypotheses about the spread of Indo-European languages to the east. While obtaining such data is likely to be more difficult than European ancient DNA work (because of poorer conditions for preservation), it could provide crucial clues. For example, if the people who brought substantial steppe ancestry to northern Europe (via the Yamnaya and Corded Ware) also spread Indo-European languages into Iran and India, we should expect to find steppe ancestry in prehistoric samples from the Iranian plateau and South Asia, unless the mode of language transgression was different there (e.g. elite dominance).

A second direction for future genetic research is to study additional ancient European populations. Present-day populations of southern Europe have lower ancestry related to the Yamnaya than those from northern Europe (Fig. 3). This could be explained either by migration from the Yamnaya that affected southern Europe to a lesser extent than northern Europe, the later dilution of this ancestry in southern Europe, or the later introduction of such ancestry only indirectly, from northern Europe as a result of much later admixture events[35,36]. It is still possible that the steppe migration detected by our study into Late Neolithic Europe might account for only a subset of Indo-European languages in Europe, and other Indo-European languages arrived in Europe not from the steppe but from either an early "Neolithic Anatolian" or later "Armenian plateau" homeland. We highlight Anatolia and the Balkans as particularly promising places to study to make further progress toward understanding these open questions. The four different theories of Indo-European origins make very different predictions about the population history of Anatolia and southeastern Europe where two of the earliest attested Indo-European languages, Hittite and Mycenaean Greek are found. By examining ancient samples from these regions it should become possible to determine if there are genetic discontinuities prior to the appearance of these languages that may be correlated with the migration of a new population. Ancient DNA provides evidence independent of archaeology or linguistics, and thus has the potential to resolve continuing controversies and to contribute to progress in answering the centuries' old riddle of Indo-European origins.